\documentclass[12pt]{article}
\pdfoutput=1
\usepackage{graphicx}
\usepackage{float}
\usepackage{bm}
\usepackage{caption}
\usepackage{color}
\usepackage{amsmath}
\usepackage[T1]{fontenc} 

\newcommand{\be}{\begin{equation}}
\newcommand{\ee}{\end{equation}}
\newcommand{\ba}{\begin{eqnarray}}
\newcommand{\ea}{\end{eqnarray}}
\newcommand{\n}[1]{\label{#1}}

\begin{document}

\title{Properties of the distorted Kerr black hole \!\!
\thanks{Alberta-Thy-14-15}}

\author{Shohreh Abdolrahimi${}^{a}$ \!\!
\thanks{Internet address: abdolrah@ualberta.ca}
 \,  ,
 Jutta Kunz${}^{b}$ \!\! 
\thanks{Internet address:
jutta.kunz@uni-oldenburg.de}
\, ,
Petya Nedkova${}^{c}$ \!\!
\thanks{Internet address:
pnedkova@phys.uni-sofia.bg}
\\ \, and
Christos Tzounis${}^{a}$ \!\!
\thanks{Internet address:
tzounis@ualberta.ca}
\\
${}^{a}$Theoretical Physics Institute, \\ 
University of Alberta, Edmonton, AB, Canada,  T6G 2G7 \\
${}^{b,c}$Institut f\"{u}r Physik, Universit\"{a}t Oldenburg, \\
D-26111 Oldenburg, Germany \\
${}^{c}$Faculty of Physics, Sofia University, \\
5 James Bourchier Boulevard, Sofia 1164, Bulgaria}

\maketitle

\begin{abstract}
We investigate the properties of the ergoregion and the location of the curvature singularities for the Kerr black hole distorted by the gravitational field of external sources. The particular cases of quadrupole and octupole distortion are studied in detail. We also investigate the scalar curvature invariants of the horizon and compare their behaviour with the case of the isolated Kerr black hole. In a certain region of the parameter space the ergoregion consists of a compact region encompassing the horizon and a disconnected part extending to infinity. The curvature singularities in the domain of outer communication, when they exist, are always located on the boundary of the ergoregion. We present arguments that they do not lie on the compact ergosurface. For quadrupole distortion the compact ergoregion size is negatively correlated with the horizon angular momentum when the external sources are varied. For octupole distortion infinitely many ergoregion configurations can exist for a certain horizon angular momentum. For some special cases we can have $J^2/M^4 > 1$ and yet avoid the naked singularity.
\end{abstract}

\normalsize
\baselineskip 17 pt
\newpage

\section{Introduction}

The main part of our knowledge about the structure and behaviour of black holes comes from the study of two classes of solutions - the Schwarzschild and the Kerr black holes. These solutions represent isolated black holes. Here, we address the question, to what extent the properties of these solutions get modified when the black holes are not isolated. In order to understand the properties of  black holes as gravitational objects predicted by the theory of general relativity, one needs to study the interaction of black holes with matter and sources. A black hole in a binary system is distorted by the presence of its companion. However, this situation is highly dynamical and can hardly be approached by analytical techniques. Great progress has been achieved in the numerical investigation of such systems and black hole mergers (see \cite{Centrella} for a review). On the other hand, quasi-stationary systems, or dynamical black holes which relax on a time scale much shorter than that of the external matter, can be reasonably well approximated by stationary and axisymmetric solutions. A class of exact solutions called distorted black holes are considered in order to describe such situations. By their investigation we can gain insights into some quasi-stationary processes of interaction of black holes with external matter sources. Moreover, study of the distorted black holes help us to see how universal are the black hole properties.

The study of distorted black holes goes back to the 50's. Originally, the term was used to describe an asymptotically flat black hole solution possessing higher mass multipole moments. Such is for example the Erez and Rosen solution \cite{ER1}, which represents a generalization of the Schwarzschild black hole with a quadrupole moment. However, by Israel's theorem \cite{Israel} these solutions suffer from the appearance of curvature singularities on the horizon or in its vicinity. The term distorted black hole can also refer to an asymptotically non-flat black hole solution which is considered to be valid only locally in a certain neighbourhood of the black hole horizon. Such solutions are interpreted as describing a black hole located in the gravitational field of external sources. Although the matter sources are not explicitly included, the solution contains information about their influence on the black hole properties. One of the first solutions of this class was constructed in 1965 by Doroshkevich, Zel'dovich, and Novikov \cite{DZN} who considered the Schwarzschild black hole in an external quadrupole gravitational field. Chandrasekhar obtained the equilibrium condition for a black hole in a static external gravitational field \cite{Chandrasekhar}. Geroch and Hartle  \cite{Geroch} considered general static black holes in four dimensions in the presence of external matter fields, and performed a fundamental analysis of the global characteristics of these solutions.

The study of rotating distorted black holes was initiated with the construction of the Kerr black hole surrounded by an external static field by Tomimatsu \cite{Tomimatsu}. He applied solitonic techniques to the static potential considered by Chandrasekhar \cite{Chandrasekhar}, which is a particular case of the Erez-Rosen potential \cite{ER1}, but did not compute all the metric functions explicitly. The complete solution describing the Kerr black hole in an external field was obtained in a concise explicit form in \cite{Breton:1997}. Distorted electrically charged black holes were constructed in \cite{Fairhurst}-\cite{Yazadjiev}. The generalization to the case of the Kerr-Newman black hole in an external gravitational field was achieved in  \cite{Breton:1998}. As it was done in the case of four dimensions, one can use the generalized Weyl solution \cite{EmR} to construct a distorted five dimensional Schwarzschild-Tangherlini black hole \cite{ASP}. Moreover, a new exact solution of the five dimensional Einstein equations in vacuum was constructed recently \cite{AKN} describing a distorted Myers-Perry black hole with a single angular momentum.

Distorted black holes can show some remarkable properties. In \cite{Breton:1997} Bret{\'o}n, Denisova and Manko have formulated the sufficient condition of the regularity of the metric in the region exterior to the black hole. The Smarr mass formula \cite{Smarr} holds for a black hole in the presence of an external gravitational field \cite{Tomimatsu}.  There exists a duality between the properties of the outer and inner horizons of a distorted charged black hole \cite{ASF}. It was shown that the inner (Cauchy) horizon of such a black hole remains regular, provided the distortion is regular at the event horizon \cite{ASF}. These properties persist even in the case of higher dimensional black holes \cite{AS}. In the case of a distorted rotating black hole, there is a duality transformation between the outer and inner horizons of the black hole, which is different from that of an electrically charged static distorted black hole \cite{Shoom}. Moreover, it is demonstrated that, in the case of a distorted five-dimensional Myers-Perry black hole, the ratio of the horizon angular momentum and the mass $J^2/M^3$ is unbounded, and can grow arbitrarily large \cite{AKN}. Thermodynamic properties of distorted black holes have been studies by various authors \cite{Geroch, Breton:1997, Breton:1998, AS, Ab}. Lastly, the propagation of light in the background of a distorted Schwarzschild spacetime and the corresponding apparent shape of the black hole was studied in \cite{AMT}.

In this paper we consider the distorted Kerr solution constructed by Bret{\'o}n, Denisova and Manko \cite{Breton:1997}. It represents a stationary and axisymmetric solution to the Einstein equations in vacuum possessing non-flat asymptotics. Considered as a local solution in some neighbourhood of the black hole horizon, it describes the Kerr black hole in the presence of external gravitational sources. In this interpretation it is assumed that a vacuum region exist between the horizon and the location of the external sources, which represents the region of validity of  the solution. The purpose of this work is to investigate the influence of the external fields on some basic properties of the Kerr black hole. We study possible ergoregion configurations, the presence and the location of the curvature singularities in the domain of outer communication, and the four-dimensional scalar curvature invariants on the black hole horizon.

In general relativity a system with very strong self-gravity and sufficiently large angular momentum is characterized by an ergoregion. Rotating black holes typically possess an ergoregion. Moreover, rotating compact objects without event horizons can have an ergoregion, such as dense stellar models \cite{Butterworth}, boson stars \cite{Kleihaus} or relativistic discs \cite{Meinel}. The ergosurface is the boundary at which the dragging of inertial frames is too strong for counter-rotating time-like or null geodesics to exist. Thus, in the ergoregion static (non rotating stationary) observers are forced to corotate with the black hole. Exotic effects such as the Penrose process or super-radiance \cite{Chandrasekhar} are associated with the ergosurface. It can be said that, like the event horizon, the ergosurface is an important surface. There might even be a ``Hawking'' radiation associated to the ergosurface, due to ``ergoregion instability'', for example in the case of fuzzball geometries \cite{fuzzball1,fuzzball2,fuzzball3,fuzzball4}. Recently, some black hole solutions with interesting ergosurface configurations were considered. Kerr black holes with scalar hair were obtained numerically in general relativity minimally coupled to a massive complex scalar field \cite{Herdeiro:2014a}. In a certain region of the parameter space these solutions can develop an ergo-Saturn \cite{Herdeiro:2014b}, i.e. an ergosurface which consists of a surface with $S^2$ topology, encompassing the horizon, and another one with toroidal topology ($S^1\times S^1$), which is disjoint from it. Similar ergosurface configurations are expected in the case of  hairy black hole generalizations of spinning gravitating Q-balls, vortons, and other soliton solutions. Disconnected compact ergoregions are also observed for boson stars \cite{Kleihaus}, where in the case of negative parity the ergosurface can consist of two topological tori. We also study the scalar curvature invariants, which are considered in quantum gravity and for effective theories of gravity applied to cosmology, the one-loop level renormalization of gravity \cite{Birrel}, and for calculating the energy density and stresses of a conformal scalar field in the Hartle-Hawking state \cite{Don, Brown, Frolov}.

The paper is organized as follows. In Section 2, we review the solution representing the distorted Kerr black hole. In Section 3, we study in detail the possible configurations of the ergoregion in the particular cases of quadrupole and octupole external fields. Some observations are made for more general types of even and odd distortion. We examine the correlation between the size of the compact part of the ergoregion and the angular momentum on the horizon. The correlation with the parameters characterizing the external field is also investigated.  In Section 4, we discuss the presence and the location of the curvature singularities for quadrupole and octupole fields. In Section 5, we calculate the scalar curvature invariants on the horizon. We finish with Section 6, where we sum up our results. In what follows, we work in geometrical units $c=G=1$.

\section{ Distorted Kerr black hole}
The metric of the distorted Kerr black hole is given in the following form \cite{Breton:1997}
\ba
d\mathcal{S}^2&=&-e^{2U}\frac{A}{B}(d\mathcal{T}-\tilde{\omega}' d\phi)^2
+ \frac{m^2}{(1-\alpha^2)^2}Be^{-2U+2V}\left(\frac{dx^2}{x^2-1}
+\frac{dy^2}{1-y^2}\right)\nonumber\\[1mm]
&+&m^2\frac{B}{A}e^{-2U}(x^2-1)(1-y^2)d\phi^2, \label{metric1}
\ea
where the metric functions are expressed by
\ba
A&=&(x^2-1)(1+ab)^2-(1-y^2)(b-a)^2, \label{A} \\[2mm]
B&=&[x+1+(x-1)ab]^2+[(1+y)a+(1-y)b]^2, \label{B} \\[2mm]
\tilde{\omega}'&=&2m e^{-2U}\frac{C}{A} -\frac{4m\alpha}{1-\alpha^2} \exp (-2 \sum_{n=0}^{\infty} c_{2n}),\nonumber\\[2mm]
C&=&(x^2-1)(1+ab)[b-a-y(a+b)] \nonumber\\[1mm]
&+&(1-y^2)(b-a)[1+ab+x(1-ab)],\nonumber\\[2mm]
a&=&-\alpha \exp\left (2 \sum_{n=1}^{\infty} c_n (x-y) \sum_{l=0}^{n-1} R^l P_l \right), \nonumber\\[2mm]
b&=& \alpha \exp \left(2 \sum_{n=1}^{\infty} c_n (x+y) \sum_{l=0}^{n-1} (-1)^{n-l} R^l P_l \right), \nonumber\\[2mm]
U&=&\sum_{n=0}^{\infty} c_n R^n P_n\left(\frac{xy}{R}\right),~~~R\equiv \sqrt{x^2+y^2-1},\nonumber\\[2mm]
V&=&\sum^{\infty}_{n,k=1}\frac{nk}{n+k}c_nc_k R^{n+k}\left(P_n P_k - P_{n-1}P_{k-1}\right)\nonumber\\[1mm]
&+&\sum^{\infty}_{n=1}c_n\sum_{k=0}^{n-1}[(-1)^{n-k+1}(x+y)
-x+y]R^k P_k\left(\frac{xy}{R}\right). \nonumber
\ea
We denote by $P_n$ the Legendre polynomials, which depend on the argument $xy/R$ in all the expressions. The solution is represented in the prolate spheroidal coordinates $\{x,y\}$, taking the ranges $x\geq 1$, $-1\leq y\leq 1$ in the domain of outer communication, and $m>0$, $\alpha\in (0,1)$ and $c_n$, $n\in\mathcal{N}$, are real constants. The physical infinity corresponds to the limit $x\rightarrow\infty$. The horizon of the distorted Kerr black hole is located at $x=1$, while the rotational axis corresponds to $y=\pm 1$. The prolate spheroidal coordinates can be extended to the interior region of the horizon, and then we observe an inner Cauchy horizon located at $x=-1$.
It is convenient to write the metric (\ref{metric1}) in the dimensionless form,
\ba
dS^2&=&-e^{2U}\frac{A}{B}(dT-\omega' d\phi)^2
+ \frac{1}{(1-\alpha^2)^2}Be^{-2U+2V}\left(\frac{dx^2}{x^2-1}
+\frac{dy^2}{1-y^2}\right)\nonumber\\
&+&\frac{B}{A}e^{-2U}(x^2-1)(1-y^2)d\phi^2, \label{metric2}
\ea where,
\ba
d\mathcal{S}^{2}&=&m^2dS^{2},~~~ T=\frac{\mathcal{T}}{m},~~~\omega'=\frac{\tilde{\omega}'}{m}\n{con1} \, .
\ea Moreover, we can make one more conformal transformation by setting,
\ba
U=\mathcal{U}+u_0 \, , ~~~~~~~ u_0=c_0
\ea then, the metric can be written in the form
\ba
ds^2&=&-e^{2\mathcal{U}}\frac{A}{B}(dt-\omega d\phi)^2
+ \frac{1}{(1-\alpha^2)^2}Be^{-2\mathcal{U}+2V}\left(\frac{dx^2}{x^2-1}
+\frac{dy^2}{1-y^2}\right)\nonumber\\
&+&\frac{B}{A}e^{-2\mathcal{U}}(x^2-1)(1-y^2)d\phi^2, \label{metric3}
\ea where,
\ba
dS^{2}&=&e^{2u_0}ds^{2},~~~ t=\frac{T}{e^{-2u_0}},~~~\omega=\frac{\omega'}{e^{-2u_0}} \, . \label{con2}
\ea Therefore, the metric function $\omega$ can be represented as
\ba
\omega=2e^{-2\mathcal{U}}\frac{C}{A}-\frac{4\alpha}{1-\alpha^2} \exp (-2 \sum_{n=1}^{\infty} c_{2n}) \, . \label{omega}
\ea Let us note that, the total conformal factor, $\Omega$, of the metric is,
\ba
d\mathcal{S}^{2}&=&\Omega^2ds^{2}\, , ~~~ \Omega^2\equiv e^{2u_0}m^2 \, . \label{con3}
\ea To satisfy the no-conical singularity condition on the axis we should set,
\be\label{reg}
\sum_{n=0}^{\infty}c_{2n+1}=0 \, .
\ee
The parameters $m$ and $\alpha$ are related to the Komar mass $M$ and angular momentum $J$ of the black hole horizon \cite{Breton:1997}
\ba\label{Mass}
M=\frac{m(1+\alpha^2)}{1-\alpha^2},~~~ J=-\frac{2m^2\alpha(1+\alpha^2)}{(1-\alpha^2)^2}\exp(-2 \sum_{n=1}^{\infty} c_{2n}),
\ea
while the parameters $c_n$ characterize the external field. We refer to them as multipole moments or distortion parameters interchangeably. In the particular case $c_n=0$ for all natural $n$ the solution reduces to the isolated Kerr black hole.
We can express the angular momentum per unit mass $a=J/M$ in the following way
\ba
a=-\frac{2\alpha m}{1-\alpha^2}\exp(-2 \sum_{n=1}^{\infty} c_{2n}) \, ,
\ea and for the spin parameter $a_{*}=J/M^2$ we have
\ba\label{spin_param}
a_{*}=-\frac{2\alpha}{1+\alpha^2}\exp(-2 \sum_{n=1}^{\infty} c_{2n}) \,.
\ea
Let us consider the simplest case of the metric which describes the Kerr black hole in an external quadrupole field ($c_2\neq 0$ and $c_{n\neq2}=0$). Then, we obtain the following expressions for the metric functions
\ba
a&=&-\alpha \exp\left[2c_2 (x-y)(1+xy)\right],\nonumber\\[1mm]
b&=&\alpha \exp\left[2c_2 (x+y)(1-xy)\right],\nonumber\\[1mm]
\mathcal{U}&=&-\frac{c_2}{2}(x^2+y^2)+\frac{c_2}{2}(3x^2y^2+1), \nonumber\\[1mm]
V&=&2c_2x(y^2-1)+\frac{1}{4}c_2^2\left[(x^2+y^2-1)(x^2+y^2-10x^2y^2)\right. \nonumber\\[1mm]
&&\left.-x^2-y^2+9x^4y^4+1\right].
\ea For octupole distortion ($c_1=-c_3$, $c_{n=2}=0$ and $c_{n>3}=0$) we have
\ba
a&=&-\alpha \exp\left(\left[(x-y)^2-3x^2y^2-1\right]c_1(x-y)\right),\nonumber\\[1mm]
b&=&\alpha\exp\left(\left[3x^2y^2+1-(x+y)^2\right]c_1(x+y)\right), \nonumber\\[1mm]
\mathcal{U}&=&\frac{c_1}{2}xy\left[3(x^2+y^2)-5x^2y^2-1\right], \nonumber\\[1mm]
V&=&c_1y(y^2-1)(1-3x^2)+\frac{1}{8}c_1^2\left[\left(1-(x^2+y^2)\right)\left(3(x^2+y^2)^2+1\right)\right. \nonumber\\[1mm]
&&\left.+x^2y^2\left(45(x^4+y^4)-54(x^2+y^2)+13\right)\right. \nonumber\\[1mm]
&&\left.+x^4y^4\left(75x^2y^2-117(x^2+y^2)+177\right)\right].
\ea

\section{Ergoregion}
In this section, we study the ergoregion of the distorted Kerr black hole for different values of the distortion parameters $c_n$ and the rotation parameter $\alpha$. We investigate in detail two particular cases of external sources - quadrupole distortion ($c_2\neq 0$, $c_{n\neq2}=0$), and octupole distortion ($c_1=-c_3$, $c_{n=2}=0$, $c_{n>3}=0$). According to the condition for the absence of conical singularities (\ref{reg}), these are the two most simple cases satisfying it. We also draw some conclusions on the behaviour of the ergoregion for more general external fields. Let us start from the definition of the ergoregion. The metric functions of the distorted Kerr black hole (\ref{metric3}), are independent of the time coordinate $t$ and the angular coordinate $\phi$. Therefore, we have the following two Killing vectors:
\be
\boldsymbol{\xi}_{(t)}\equiv \frac{\partial}{\partial t} \, , ~ \boldsymbol{\xi}_{(\phi)}\equiv \frac{\partial}{\partial \phi}
\ee which are associated with the stationarity and the axial symmetry of the black hole. An observer moving along a worldline of constant $(x,y)$ with uniform angular velocity,
\be
\Omega\equiv \frac{d \phi}{d t} \, ,
\ee can be viewed as a stationary observer. If the angular velocity of the observer is zero i.e. he/she moves along a worldline of constant $(x,y, \phi)$, then he/she is a static observer. For a stationary observer the four-velocity is
\be
\boldsymbol{u}= u^t\left(\frac{\partial}{\partial t}+\Omega\frac{\partial}{\partial \phi}\right)= \frac{\boldsymbol{\xi}_{(t)}+\Omega\boldsymbol{\xi}_{(\phi)}}{\sqrt{-g_{tt}-2\Omega g_{t\phi}-\Omega^2 g_{\phi\phi}}} \, .
\ee Static observers can exist when $g_{tt}<0$. We call this the ``static'' region. The region where, $g_{tt}>0$ is called ergoregion, while its boundary $g_{tt}=0$ is the ergosurface. In our case the static region, the ergoregion and the ergosurface are defined by the following conditions $A>0$, $A<0$ and $A=0$ on the metric function (\ref{A}), respectively.

In the case of the undistorted Kerr black hole the ergosurface is always a compact two dimensional surface which touches the horizon on the axis $y = \pm1$. In the case of the Kerr black hole in an external gravitational field the ergoregion has a more complicated behaviour. In  general we have two cases. In the first one, there is a compact ergoregion which encompasses the horizon of the black hole and touches the horizon on the axis. In addition, we observe other non-compact parts of the ergoregion, which are disconnected from the ergoregion in the vicinity of the horizon. In the second case, there is no compact ergoregion, but only a connected ergoregion extending to infinity.
\begin{figure}[htp]
\setlength{\tabcolsep}{ 0 pt }{\scriptsize\tt
		\begin{tabular}{ ccc }
	\includegraphics[width=5 cm]{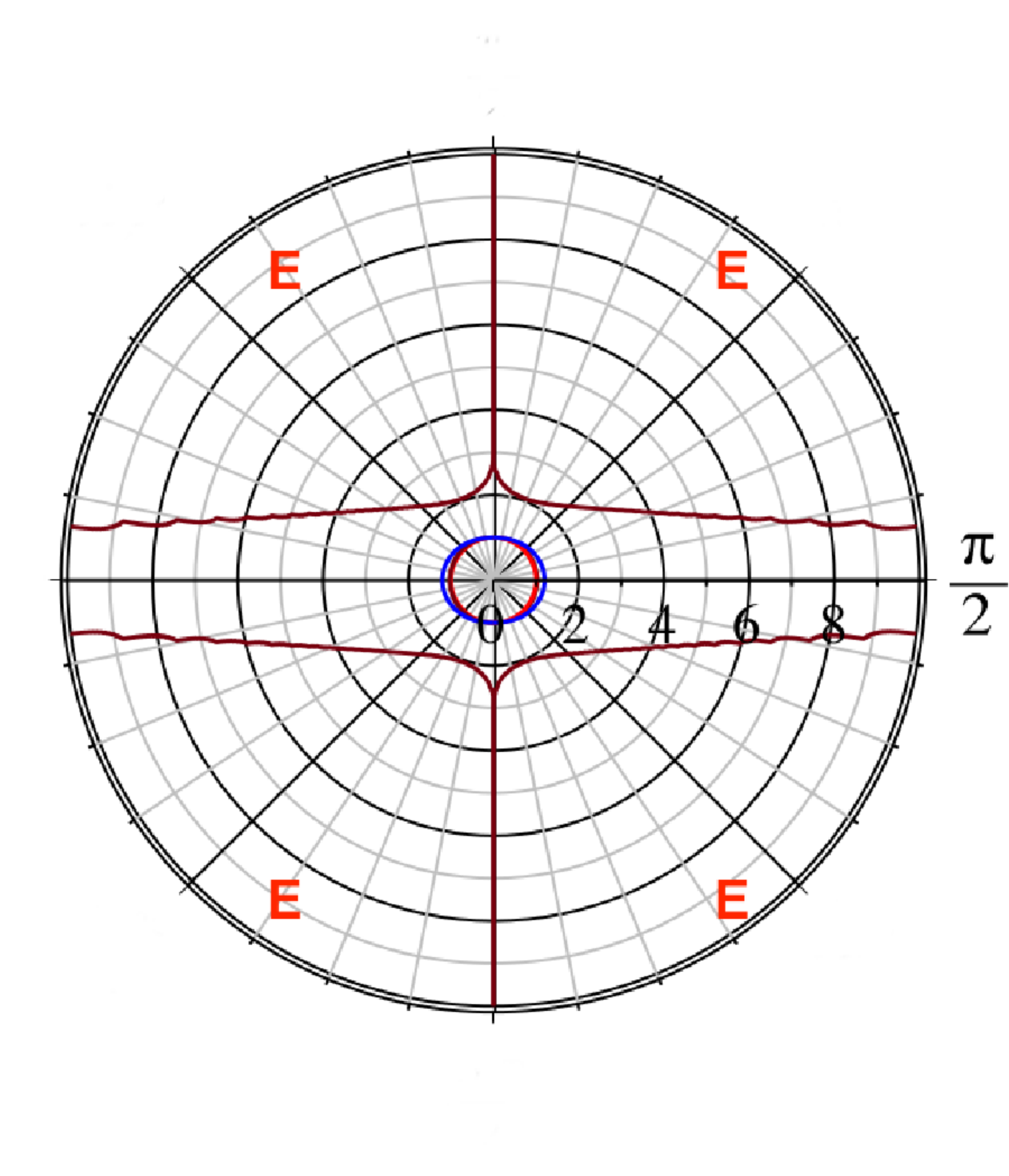} &
            \includegraphics[width=5 cm]{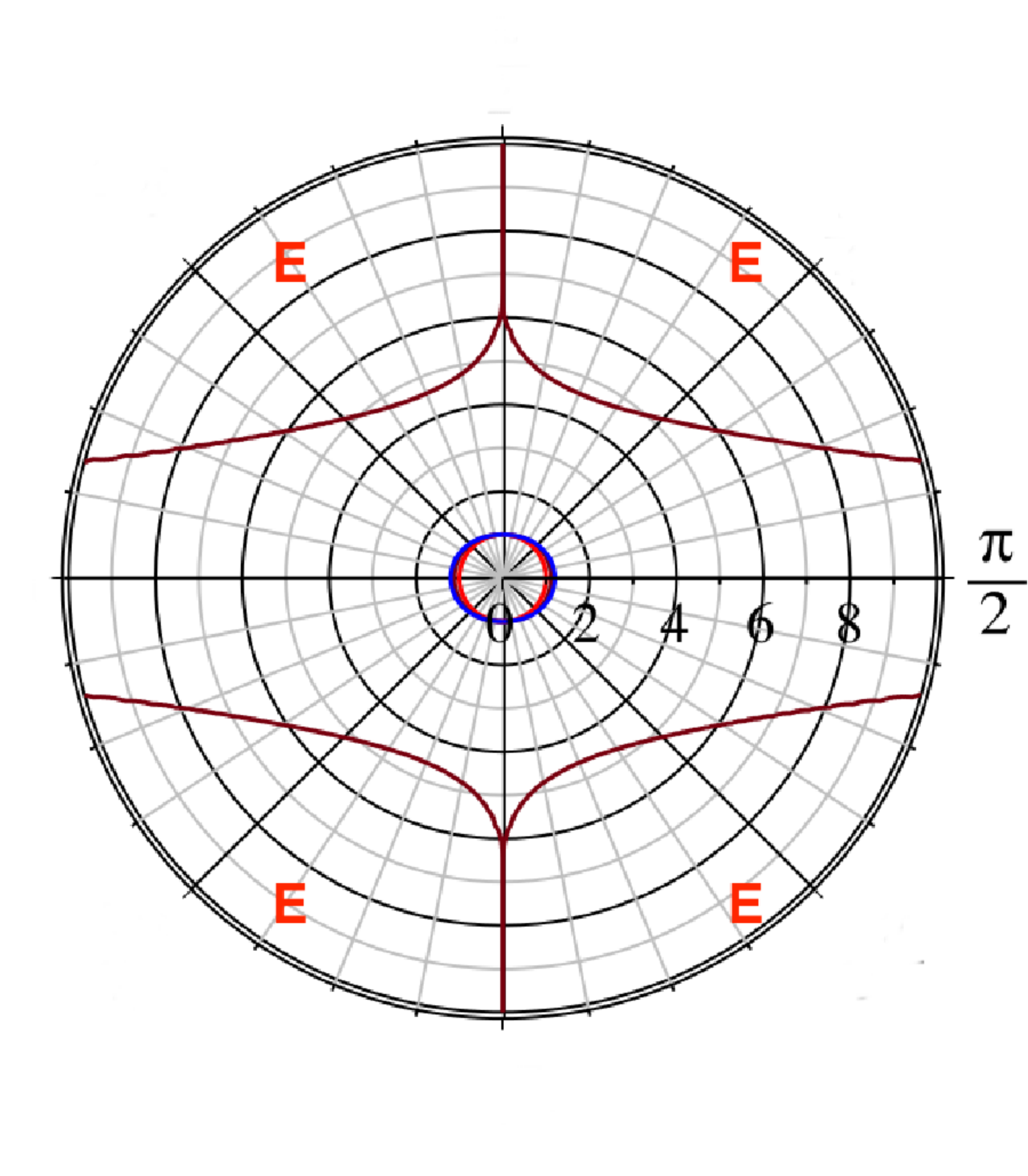} &
            \includegraphics[width=5 cm]{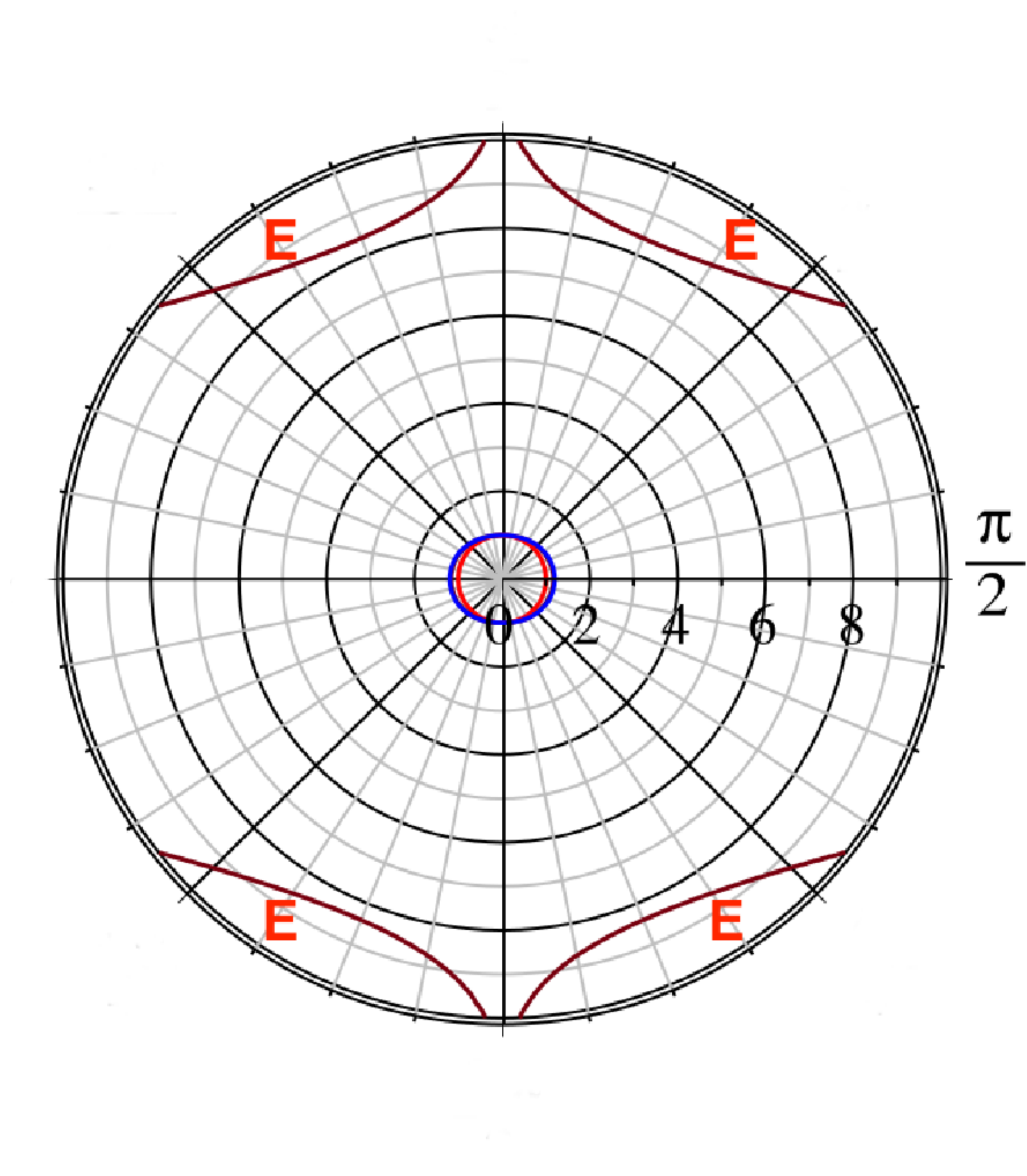} \\
            $c_2=-1/2$ & $c_2=-1/10$ & $c_2=-1/30$    \\
            \includegraphics[width=5 cm]{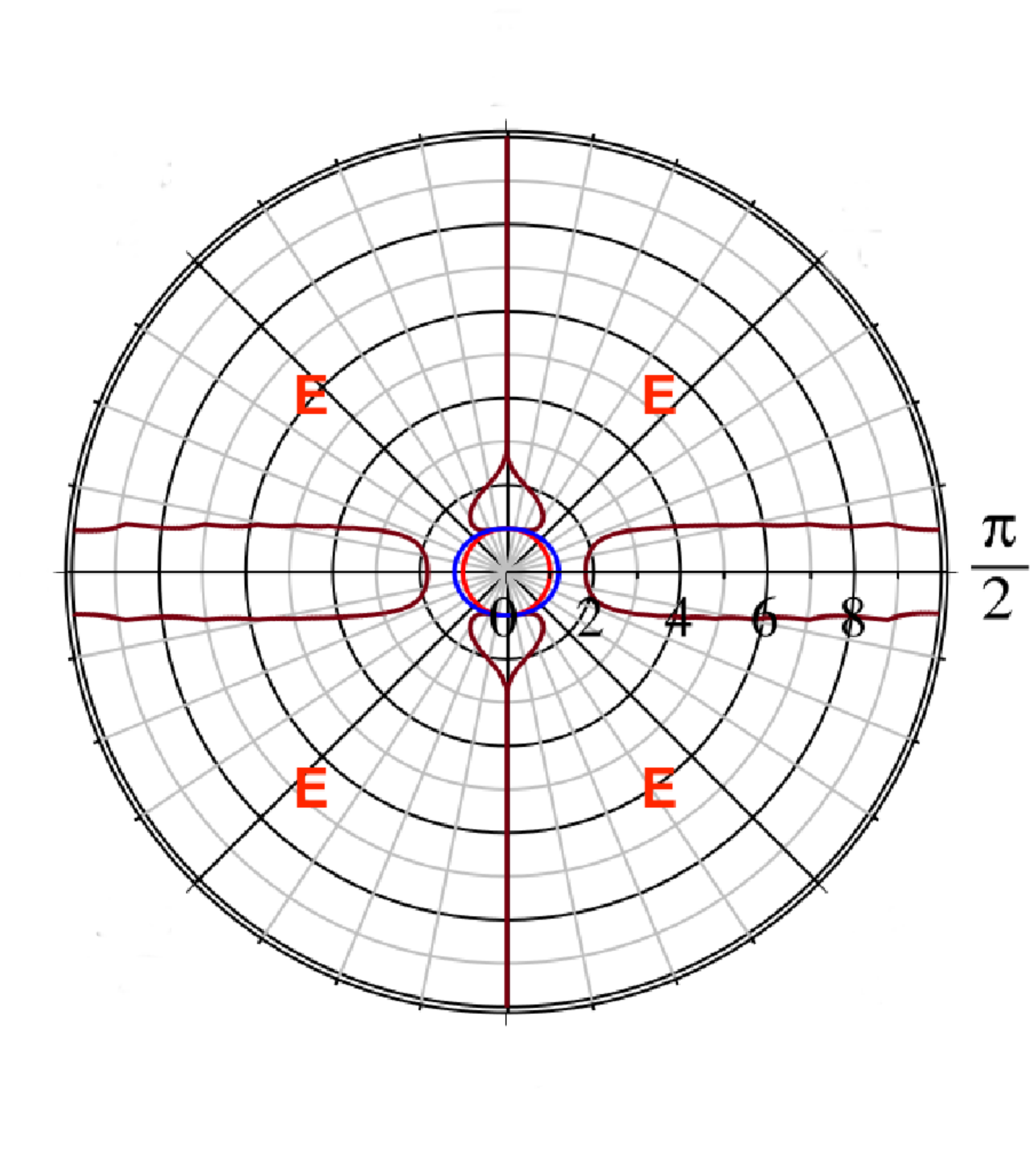} &
            \includegraphics[width=5 cm]{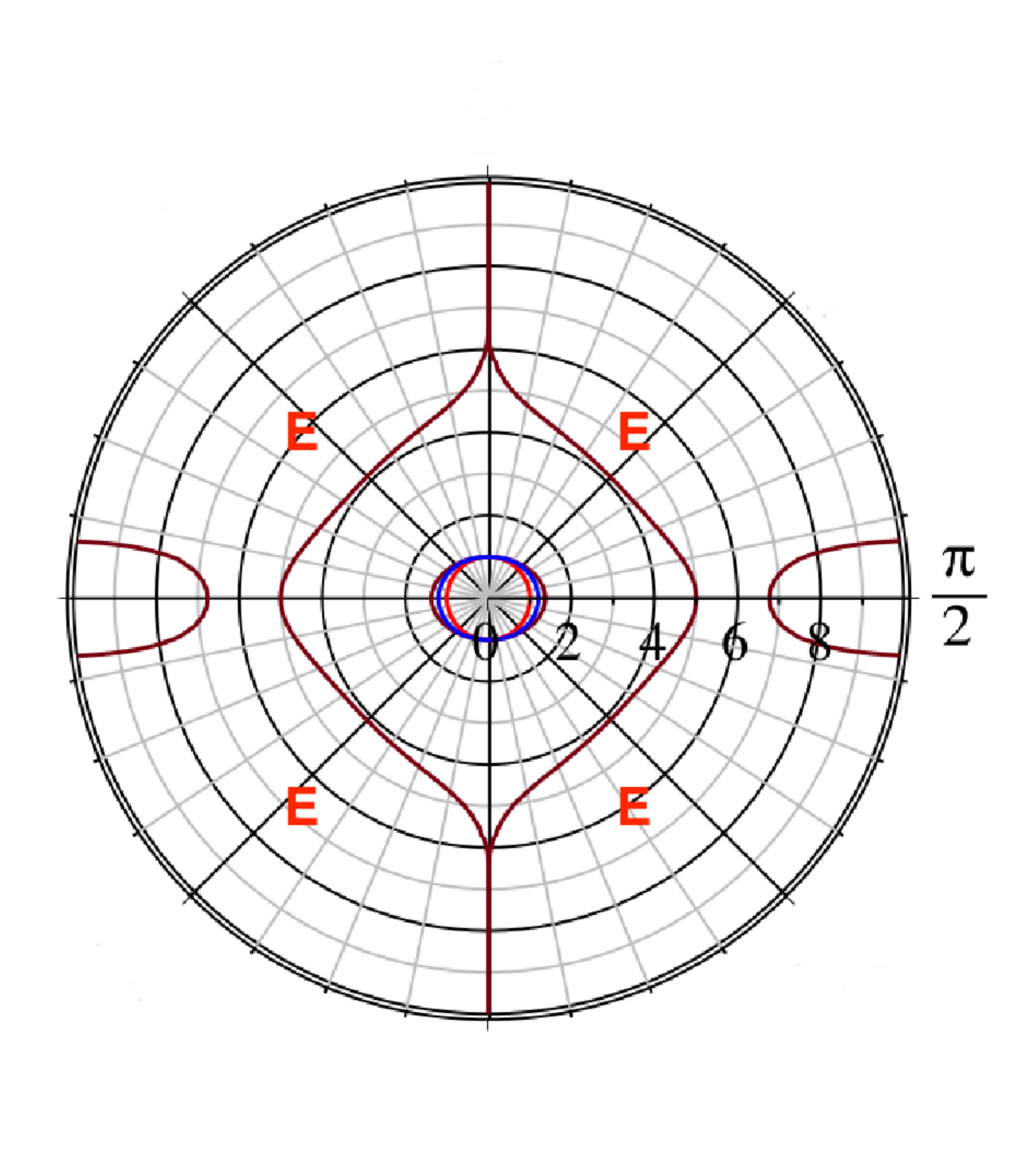} &
            \includegraphics[width=5 cm]{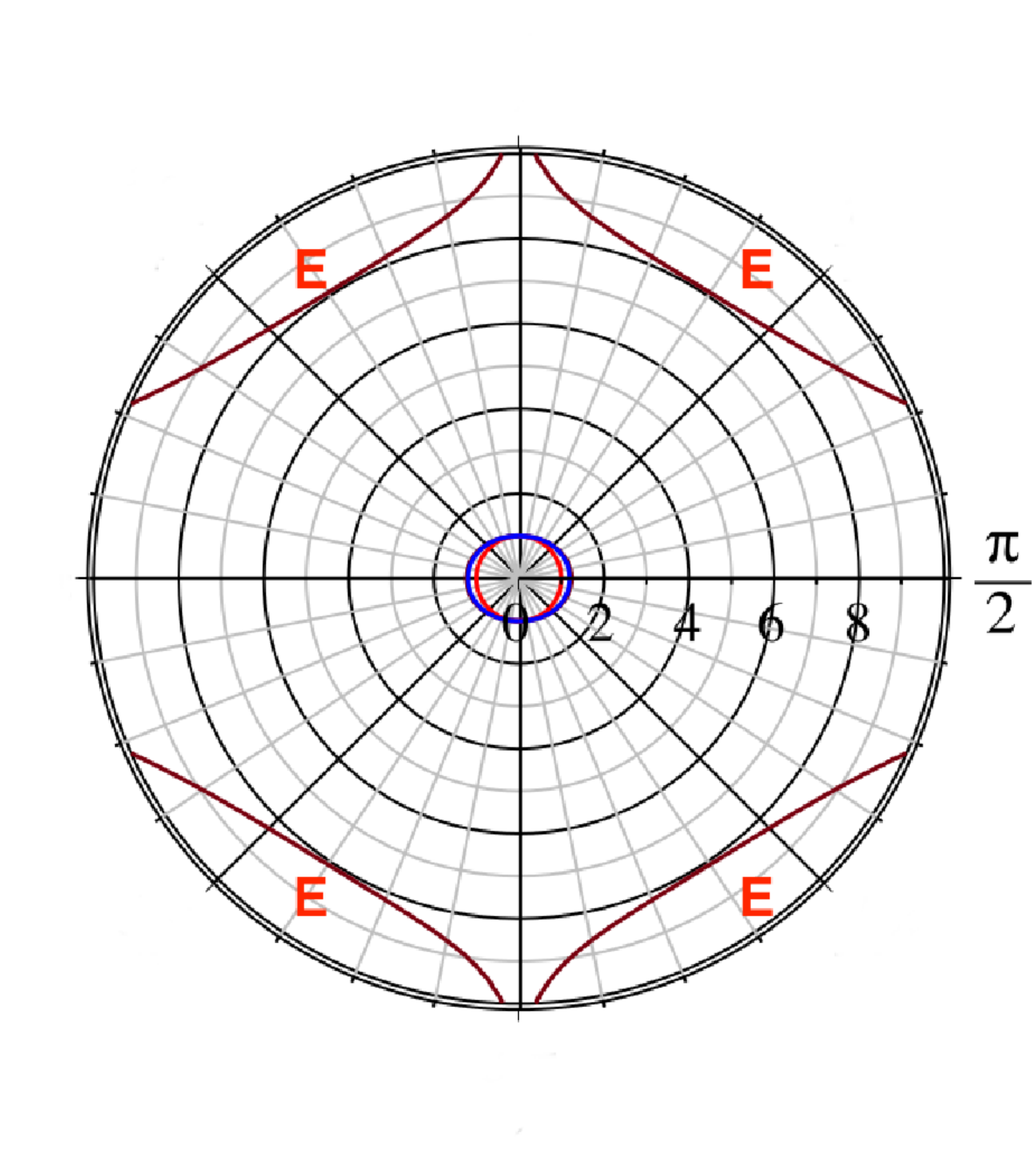} \\
            $c_2=1/2$ & $c_2=1/10$ & $c_2=1/30$
           \end{tabular}}
           \caption{\footnotesize{ Ergoregion for quadrupole distortion and $\alpha=0.3$}}\label{even03}
\end{figure}
In this work, we consider the distorted Kerr black hole as a local solution, which is valid only in a certain neighbourhood of the horizon. We assume that there are external sources which distort the black hole and in the region where the sources are located the spacetime is not vacuum and the solution is not valid. Thus, even though our metric represents a vacuum spacetime, some matter sources should exist which are exterior to the region of validity of the solution and cause distortion of the black hole. Provided these sources are located in some finite-sized region we can extend the solution beyond this non-vacuum region to a yet more distant vacuum region. A global solution can be constructed if (\ref{metric3}) is extended to an asymptotically flat solution by some sewing technique. This can be realized by cutting the spacetime manifold in the region where the metric (\ref{metric3}) is valid and attaching to it another spacetime manifold where the solution is not vacuum anymore, but the sources of the distorting matter are also included. In the cases, when the ergoregion consists of a compact region close to the horizon and another non-compact region, one can always choose to cut the manifold at such a value of the $x$ and $y$ coordinates that the other ergoregion is not included. In this way one may be able to construct an extension to an asymptotically flat solution with a compact ergoregion.

\subsection{Quadrupole distortion}
\begin{figure}[htp]
\setlength{\tabcolsep}{ 0 pt }{\scriptsize\tt
		\begin{tabular}{ ccc }
	\includegraphics[width=5 cm]{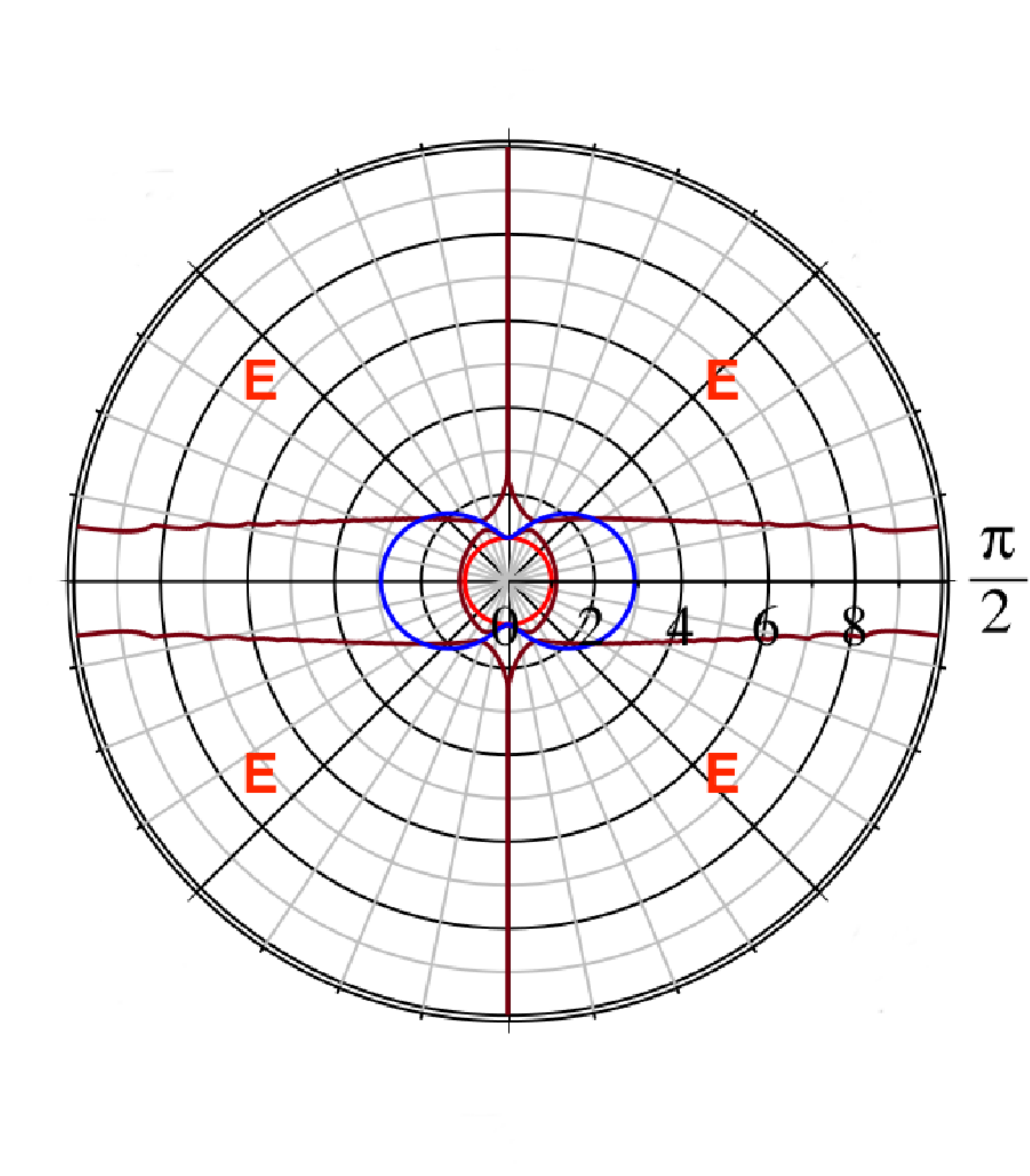} &
            \includegraphics[width=5 cm]{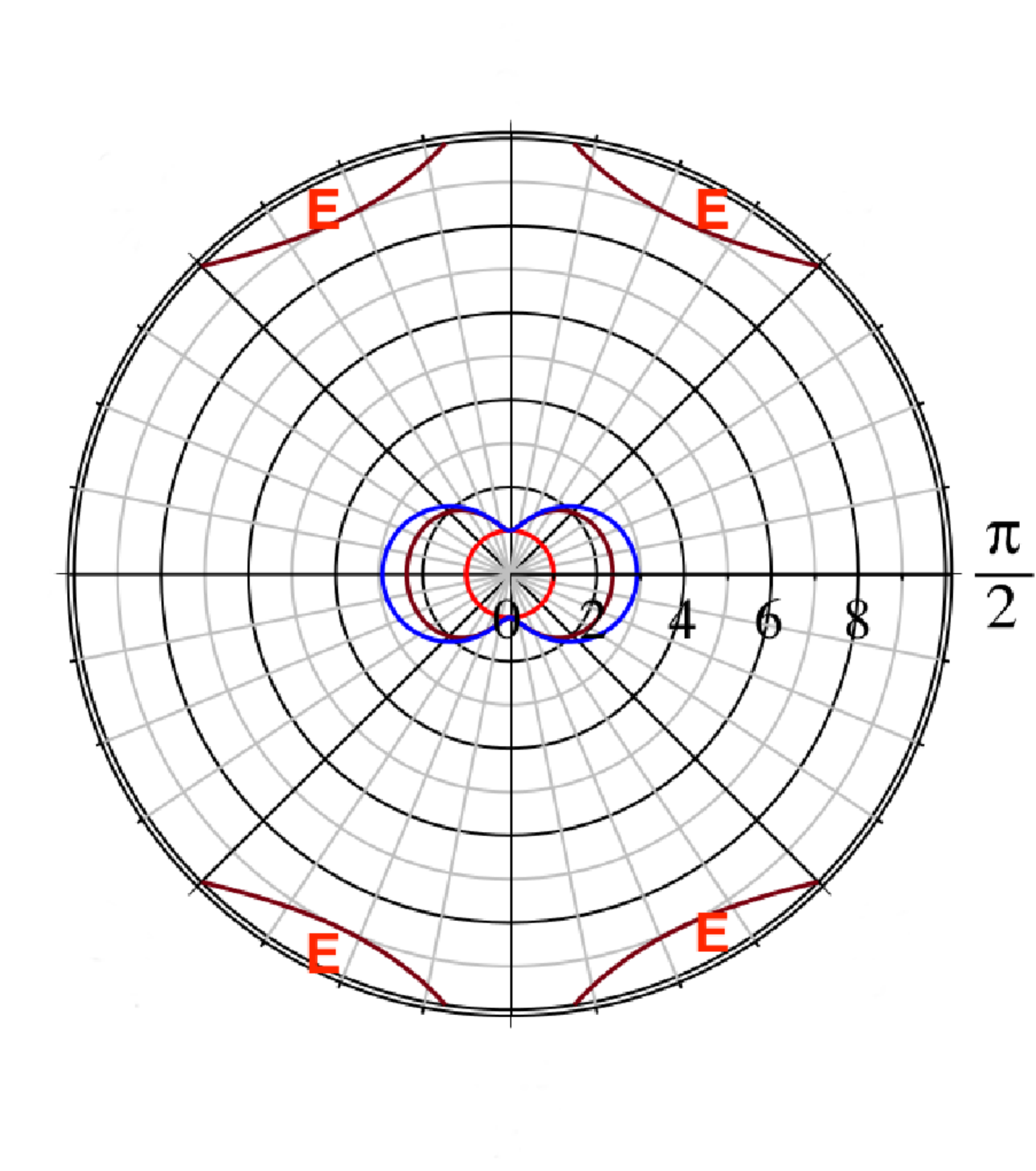} &
            \includegraphics[width=5 cm]{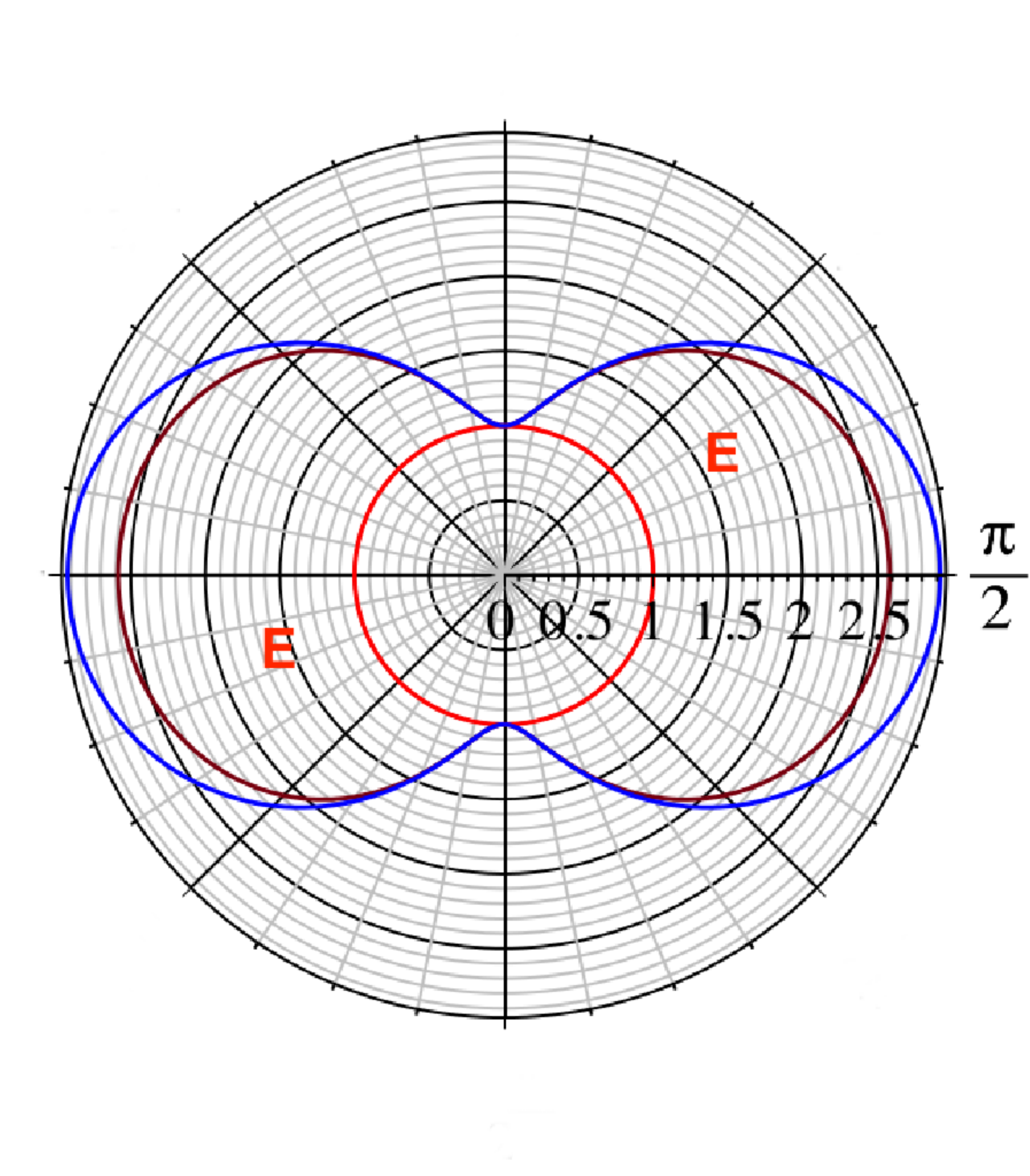} \\
            $c_2=-1/2$ & $c_2=-1/50$ & $c_2=-1/100$    \\
            \includegraphics[width=5 cm]{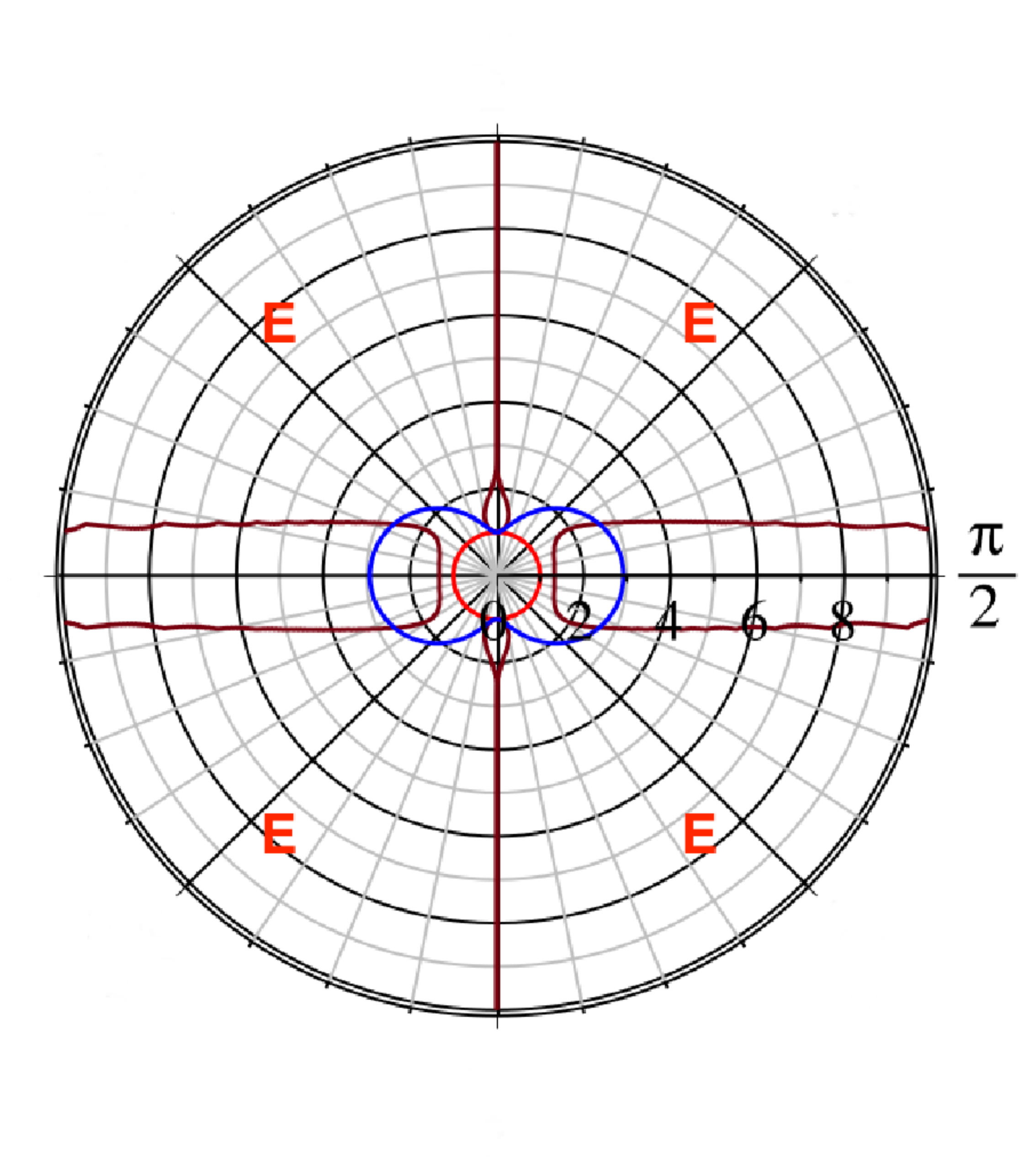} &
            \includegraphics[width=5 cm]{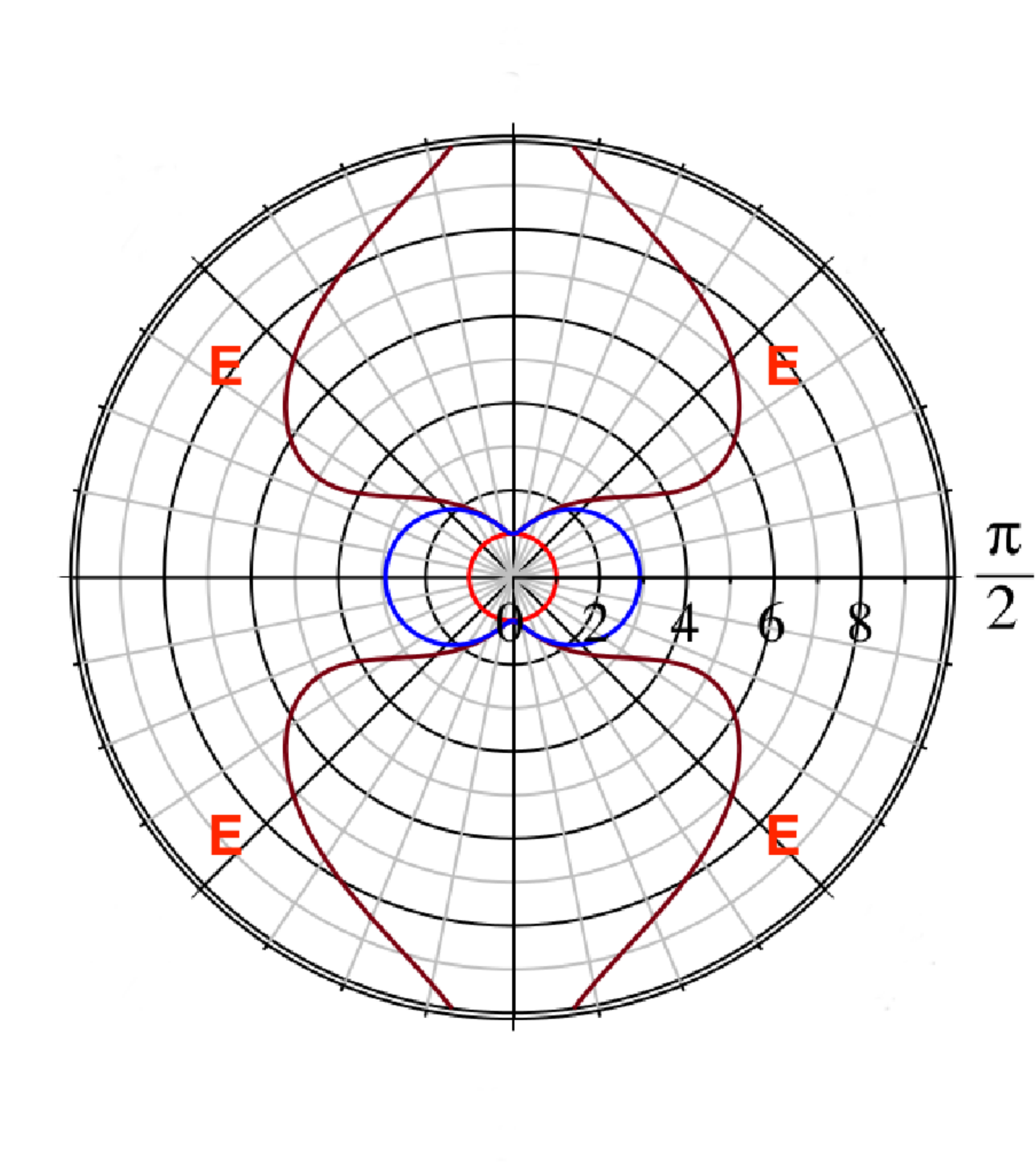} &
            \includegraphics[width=5 cm]{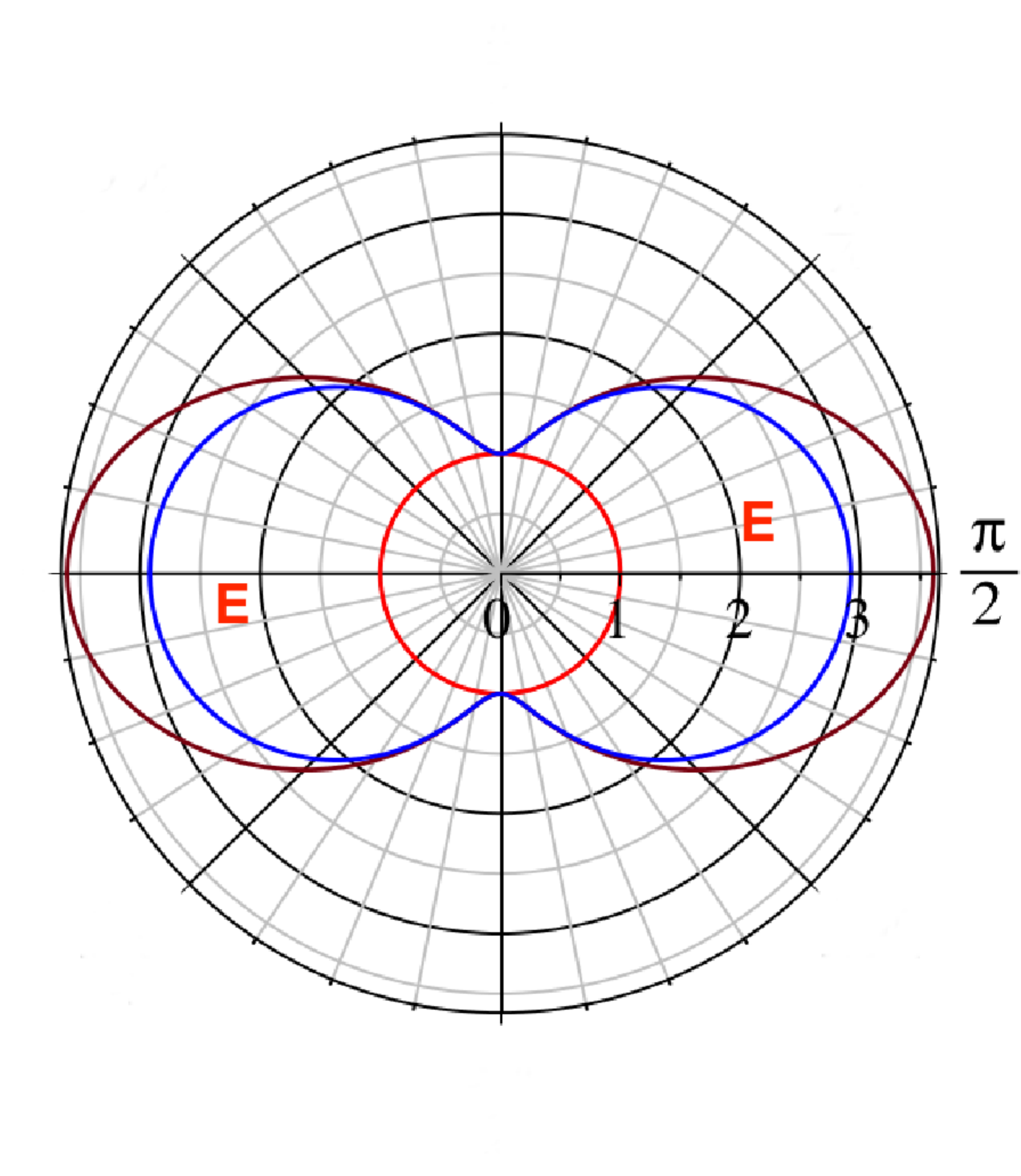} \\
            $c_2=1/2$ & $c_2=1/50$ & $c_2=1/100$
           \end{tabular}}
           \caption{\footnotesize{Ergoregion for quadrupole distortion and $\alpha=0.7$}}
		\label{even07}
\end{figure}

\begin{figure}[htp]
\setlength{\tabcolsep}{ 0 pt }{\scriptsize\tt
		\begin{tabular}{ ccc }
	\includegraphics[width=5 cm]{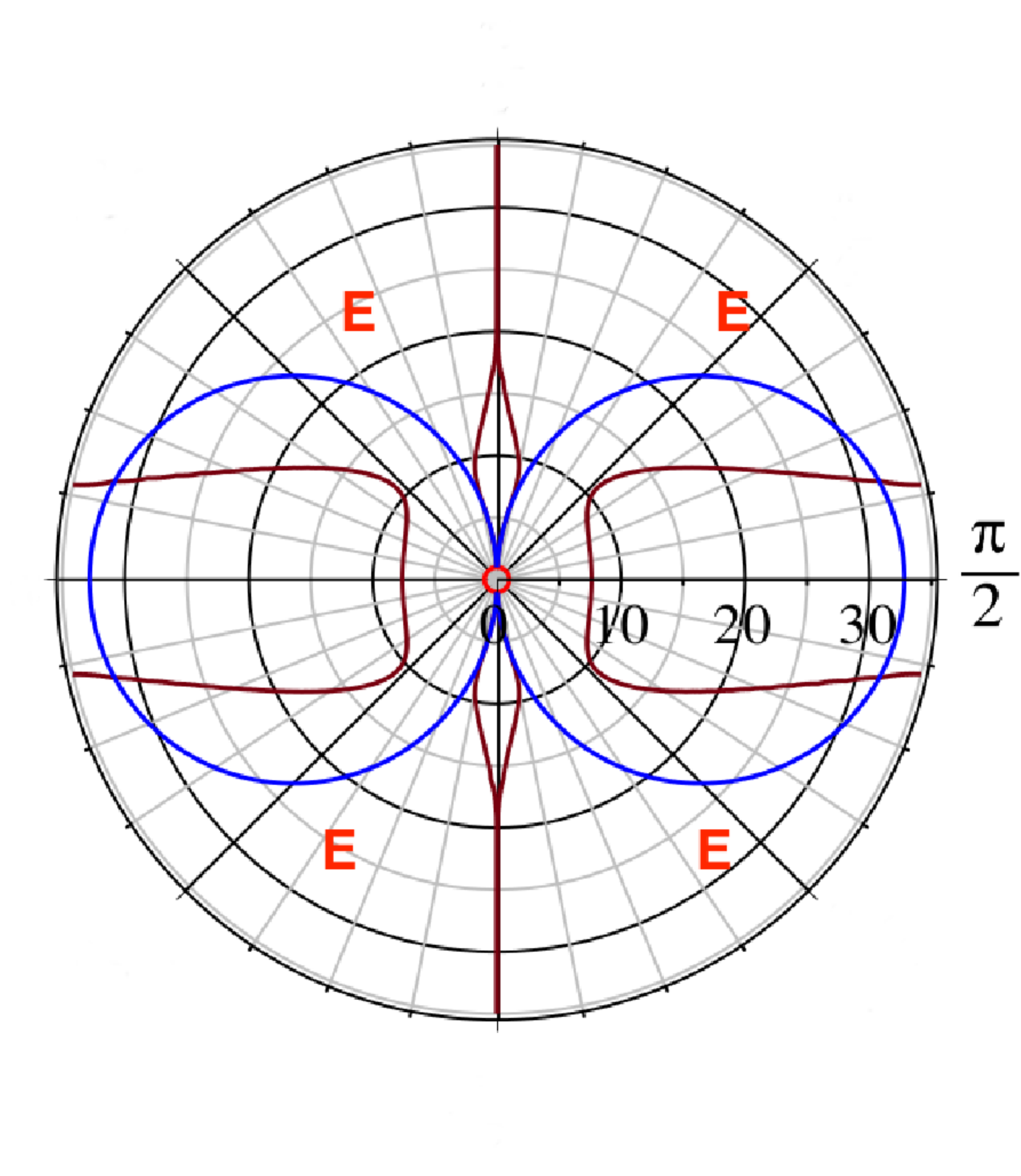} &
            \includegraphics[width=5 cm]{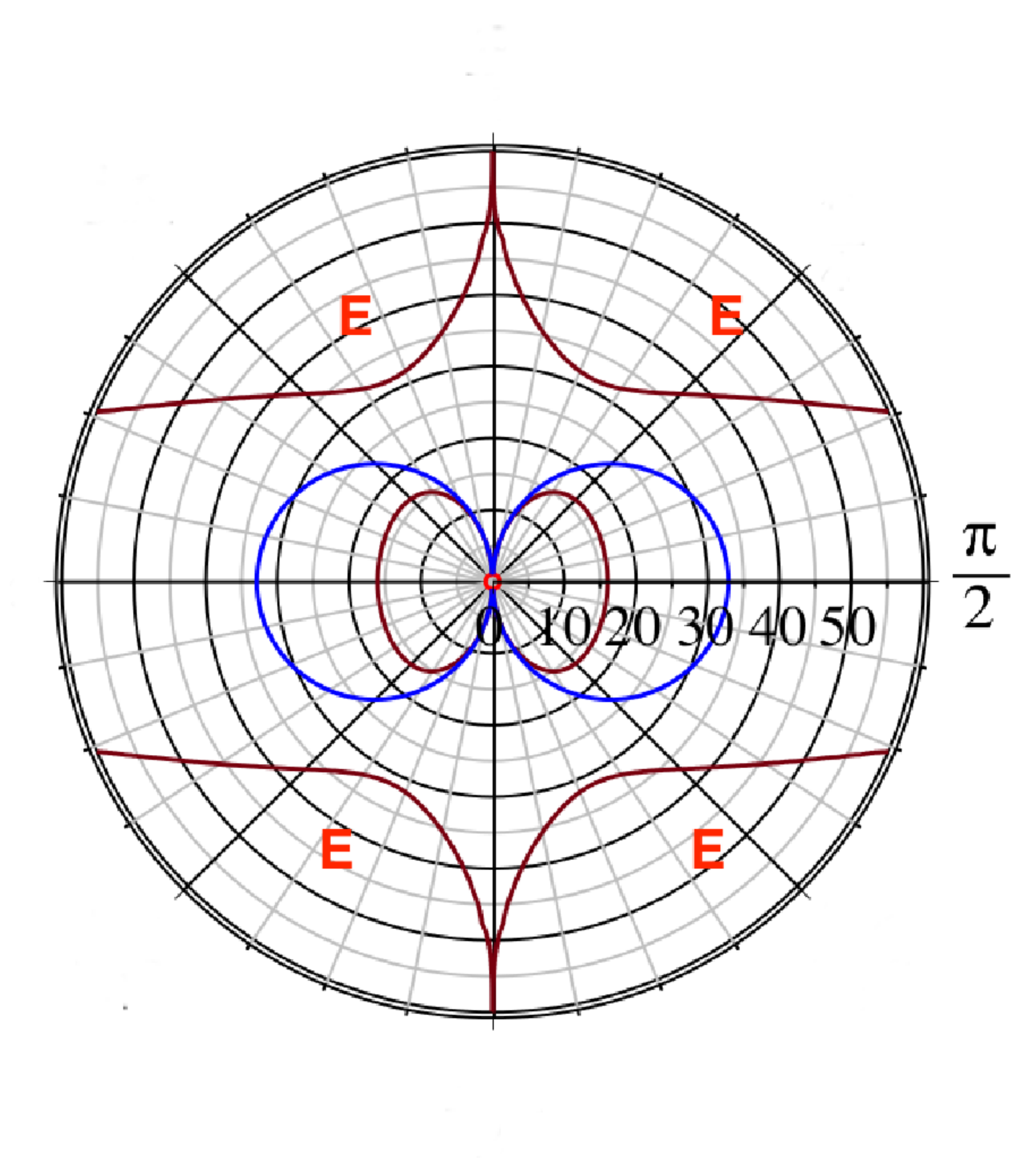} &
            \includegraphics[width=5 cm]{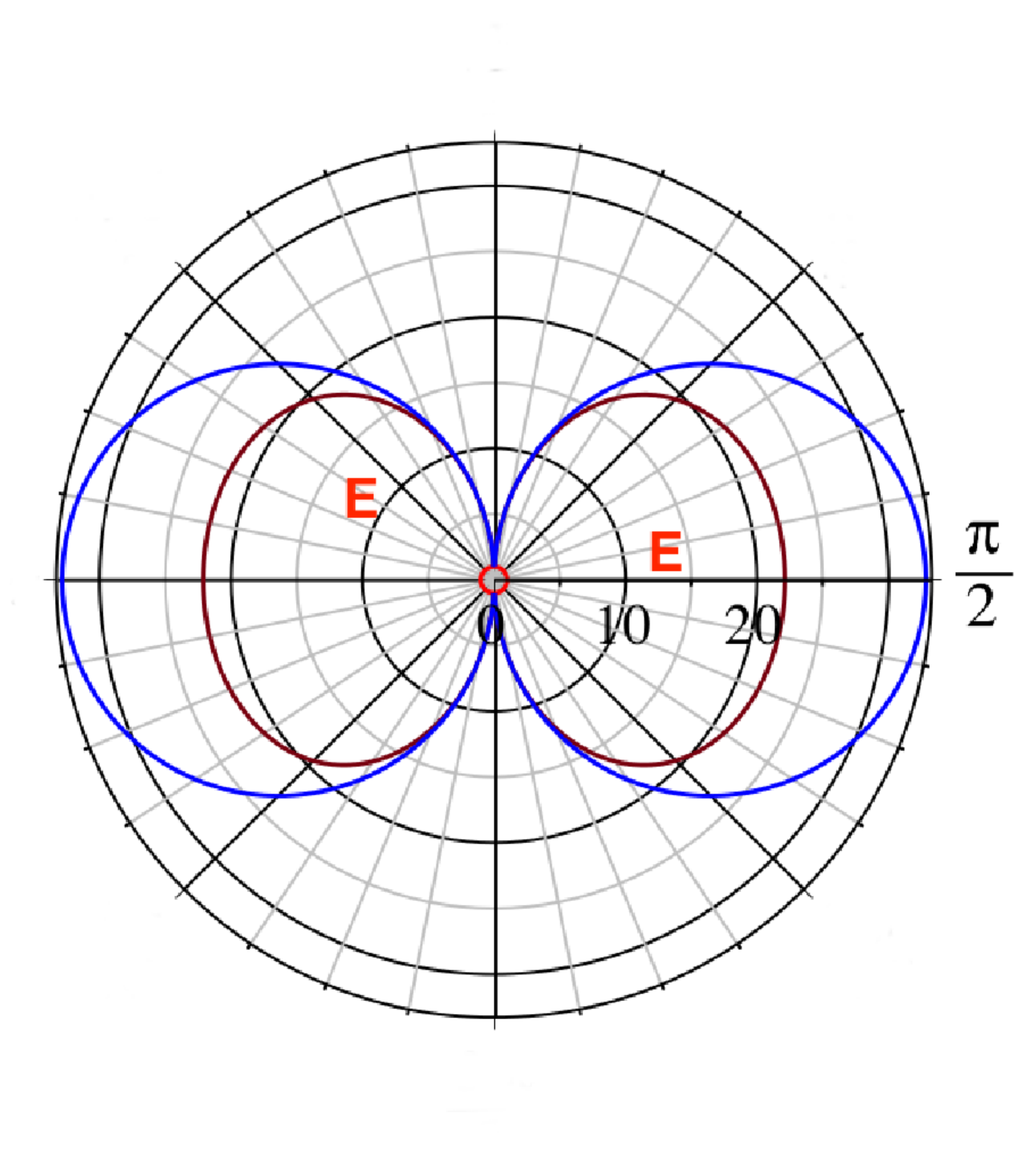} \\
            $c_2=-1/150$ & $c_2=-1/1000$ & $c_2=-1/3000$    \\
            \includegraphics[width=5 cm]{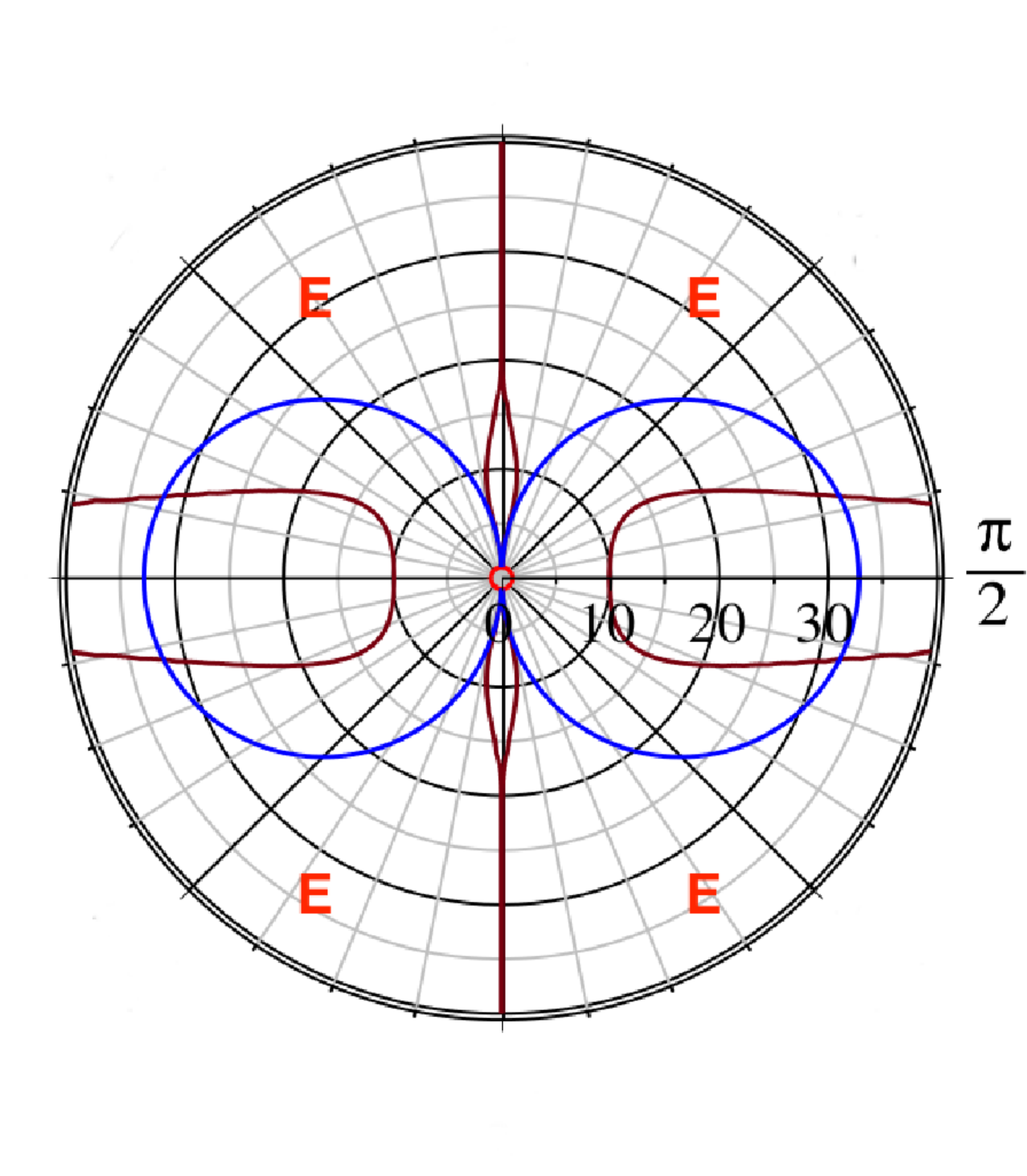} &
            \includegraphics[width=5 cm]{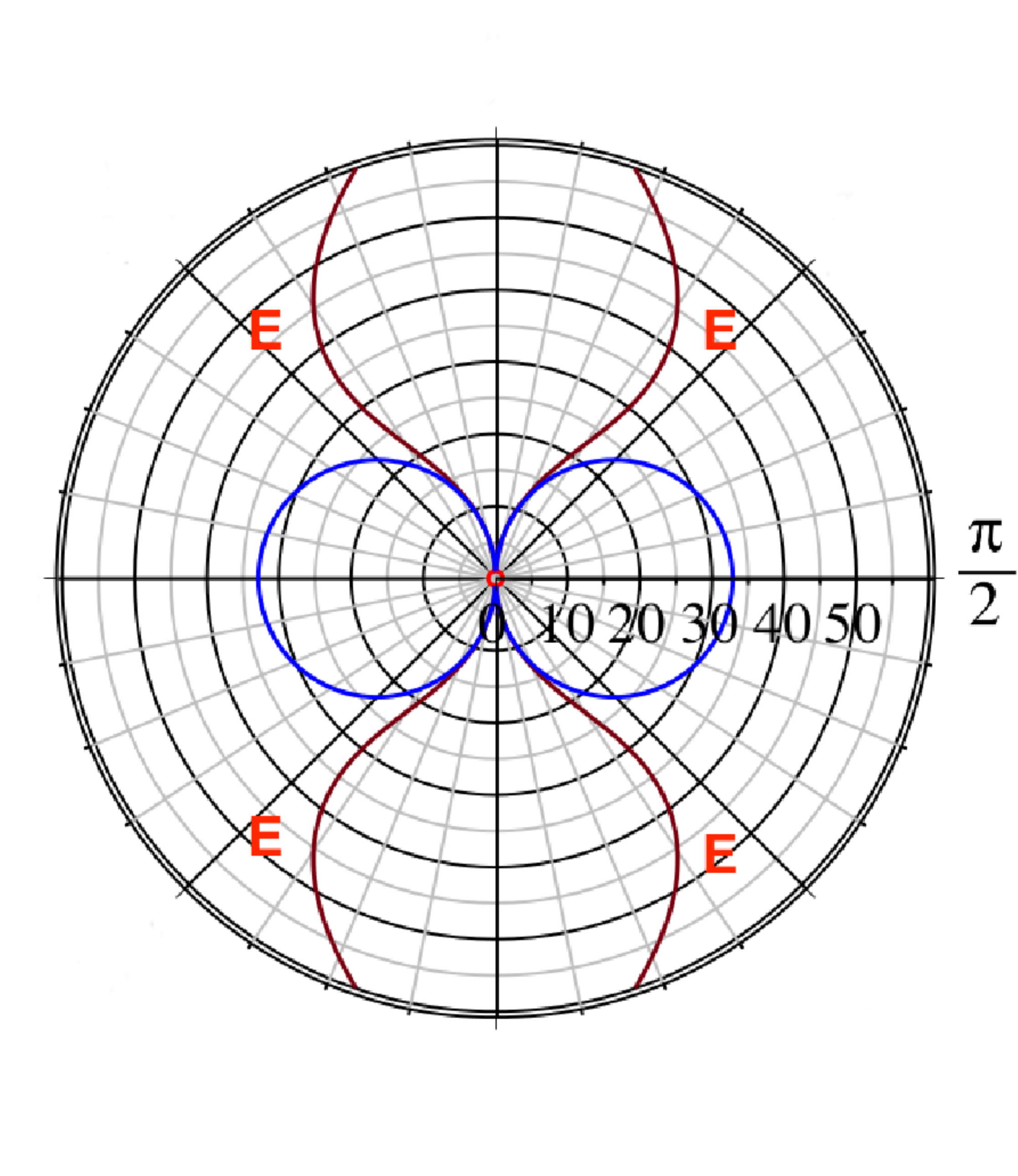} &
            \includegraphics[width=5 cm]{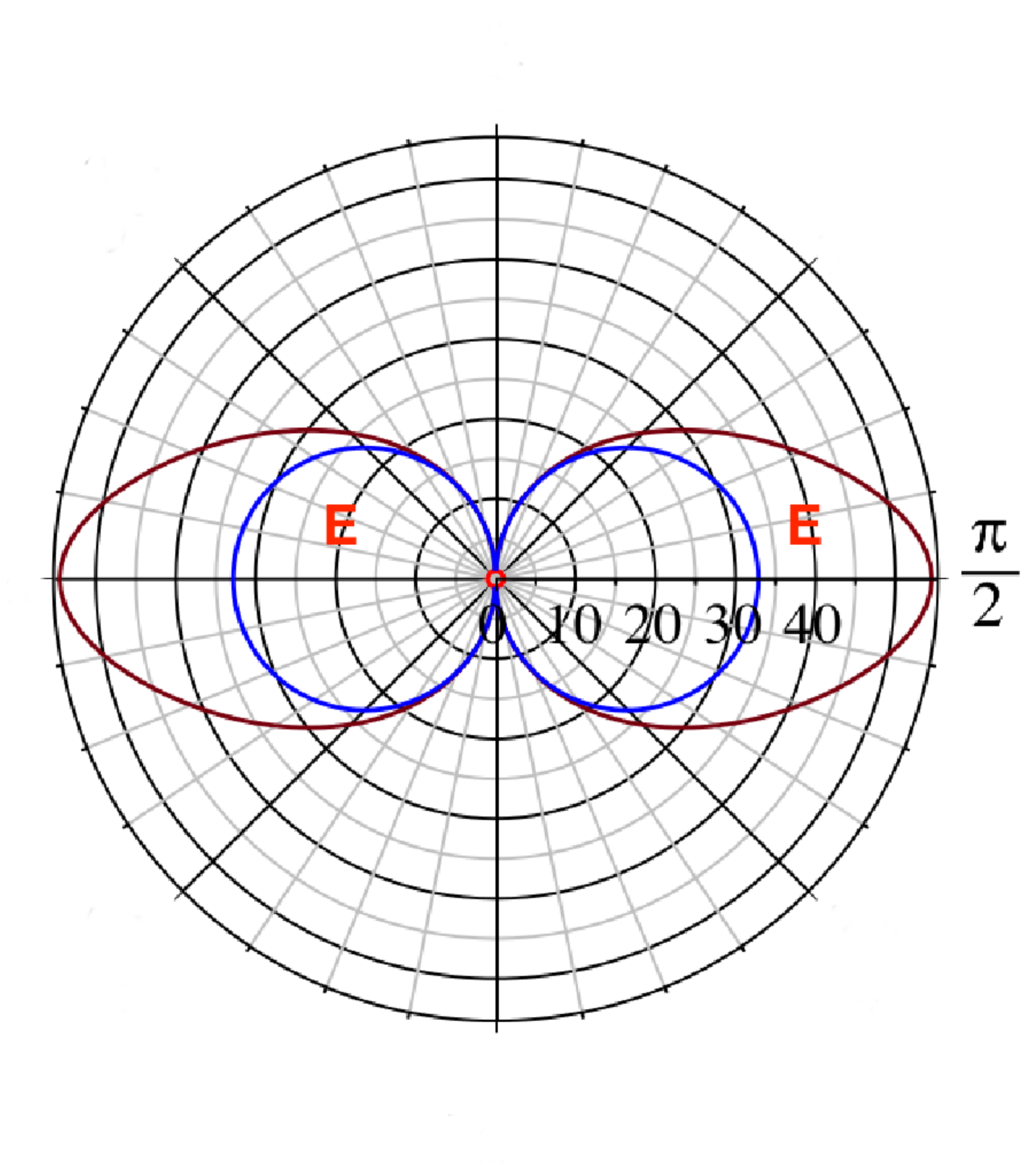} \\
            $c_2=1/150$ & $c_2=1/3000$ & $c_2=1/9000$
           \end{tabular}}
           \caption{\footnotesize{ Ergoregion for quadrupole distortion and $\alpha=0.97$}}
		\label{even097}
\end{figure}

In this section, we investigate the influence of an external quadrupole field on the behaviour of the ergoregion. The analysis is organized in the following way. First we describe the configurations which are observed for three values of the rotation parameter $\alpha$ and illustrate the results in Fig. \ref{even03}-\ref{even097}. Then, we deduce some general properties of the ergoregion for quadrupole distortion and summarize the dependance of its features on the parameters $c_2$ and $\alpha$. In the analysis presented in Fig. \ref{even03}-\ref{even097}, we distinguish two cases - of positive and negative values of the distortion parameter $c_2$ located in the first and the second row in each figure, respectively. 
 For  better visualisation, we introduce  a new coordinate $\theta\in[0,\pi]$, which is related to the coordinate $y$ as $y=\cos(\theta)$. Consequently, in each of the figures \ref{even03}-\ref{even097}, we observe the cross-section of the ergoregion with the $(x, \theta)$ - plane. The rotation axis is located at $\theta =0$ and $\theta = \pi$. Thus, the full ergoregion is obtained by rotating the plots around the vertical axis. In all the figures the red colour represents the horizon, the blue corresponds to the ergoregion of the undistorted Kerr black hole and the dark red to the ergoregion of the distorted case. All the non-compact parts of the ergoregion and some compact ergoregions  of the distorted Kerr black hole are denoted by $E$ in order to be distinguished from the static regions.

In Fig. \ref{even03}, we see the ergoregion for $\alpha = 0.3$.   For $c_2=-1/2$, we have a compact ergoregion which encompasses the horizon of the black hole and touches the horizon on the axis. This region gets larger when the absolute value of the distortion parameter $c_2$ decreases. For $c_2=-1/50$, it coincides approximately with the ergoregion of the undistorted case. At the same time the non-compact parts of the ergoregion, which do not touch the horizon, are pushed further away from the black hole.

For $c_2=1/2$, we do not have a compact ergoregion which encompasses the horizon. However, for $c_2\approx1/6.5$ a compact ergoregion appears which is very close to the undistorted case. This happens when the two static regions on the axis approach each other in order to finally form the shape that we can see in Fig. \ref{even03} for $c_2=1/10$. When the values of the distortion parameter decrease, the static region around the black hole increases, as we see in Fig. \ref{even03} for $c_2=1/30$. At the same time the static region situated on and around the equator is pushed further away from the black hole and gets more narrow (see Fig. 1 for $c_2=1/2$, and $c_2=1/10$).

In Fig. \ref{even07}, we see the ergoregion for $\alpha = 0.7$. The evolution of the ergoregion as the distortion parameter varies is similar to the case $\alpha = 0.3$. For $c_2=-1/2$, we observe a non-compact part of the ergoregion disconnected from the horizon, and  a compact one which encompasses the horizon and touches the horizon on the axis. This region gets larger when the absolute value of the distortion parameter decreases. At the same time the non-compact parts of the ergoregion are pushed further away from the black hole. The compact ergoregion in the vicinity of the horizon is illustrated for low absolute values of the distortion parameter ($c_2= -1/100$). We observe that it is smaller than the ergoregion of the isolated Kerr black hole for the same rotation parameter $\alpha$. We investigate further this effect in section 3.3.

For $c_2=1/2$, we do not have a compact ergoregion which encompasses the horizon. However, for $c_2\approx1/63.6$ a compact ergoregion appears which is very similar in shape to the undistorted case. This happens when the two static regions on the axis approach each other in order to finally form a shape similar to the one we observe in Fig. \ref{even03} for $c_2=1/10$. The static region around the black hole increases when the distortion parameter decreases, while the static region located on the equator is pushed further away from the black hole and gets more narrow (see Fig. \ref{even07} for $c_2=1/2$). The compact ergoregion in the vicinity of the horizon for low values of the distortion parameter is larger than the ergoregion of the isolated Kerr black hole for the same rotation parameter $\alpha$ (see Fig. \ref{even07} for $c_2=1/100$).

In Fig. \ref{even097}, we see the ergoregion for $\alpha = 0.97$. For $c_2=-1/2$ we do not have a compact ergoregion which encompasses the horizon. Moreover, the static regions around the axis are very small. When the absolute value of the distortion parameter decreases, the static regions on the axis increase, while the static region on the equator is pushed further away from the black hole. Thus, for $c_2=-1/150$ we observe the configuration presented in Fig. \ref{even097}. The static regions on the equator and on the axis get larger, and for $c_2\approx-1/431$ touch each other forming the first compact ergoregion which encompasses the horizon. When the absolute value of the distortion parameter further decreases, the compact ergoregion increases and its shape approaches that of the undistorted case. At the same time the non-compact part of the ergoregion is pushed further away (see Fig. \ref{even097} for $c_2=-1/1000$). The compact ergoregion in the vicinity of the horizon is illustrated for low absolute values of the distortion parameter ($c_2=-1/3000$) in comparison with the ergoregion of the non-distorted Kerr black hole for the same value of the rotation parameter.

For positive distortion parameter $c_2=1/150$, we observe no compact ergoregion. When $c_2$ decreases, the static regions on the axis get larger and  touch each other for $c_2\approx1/8673$ . Thus, we have the appearance of the first compact ergoregion which encompasses the horizon. This ergoregion gets smaller when the value of the distortion parameter decreases. We have illustrated this effect by presenting the vicinity of the horizon for $c_2=1/9000$ in Fig. \ref{even097}. Globally, the ergoregion is similar to the configuration presented in Fig. \ref{even03} for $c_2=1/10$.

\paragraph{}
We can summarize the observed characteristics of the ergoregion in the following way. In the case of quadrupole distortion the ergoregion is determined by the function $A(x,y;c_2,\alpha)$ defined in eq. (\ref{A}),  which depends on two parameters $c_2$ and $\alpha$. It can be presented explicitly in the form
\begin{eqnarray}\label{A_c2}
A(x,y;c_2, \alpha) &=& (x^2-1)\left(1-\alpha^2e^{4c_2x(1-y^2)}\right)^2 \nonumber \\
 &-& 4\alpha^2(1-y^2)e^{4c_2x(1-y^2)}\cosh^2\left[2c_2y(1-x^2)\right].
\end{eqnarray}
Analyzing its behaviour, we can deduce the following general properties, which characterize the ergoregion:
\begin{enumerate}
\item The ergoregion is symmetric with respect to the cross-section $y=0$ (or equivalently $\theta=\pi/2$). Indeed, if we examine the inequality $A(x,y;c_2, \alpha)<0$, which determines the ergoregion, we observe that it is invariant with respect to the shift $y\longleftrightarrow -y$. Consequently, if $y=\tilde{y}$ is a solution to $A(x,y;c_2, \alpha)<0$, $y=-\tilde{y}$ is a solution as well, and the corresponding point belongs also to the ergoregion.
\item The ergoregion always extends to infinity, if we consider the distorted Kerr black hole as a global solution. This is evident from the behaviour of the function $A(x,y;c_2, \alpha)$ at $x\rightarrow\infty$. In particular, it implies that there are no cases (for any values of the parameters $c_2$ and $\alpha$) when the ergoregion consists of a single compact region in the vicinity of the horizon, in contrast to the non-distorted Kerr black hole.
\item The axis $y=\pm1$ does not belong to the ergoregion for any values of the parameters $c_2$ and $\alpha$ except on the horizon. If we consider the equation $A(x,y;c_2, \alpha)=0$ for $y=\pm1$, the only solution in the interval $x\in(-1,+\infty)$ is $x=1$, where the ergoregion touches the horizon.
\item The cross-section $y=0$ does not belong to the ergoregion for large values of $x$ for any values of the parameters $c_2$ and $\alpha$. Indeed, we see that the function $A(x,y=0;c_2, \alpha)$ is positive for $x\rightarrow\infty$. Consequently, there always exists some value $\tilde{x}$, which depends on the parameters $c_2$ and $\alpha$, such that for $x>\tilde{x}$ a static region is formed around the cross-section $y=0$ and it extends to infinity.
\item The ergosurface intersects the cross-section $y=0$ an odd number of times. Since the function $A(x,y=0;c_2, \alpha)<0$ on the horizon, and $A(x,y=0;c_2, \alpha)>0$ at infinity for any values of $c_2$ and $\alpha$, it should possess an odd number of real zeros located in the interval $x\in (1, +\infty)$.
\end{enumerate}
\begin{figure}[htp]
\setlength{\tabcolsep}{ 0 pt }{\scriptsize\tt
		\begin{tabular}{ ccccc }
	\includegraphics[width=0.2\textwidth]{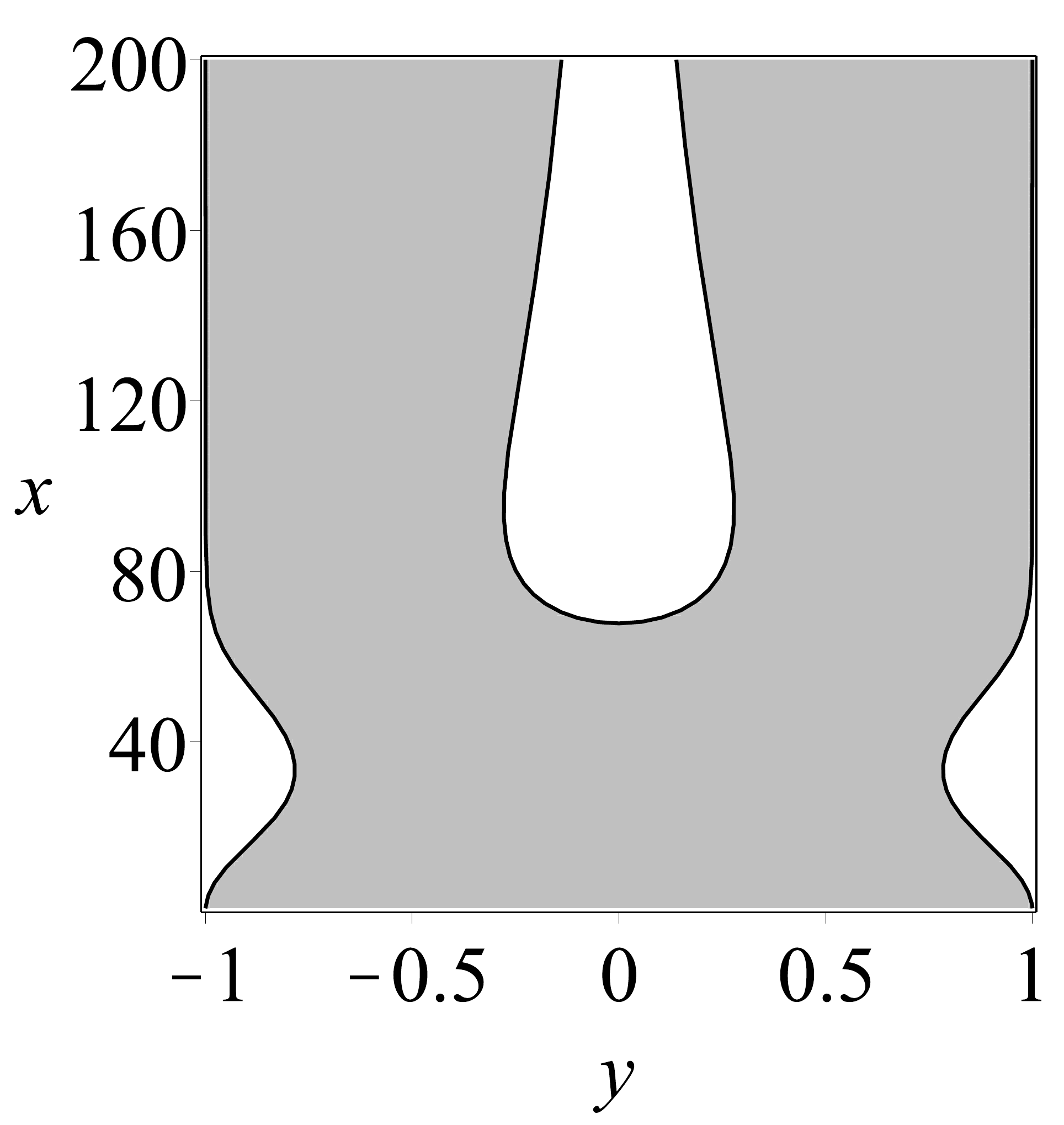} &
            \includegraphics[width=0.2\textwidth]{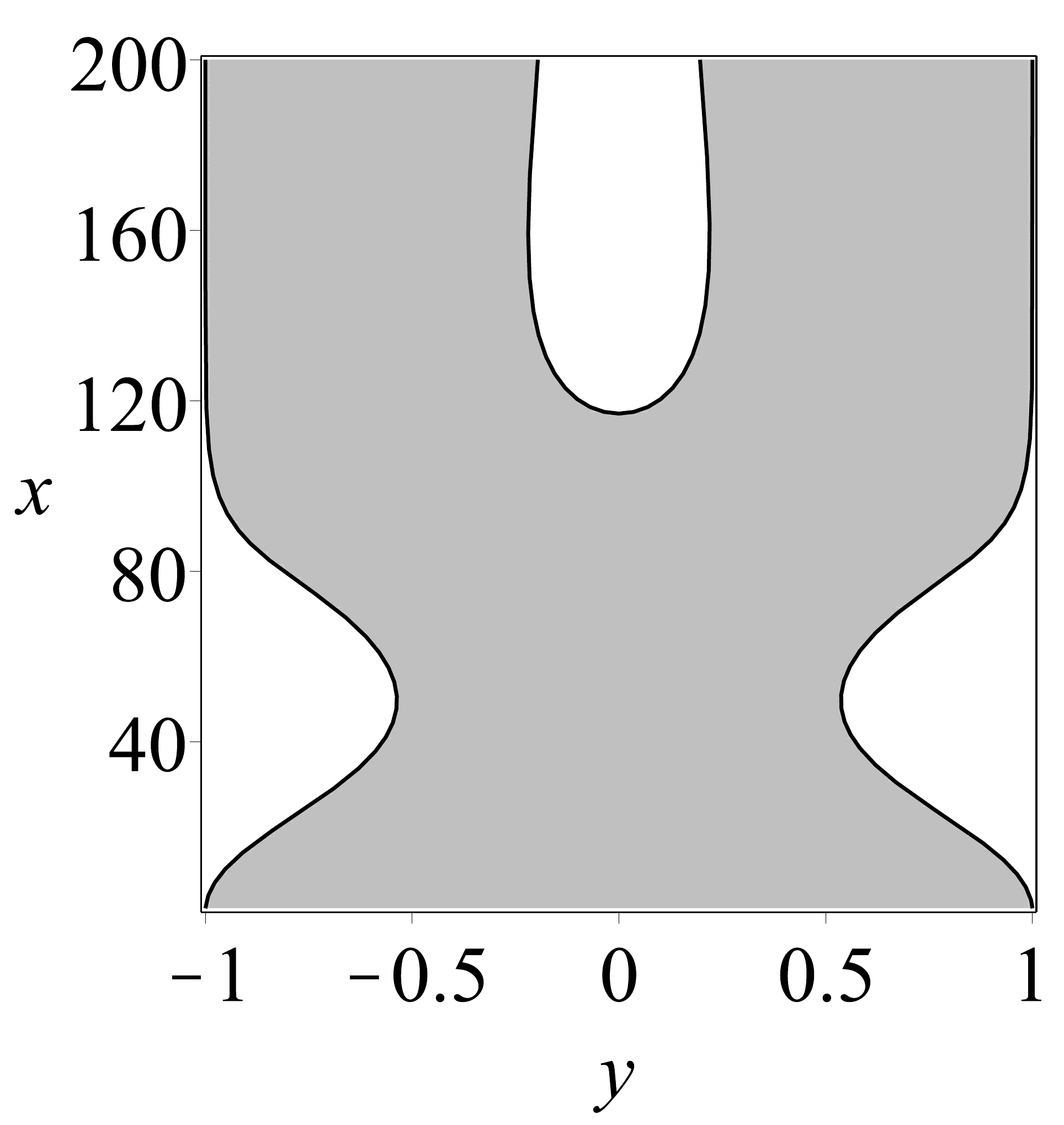} &
            \includegraphics[width=0.2\textwidth]{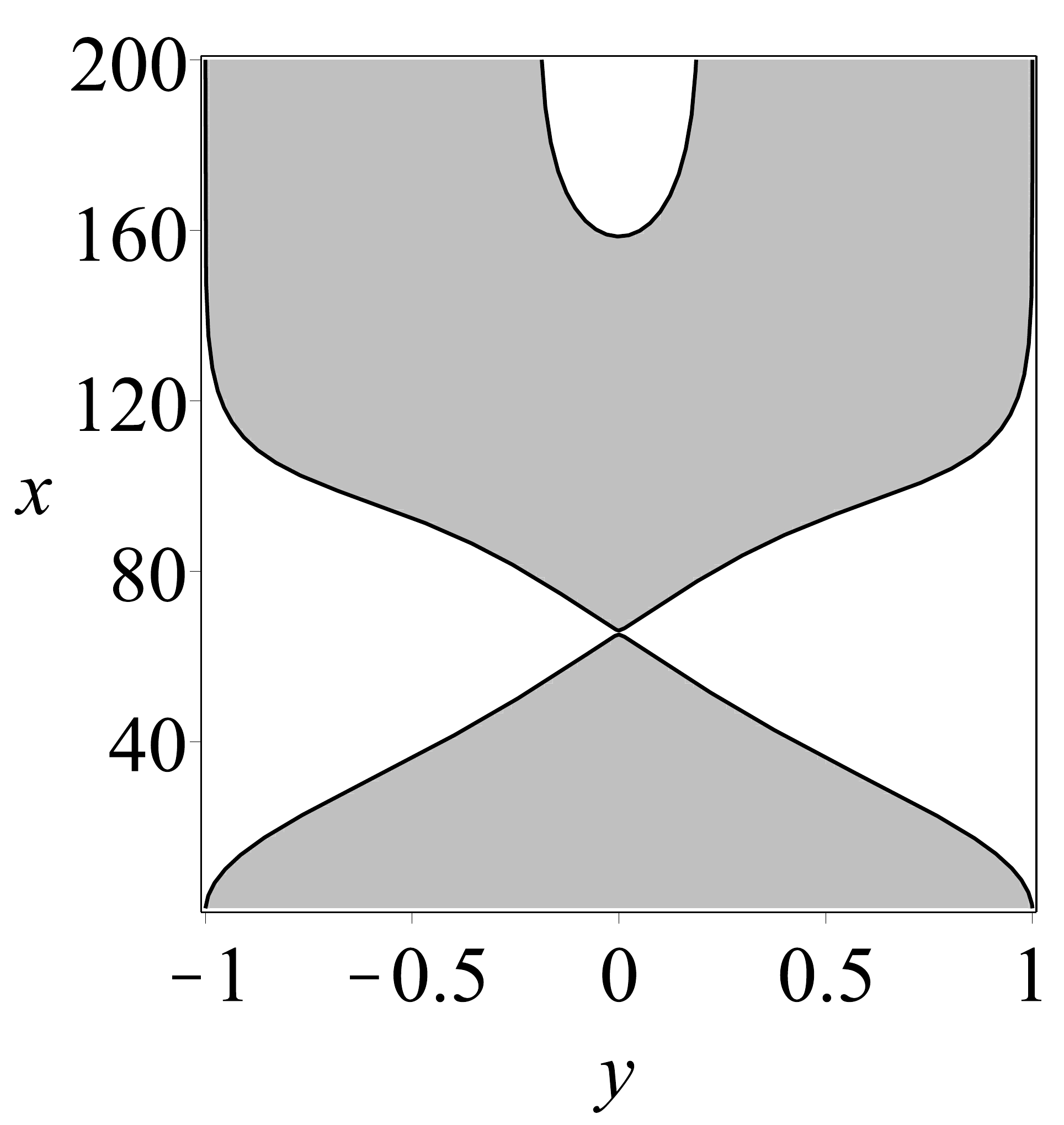} &
            \includegraphics[width=0.2\textwidth]{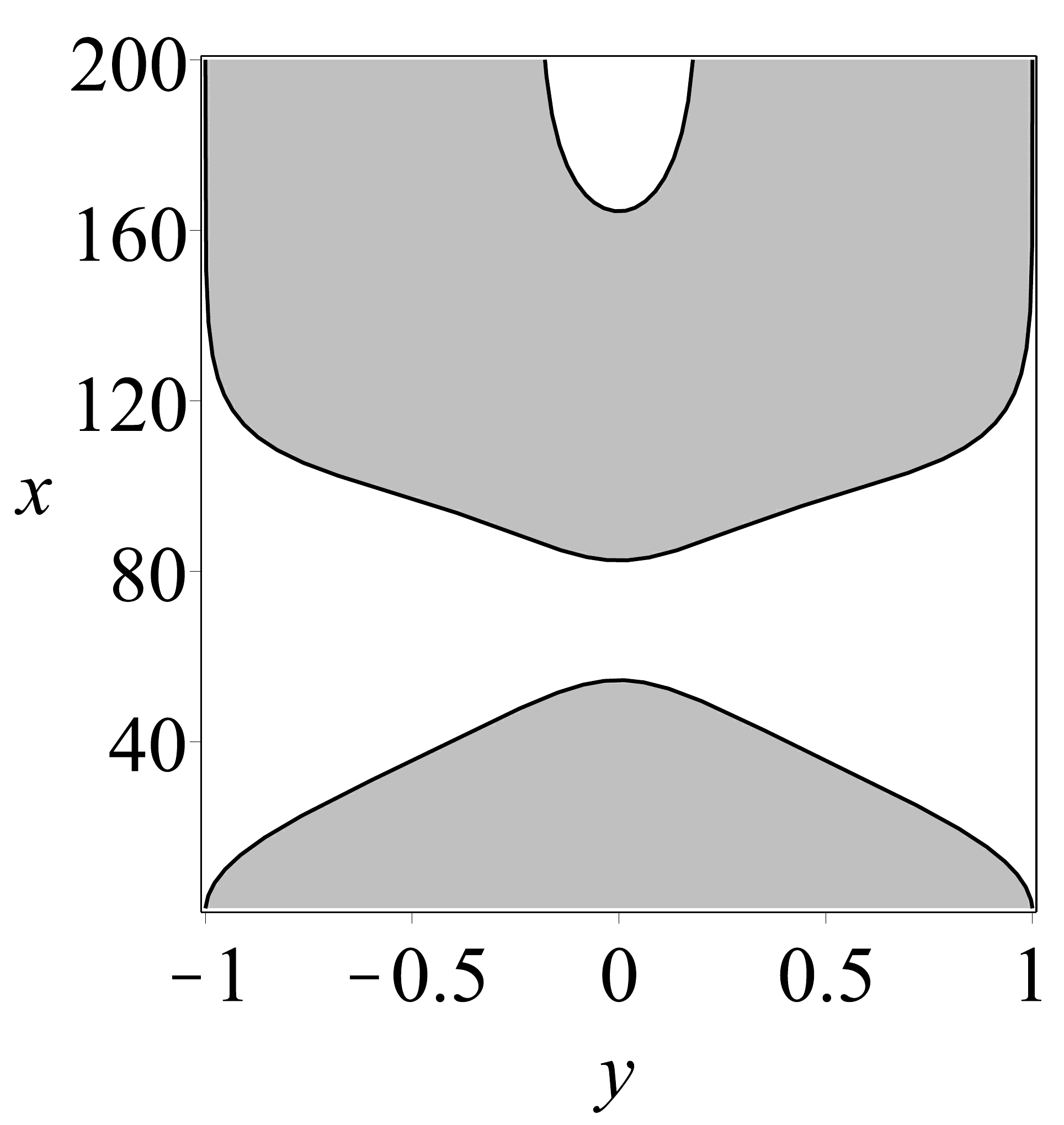}&
            \includegraphics[width=0.2\textwidth]{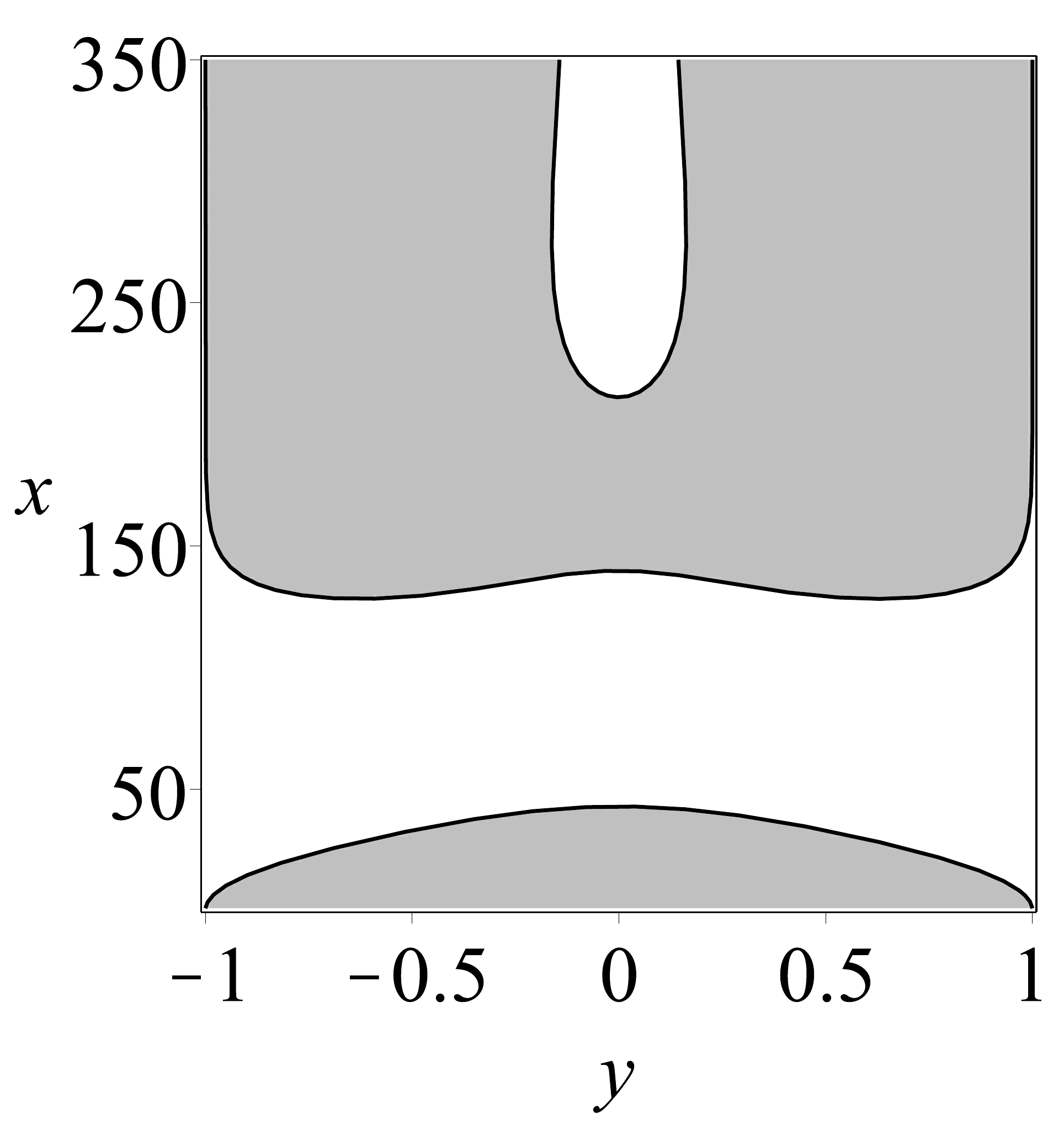} \\
             $c_2=1/3000$ & $c_2=1/6000$ & $c_2=1/8624$& $c_2=1/9000$ &$c_2=1/12000$
                       \end{tabular}}
           \caption{\footnotesize{behaviour of the ergoregion (grey area) as a function of $c_2$ for positive distortion parameter. The rotation parameter is fixed to $\alpha=0.97$.}}
		\label{c2gen}
\end{figure}
These general properties are relevant both for positive and negative values of the distortion parameter $c_2$, but the ergoregion behaves qualitatively different in the two cases. The behaviour of the ergoregion as a function of $c_2$ for positive distortion parameter is summarized in Fig. \ref{c2gen}. For high values of $c_2$ the ergoregion consists of a single connected region extending to infinity (if we consider the solution as a global one). The static regions are non-compact as well and consist of two types: the first one is located in the vicinity of the cross-section $y=0$, and the other one contains the axis.  As $c_2$ decreases the ergoregion pinches symmetrically with respect to the cross-section $y=0$, until for a certain value of $c_2= c_{crit}$ two parts of the ergoregion are formed touching at a single point in the $(x,y)$-plane located at $y=0$. For values of $c_2 < c_{crit}$ two disconnected parts of the ergoregion develop - a compact and a non-compact one. The compact one is located in the vicinity of the horizon, and closely resembles the ergoregion of the non-distorted Kerr black hole for the same value of the rotation parameter $\alpha$. When $c_2$ decreases, the size of the compact ergoregion decreases. At the same time the distance between the two disconnected parts of the ergoregion increases, and the static region located around the cross-section $y=0$ becomes more remote with respect to the horizon.
\begin{figure}[htp]
\setlength{\tabcolsep}{ 0 pt }{\scriptsize\tt
		\begin{tabular}{ cccc }

	\includegraphics[width=0.25\textwidth]{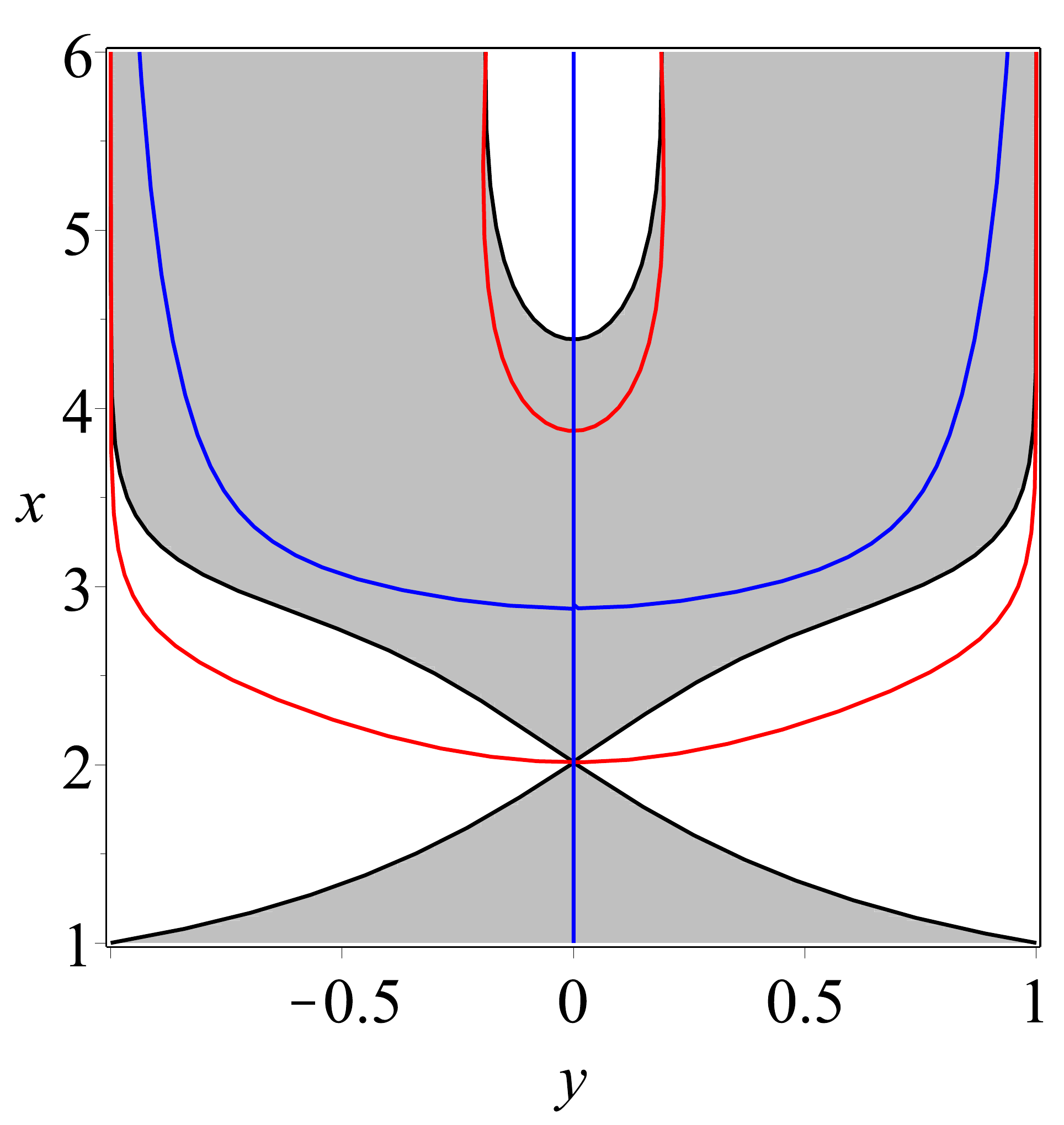} &
            \includegraphics[width=0.25\textwidth]{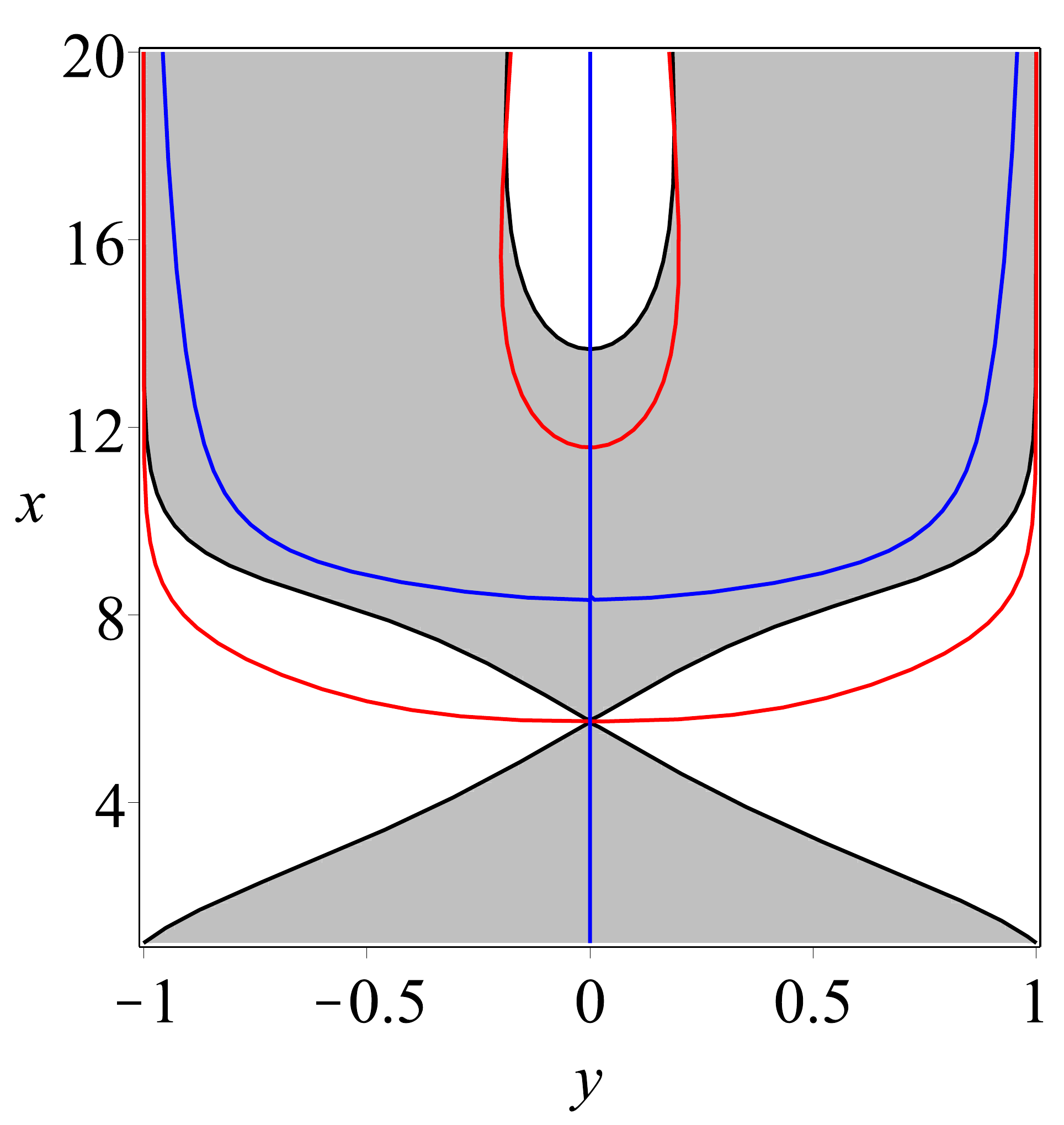} &
            \includegraphics[width=0.25\textwidth]{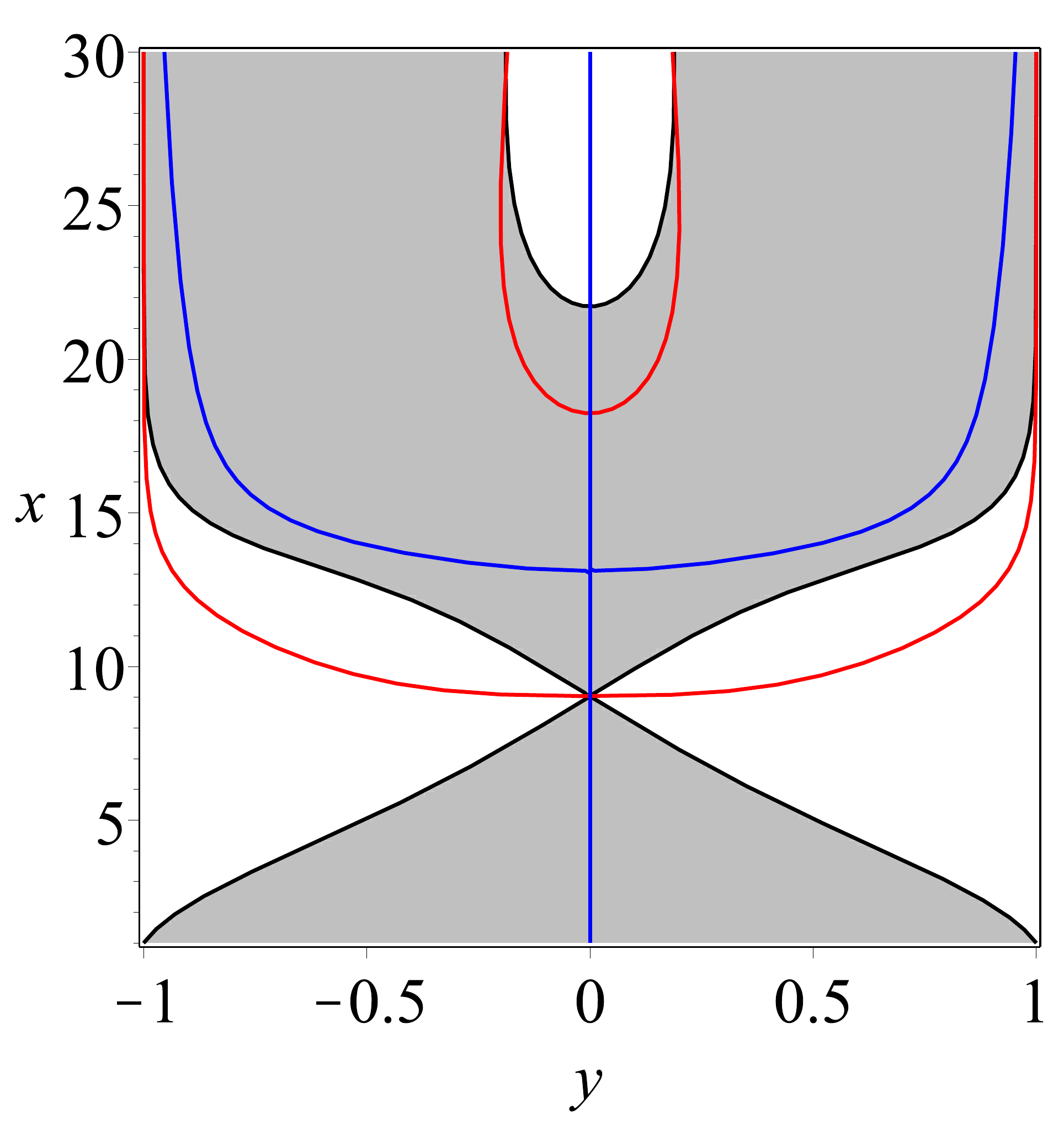} &
            \includegraphics[width=0.25\textwidth]{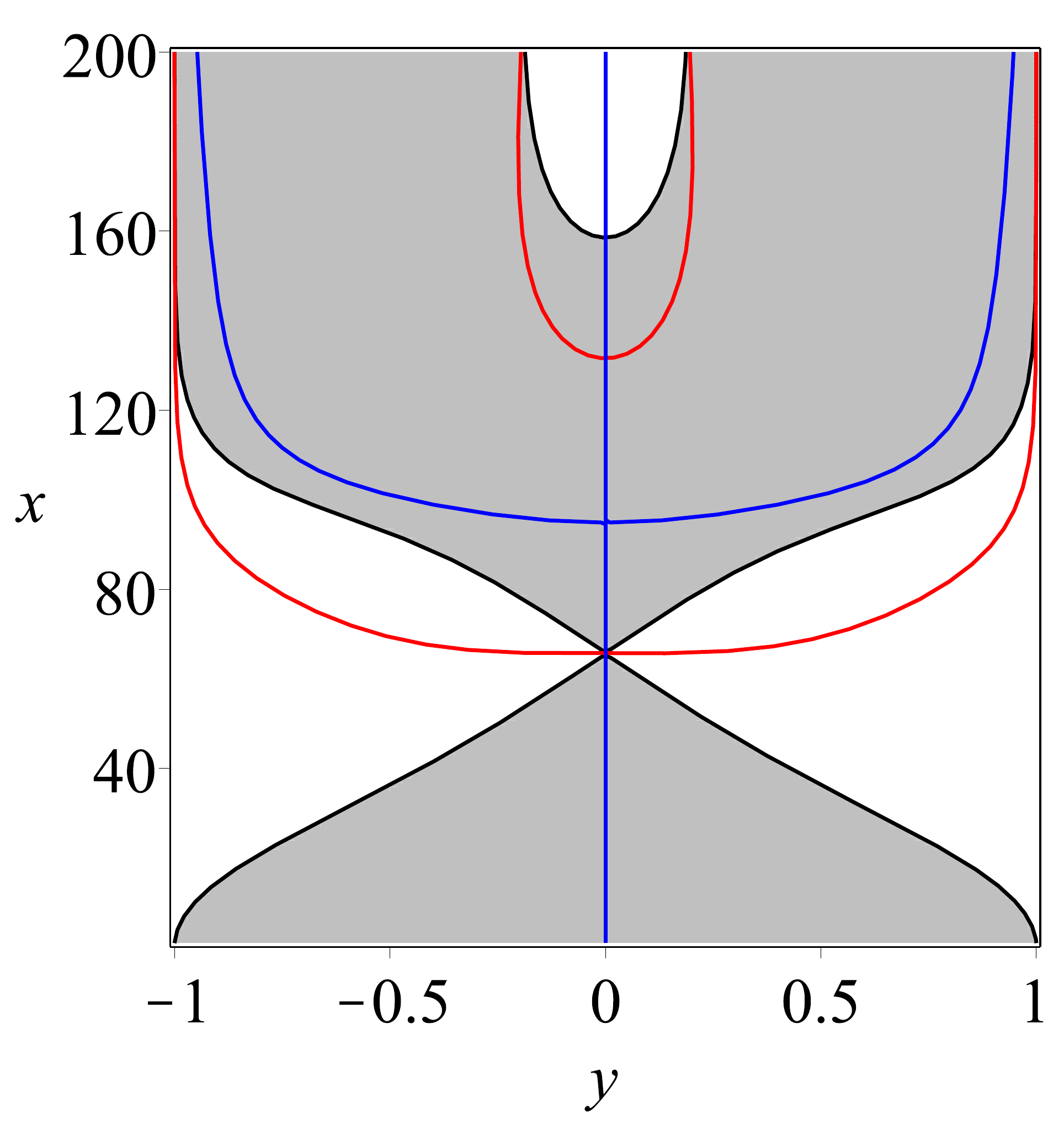}\\
	\includegraphics[width=0.25\textwidth]{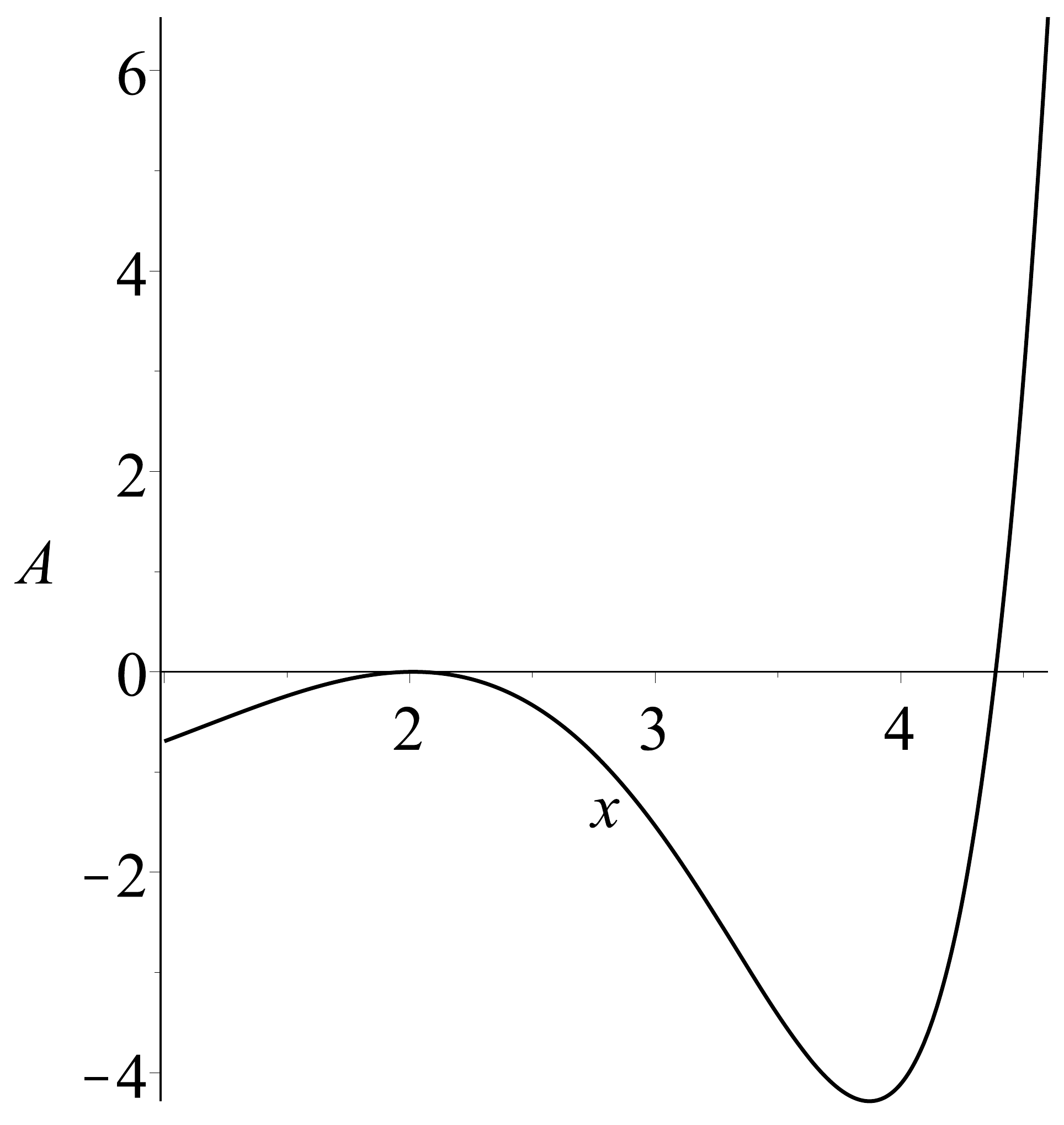} &
            \includegraphics[width=0.25\textwidth]{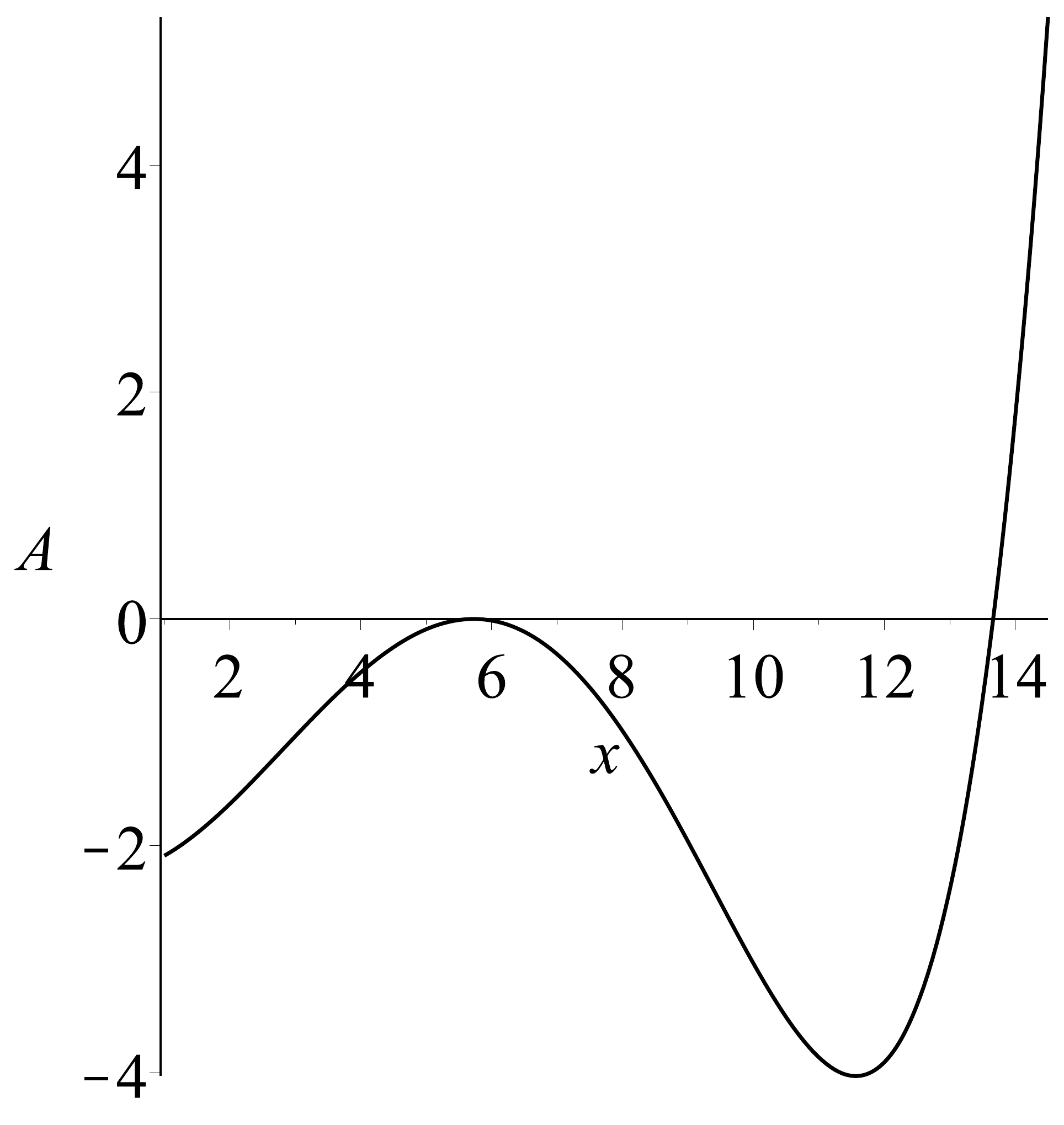} &
            \includegraphics[width=0.25\textwidth]{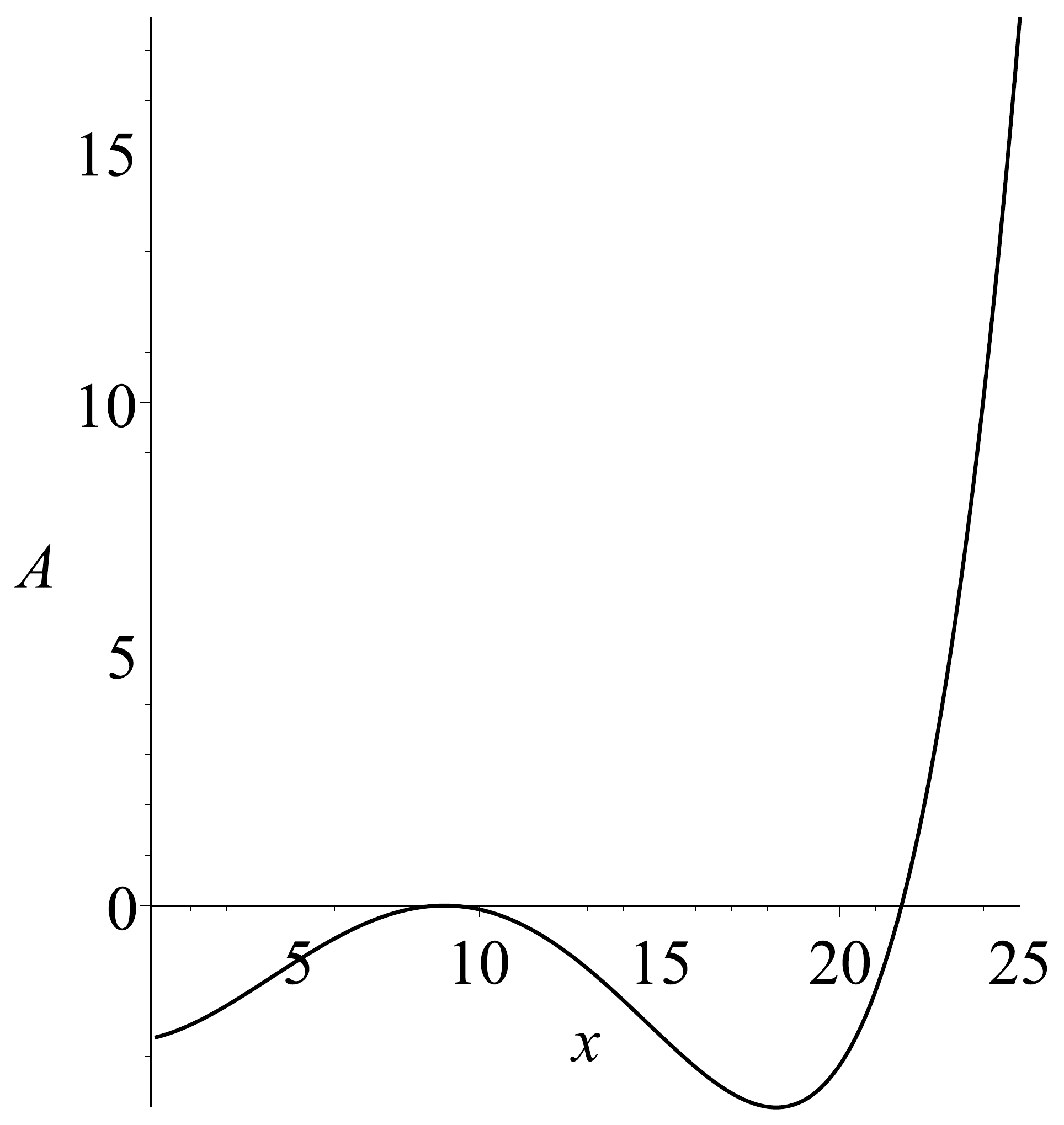} &
            \includegraphics[width=0.25\textwidth]{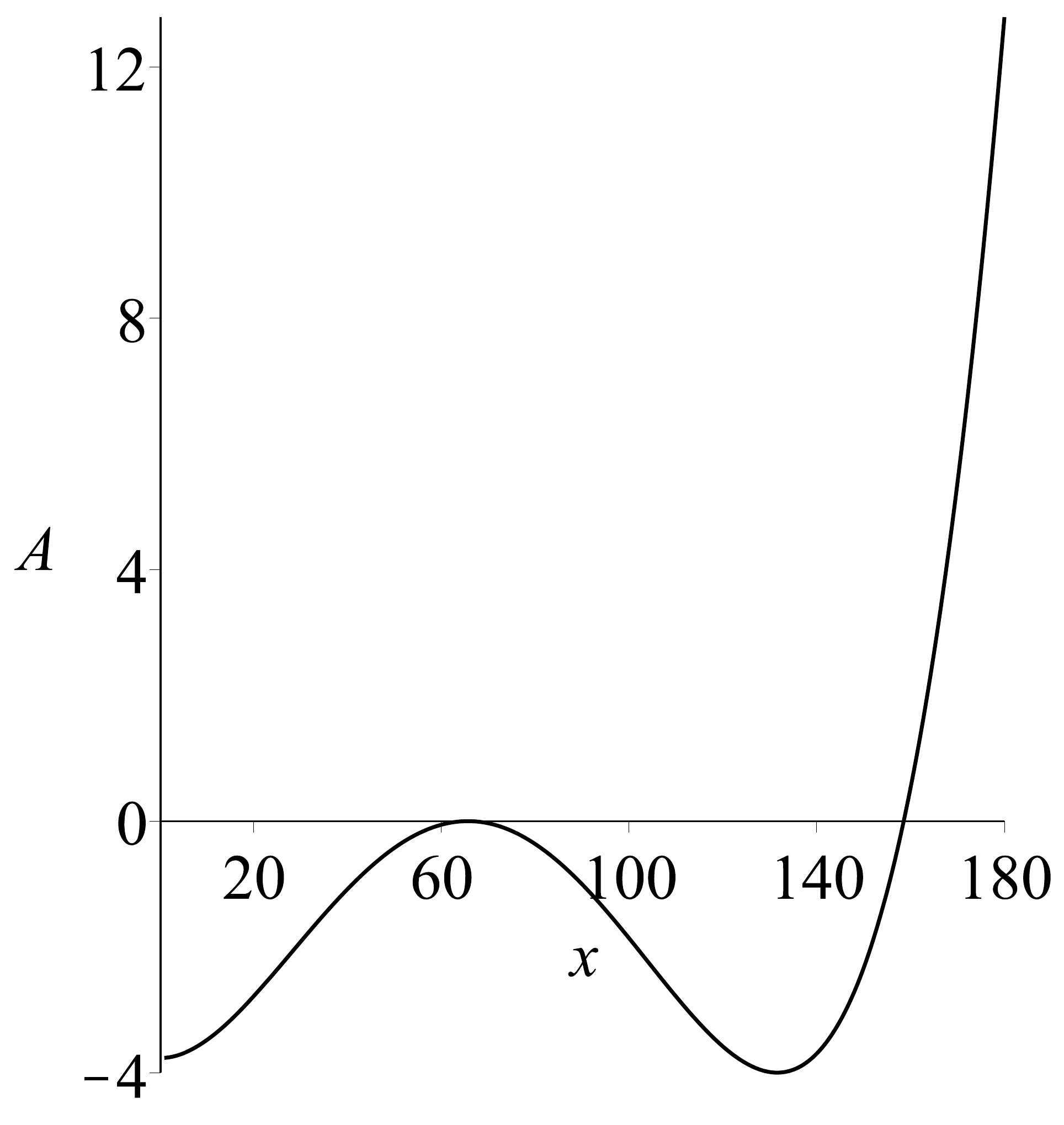}\\
            $\alpha=0.3$ & $\alpha=0.7$ & $\alpha=0.8$ & $\alpha=0.97$\\
            $c_{crit}\approx 0.164$ & $c_{crit}\approx 0.0157$ & $c_{crit}\approx 0.00619$ & $c_{crit}\approx0.1159\times10^{-3}$
                                   \end{tabular}}
           \caption{\footnotesize{Dependence of the critical point location on the rotation parameter $\alpha$ (top panel) and behaviour of the metric function $A(x,y)$ in the equatorial plane $y=0$ (bottom panel). The red and blue lines correspond to the curves $\partial_x A(x,y)=0$ and $\partial_y A(x,y)=0$, respectively, for each set of parameters $c_{crit}$ and $\alpha$.}}
		\label{c2pin}
\end{figure}

The behaviour of the ergoregion as a function of $c_2$, which is presented in Fig. \ref{c2gen}, is considered for the fixed value of the rotation parameter $\alpha=0.97$. However, it is representative for a general rotation parameter. For a different value of $\alpha$ we observe qualitatively the same configurations and the same type of evolution between them with only quantitative differences. These concern the value of $c_{crit}$, at which the transition between the connected and disconnected type of ergoregion occurs, and the location of the critical point, in which the ergoregion pinches off. The size of the compact part of the ergoregion and the separation between the compact and non-compact parts also depend on the value of the rotation parameter $\alpha$. In Fig. \ref{c2pin}, we illustrate the dependence of the critical point location $(x_c,y_c)$ in the $(x,y)$ - plane on $\alpha$. The value of $c_{crit}$, at which the transition occurs, is also specified for each rotation parameter. We observe that the critical point is always located on the $y=0$ cross-section. The value $x_c$ of the $x$-coordinate increases as $\alpha$ increases. On the other hand, the value of $c_{crit}$ decreases when the rotation parameter grows.

Mathematically, the pinching off of the ergoregion at a single point in the $(x,y)$-plane corresponds to the following situation. The boundary of the ergoregion, i.e. the ergosurface, is determined by the equation $A(x,y;c_2, \alpha)=0$. Thus, for each fixed value of $\alpha$ it defines a $1$-parameter family of curves in the $(x,y)$ - plane. If the conditions of the implicit function theorem are satisfied at some point $(x_0,y_0)$ for some value of $c_2$,  there exists a unique function $x(y)$ in some neighbourhood of $(x_0,y_0)$, determined by the equation $A(x,y;c_2, \alpha)=0$. Consequently, the critical points correspond to points in the $(x,y)$ - plane, in which the conditions of the implicit function theorem (see e.g. \cite{Zorich}) are violated for some value of $c_2$ and $\alpha$, and the equation $A(x,y;c_2, \alpha)=0$ does not define a unique function $x(y)$ (nor $y(x)$) even in an arbitrary small neighbourhood. For a fixed value of $\alpha$ such points are determined by the system of equations
\begin{eqnarray}
A(x,y;c_2, \alpha)=0, \quad~~~ \partial_x A(x,y;c_2, \alpha)=0, \quad~~~ \partial_y A(x,y;c_2, \alpha)=0.
\end{eqnarray}

\noindent
The solution $(x_c,y_c;c_{crit})$ provides the coordinates of the critical point $(x_c,y_c)$, and the value of $c_2=c_{crit}$, at which  the transition between connected/disconnected type of ergoregion occurs. In Fig. \ref{c2pin}, we have presented the curves defined by the equations $\partial_xA(x,y;c_2, \alpha)=0$ and $\partial_y A(x,y;c_2, \alpha)=0$  with red and blue lines, respectively. We observe that they intersect simultaneously the ergosurface $A(x,y;c_2, \alpha)=0$ only in the critical point.

\begin{figure}[htp]
\setlength{\tabcolsep}{ 0 pt }{\scriptsize\tt
		\begin{tabular}{ ccc}
	       \includegraphics[width=4.1 cm]{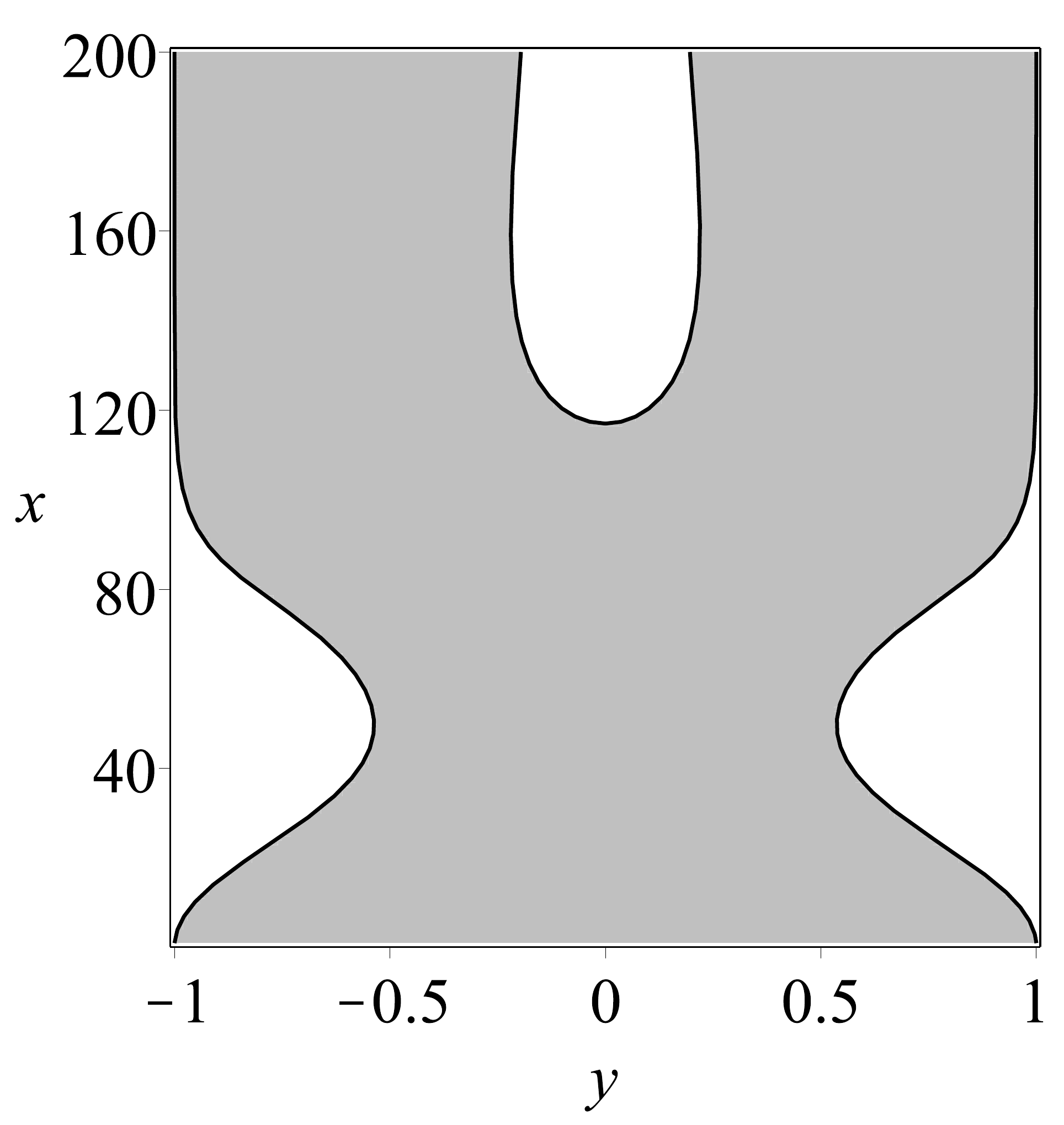} &
            \includegraphics[width=4.1 cm]{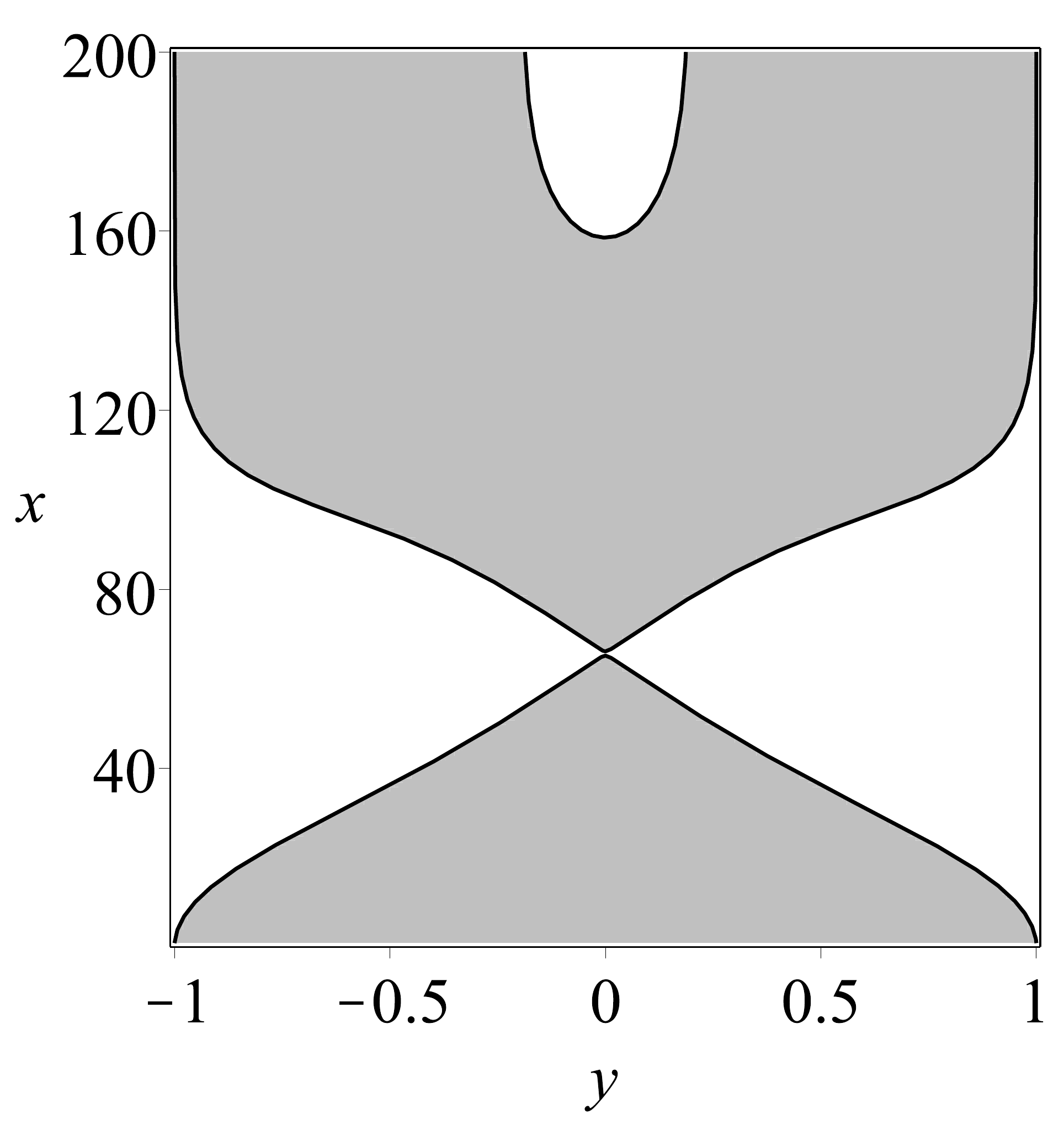} &
            \includegraphics[width=4.1 cm]{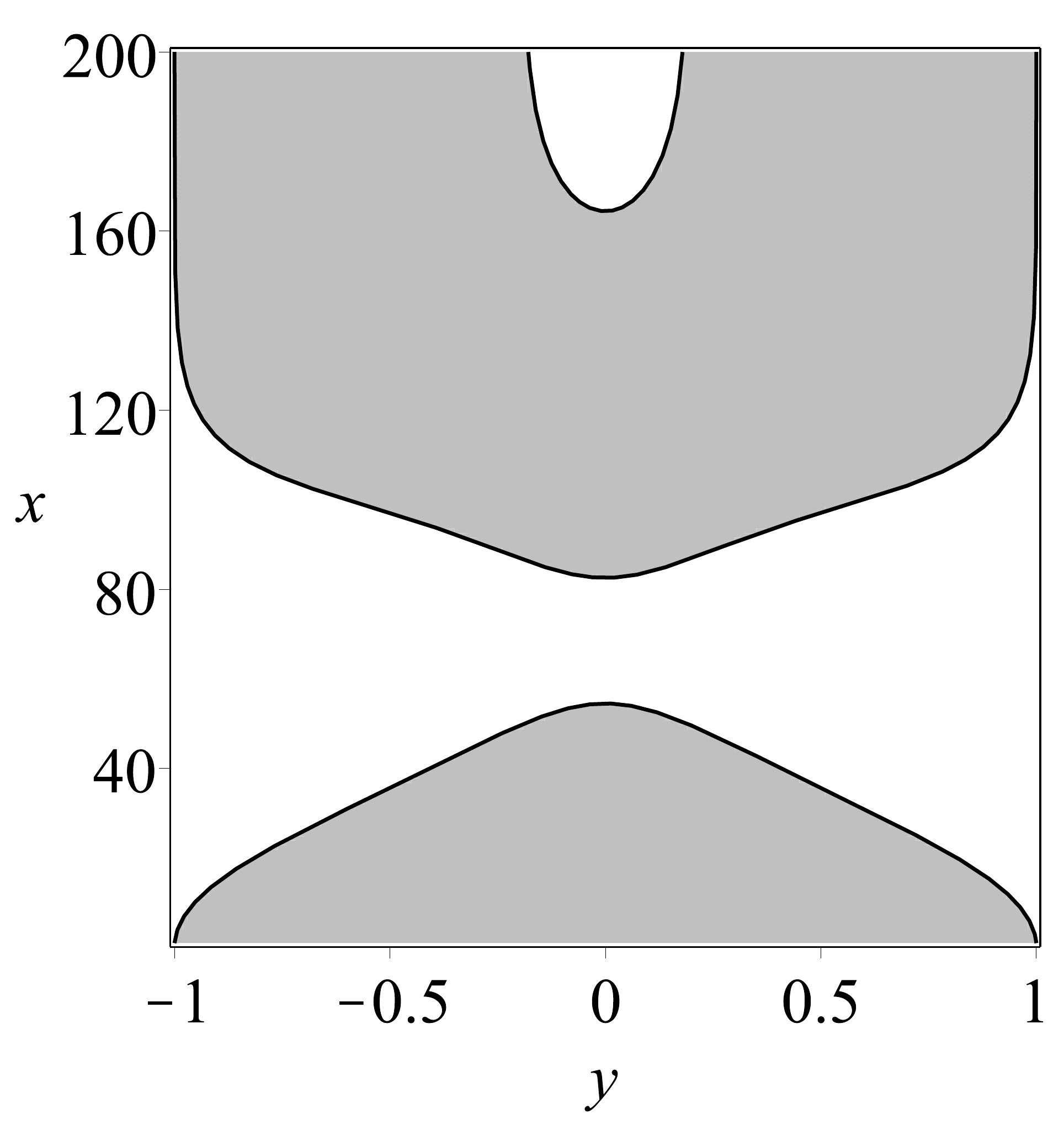} \\
	       \includegraphics[width=4.1 cm]{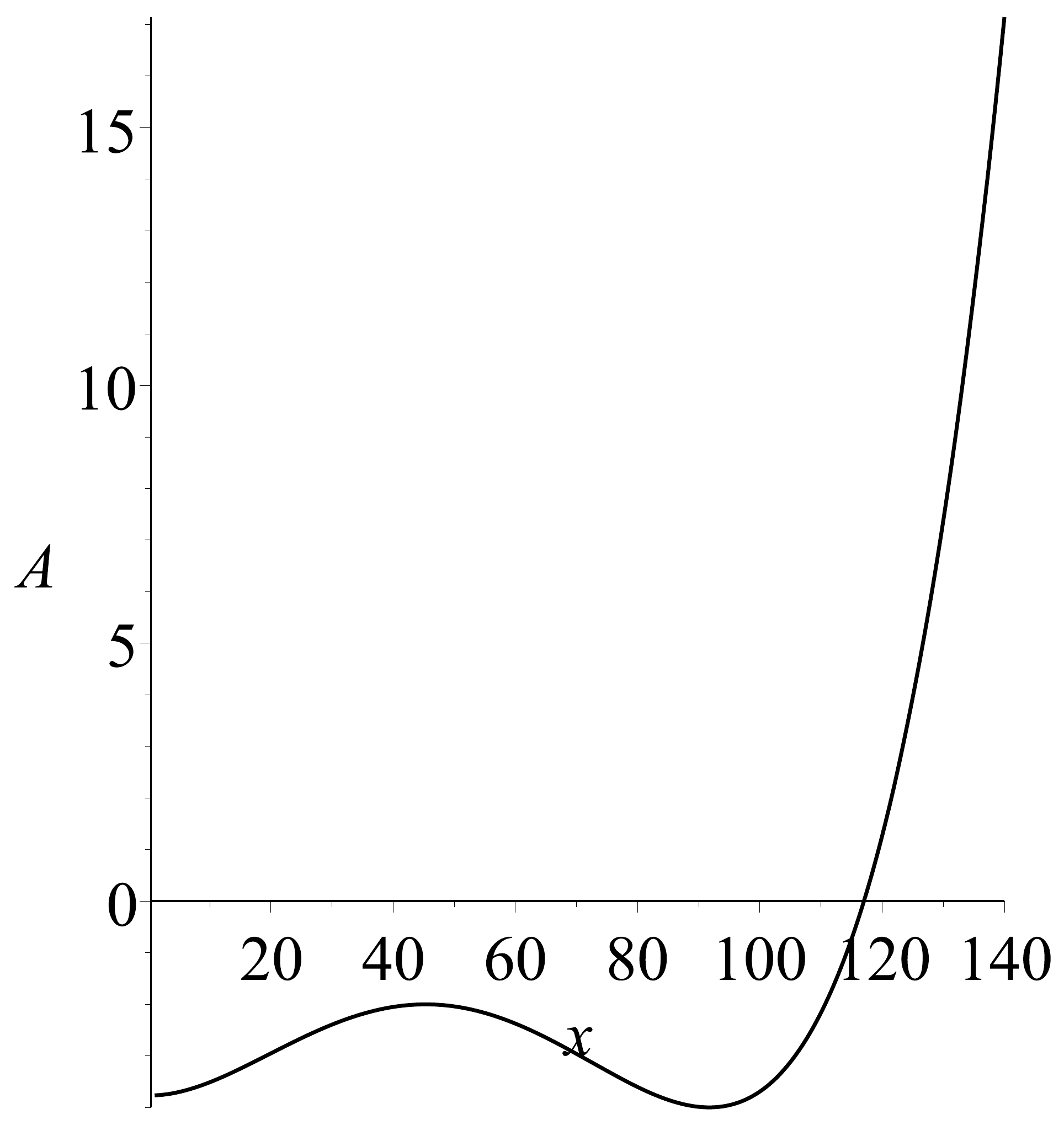} &
            \includegraphics[width=4.1 cm]{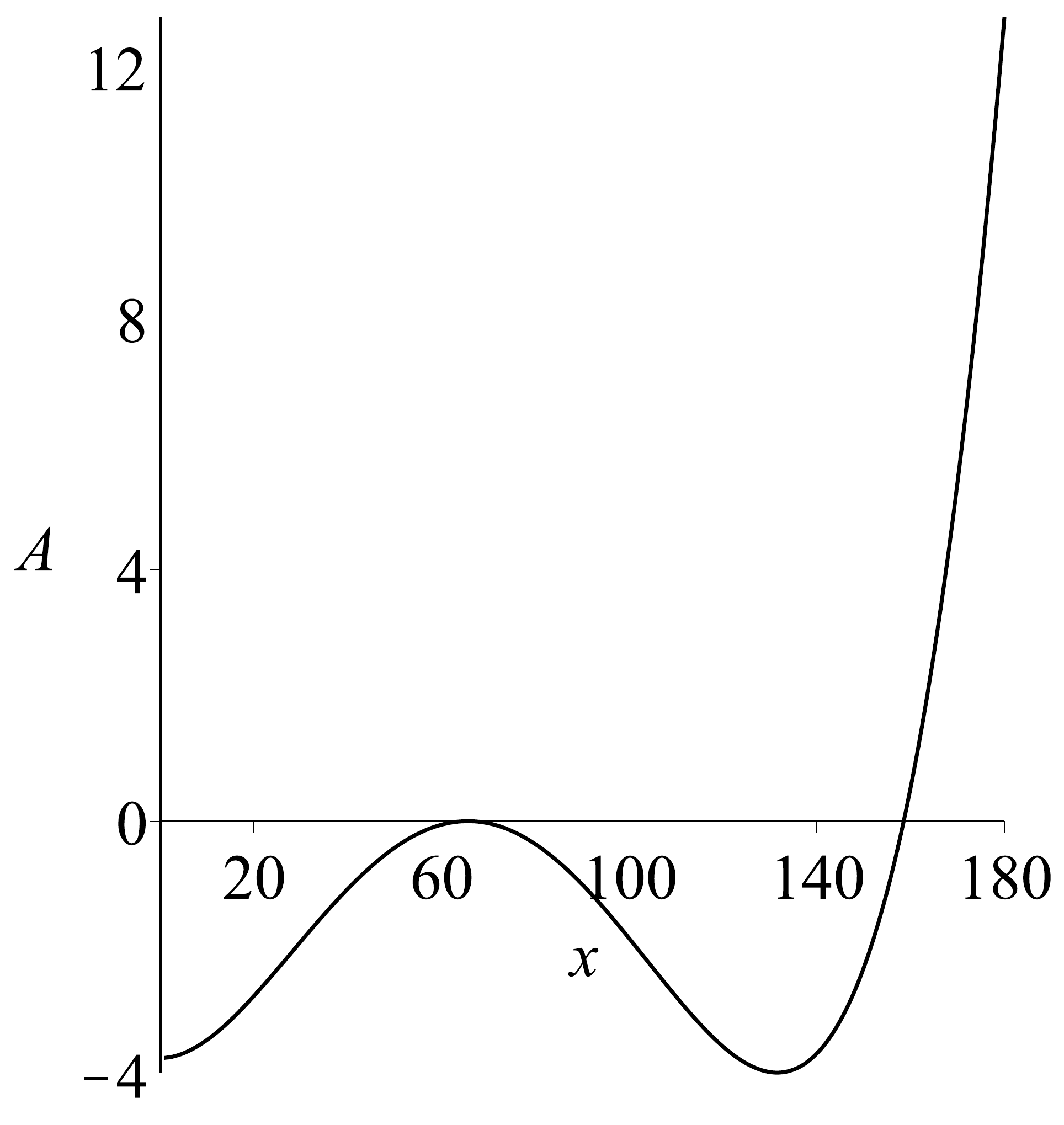} &
            \includegraphics[width=4.1 cm]{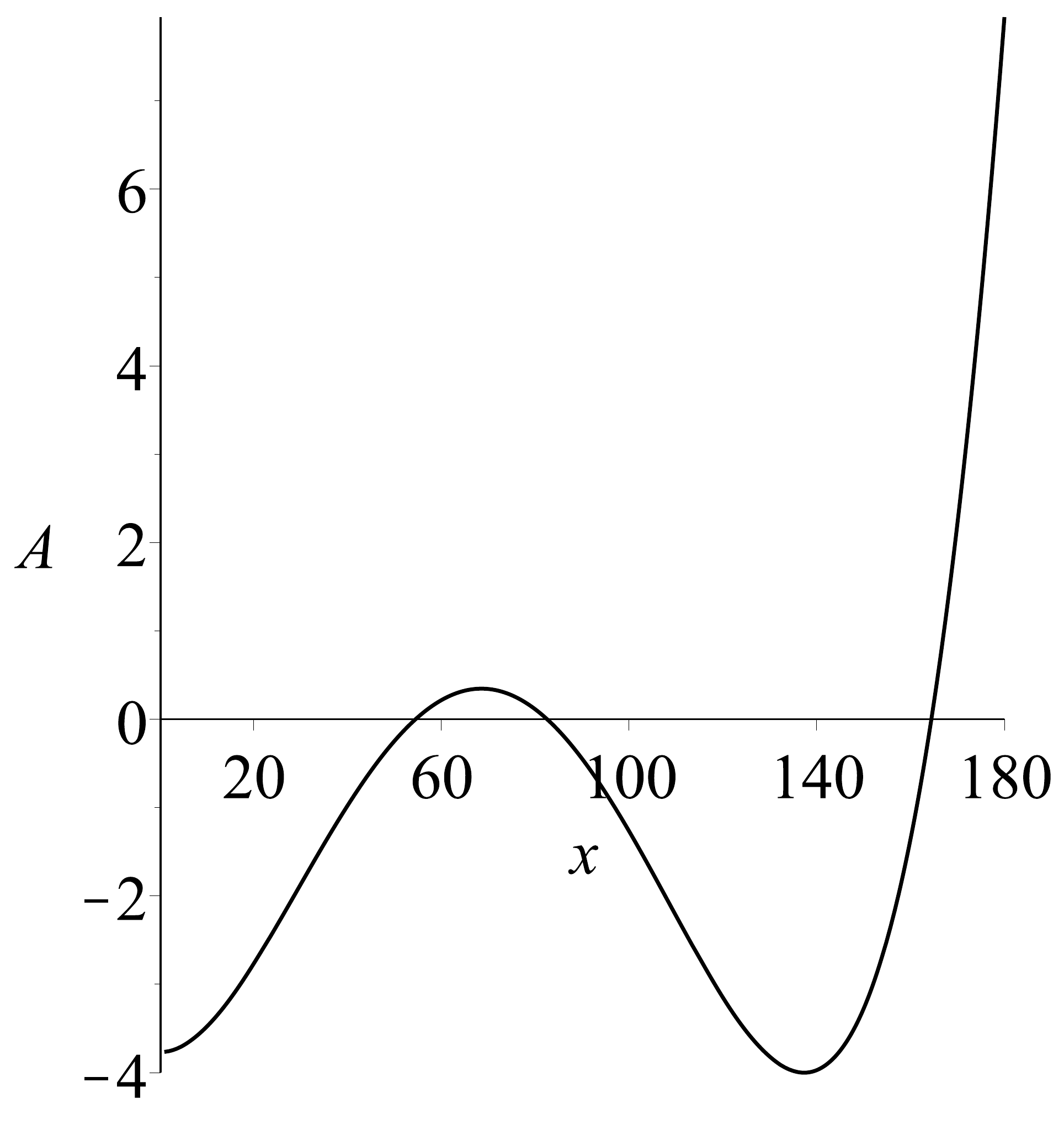} \\
             $c_2=1/6000$ & $c_2=1/8624$ & $c_2=1/9000$
                       \end{tabular}}
           \caption{\footnotesize{Correlation between the connected/disconnected type of the ergoregion (grey area) and the root structure of the metric function $A(x,y;c_2, \alpha)=0$ restricted to the equatorial plane $y=0$. The value of the rotation parameter is fixed to $\alpha=0.97$.}}
		\label{c2pin1}
\end{figure}

For positive values of $c_2$ the critical points always occur at $y=0$. Therefore, we can determine whether the ergoregion is connected or disconnected for some fixed values of $c_2$ and $\alpha$ by examining the root structure of the equation $A(x,y=0;c_2, \alpha)=0$. If it possesses a single real root $x=x_0$ located in the interval $x\in (1, +\infty)$, we observe a non-compact ergoregion extending to infinity. The root $x=x_0$ corresponds to the point, in which the static region located around the cross-section $y=0$ begins. If the equation possesses three reals roots in the interval $x\in (1, +\infty)$, we observe two disconnected parts of the ergoregion - a compact one and a non-compact one. The critical points correspond to the points in which the function $A(x;c_2, \alpha)=0$ is tangent to the $x$-axis. In Fig. \ref{c2pin1}, we illustrate the described three cases for the fixed value of the rotation parameter $\alpha=0.97$. The correlation between the root structure of the function $A(x;c_2, \alpha)=0$  and the location of the critical points is presented in Fig. \ref{c2pin} for different values of $\alpha$.
\begin{figure}[htp]
\setlength{\tabcolsep}{ 0 pt }{\scriptsize\tt
		\begin{tabular}{ ccccc }
	\includegraphics[width=0.2\textwidth]{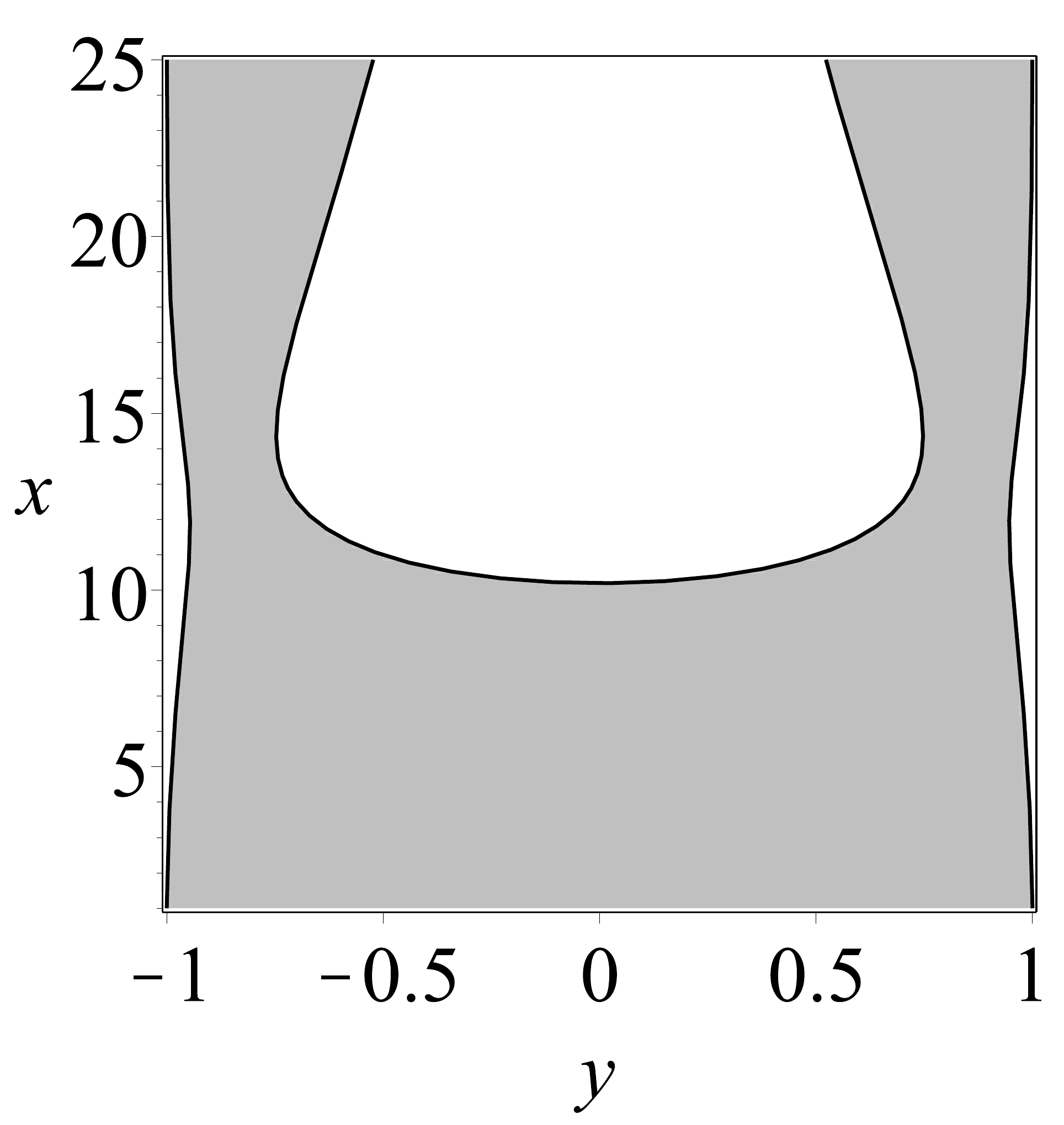} &
            \includegraphics[width=0.2\textwidth]{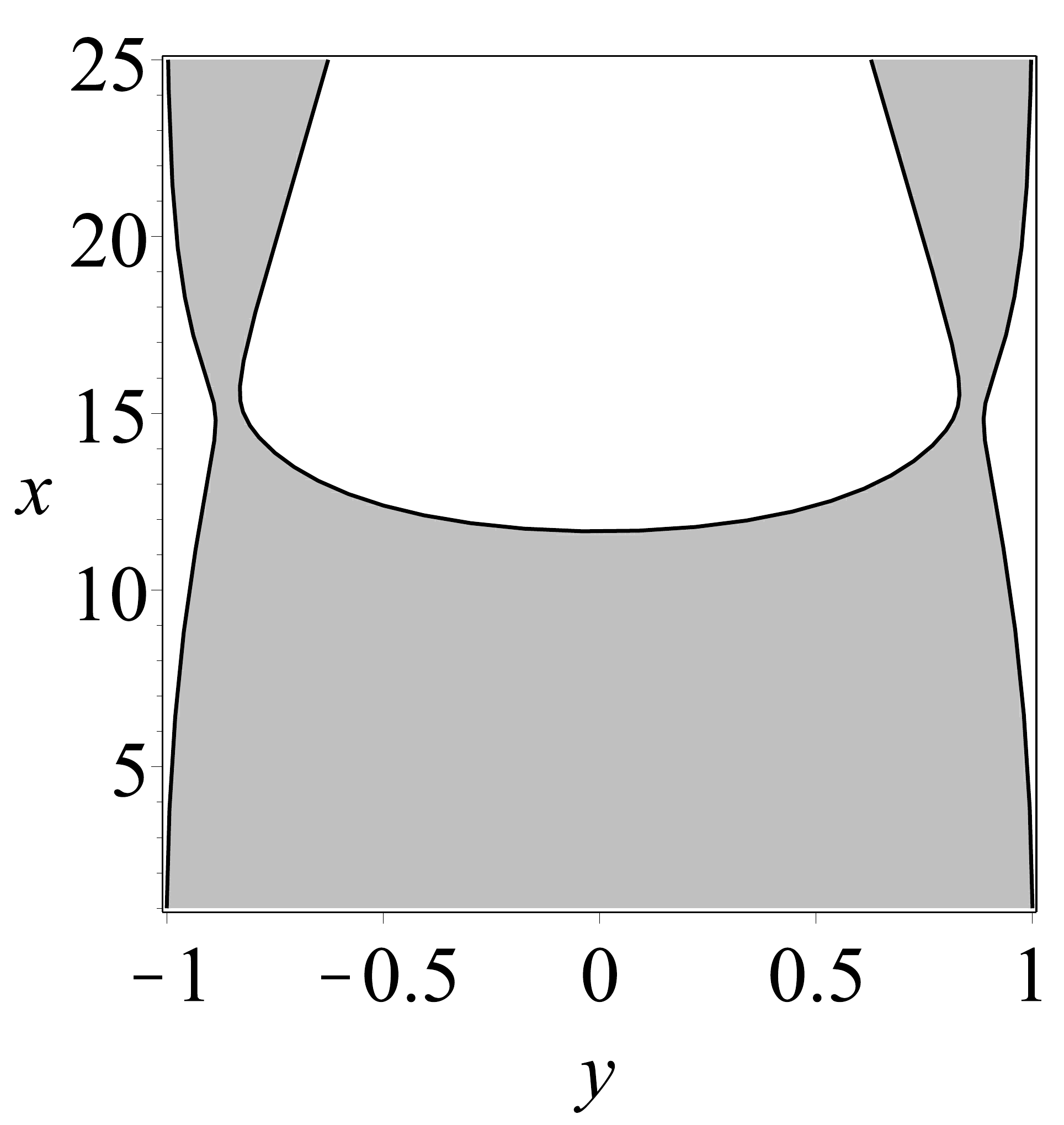} &
            \includegraphics[width=0.2\textwidth]{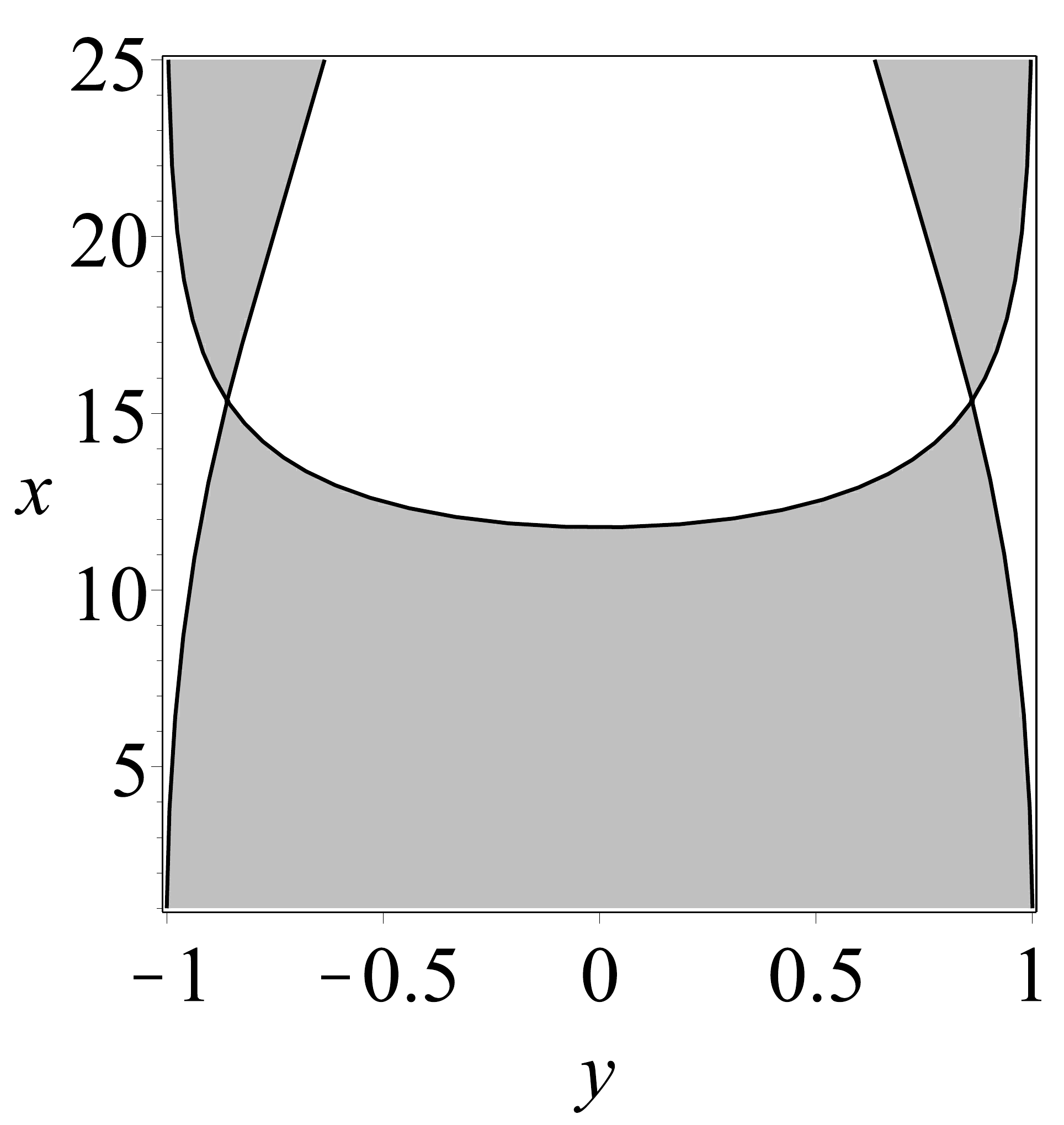} &
            \includegraphics[width=0.2\textwidth]{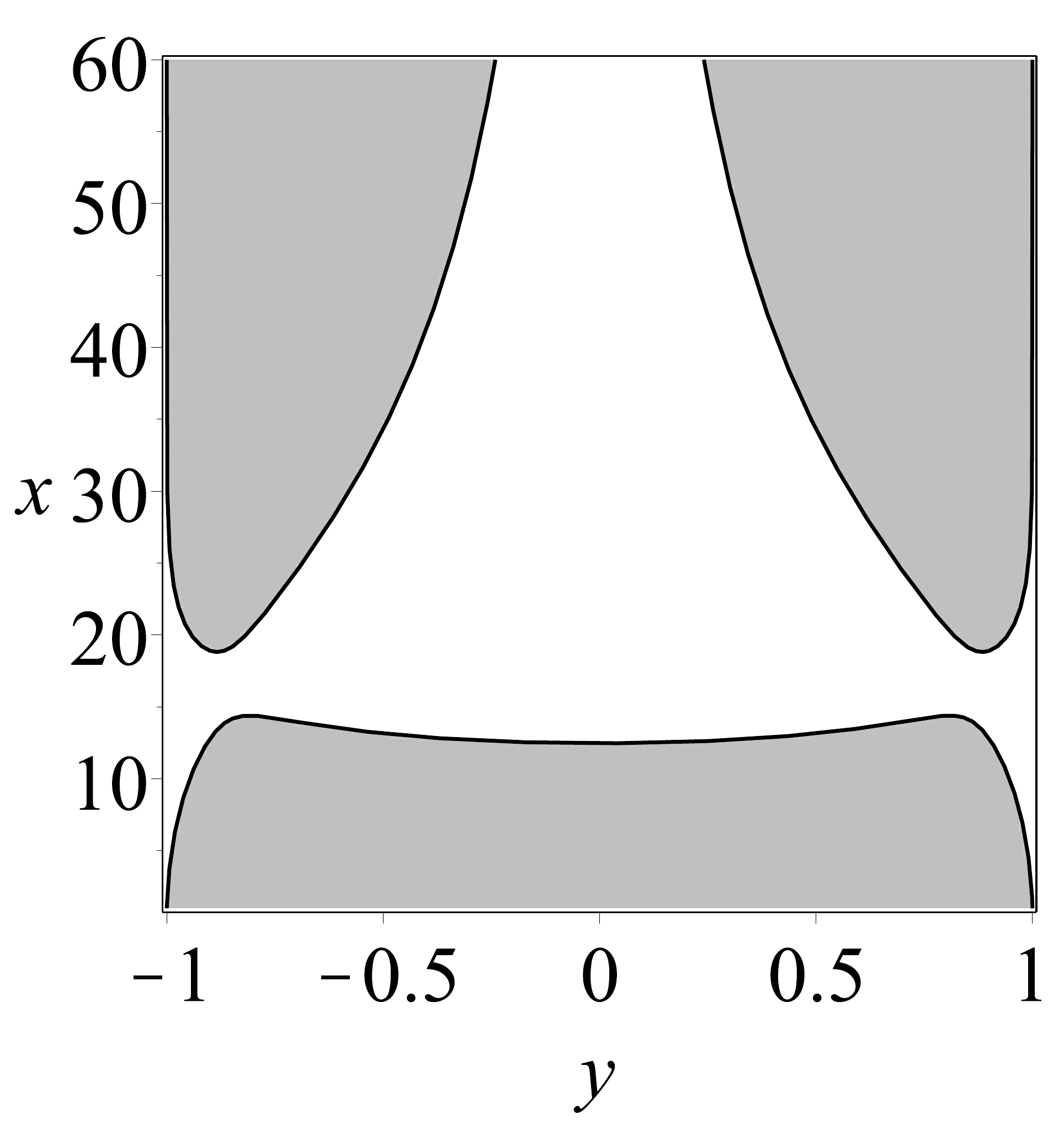}&
            \includegraphics[width=0.2\textwidth]{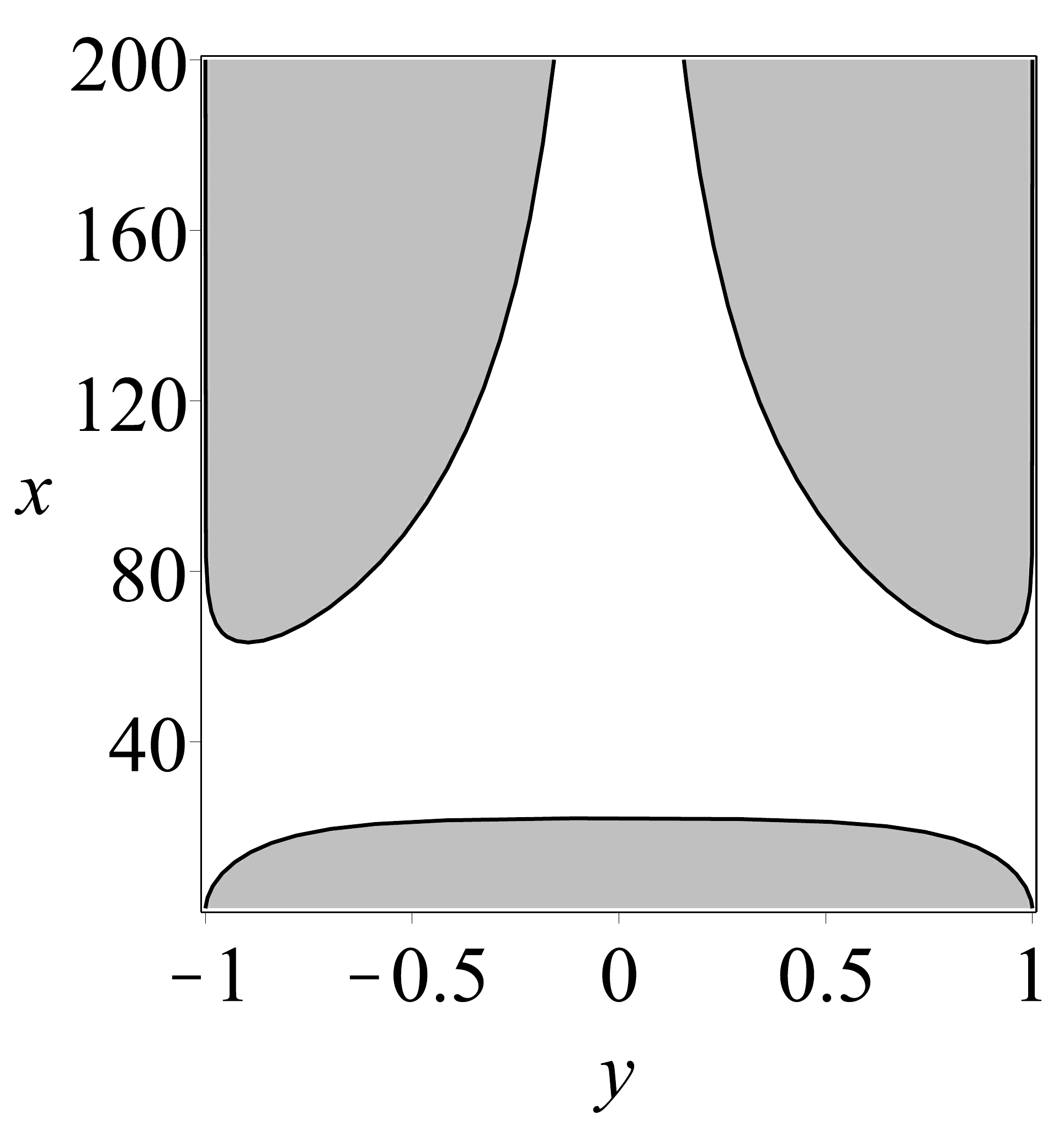}\\
            $c_2=-1/300$ & $c_2=-1/420$ & $c_2=-1/430$ & $c_2=-1/500$
            & $c_2=-1/3000$
                       \end{tabular}}
           \caption{\footnotesize{Behaviour of the ergoregion (grey area) as a function of $c_2$ for negative distortion parameter. The rotation parameter is fixed to $\alpha=0.97$.}}
		\label{c2gen_neg}
\end{figure}

For negative values of the distortion parameter, we observe some qualitative differences in the behaviour of the ergoregion as a function of $c_2$. Its general evolution as $c_2$ varies is summarized in Fig. \ref{c2gen_neg}. For high absolute values of the distortion parameter a single connected ergoregion is present which extends to infinity. Again, two types of static regions are observed - located around the axis, or containing the cross-section $y=0$, both of which are non-compact. However, the static regions containing the cross-section $y=0$ are considerably broader than the corresponding ones for positive values of $c_2$. As the absolute value of $c_2$ decreases, the ergoregion pinches in two areas located symmetrically with respect to the cross-section $y=0$. For a certain value of $c_2=c_{crit}$ two parts of the ergoregion develop which touch the ergoregion encompassing the horizon only in two critical points in the $(x,y)$ - plane, with coordinates $(x_c, y_c)$ and $(x_c, -y_c)$. For absolute values of the distortion parameter lower than $c_{crit}$ the ergoregion consists of three disconnected parts - a compact one encompassing the horizon, and two non-compact ones. If we consider $c_2\approx c_{crit}$, we observe that the compact part of the ergoregion is deformed in the vicinity of the critical points. Thus, its shape differs from the ergoregion of the non-distorted Kerr black hole. When $|c_2|$ further decreases, the distance between the compact and non-compact parts of the ergoregion increases. At the same time the ergoregion in the vicinity of the horizon gets larger, and resembles the ergoregion of the non-distorted Kerr black hole.
\begin{figure}[htp]
\setlength{\tabcolsep}{ 0 pt }{\scriptsize\tt
		\begin{tabular}{ ccccc }
	\includegraphics[width=0.2\textwidth]{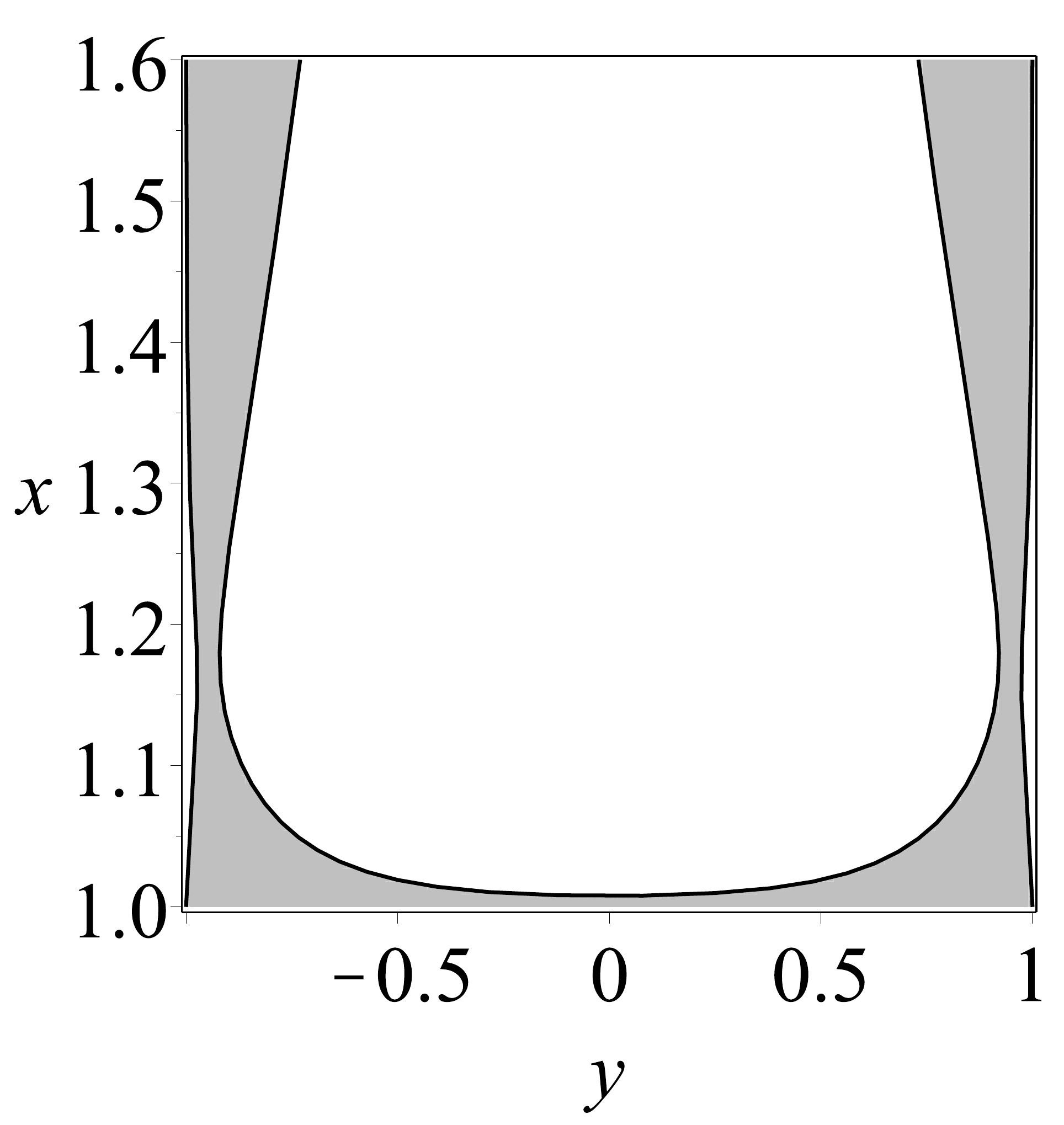} &
            \includegraphics[width=0.2\textwidth]{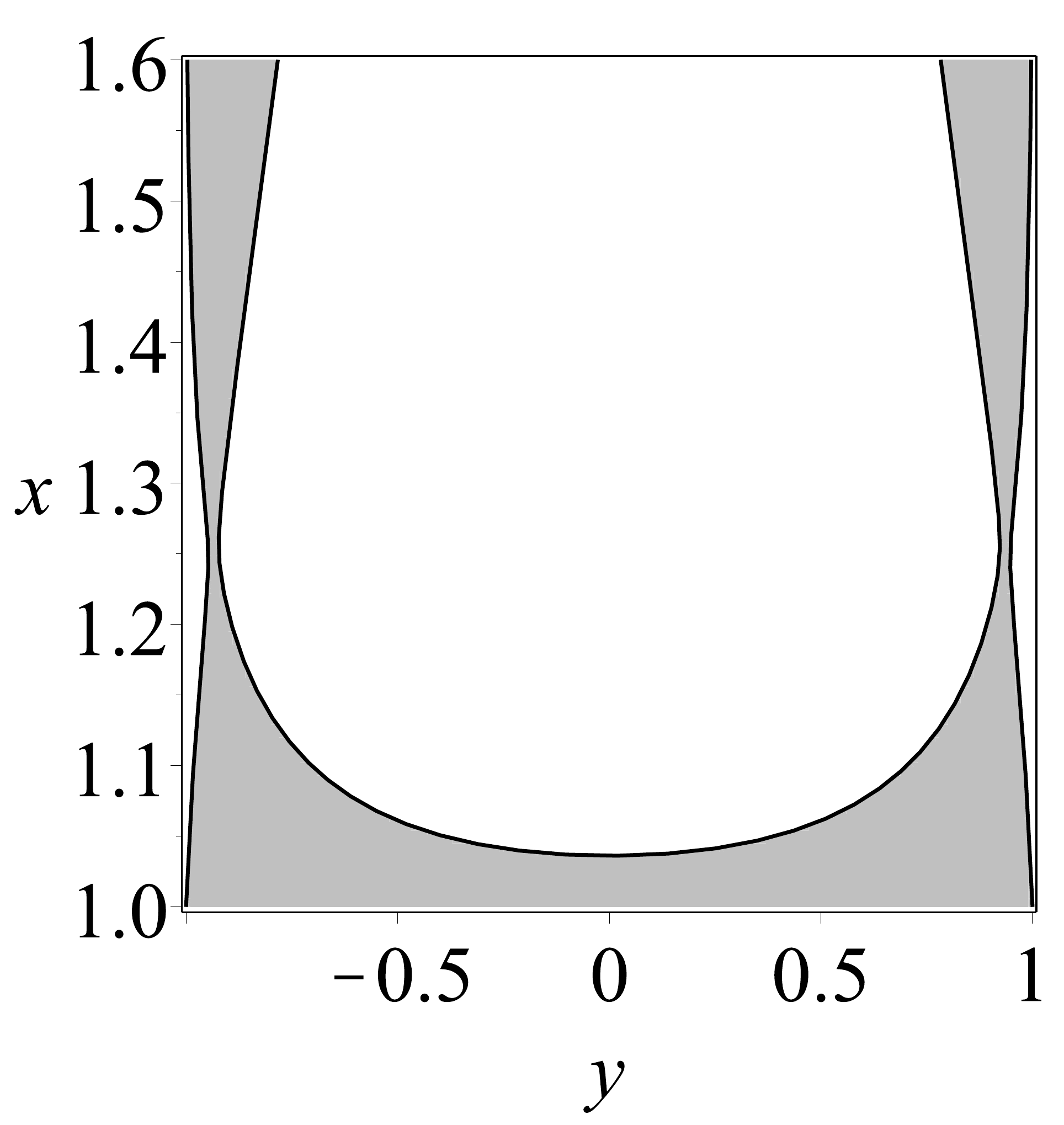} &
            \includegraphics[width=0.2\textwidth]{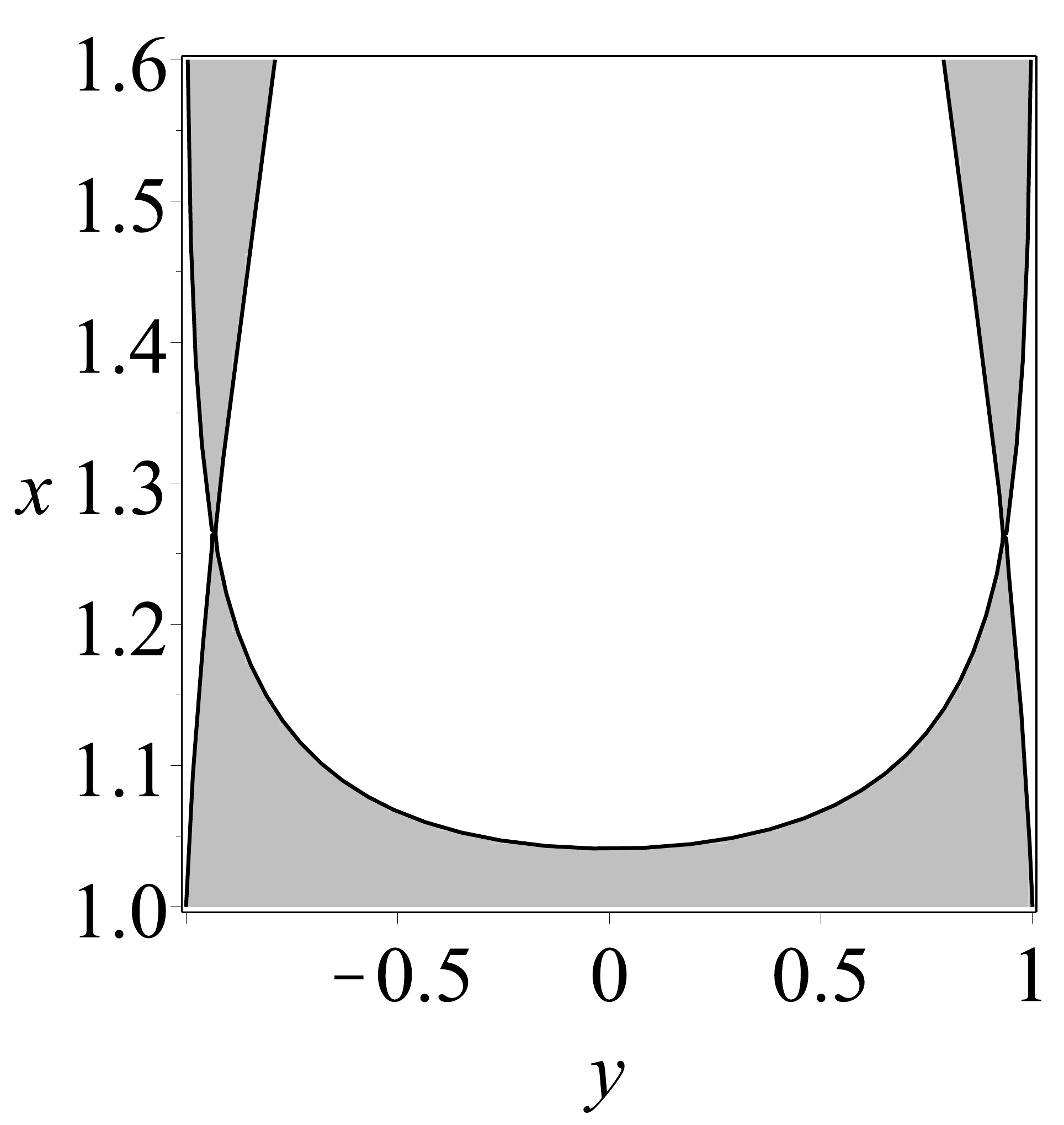} &
            \includegraphics[width=0.2\textwidth]{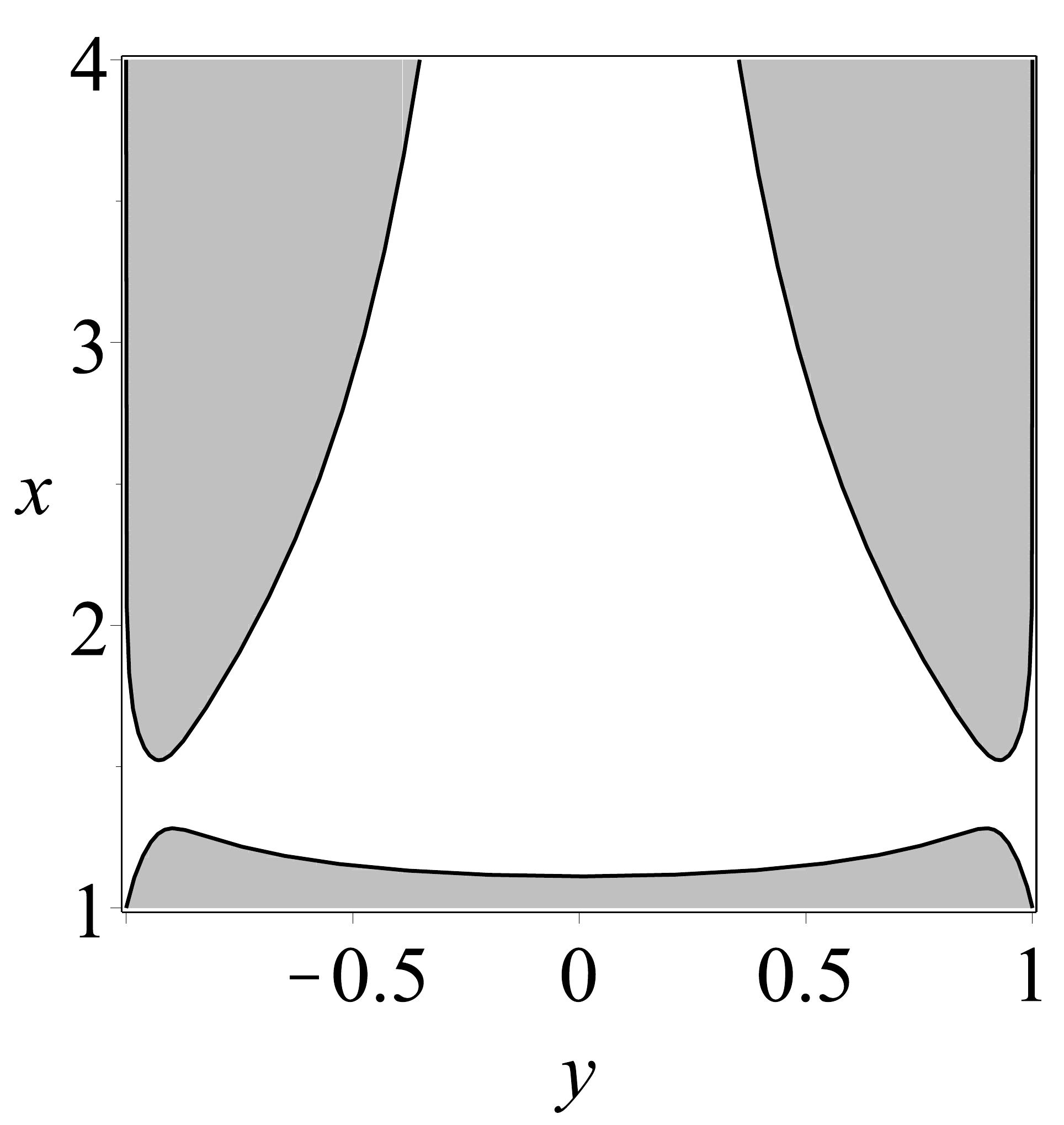}&
            \includegraphics[width=0.2\textwidth]{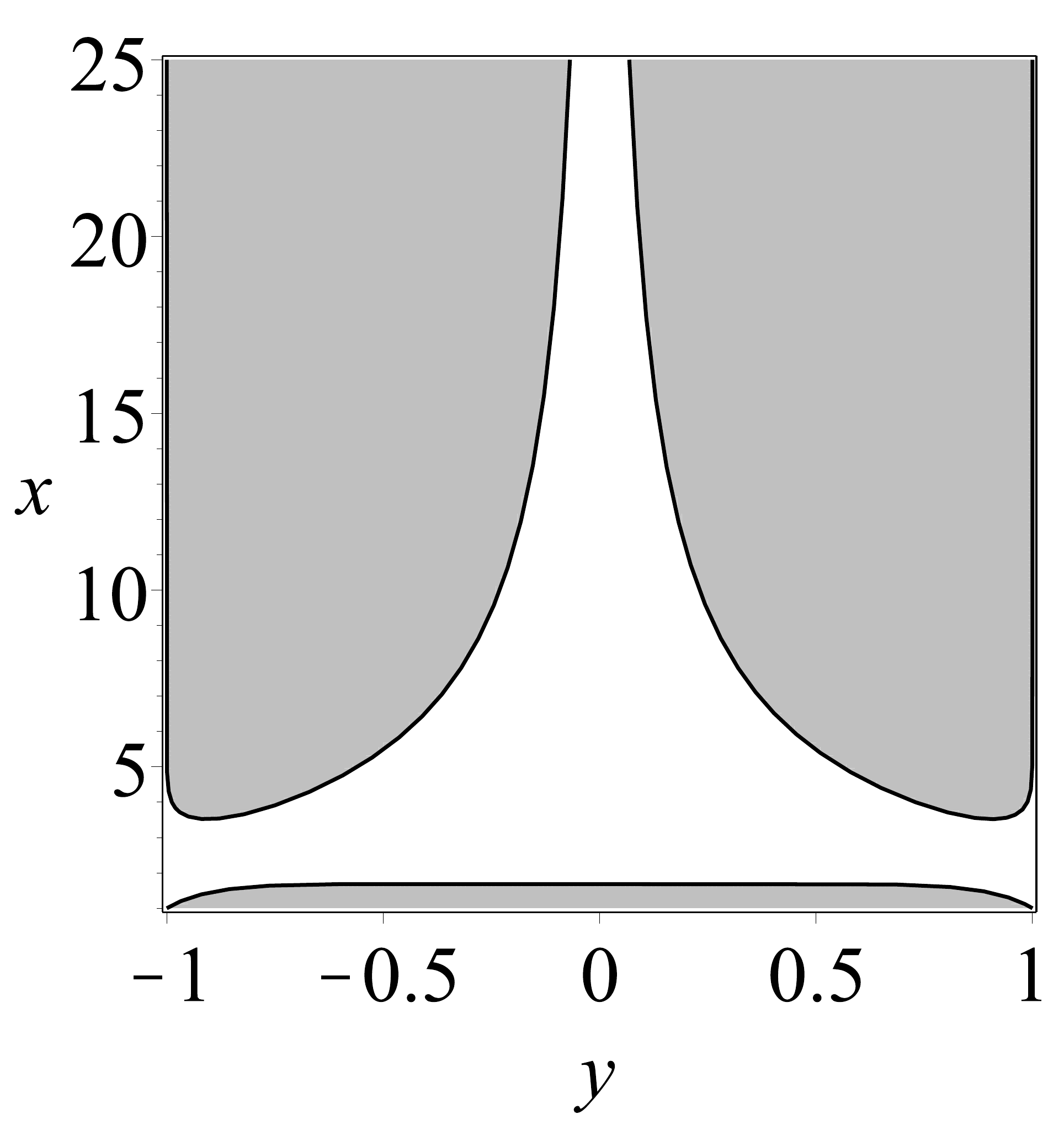}\\
            $c_2=-1.2$ & $c_2=-0.8$ & $c_2=-0.76$ & $c_2=-0.5$
            & $c_2=-0.1$
                       \end{tabular}}
           \caption{\footnotesize{ Behaviour of the ergoregion (grey area) as a function of $c_2$ for negative distortion parameter. The rotation parameter is fixed to $\alpha=0.7$.}}
		\label{c2gen_neg1}
\end{figure}

As in the case of positive distortion parameter, in Fig. \ref{c2gen_neg} we illustrate the behaviour of the ergoregion as a function of $c_2$ for the fixed value of $\alpha=0.97$. If we consider a different rotation parameter, the dependence of the ergoregion on $c_2$ is qualitatively the same. The properties which depend on the value of $\alpha$ are the location of the critical points in the $(x,y)$ - plane, the value of $c_{crit}$, the size of the compact part of the ergoregion and the distance between the compact and non-compact parts. For comparison, in Fig. \ref{c2gen_neg1} we present the behaviour of the ergoregion as a function of $c_2$ for the fixed value of $\alpha=0.7$.

\begin{figure}[htp]
\setlength{\tabcolsep}{ 0 pt }{\scriptsize\tt
		\begin{tabular}{ cccc }
	\includegraphics[width=0.25\textwidth]{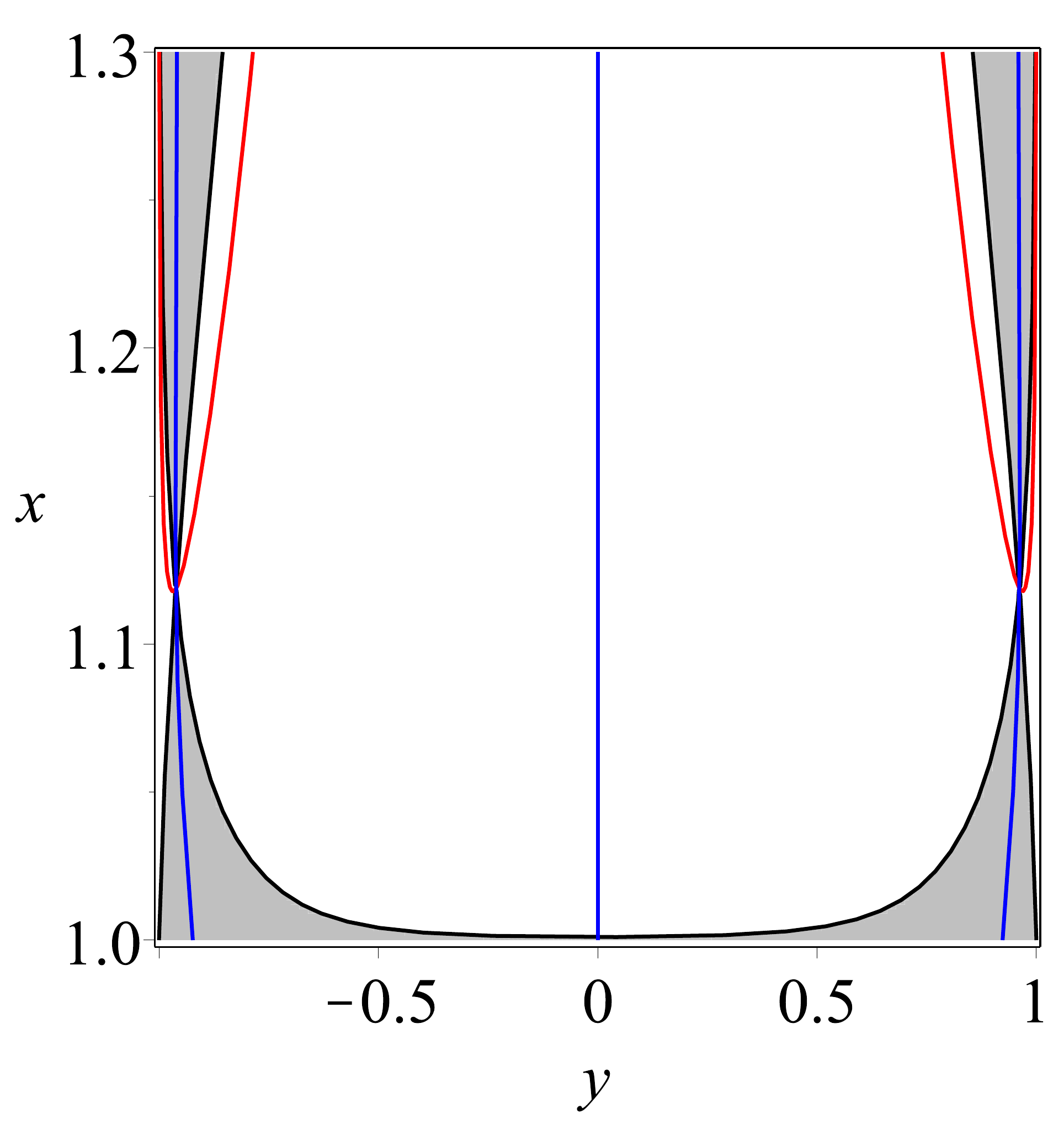} &
            \includegraphics[width=0.25\textwidth]{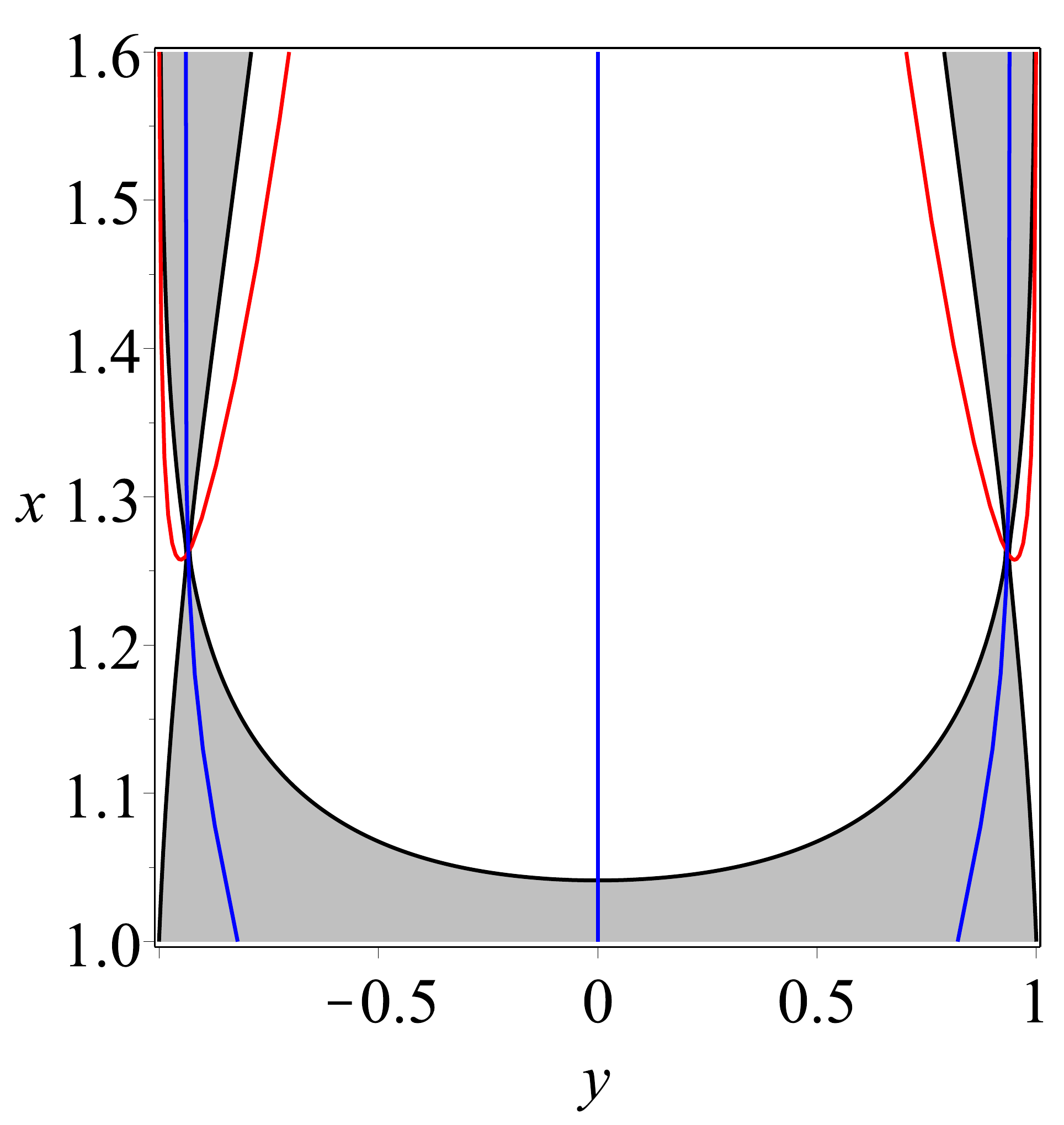} &
            \includegraphics[width=0.25\textwidth]{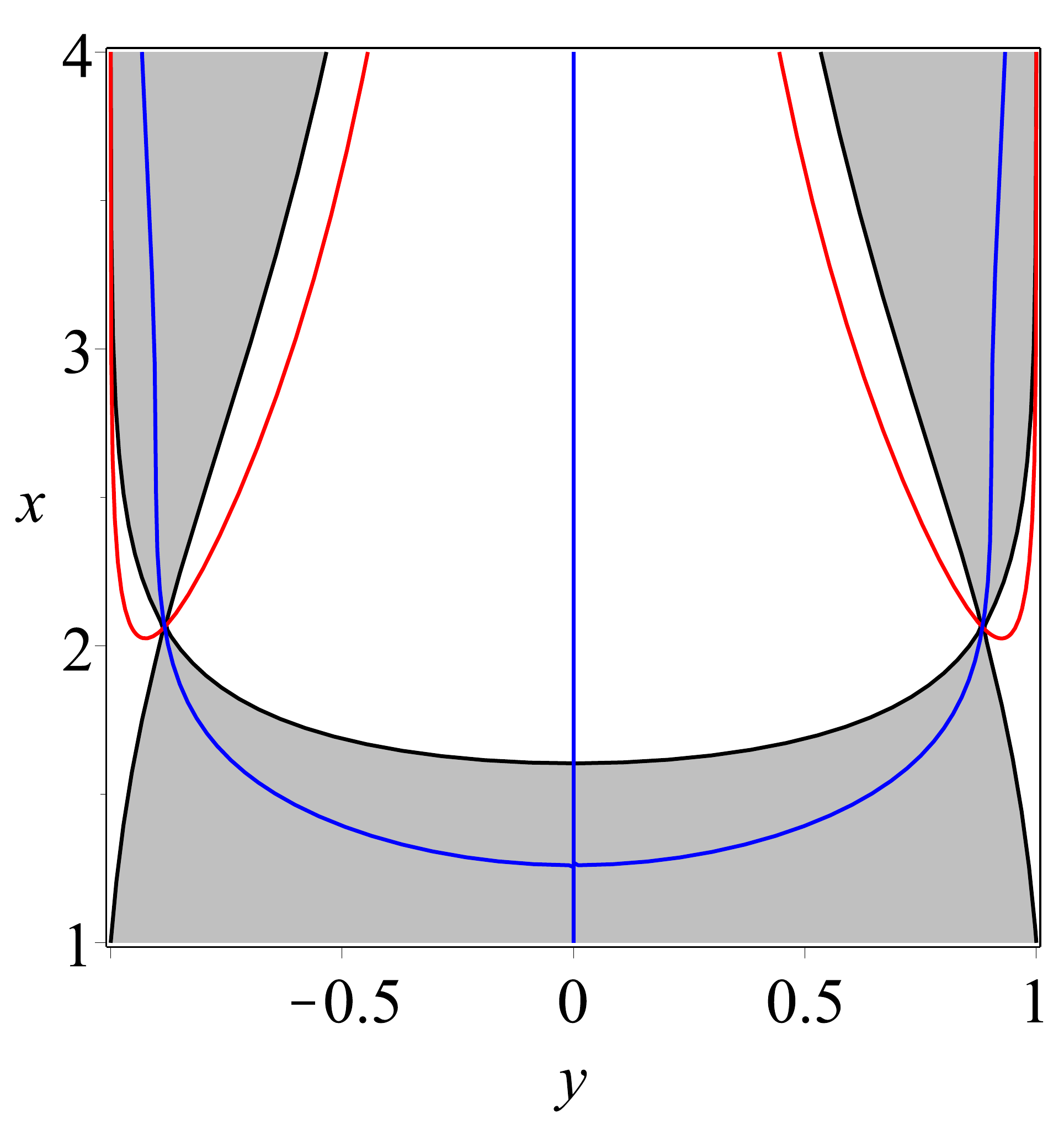} &
            \includegraphics[width=0.25\textwidth]{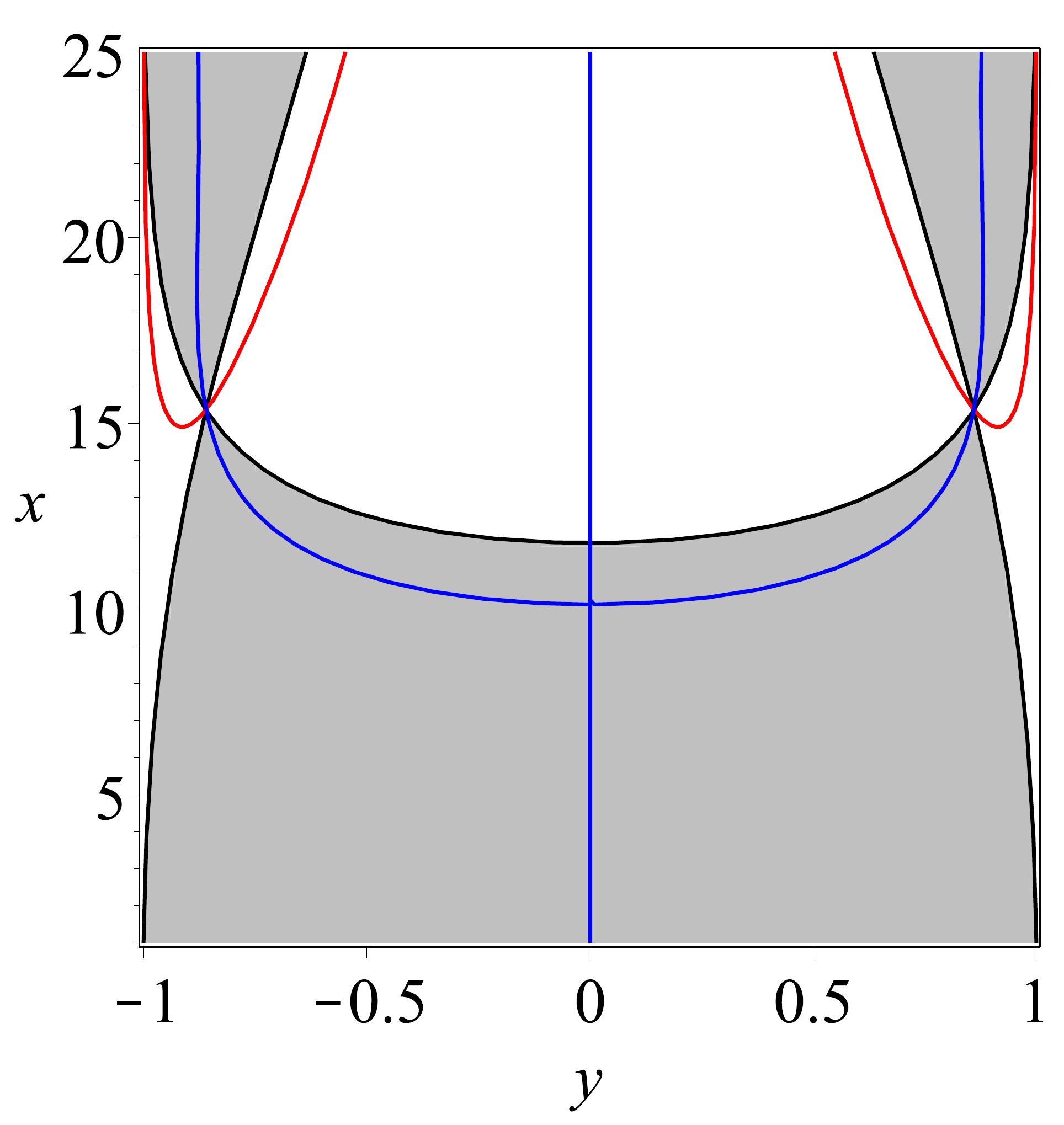}\\
            $\alpha=0.67$ & $\alpha=0.7$ & $\alpha=0.8$ & $\alpha=0.97$ \\
            $c_{crit}\approx -1.687$ & $c_{crit}\approx -0.7655$ & $c_{crit}\approx -0.1581$ & $c_{crit}\approx -0.23218\times10^{-2}$
                                   \end{tabular}}
           \caption{\footnotesize{Dependence of the critical points location on the rotation parameter $\alpha$ for negative $c_2$. The red and blue lines correspond to the curves $\partial_x A(x,y)=0$ and $\partial_y A(x,y)=0$, respectively, for each couple of parameters $c_{crit}$ and $\alpha$.}}
		\label{c2pin_neg}
\end{figure}

In Fig. \ref{c2pin_neg}, we illustrate the dependence of the critical points location in the $(x,y)$-plane on $\alpha$. We also specify for each rotation parameter the value of $c_{crit}$, at which the transition between the connected and disconnected type of the ergoregion occurs. We observe that the critical points are always located symmetrically with respect to the $y=0$ cross-section, i.e. they have coordinates $(x_c,y_c)$ and $(x_c,-y_c)$. The value $x_c$ of the $x$-coordinate increases as $\alpha$ increases. On the other hand, the absolute value of $c_{crit}$ decreases when the rotation parameter grows. For each set of parameters $\alpha$ and $c_{crit}$ we have presented the curves defined by the equations $\partial_xA(x,y;c_2, \alpha)=0$ and $\partial_y A(x,y;c_2, \alpha)=0$  with red and blue lines, respectively. As we argued in the analysis for positive distortion parameter $c_2$, the points in which they simultaneously intersect the ergosurface $A(x,y;c_2, \alpha)=0$ correspond to the critical points.

\subsection{Octupole distortion}
\begin{figure}[htp]
\setlength{\tabcolsep}{ 0 pt }{\scriptsize\tt
		\begin{tabular}{ ccc }
	\includegraphics[width=5 cm]{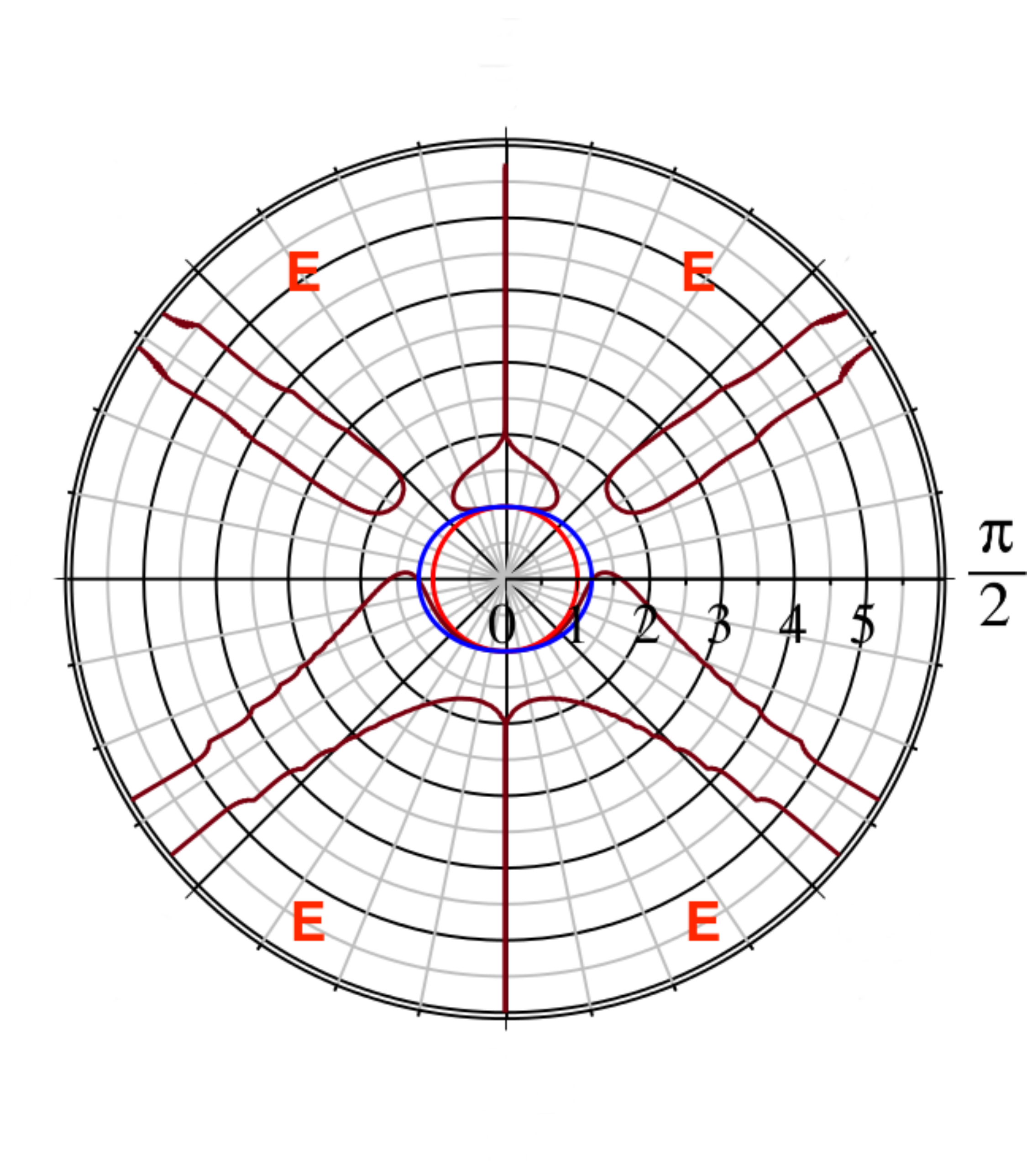} &
            \includegraphics[width=5 cm]{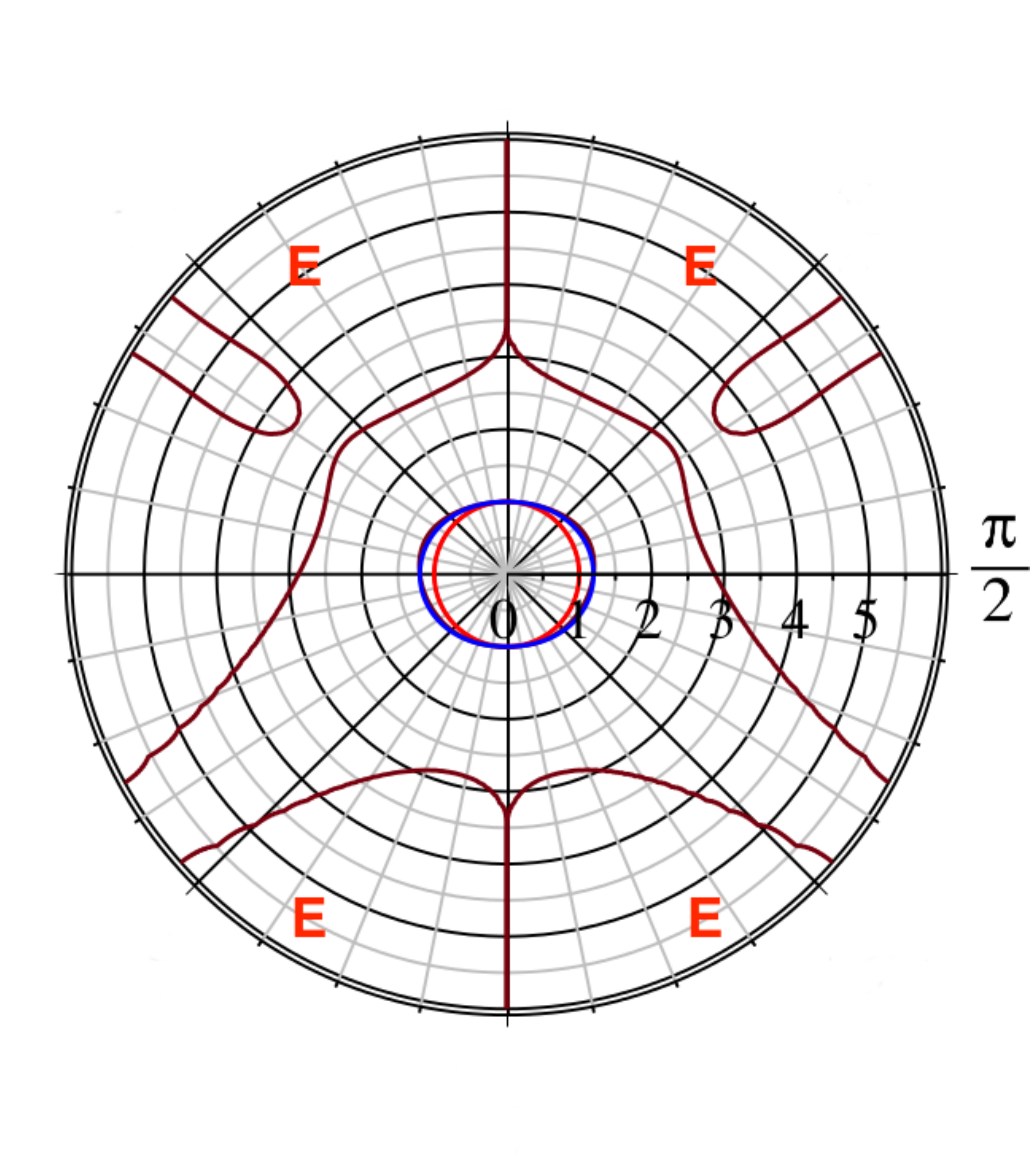} &
            \includegraphics[width=5 cm]{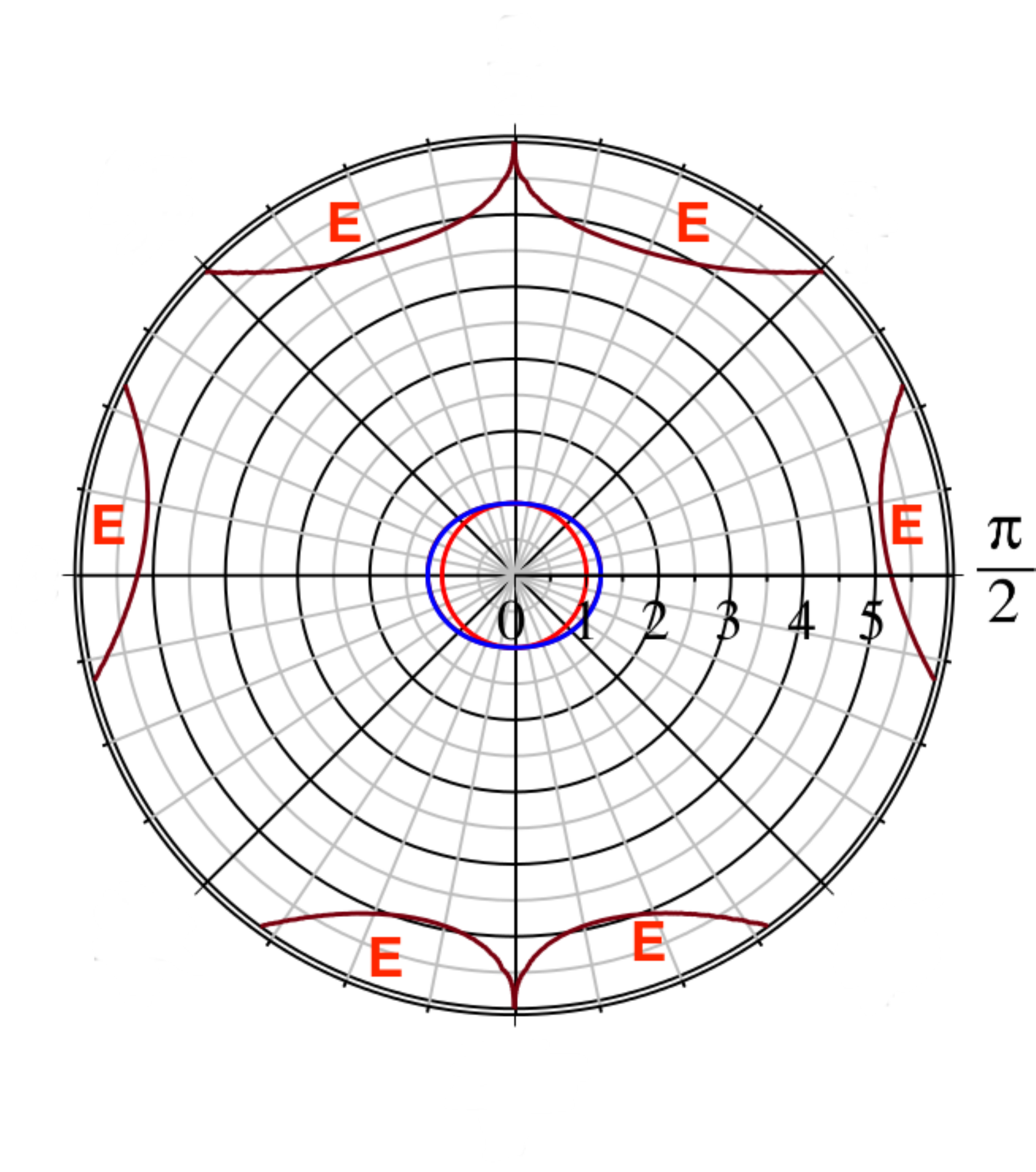} \\
            $c_1=-1/2$ & $c_1=-1/10$ & $c_1=-1/50$    \\
            \includegraphics[width=5 cm]{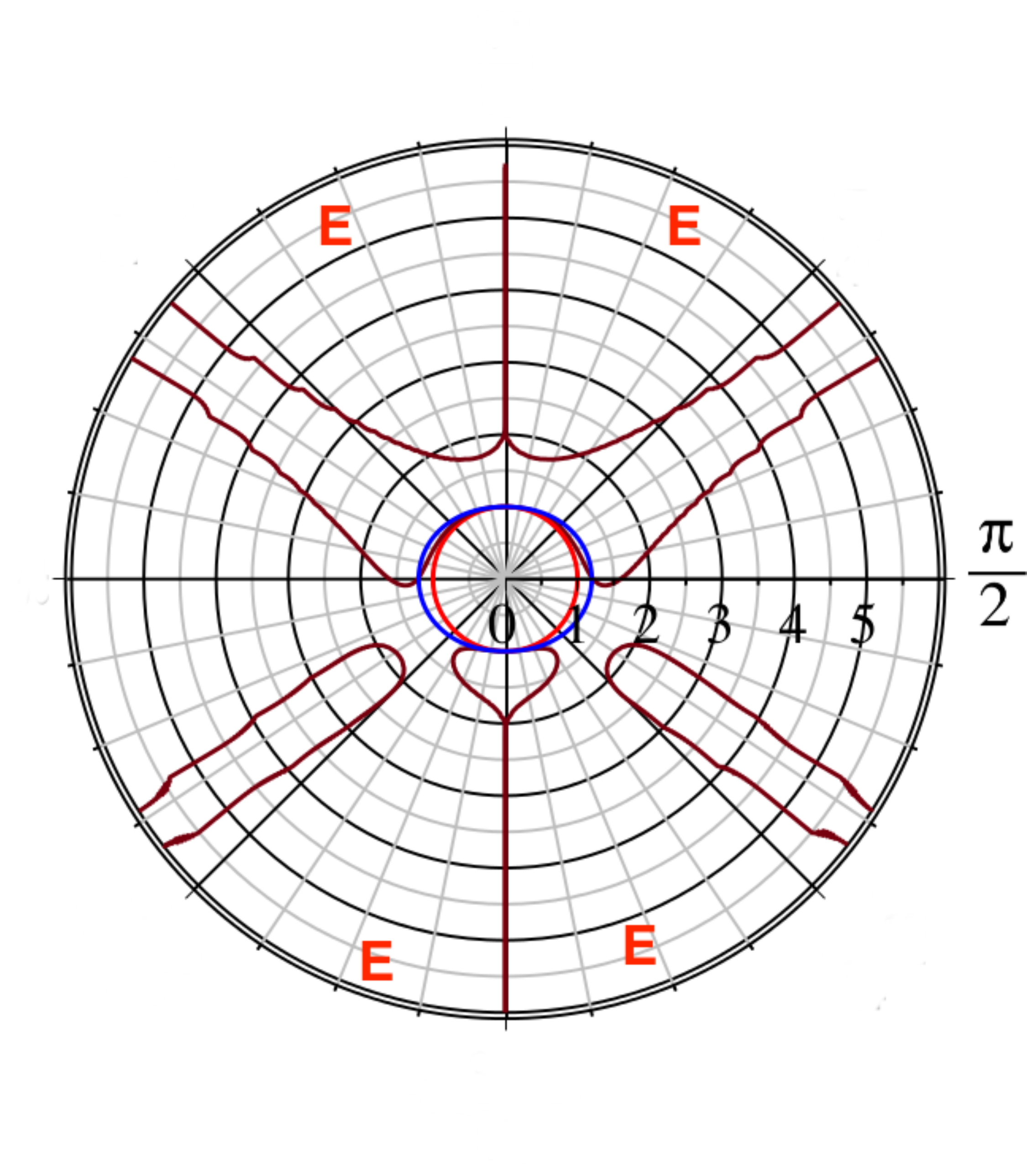} &
            \includegraphics[width=5 cm]{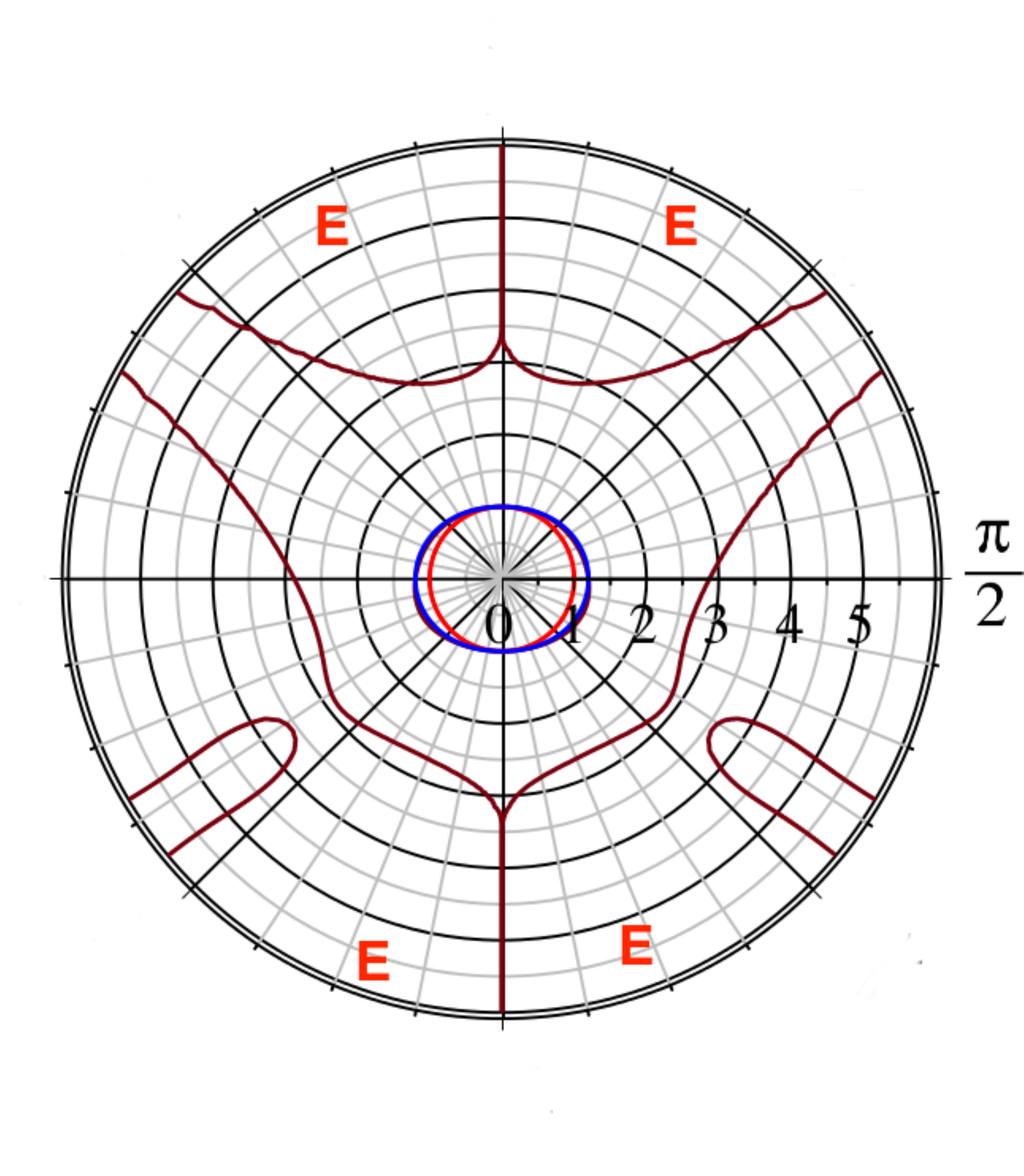} &
            \includegraphics[width=5 cm]{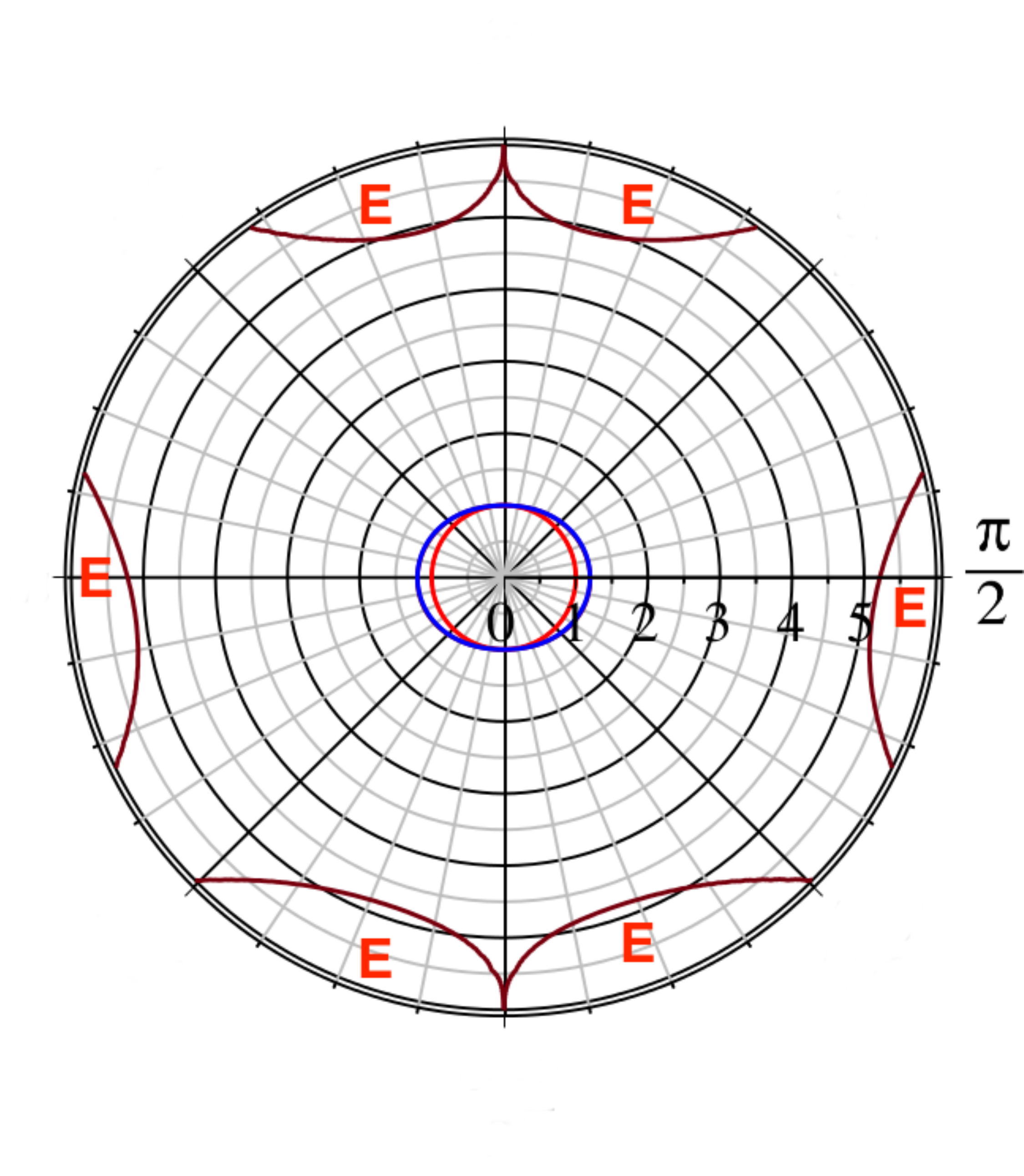} \\
            $c_1=1/2$ & $c_1=1/10$ & $c_1=1/50$
           \end{tabular}}
           \caption{\footnotesize{Ergoregion for octupole distortion and $\alpha=0.3$}}
		\label{odd03}
\end{figure}

\begin{figure}[htp]
\setlength{\tabcolsep}{ 0 pt }{\scriptsize\tt
		\begin{tabular}{ ccc }
	\includegraphics[width=5 cm]{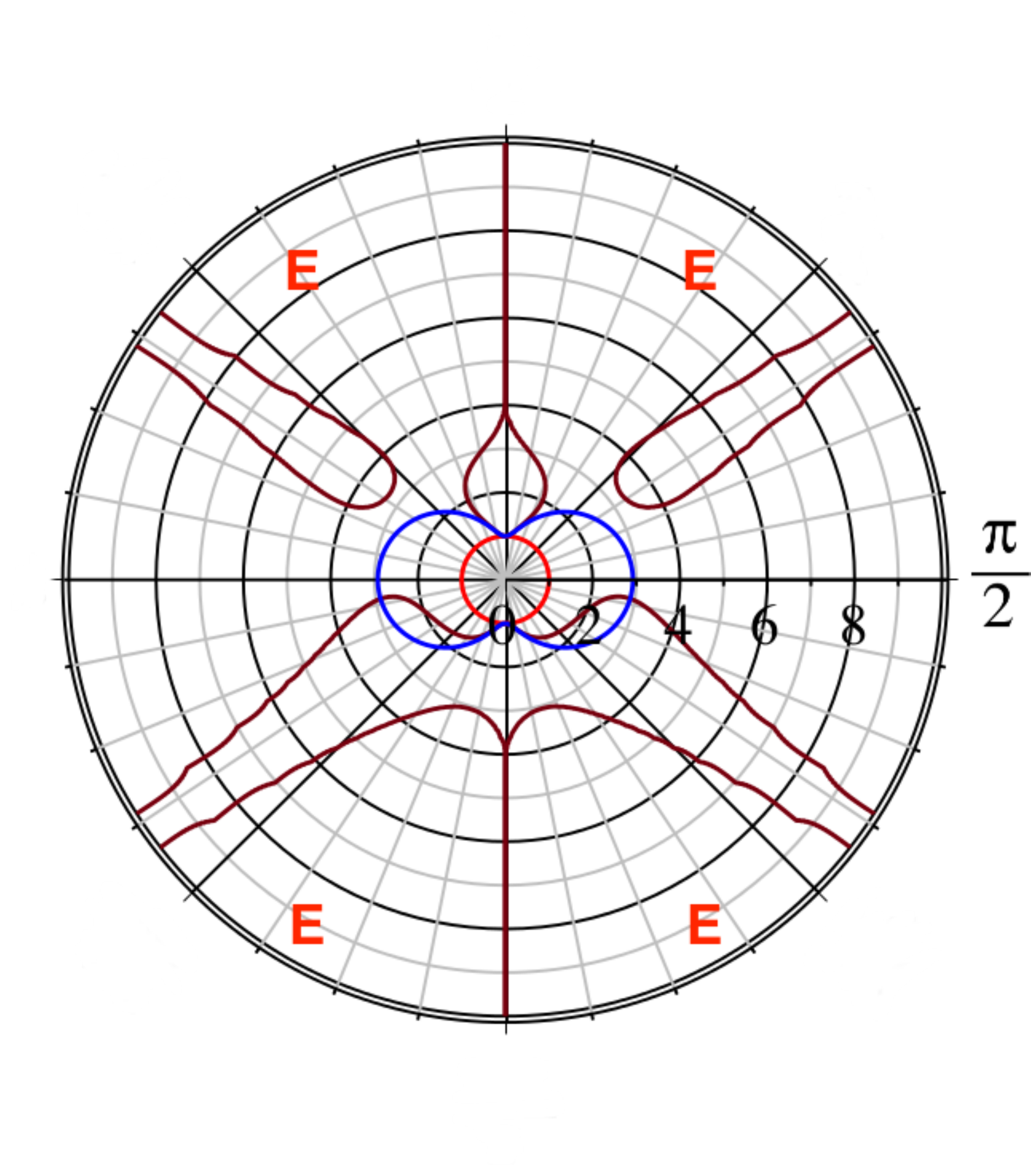} &
            \includegraphics[width=5 cm]{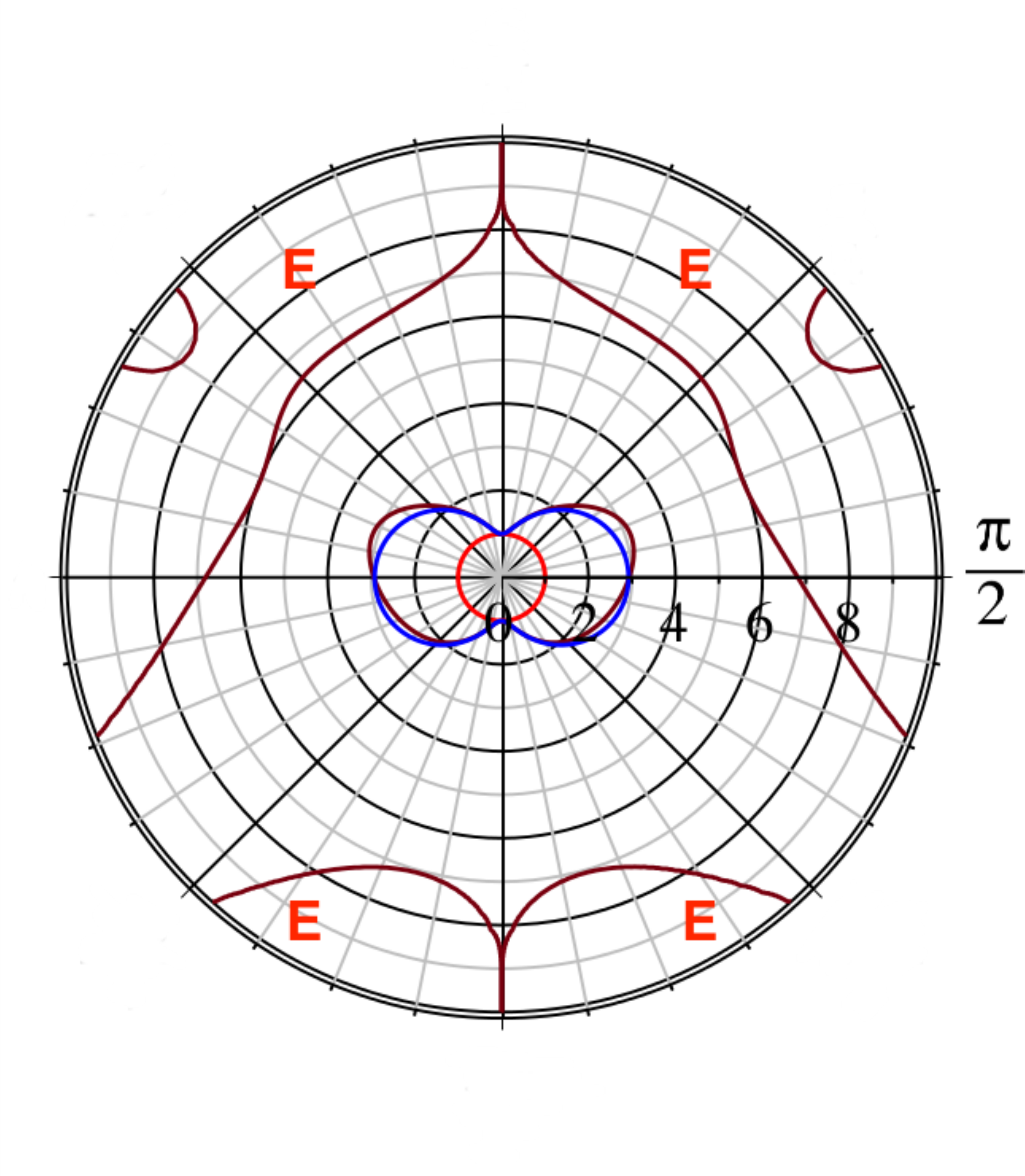} &
            \includegraphics[width=5 cm]{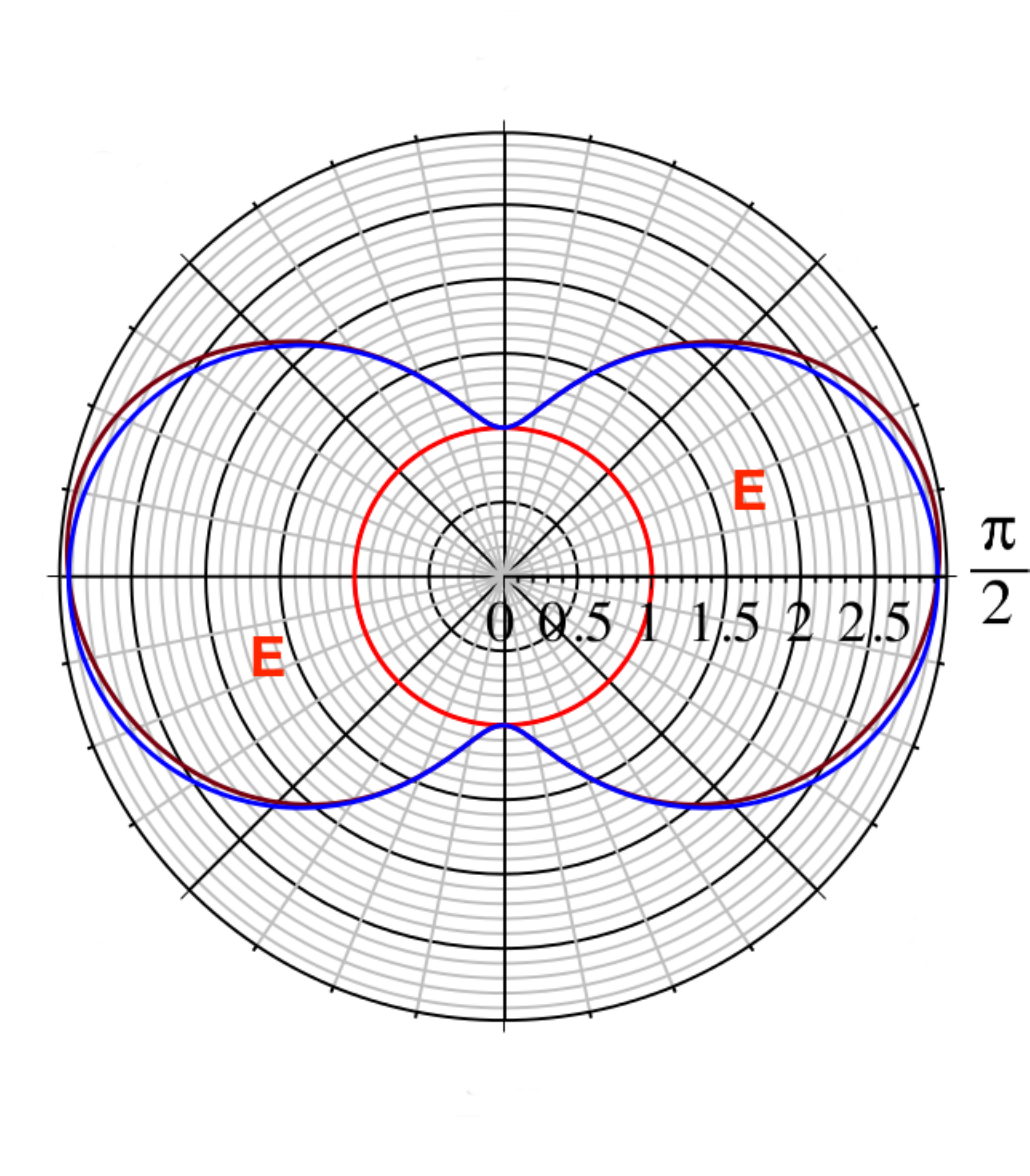} \\
            $c_1=-1/20$ & $c_1=-1/200$ & $c_1=-1/600$    \\
            \includegraphics[width=5 cm]{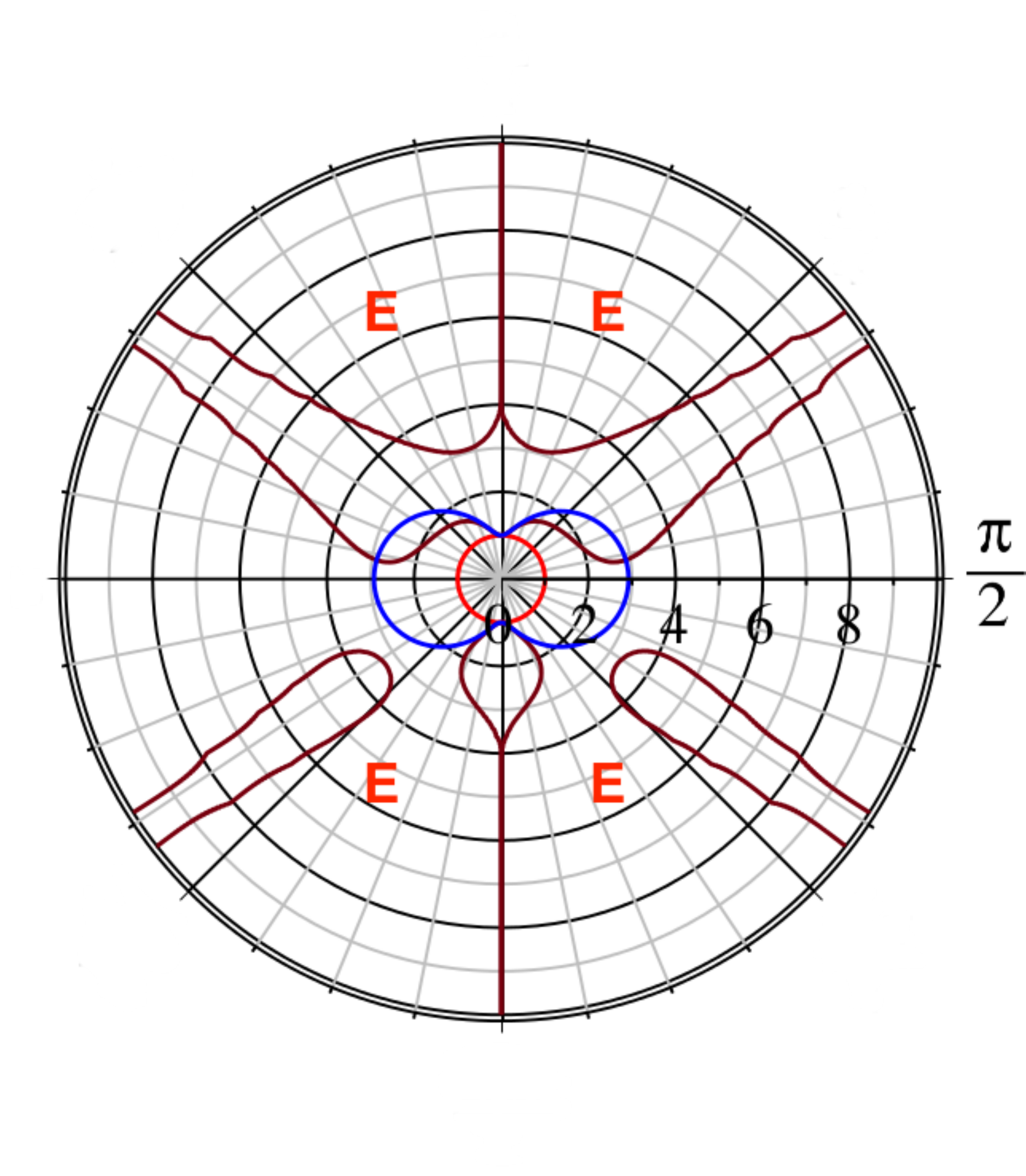} &
            \includegraphics[width=5 cm]{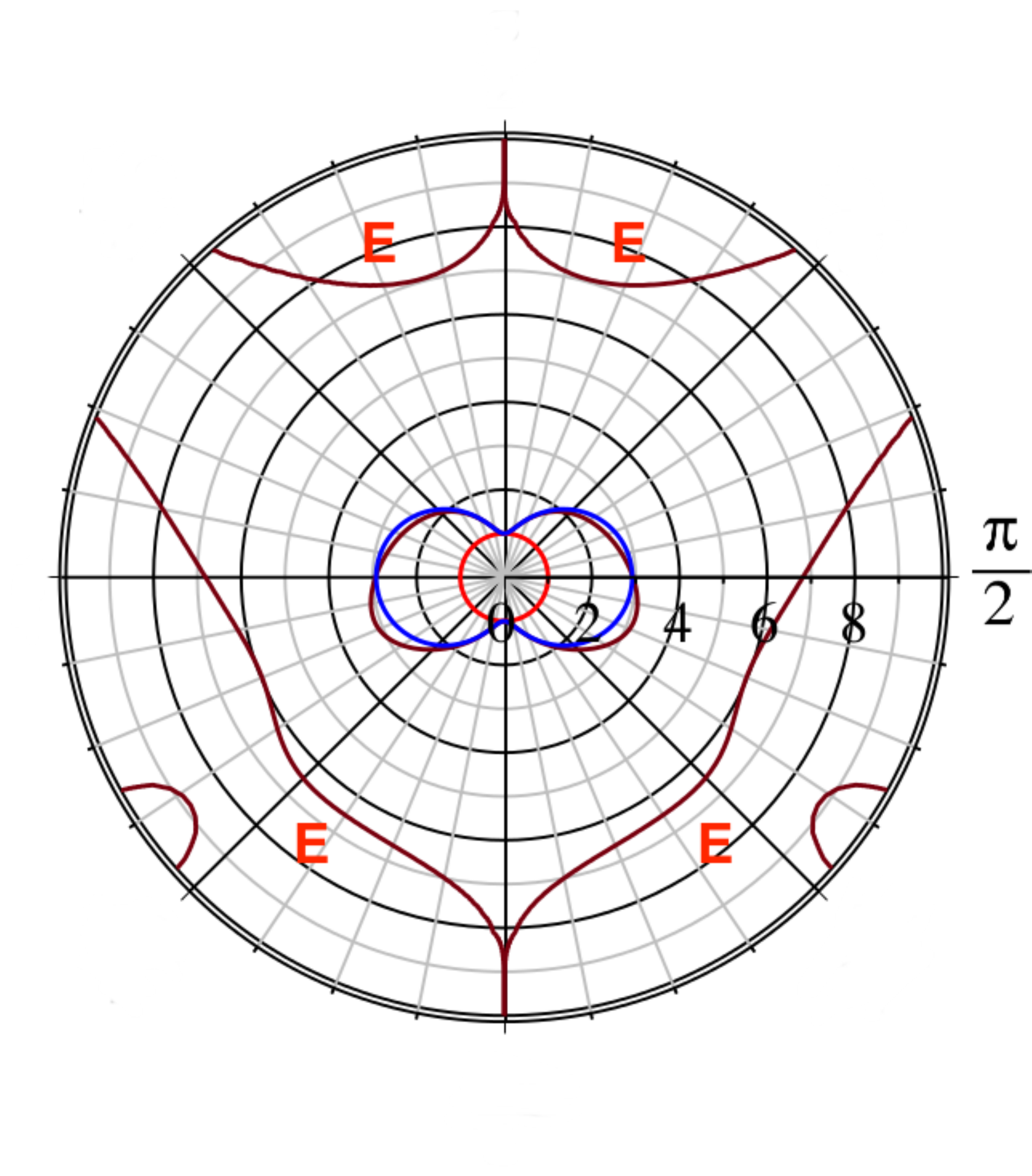} &
            \includegraphics[width=5 cm]{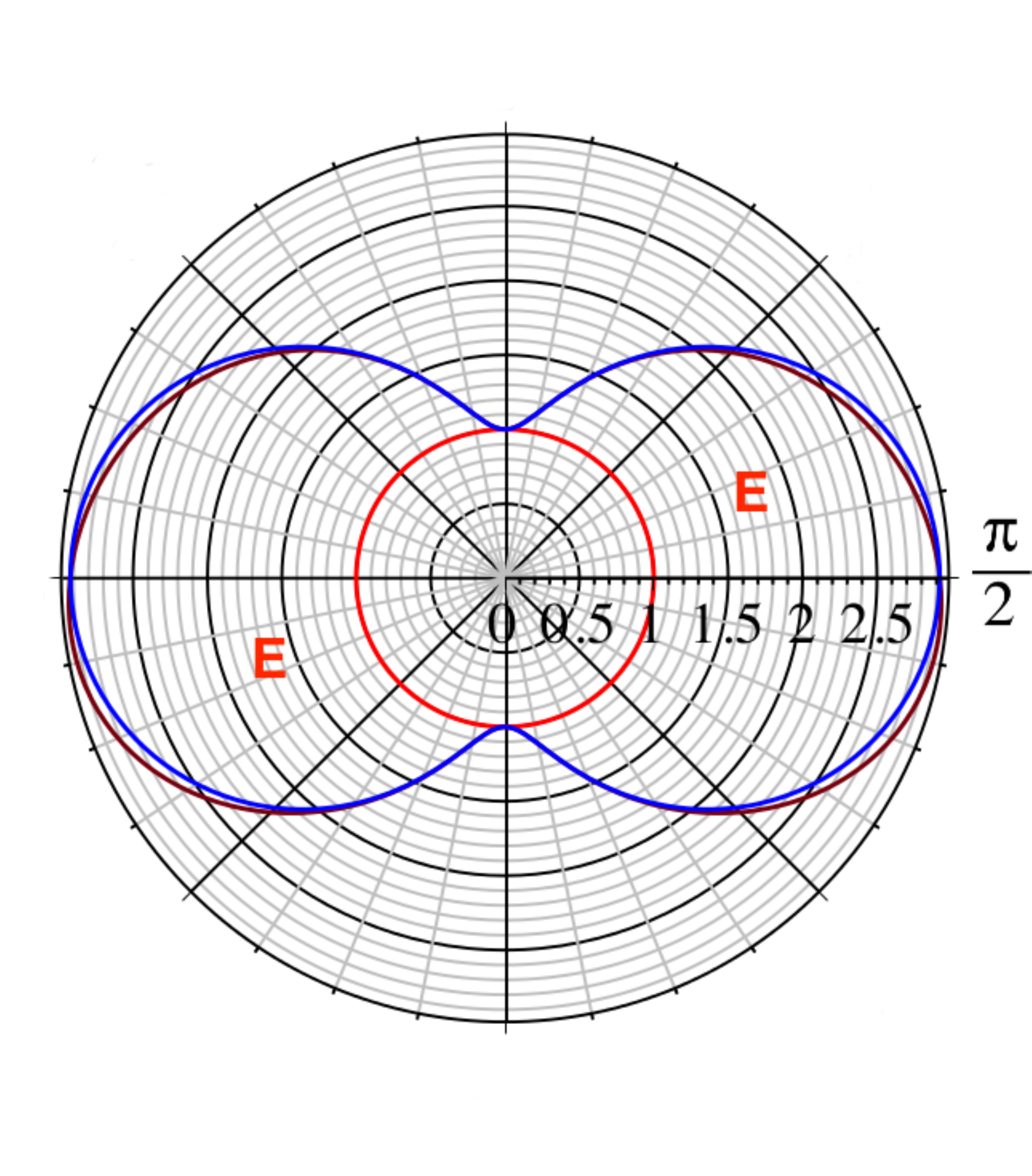} \\
            $c_1=1/20$ & $c_1=1/200$ & $c_1=1/600$
           \end{tabular}}
           \caption{\footnotesize{Ergoregion for octupole distortion and $\alpha=0.7$}}
		\label{odd07}
\end{figure}

\begin{figure}[htp]
\setlength{\tabcolsep}{ 0 pt }{\scriptsize\tt
		\begin{tabular}{ ccc }
	\includegraphics[width=5 cm]{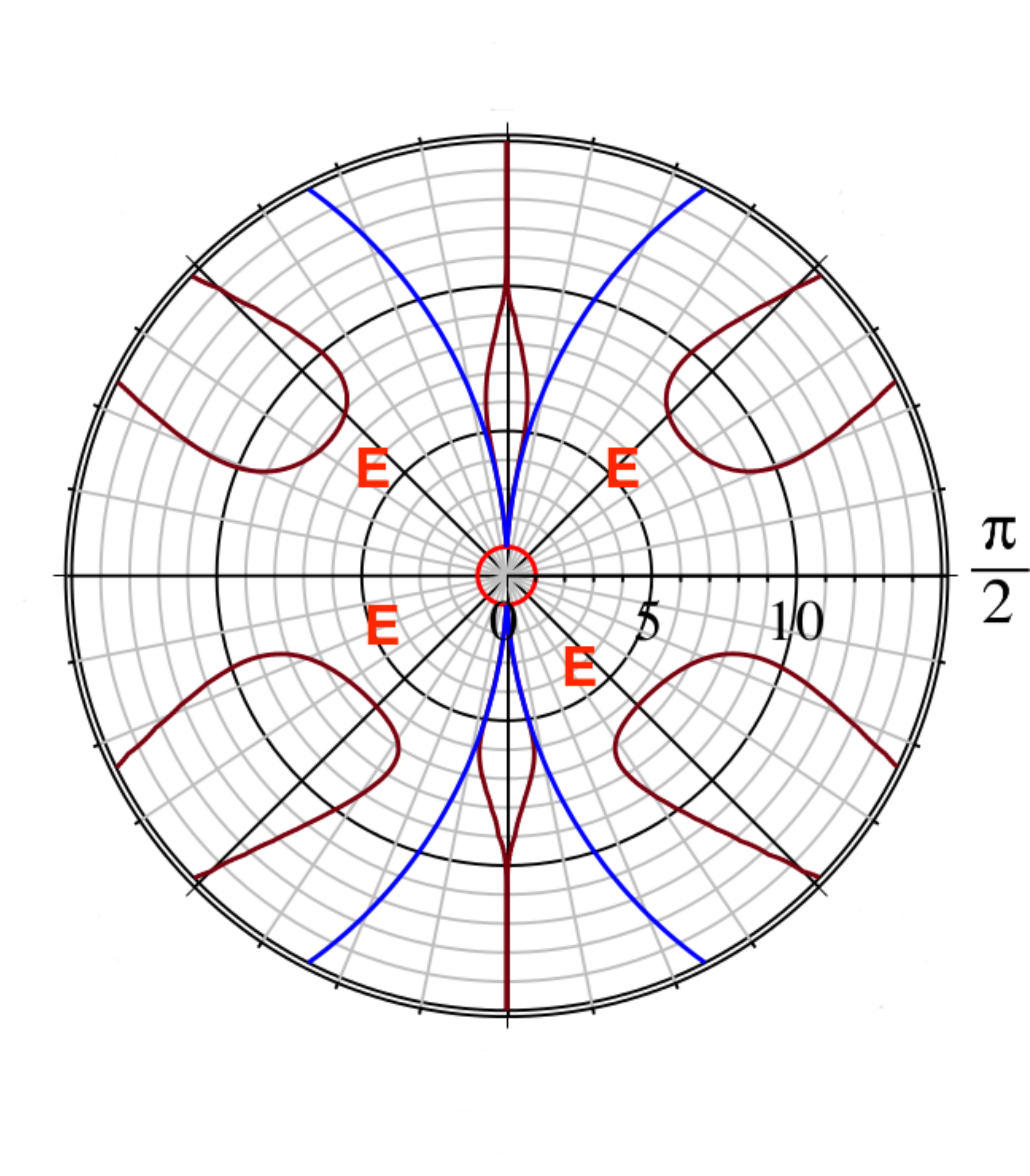} &
            \includegraphics[width=5 cm]{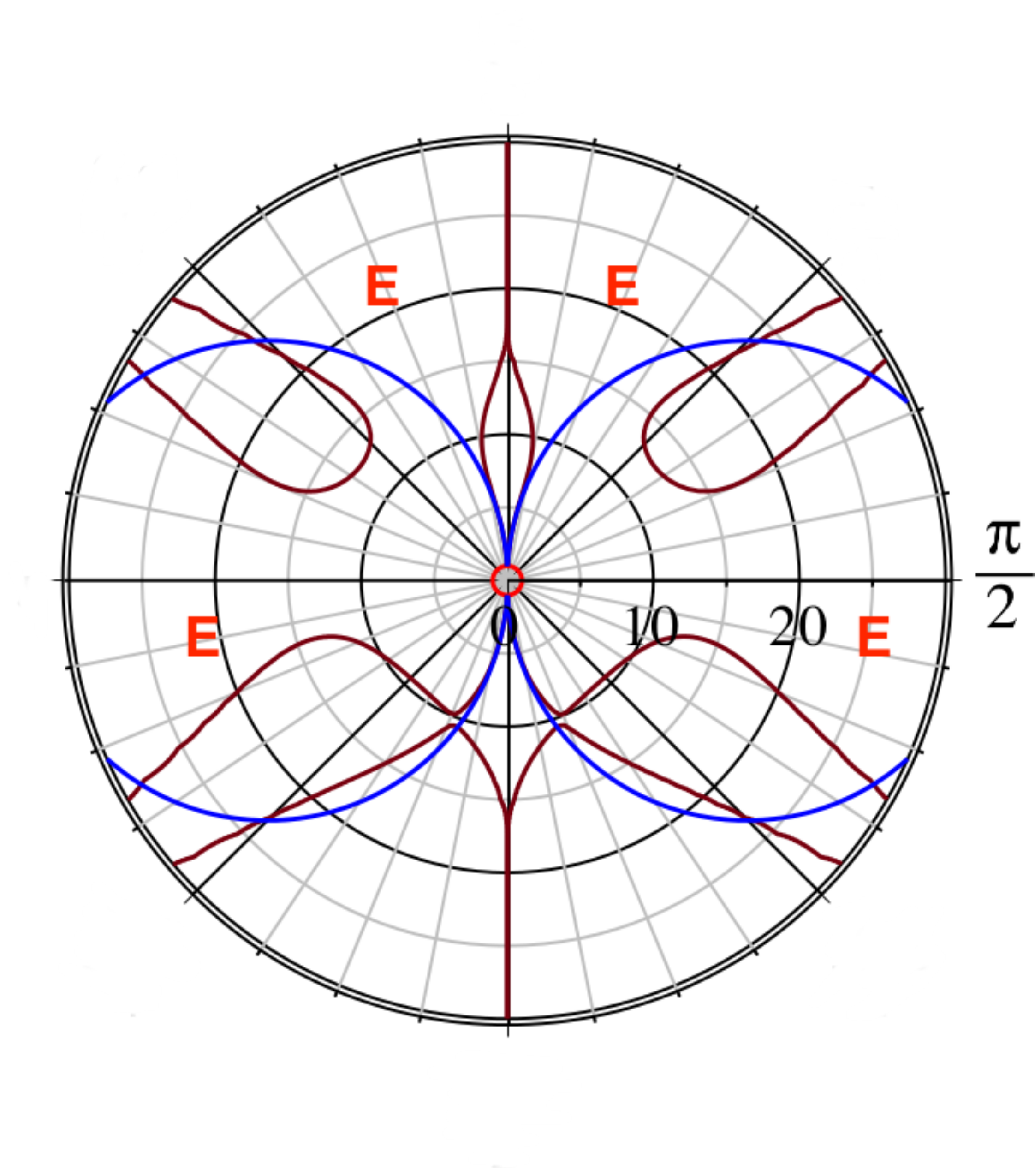} &
            \includegraphics[width=5 cm]{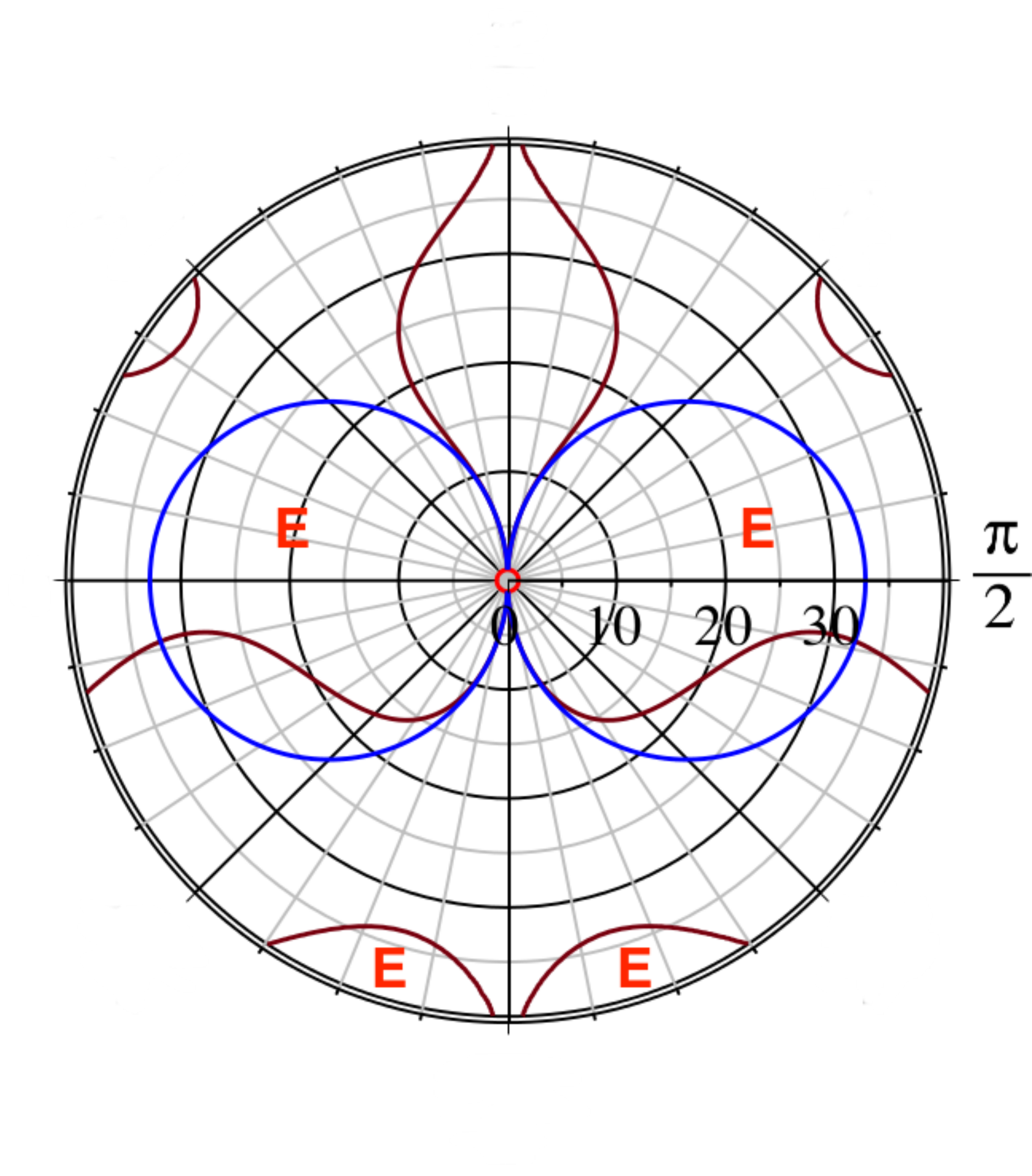} \\
            $c_1=-1/500$ & $c_1=-1/2000$ & $c_1=-1/30000$    \\
            \includegraphics[width=5 cm]{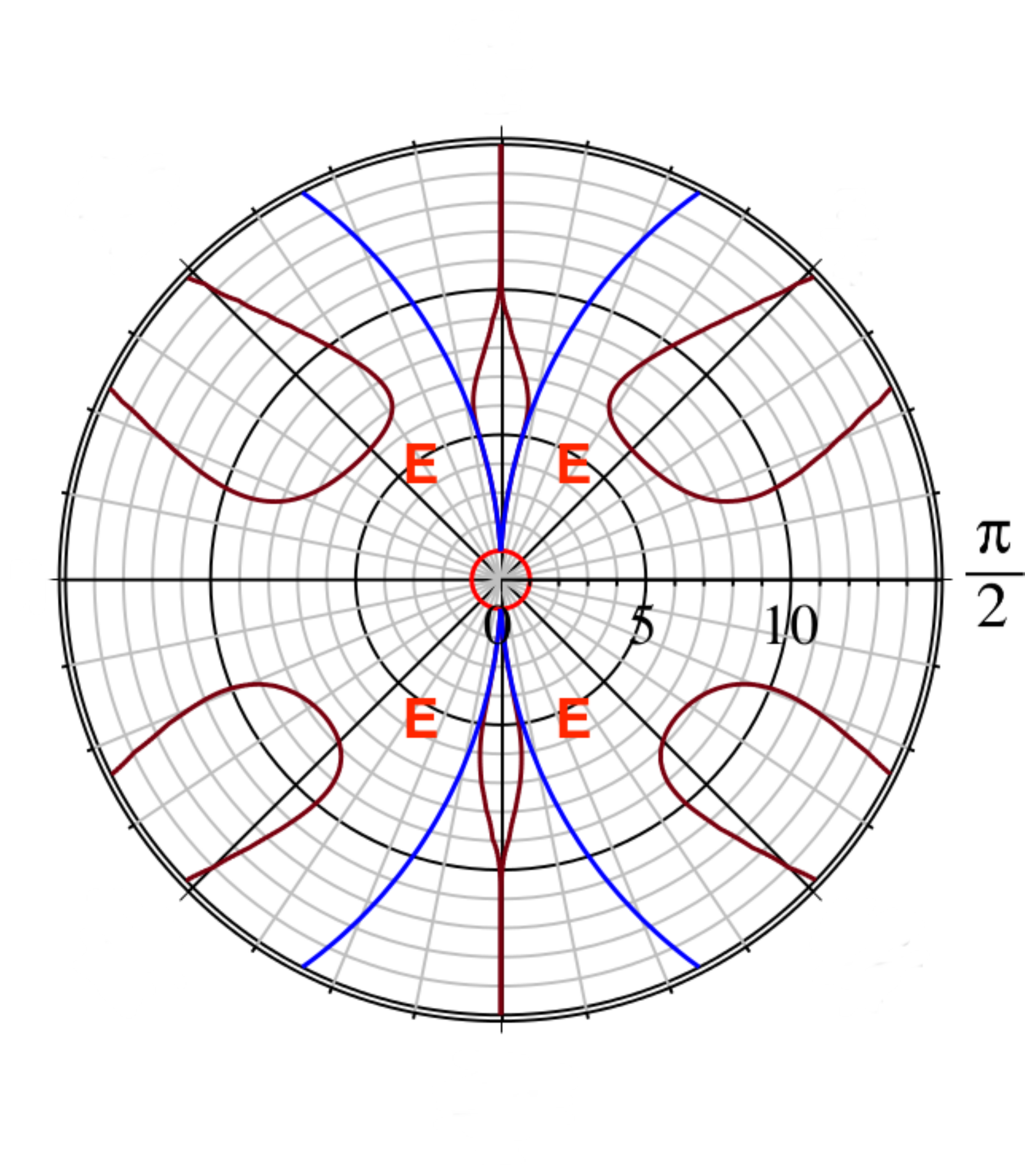} &
            \includegraphics[width=5 cm]{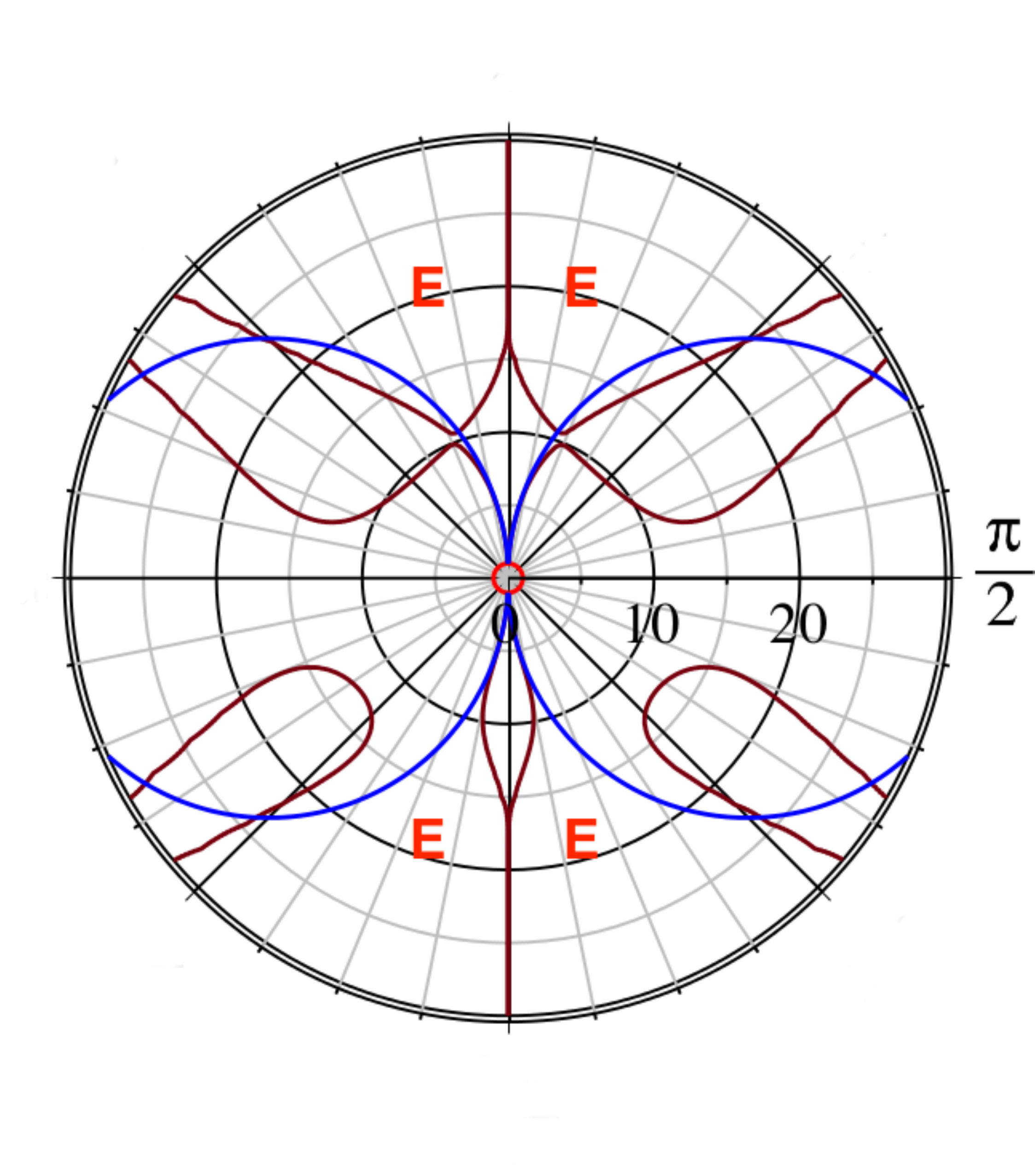} &
            \includegraphics[width=5 cm]{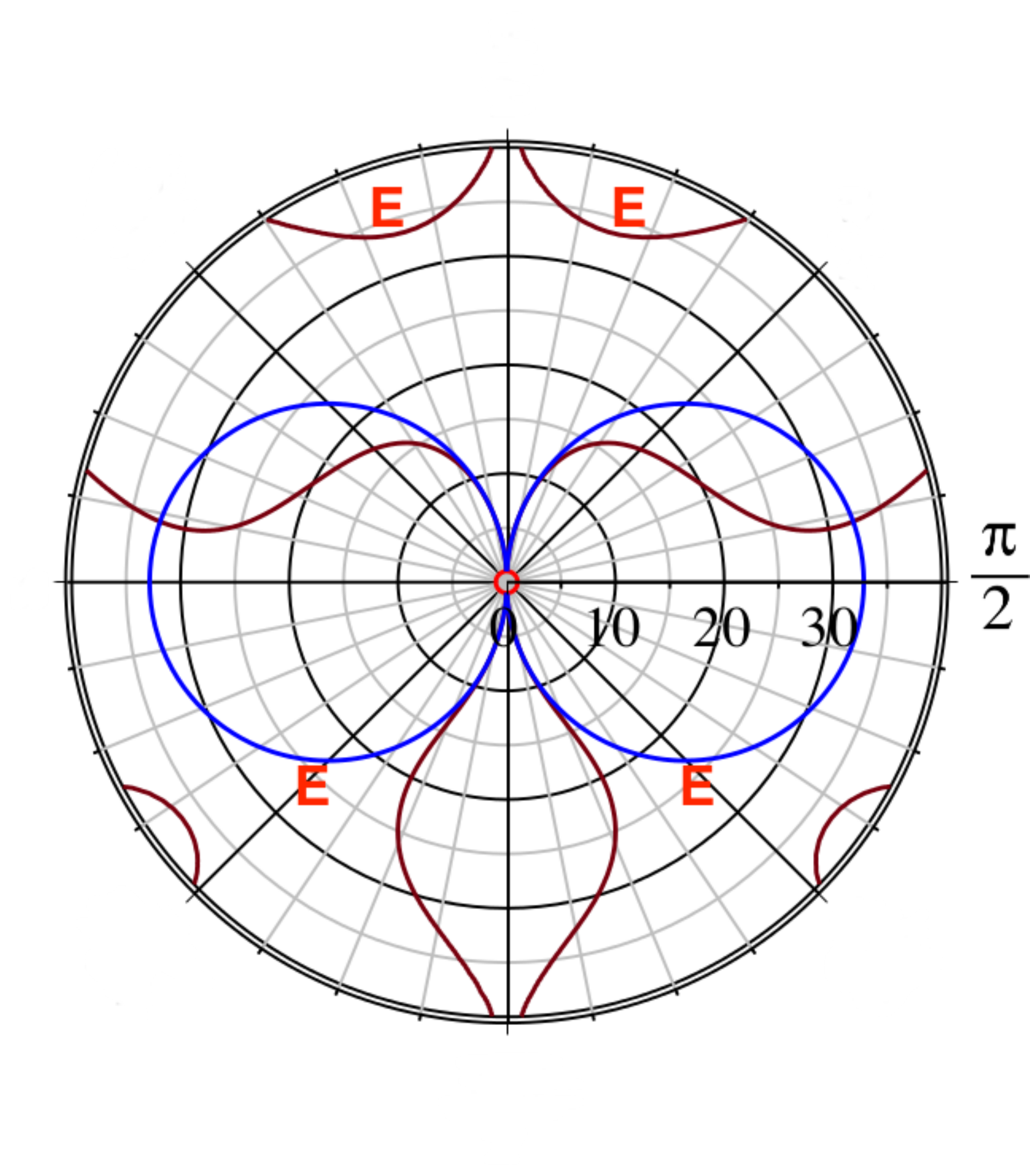} \\
            $c_1=1/500$ & $c_1=1/2000$ & $c_1=1/30000$
           \end{tabular}}
           \caption{\footnotesize{ Ergoregion for octupole distortion and $\alpha=0.97$}}
		\label{odd097}
\end{figure}

We investigate the behaviour of the ergoregion in the presence of an octupole external field in a similar way as the previous analysis for quadrupole distortion. First we describe the observed configurations for three values of the rotation parameter $\alpha$ and illustrate the results in Fig. \ref{odd03}-\ref{odd097}. Then, we deduce some general properties of the ergoregion for octupole distortion and summarize the dependance of its features on the parameters $c_1$ and $\alpha$.

The organization of the figures follows the same conventions as in the analysis for quadrupole distortion. Fig. \ref{odd03}-\ref{odd097} represent the cross-section of the ergoregion  with the $(x, \theta)$ - plane, where the coordinate $\theta$ is defined by $y=\cos(\theta)$.  The cases of positive and negative values of the distortion parameter $c_1$ are located in the first and the second row, respectively. The red contours represent the horizon, the blue ones correspond to the ergoregion of the undistorted Kerr black hole and the dark red to the ergoregion of the distorted case. All the non-compact parts of the ergoregion and some compact ergoregions are denoted by the letter $E$ in order to be distinguished from the static regions.

In Fig. \ref{odd03}, we see the ergoregion for $\alpha = 0.3$. For $c_1=-1/2$, we do not have a compact ergoregion which encompasses the horizon. Such an ergoregion appears when the static region on the axis, which gets larger as the absolute value of $c_1$ decreases, touches the ``butterfly'' static region located in the lower plane for $c_1\approx-1/4.4$. We can see the new configuration of the ergoregion in Fig. \ref{odd03} for $c_1=-1/10$. The static region which separates the compact and non-compact parts of the ergoregion increases when the absolute value of the distortion parameter decreases, as we can see in Fig. \ref{odd03} for $c_1=-1/50$. At the same time the static regions located left and right to the axis on the upper plane get more narrow. For positive values of the distortion parameter we observe the same behaviour, however all the configurations are reflected with respect to the equatorial plane. For example, the ``butterfly'' static region which is located in the lower plane for negative $c_1$ is transformed into its mirror image in the upper plane (see Fig. \ref{odd03} for $c_1=1/2$).

In Fig. \ref{odd07}, we see the ergoregion for $\alpha = 0.7$. The analysis is similar to the case $\alpha = 0.3$. Thus, for $c_1=-1/20$, we do not have a compact ergoregion which encompasses the horizon. Such an ergoregion appears when the static region on the axis touches the ``butterfly'' static region located in the lower plane for $c_1\approx-1/136$. The compact ergoregion is similar to the undistorted case with the difference that two small ``bumps'' appear between the angles $\pi/16<\theta<3\pi/16$ and $12\pi/16<\theta<15\pi/16$. However, as the absolute value of the distortion parameter decreases these ``bumps'' gradually disappear and the shape resembles the ergoregion of the undistorted Kerr black hole. We illustrate the compact ergoregion in the vicinity of the horizon for low absolute values of the distortion parameter ($c_1 = -1/600$) in comparison with the ergoregion of the undistorted Kerr black hole. For positive distortion parameter we observe the same behaviour, however all the configurations are reflected with respect to the equatorial plane. Thus, the deformation of the compact ergoregion compared to the undistorted Kerr black hole  appears between the angles $17\pi/16<\theta<19\pi/16$ and $29\pi/16<\theta<31\pi/16$ as we can see in Fig. \ref{odd07} for $c_1=1/200$, and the ``butterfly'' static region is located in the upper plane.

In Fig. \ref{odd097}, we see the ergoregions for $\alpha = 0.97$. For $c_1=-1/2$, there is no static region in the shape of ``butterfly'' but there are four disconnected static regions. Moreover, the static regions located on the axis are very narrow. When the distortion parameter $c_1$ decreases,  the static region on the axis increases and  touches the two static regions in the lower plane for $c_1\approx-1/1939$. Thus, the ``butterfly'' static region is formed, which we can see in Fig. \ref{odd097} for $c_1=-1/2000$. For $c_1\approx-1/212770$, the ``butterfly'' static region touches the static region located on the axis in the upper plane. As a result, we have the formation of a compact ergoregion similar to that presented in Fig. \ref{odd07} for $c_1=-1/200$.  The analysis for positive distortion parameter is the same, however, all the configurations are reflected with respect to the equatorial plane.  Thus, for $c_1\approx1/1939$ the two static regions in the upper plane touch the static region on the axis in order to form the ``butterfly'' static region (see Fig. \ref{odd097} for $c_1=1/2000$), and for $c_1\approx1/212770$, the static region located on the axis in the lower plane touches the ``butterfly'' static region to form the compact ergoregion in the vicinity of the horizon.

\paragraph{}
We can summarize the observed characteristics of the ergoregion in the following way. In the case of octupole distortion the ergoregion is determined by the function $A(x,y;c_1,\alpha)$ defined in eq. (\ref{A}),  which depends on two parameters $c_1$ and $\alpha$. It can be presented explicitly in the form
\begin{eqnarray}\label{A_c1}
&&A(x,y;c_1, \alpha) = (x^2-1)\left(1-\alpha^2e^{4\chi_1}\right)^2
 - 4\alpha^2(1-y^2)e^{4\chi_1}\cosh^2 2\chi_2,  \\[2mm]
 &&\chi_1 = \frac{1}{2}c_1\,y(y^2-1)(3x^2-1), \nonumber\\[2mm]
 &&\chi_2 = \frac{1}{2}c_1\,x(x^2-1)(3y^2-1). \nonumber
\end{eqnarray}
Analyzing its behaviour, we can deduce the following general properties, which characterize the ergoregion:
\begin{enumerate}
\item In contrast to the ergoregion of the non-distorted Kerr black hole, the ergoregion is not symmetric with respect to the cross-section $y=0$ (or equivalently $\theta=\pi/2$). The inequality $A(x,y;c_1, \alpha)<0$, which determines the ergoregion, is not invariant with respect to the shift $y\longleftrightarrow -y$. So, in general, if $y=\tilde{y}$ is a solution, $y=-\tilde{y}$ is not a solution. Thus, some discrete symmetries of the non-distorted case are violated.
\item The ergoregion is symmetric with respect to the shift $(y,c_1)\longleftrightarrow(-y,-c_1)$. We can see that this transformation leaves the inequality $A(x,y;c_1, \alpha)<0$  invariant. Consequently, if $y=\tilde{y}$ is a solution for some value of $c_1=\tilde{c_1}$, $y=-\tilde{y}$ is a solution as well for the corresponding negative value of the distortion parameter $c_1= -\tilde{c_1}$. This property accounts for the mirror symmetry of the ergoregion for negative and positive $c_1$ with the same absolute value with respect to the cross-section $y=0$.
\item The ergoregion always extends to infinity, if we consider the distorted Kerr black hole as a global solution. This is evident from the behaviour of the function $A(x,y;c_1, \alpha)$ at $x\rightarrow\infty$. In particular, it implies that there are no cases (for any values of the parameters $c_1$ and $\alpha$) when the ergoregion consists of a single compact region in the vicinity of the horizon, as for the non-distorted Kerr black hole.
\item The axis $y=\pm1$ (or $\theta = 0$, $\theta=\pi$) does not belong to the ergoregion for any values of the parameters $c_1$ and $\alpha$ except on the horizon. If we consider the equation $A(x,y;c_1, \alpha)=0$ for $y=\pm1$, the only solution in the interval $x\in(-1,+\infty)$ is $x=1$, where the ergoregion touches the horizon.
\item The cross-sections $y=\pm\frac{1}{\sqrt{3}}$ (or equivalently $\theta = \arccos\left(\pm\frac{1}{\sqrt{3}}\right)$) do not belong to the ergoregion for large values of $x$ for any values of the parameters $c_1$ and $\alpha$. Indeed, we see that the function $A(x,y=\pm\frac{1}{\sqrt{3}};c_1, \alpha)$ is positive for $x\rightarrow\infty$. Consequently, there always exists some value $\tilde{x}$, which depends on the parameters $c_1$ and $\alpha$, such that for $x>\tilde{x}$  static regions are formed around the cross-sections $y=\pm\frac{1}{\sqrt{3}}$, and they extend to infinity.
\end{enumerate}
\begin{figure}[htp]
\setlength{\tabcolsep}{ 0 pt }{\scriptsize\tt
		\begin{tabular}{ ccccc }
	\includegraphics[width=0.2\textwidth,height=3.2cm]{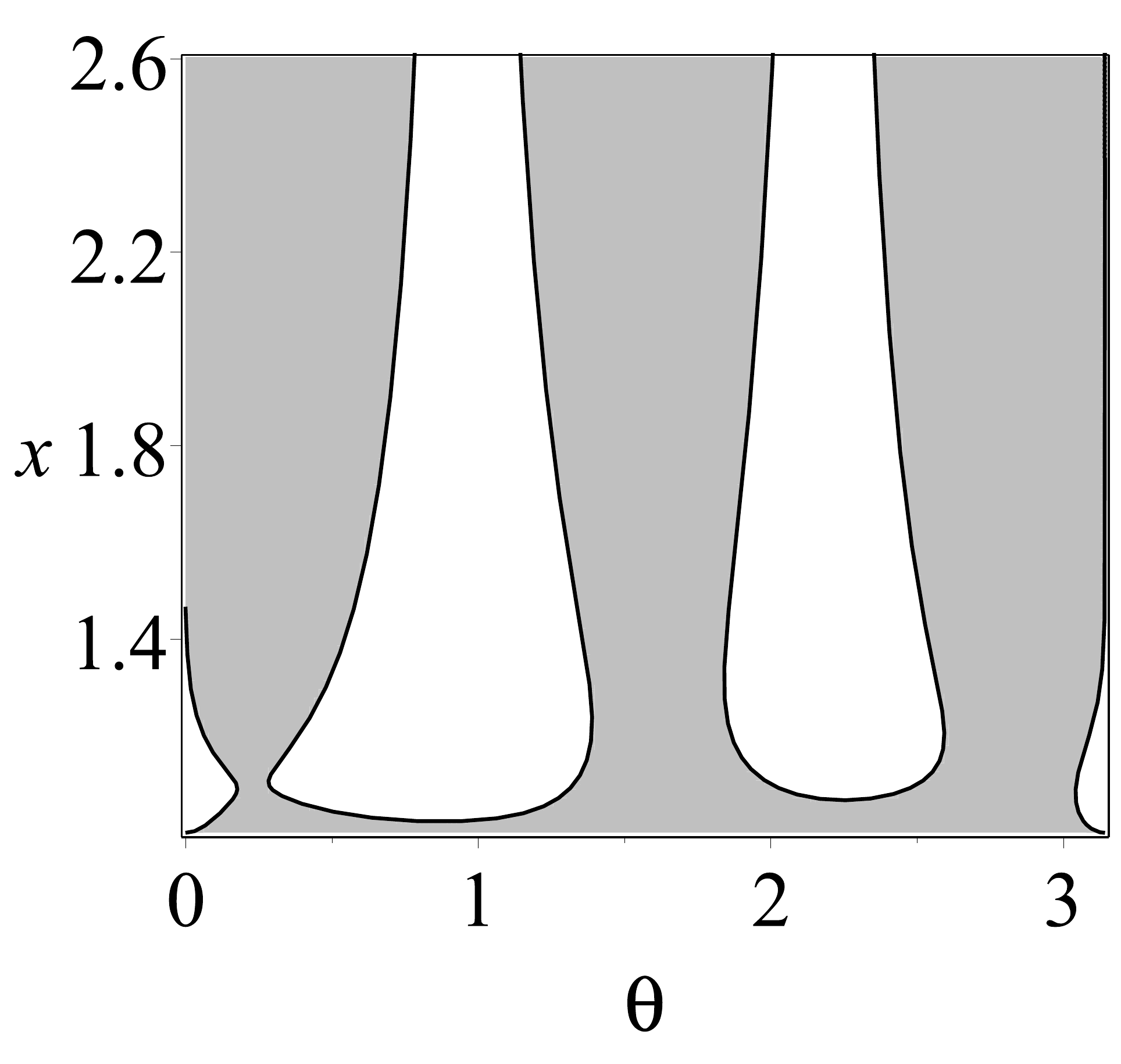} &
            \includegraphics[width=0.2\textwidth,height=3.2cm]{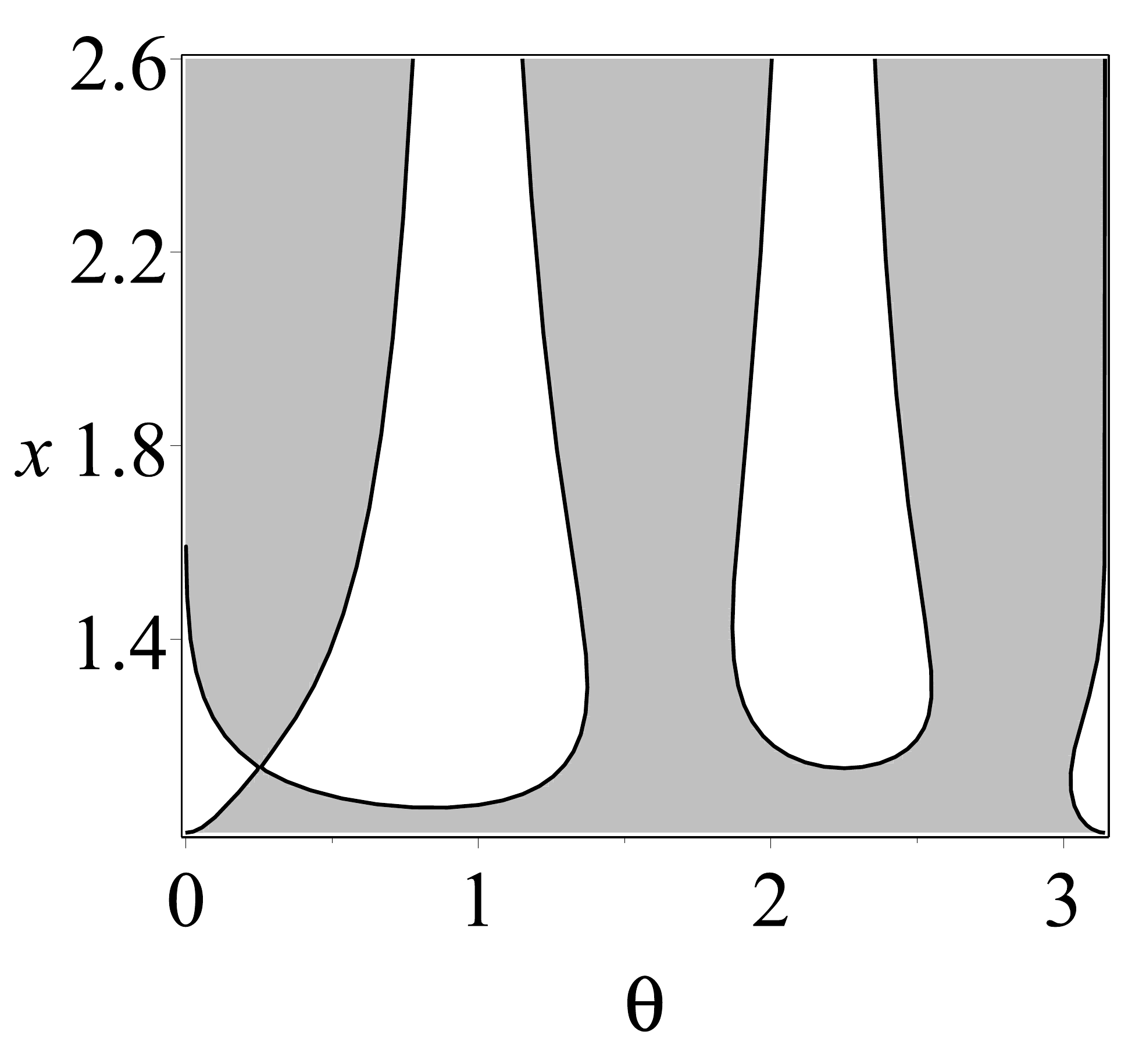} &
            \includegraphics[width=0.2\textwidth,height=3.2cm]{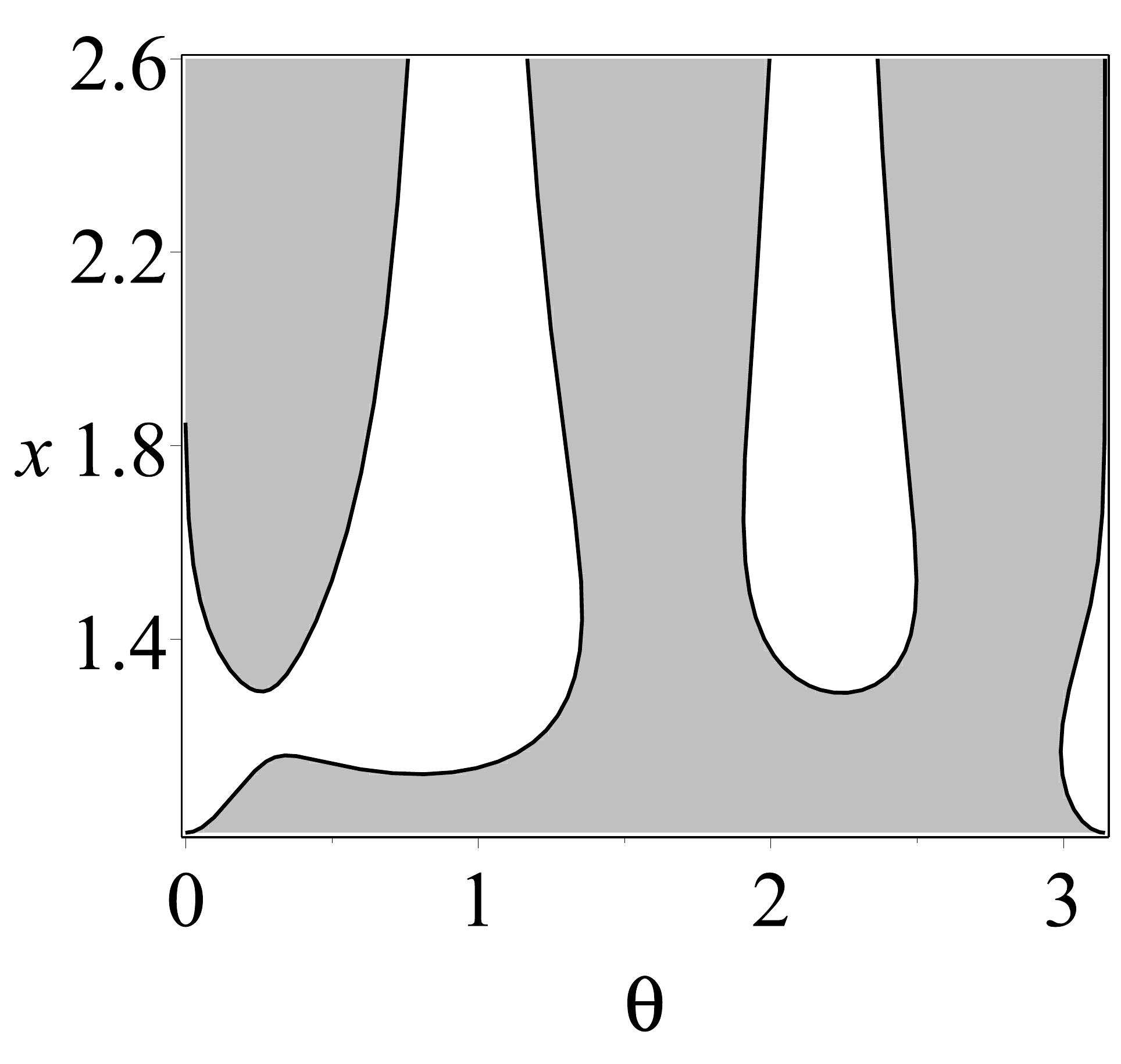} &
            \includegraphics[width=0.2\textwidth,height=3.2cm]{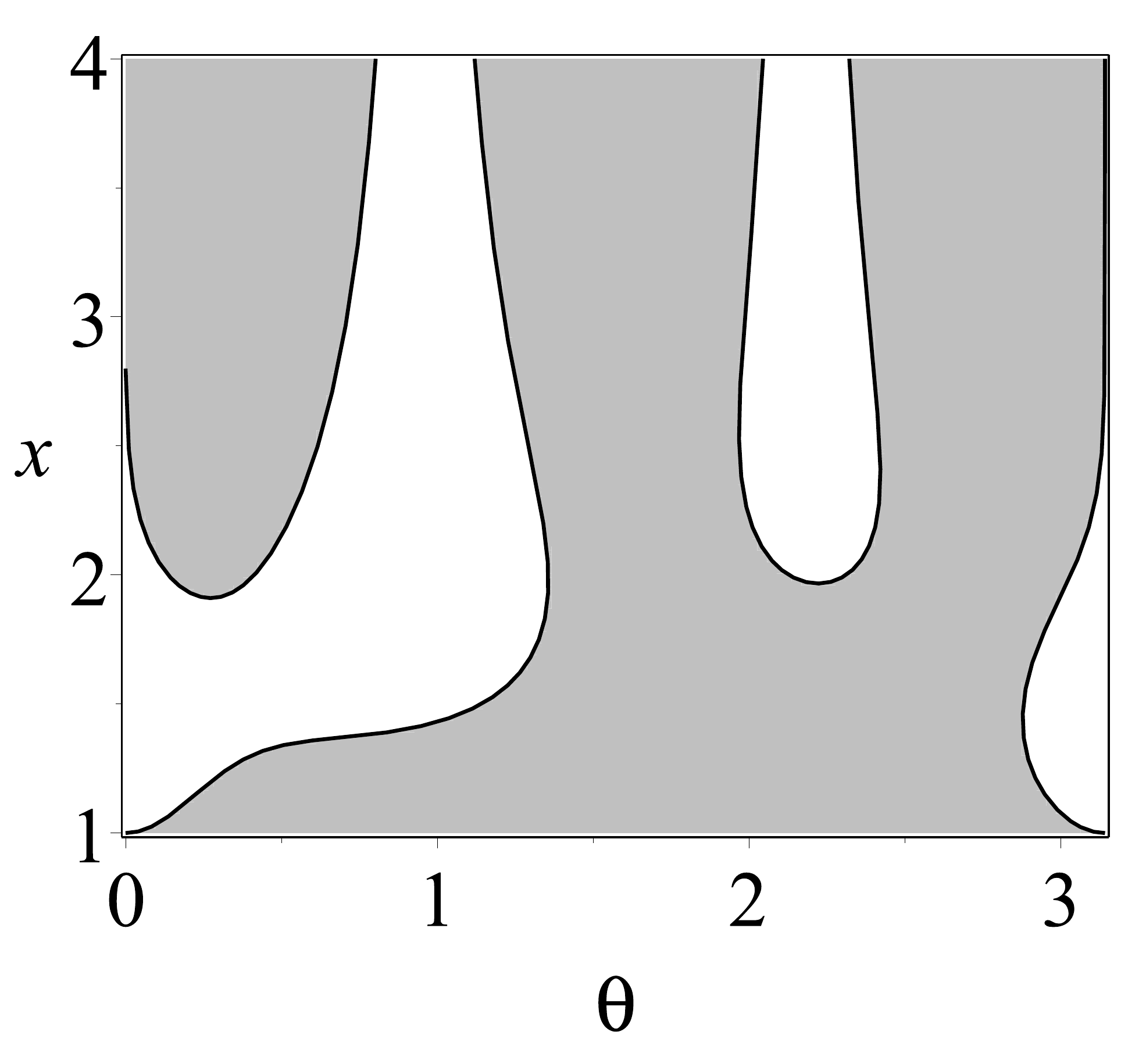}&
            \includegraphics[width=0.2\textwidth,height=3.2cm]{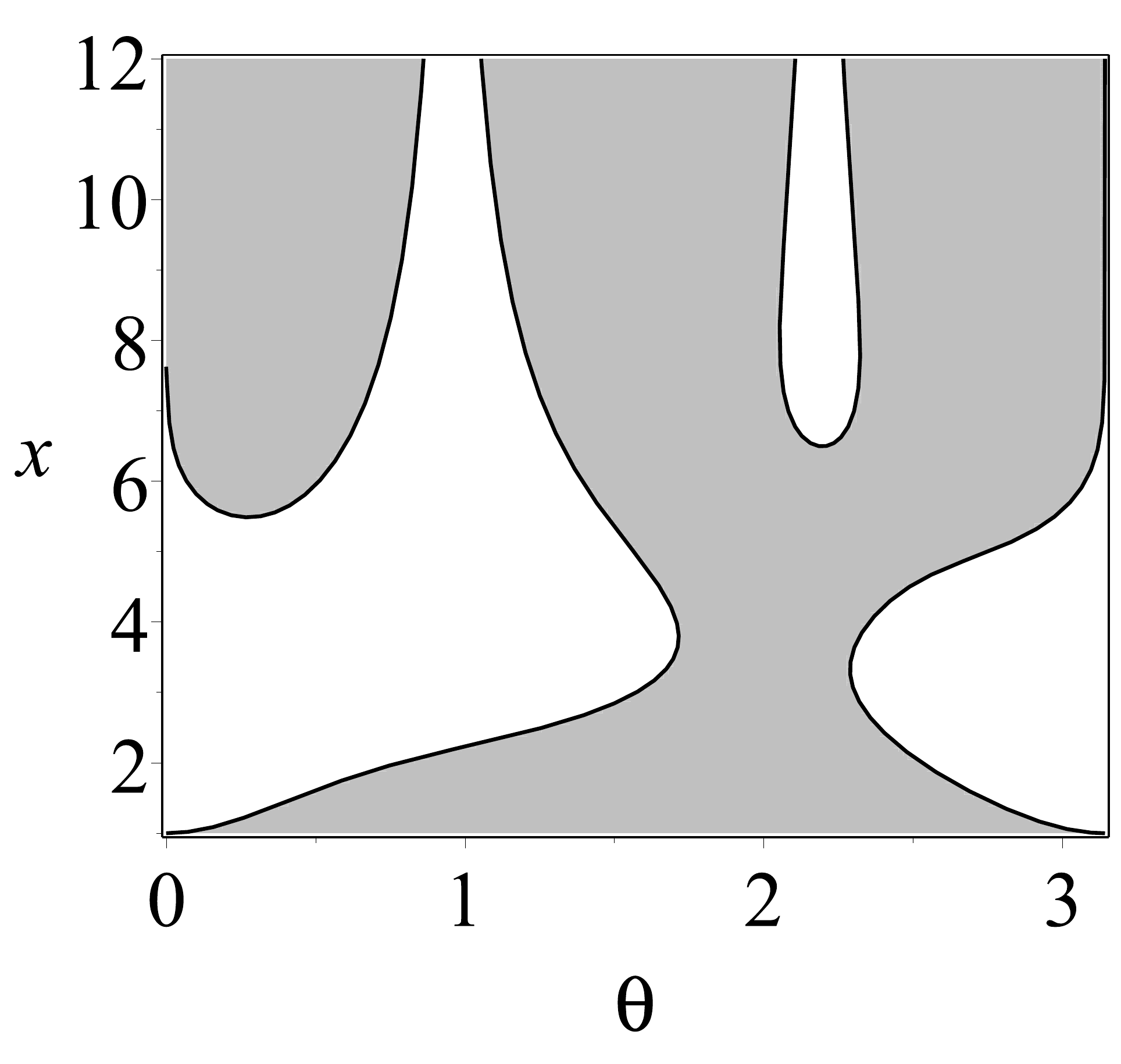}\\
            $c_1=2.0$ & $c_1=1.41$ & $c_1=0.8$ & $c_1=0.2$
            & $c_1=0.01$ \\
	        \includegraphics[width=0.2\textwidth,height=3.2cm]{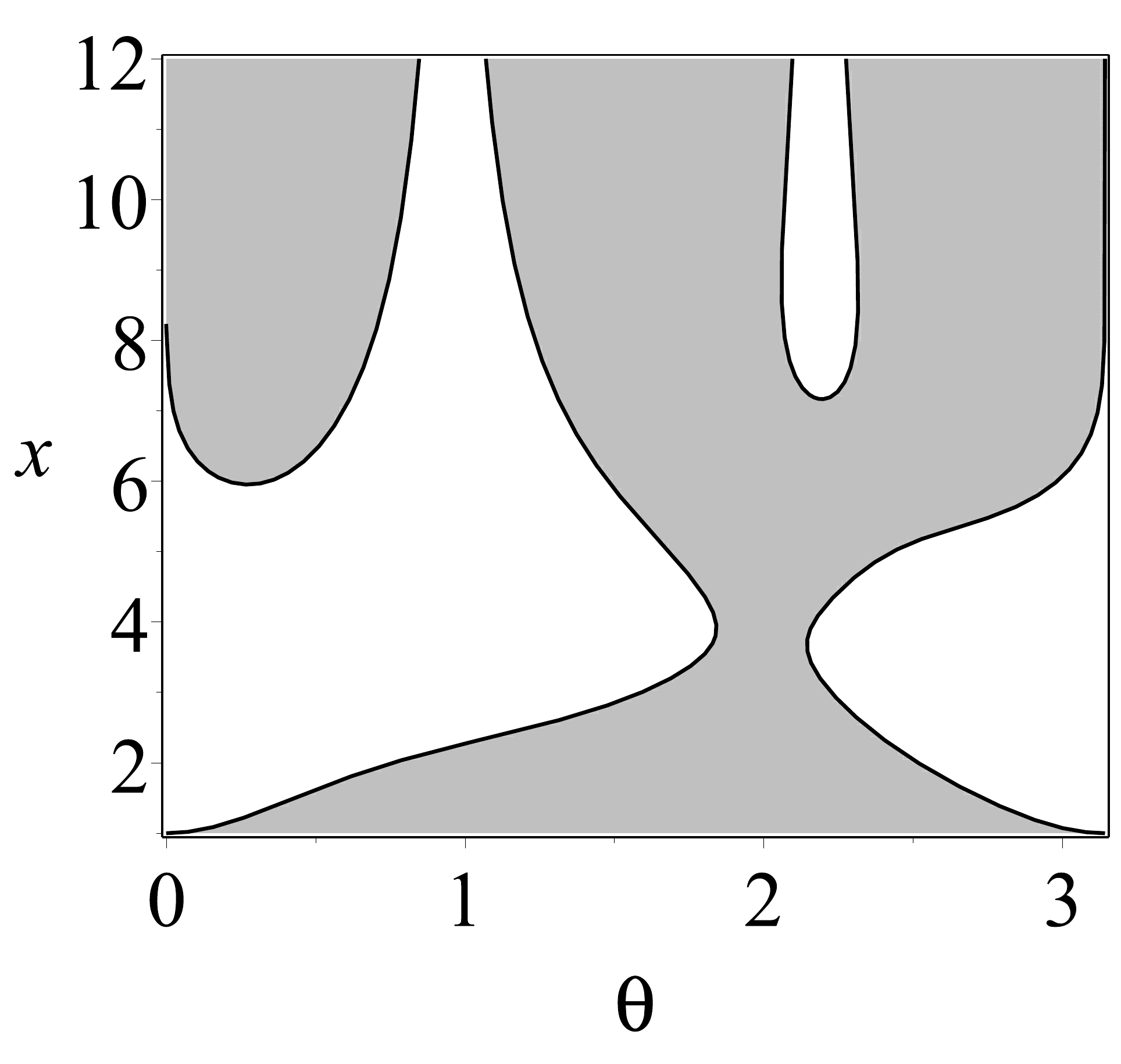} &
            \includegraphics[width=0.2\textwidth,height=3.2cm]{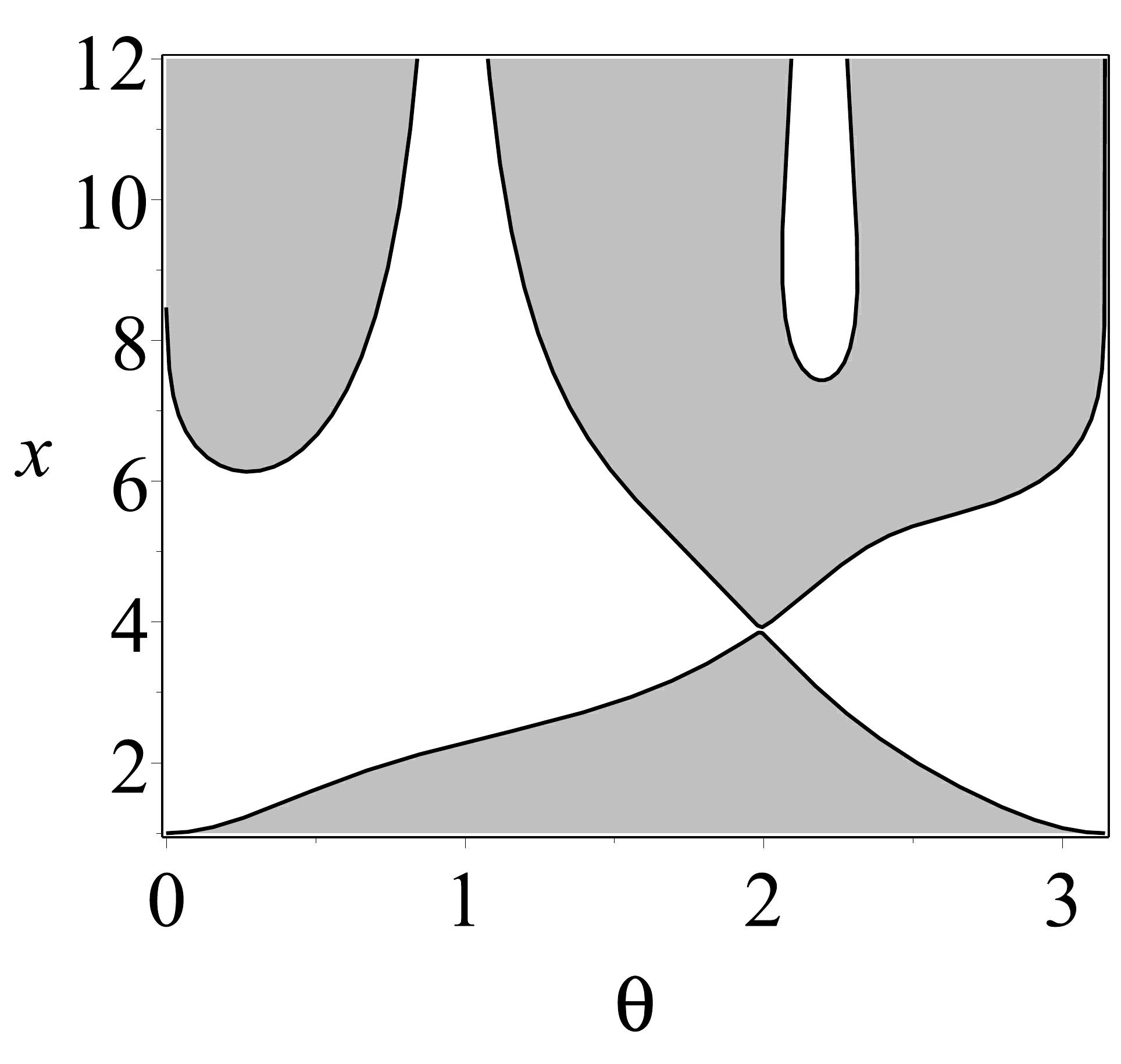} &
            \includegraphics[width=0.2\textwidth,height=3.2cm]{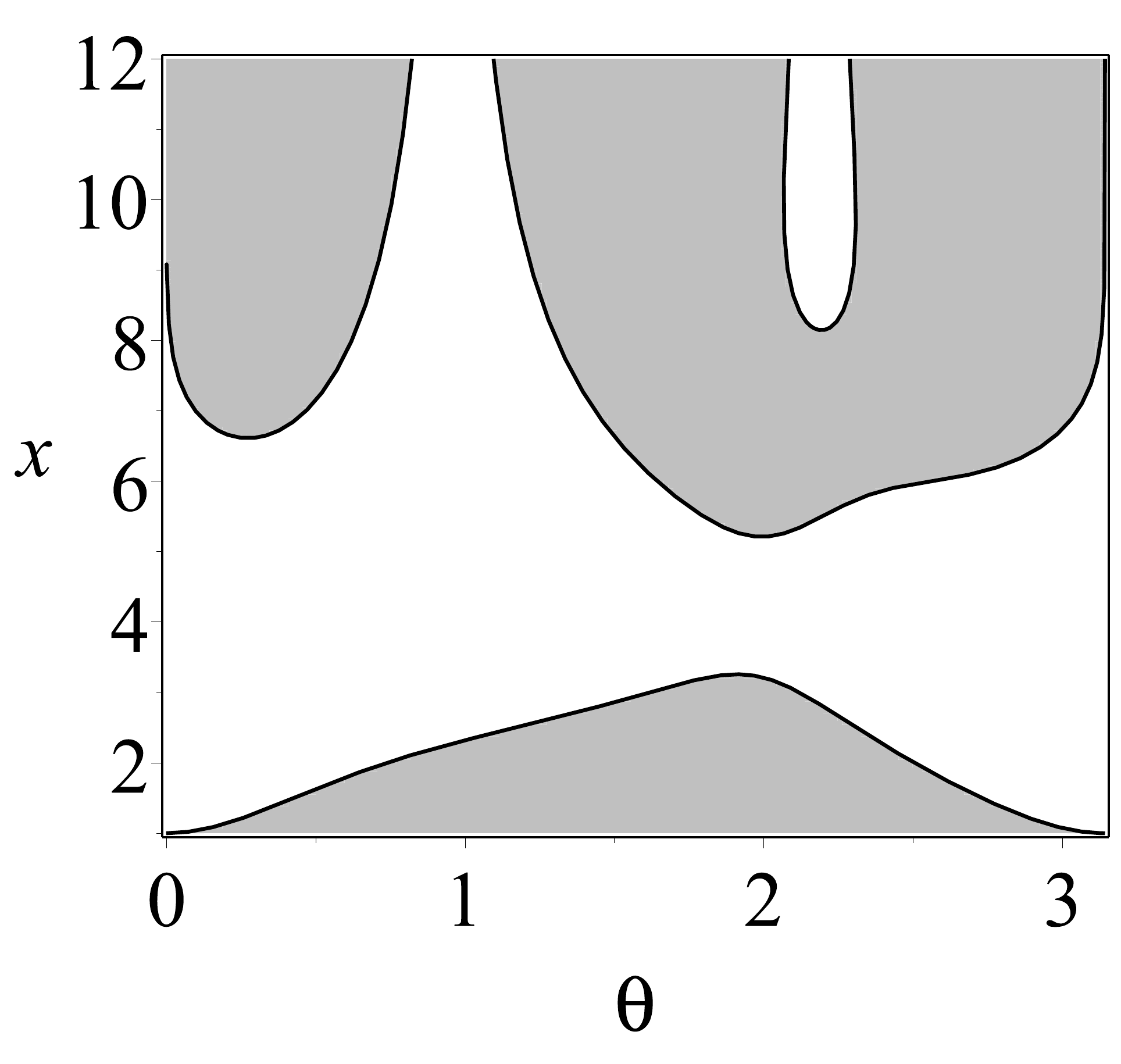}&
            \includegraphics[width=0.2\textwidth,height=3.2cm]{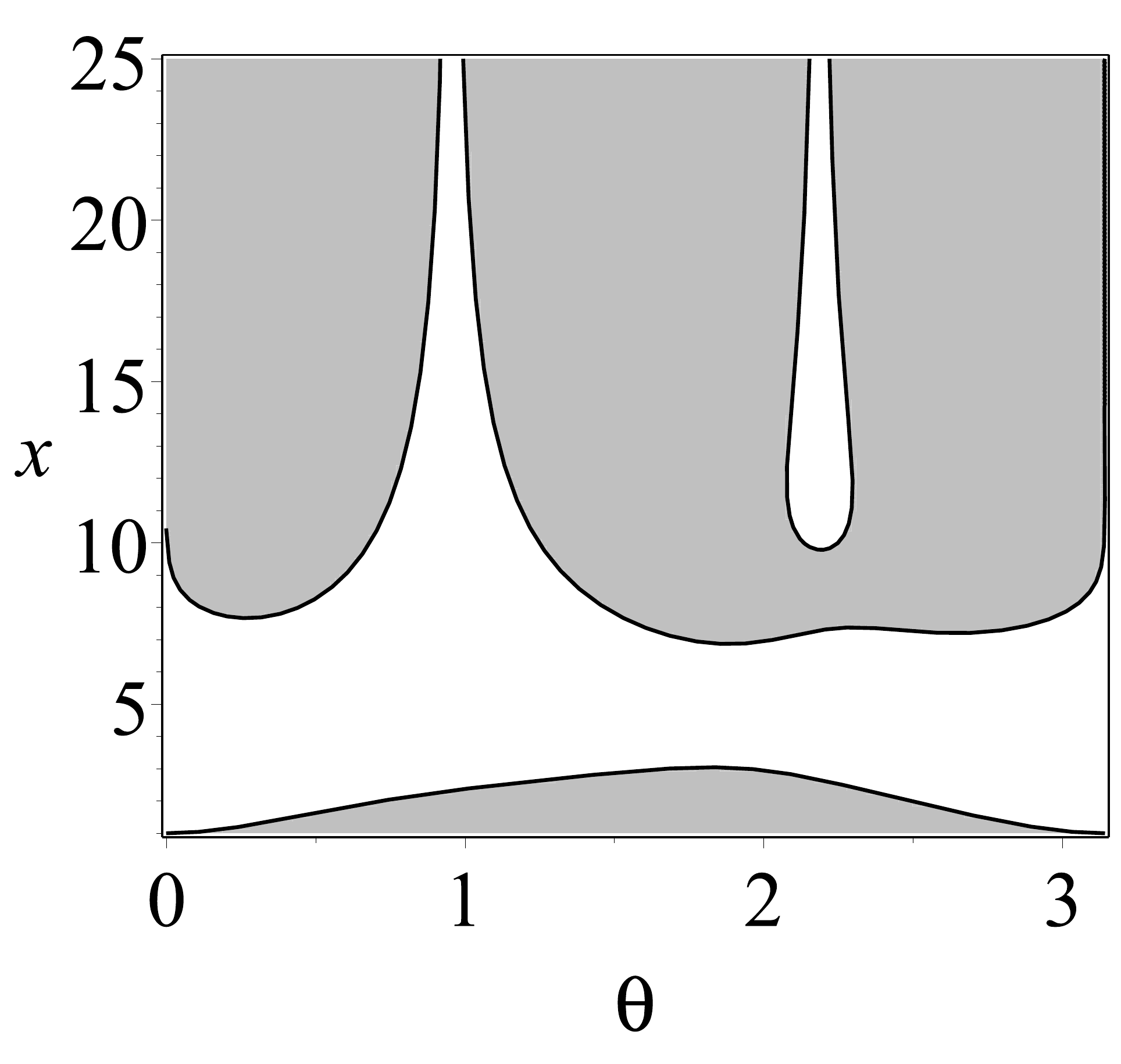}&
            \includegraphics[width=0.2\textwidth,height=3.2cm]{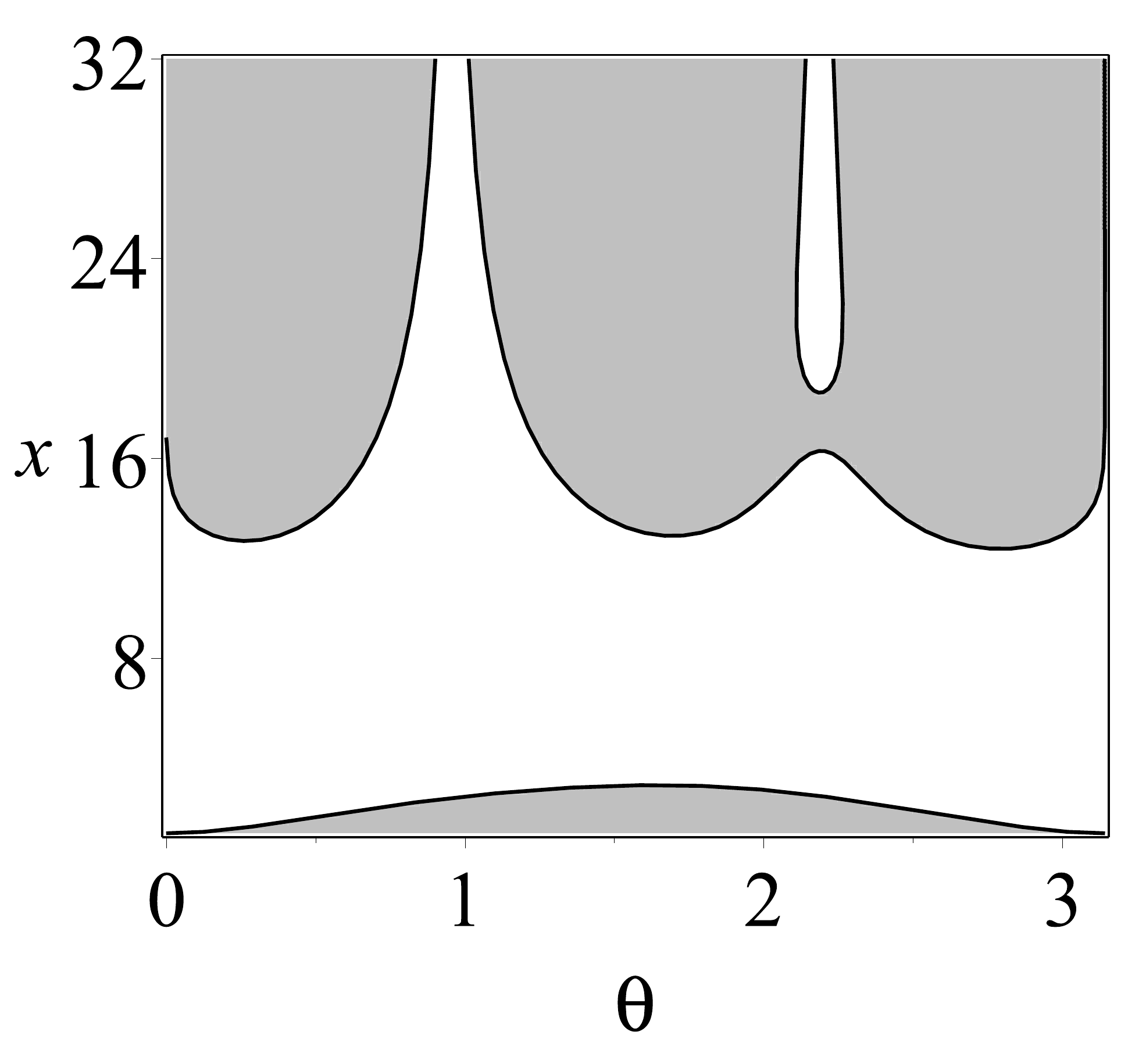} \\
            $c_1=0.008$ & $c_1=0.0073$ & $c_1=0.006$ & $c_1=0.004$
            & $c_1=0.001$
                       \end{tabular}}
           \caption{\footnotesize{ Behaviour of the ergoregion (grey area) as a function of $c_1$ for positive distortion parameter. The rotation parameter is fixed to $\alpha=0.7$.}}
		\label{c1gen}
\end{figure}

Due to the symmetry of the ergoregion with respect to the shift $(y,c_1)\longleftrightarrow(-y,-c_1)$, its analysis for positive $c_1$ completely determines the case $c_1<0$. The behaviour of the ergoregion as a function of $c_1$ for positive distortion parameter is summarized in Fig. \ref{c1gen}. For high values of $c_1$ the ergoregion consists of a single connected region extending to infinity. Three types of  static regions are observed, all of which are non-compact. The first one contains the axis, and the other two are located in the vicinity of the cross-sections $\theta=\arccos\left(\pm\frac{1}{\sqrt{3}}\right)$.  As $c_1$ decreases, the ergoregion pinches in some area located in the interval $\theta \in (0,\pi/2)$, until for a certain value of $c_1= c^{(1)}_{crit}$ two parts of the ergoregion are formed touching at a single critical point in the $(x,\theta)$-plane. For values of $c_1 < c^{(1)}_{crit}$ two disconnected parts of the ergoregion develop, both of which non-compact, and the static region containing the axis $\theta = 0$ merges with the static region around the cross-section $\theta=\arccos\left(\frac{1}{\sqrt{3}}\right)$. Within the context that we discussed before in the analysis of the Fig. \ref{odd03} - \ref{odd097}, for $c_1 < c^{(1)}_{crit}$ we observe the formation of the ``butterfly'' static region. As $c_1$ further decreases, the part of the ergoregion which touches the horizon pinches in a second area located in the interval $\theta \in (\pi/2,\pi)$. When the distortion parameter reaches a second critical value $c_1= c^{(2)}_{crit}$, two parts of the ergoregion are formed again touching at a single critical point in the $(x,\theta)$-plane. For values of $c_1 < c^{(2)}_{crit}$ they separate, and then the ergoregion consists of three disconnected regions - two non-compact ones, and a compact one encompassing the horizon. When the distortion parameter is relatively close to $c^{(2)}_{crit}$, we observe that the compact part of the ergoregion is deformed in the vicinity of the critical point, and the deformation is not symmetric with respect to the equatorial plane. Thus, its shape differs both from the ergoregion of the non-distorted Kerr black hole, and the case of quadrupole distortion. For $c_1= c^{(2)}_{crit}$ the static region on the axis touches the ``butterfly'' static region so, for $c_1< c^{(2)}_{crit}$ the latter stops to exist. When $c_1$ further decreases, the distance between the compact and non-compact parts of the ergoregion increases. The ergoregion encompassing the horizon gets smaller, and its deviation from the ergoregion of the non-distorted Kerr black hole for the same value of $\alpha$ reduces.
\begin{figure}[htp]
\setlength{\tabcolsep}{ 0 pt }{\scriptsize\tt
		\begin{tabular}{ cccc }
	\includegraphics[width=0.25\textwidth]{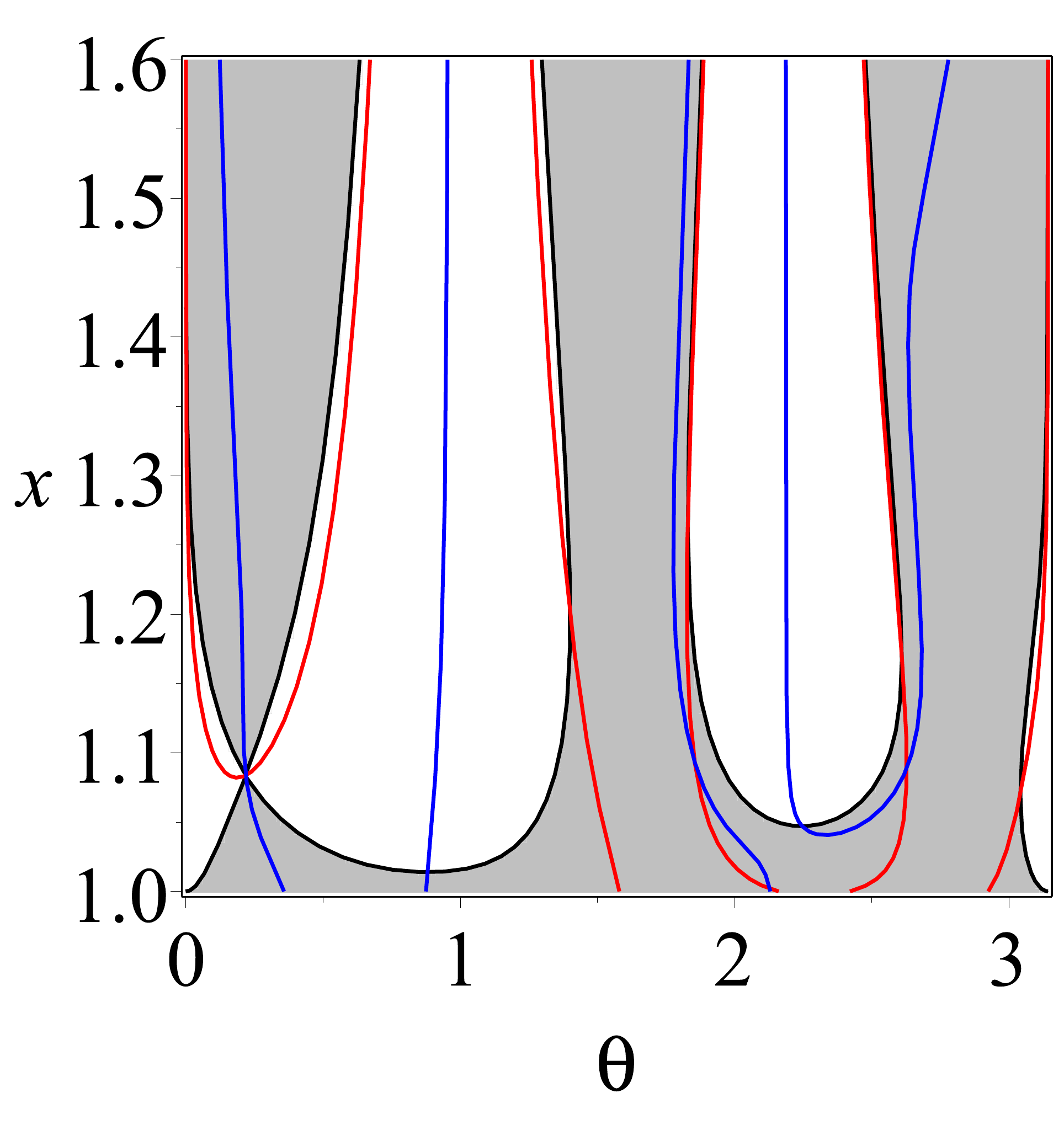} &
            \includegraphics[width=0.25\textwidth]{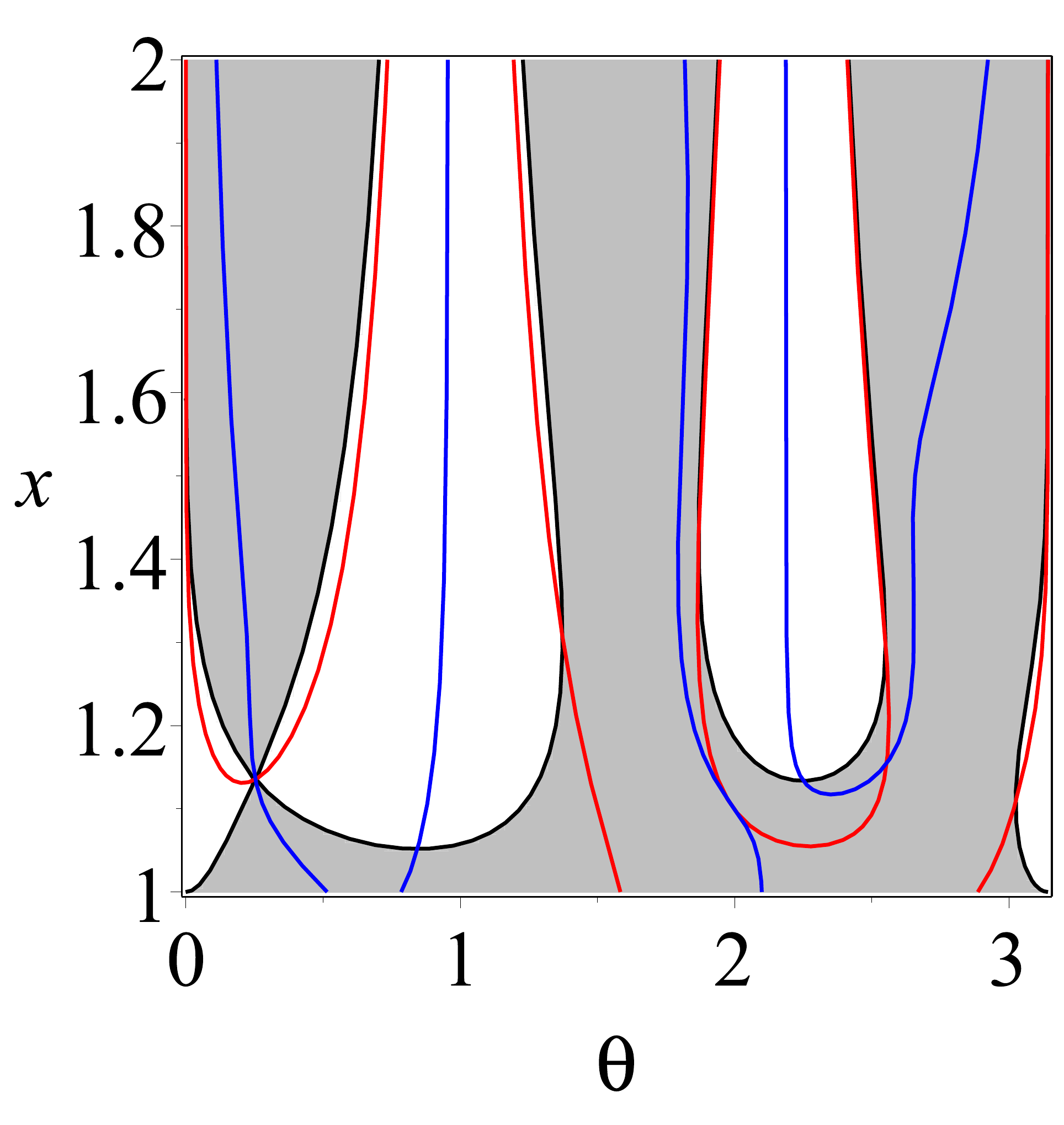} &
            \includegraphics[width=0.25\textwidth]{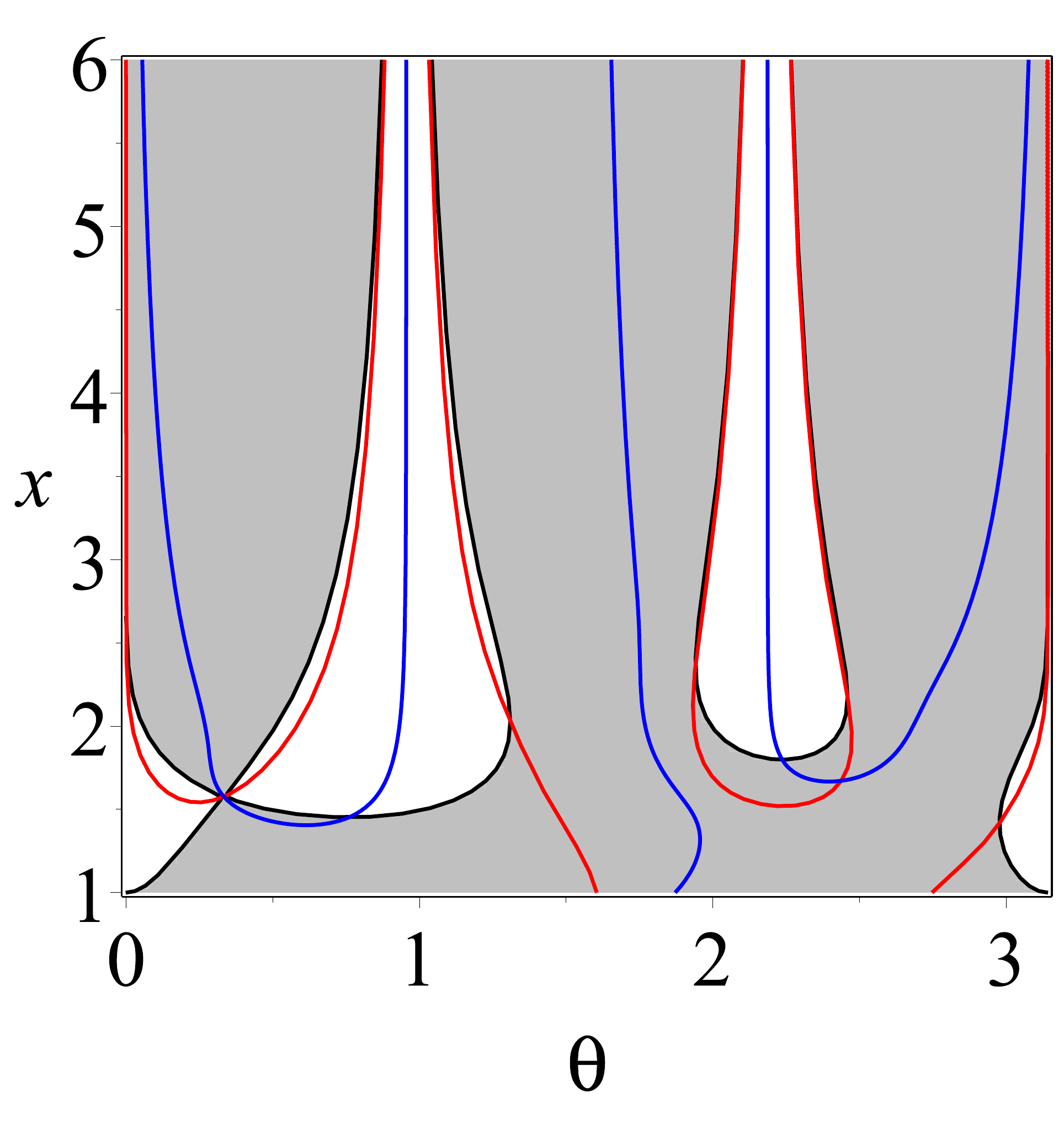} &
            \includegraphics[width=0.25\textwidth]{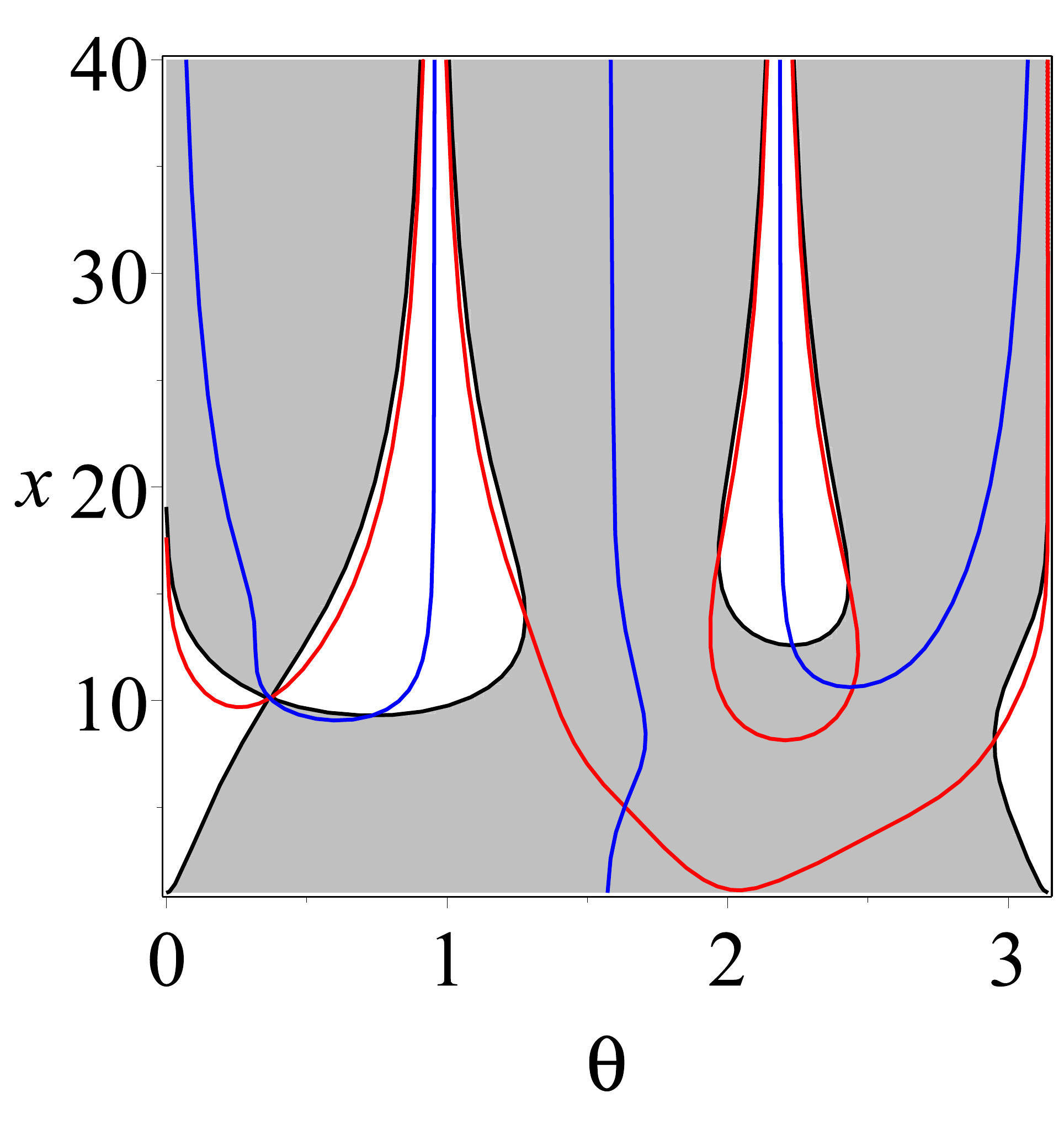}\\
            $\alpha=0.68$ & $\alpha=0.7$ & $\alpha=0.8$ & $\alpha=0.97$\\
            $c^{(1)}_{crit}\approx 2.34$ & $c^{(1)}_{crit}\approx1.41 $ & $c^{(1)}_{crit}\approx 0.22$ & $c^{(1)}_{crit}\approx 0.518\times10^{-3}$
                                   \end{tabular}}
           \caption{\footnotesize{Dependence of the first critical point location on the rotation parameter $\alpha$. The red and blue lines correspond to the curves $\partial_x A(x,y)=0$ and $\partial_y A(x,y)=0$, respectively, for each couple of parameters $c^{(1)}_{crit}$ and $\alpha$.}}
		\label{c1pin1}
\end{figure}
\begin{figure}[htp]
\setlength{\tabcolsep}{ 0 pt }{\scriptsize\tt
		\begin{tabular}{ cccc }
	\includegraphics[width=0.25\textwidth]{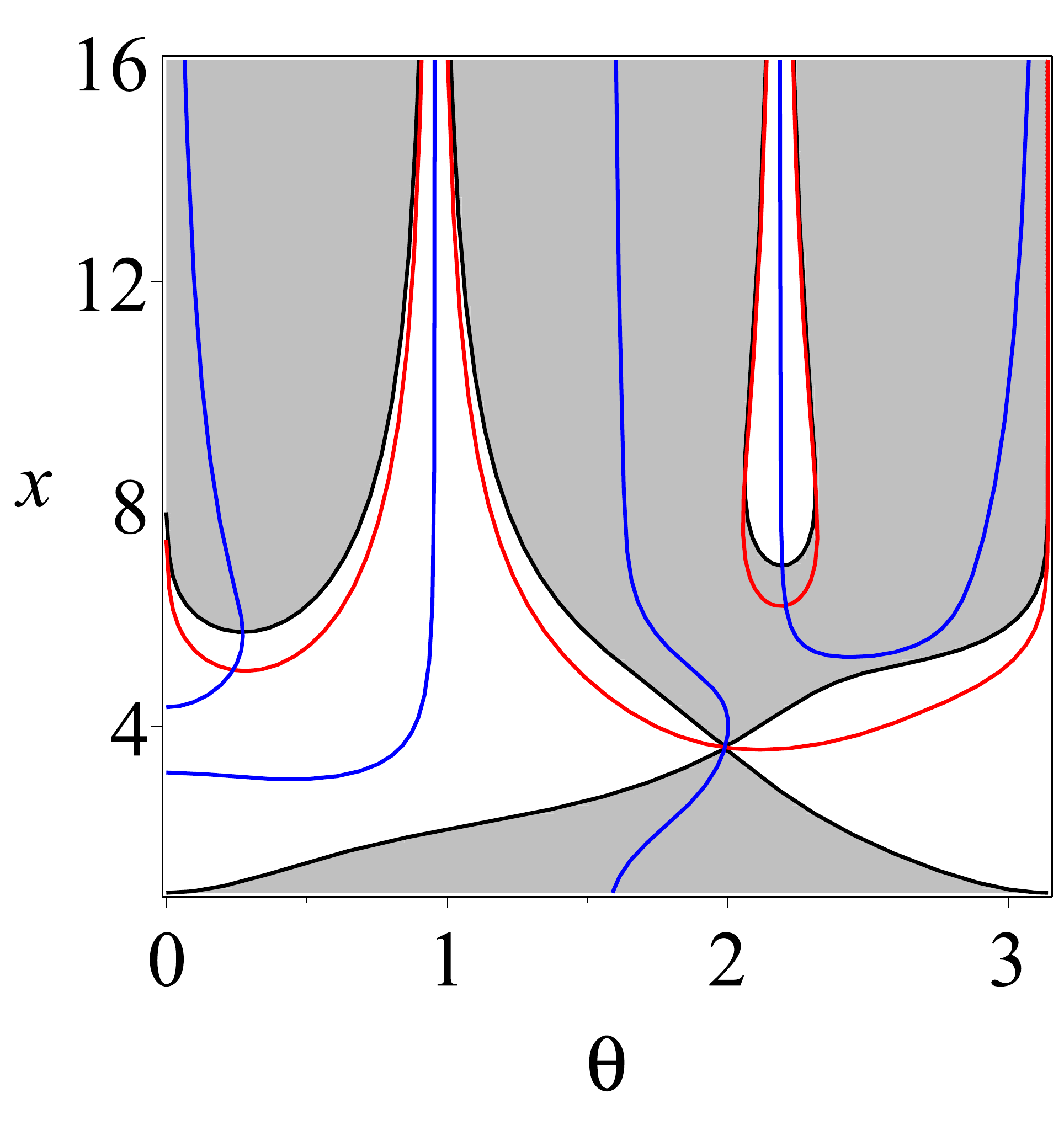} &
            \includegraphics[width=0.25\textwidth]{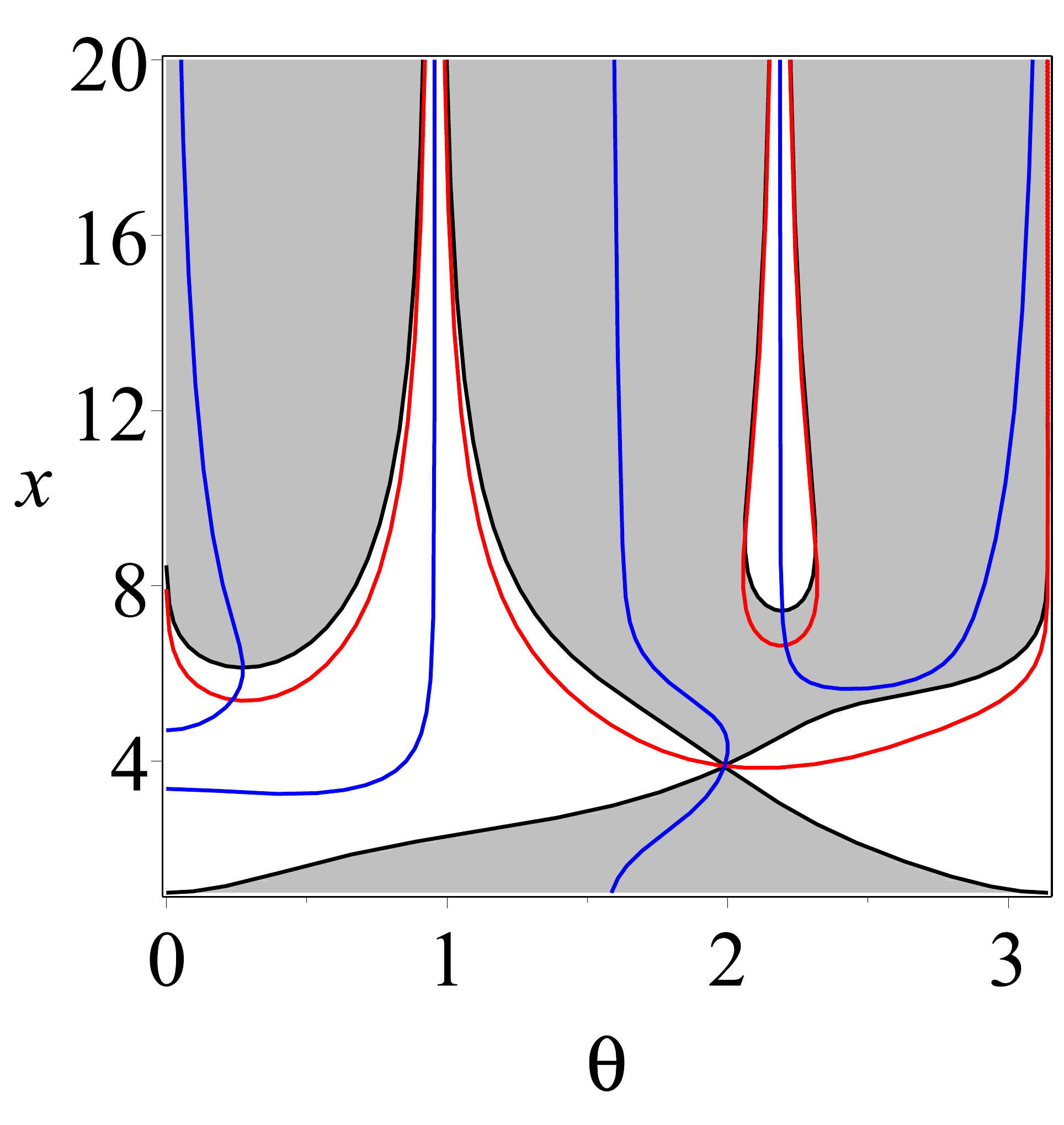} &
            \includegraphics[width=0.25\textwidth]{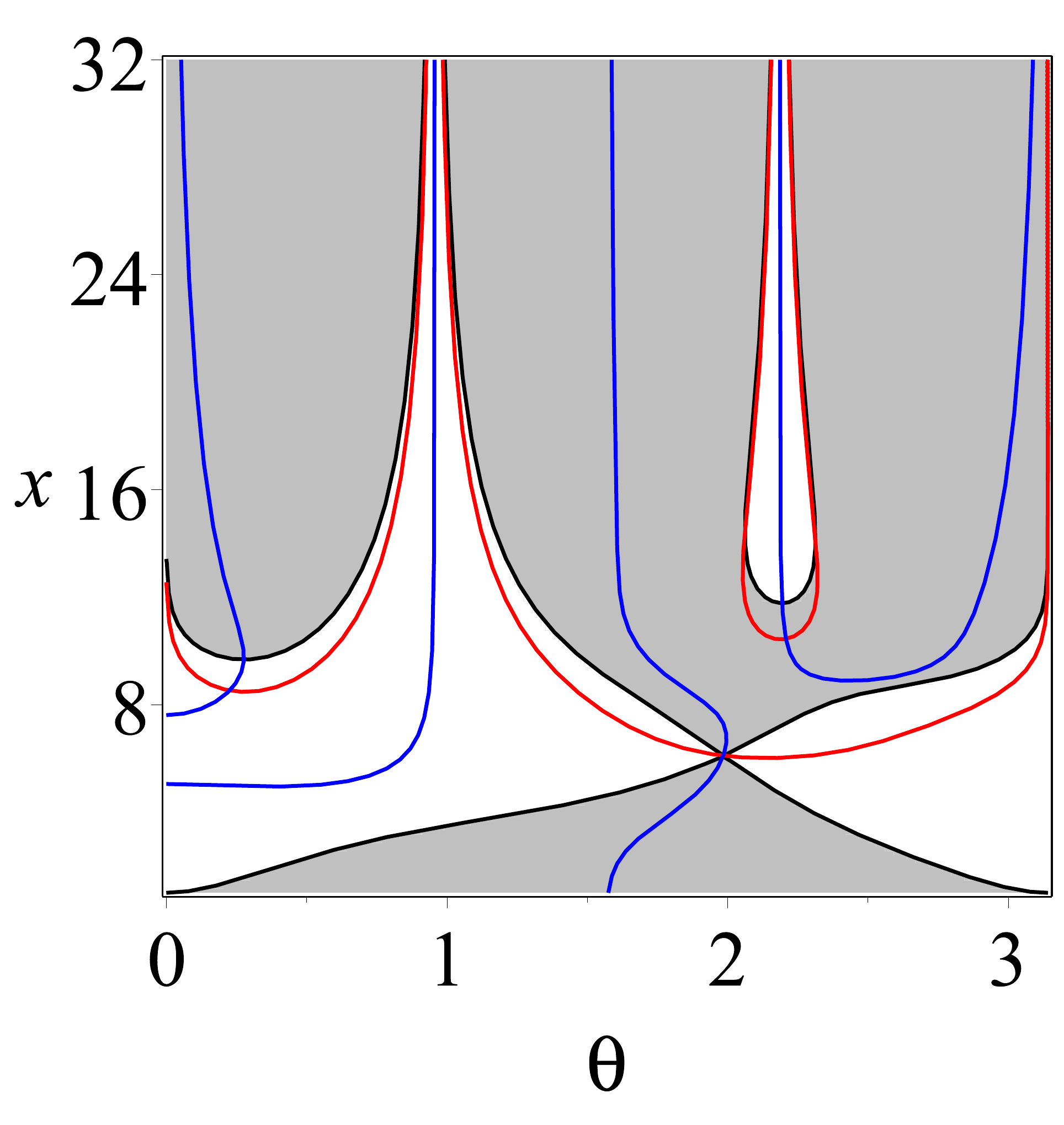} &
            \includegraphics[width=0.25\textwidth]{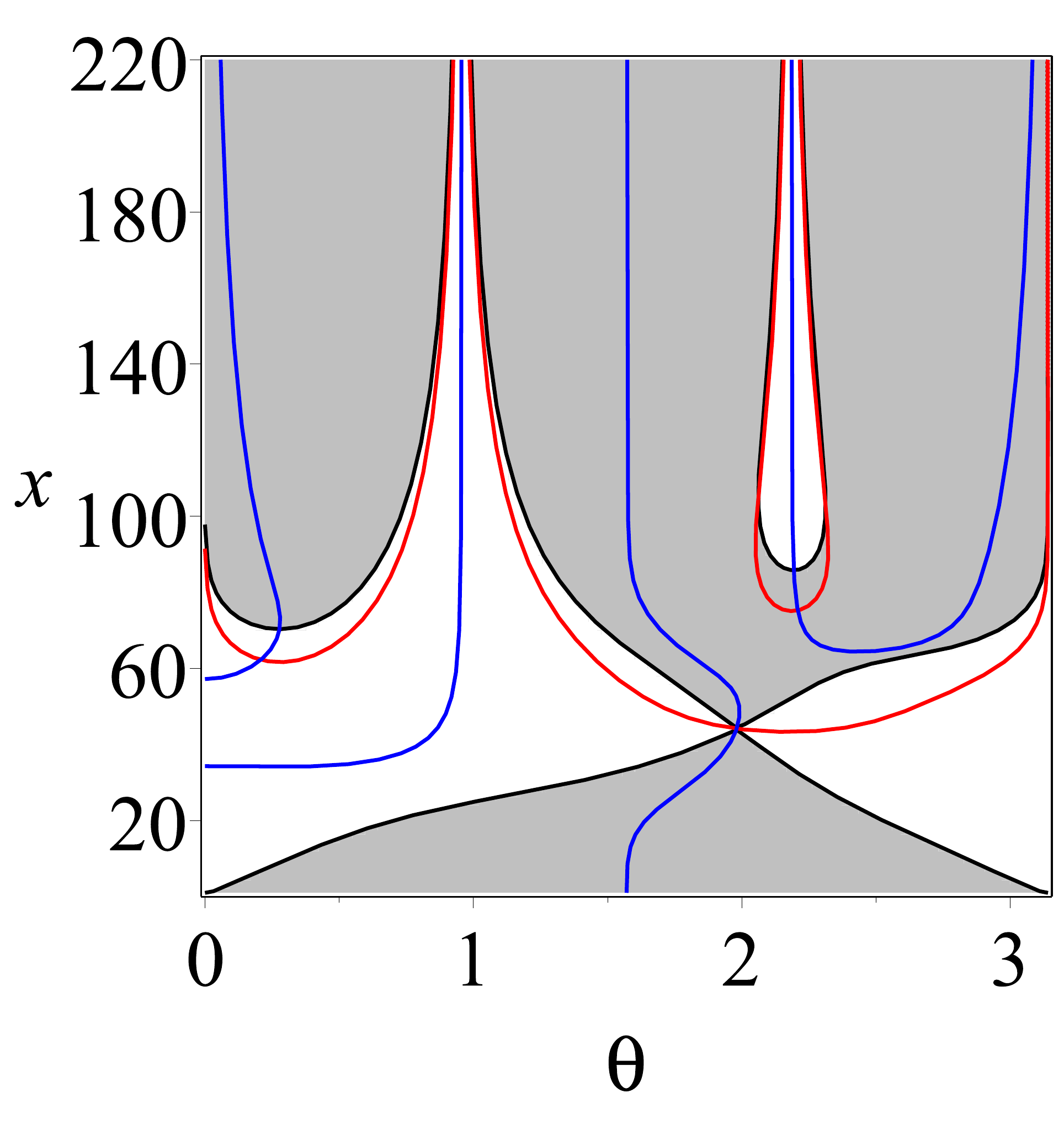}\\
            $\alpha=0.68$ & $\alpha=0.7$ & $\alpha=0.8$ & $\alpha=0.97$ \\
            $c^{(2)}_{crit}\approx 0.009275 $ & $c^{(2)}_{crit}\approx 0.007367$ & $c^{(2)}_{crit}\approx 0.001831$ & $c^{(2)}_{crit}\approx 0.47\times10^{-5}$
                                   \end{tabular}}
           \caption{\footnotesize{Dependence of the second critical point location on the rotation parameter $\alpha$. The red and blue lines correspond to the curves $\partial_x A(x,y)=0$ and $\partial_y A(x,y)=0$, respectively, for each couple of parameters $c^{(2)}_{crit}$ and $\alpha$.}}
		\label{c1pin2}
\end{figure}

The described behaviour of the ergoregion as a function of $c_1$ is illustrated in Fig. \ref{c1gen} for the fixed value of the rotation parameter $\alpha=0.7$. However, the same qualitative behaviour is observed for general rotation parameter. The value of $\alpha$ influences the location of the two types of critical points in the $(x, \theta)$ - plane, and the corresponding critical values of the distortion parameter $c_1= c^{(1)}_{crit}$ and $c_1=c^{(2)}_{crit}$, for which the ergoregion pinches off. The dependence is illustrated in Fig. \ref{c1pin1} - \ref{c1pin2}, where we have also specified the values of $c^{(1)}_{crit}$ and $c^{(2)}_{crit}$ for each $\alpha$. When $\alpha$ increases, the $x$-coordinates of the two critical points increase, while the critical values of the distortion parameter $c^{(1)}_{crit}$ and $c^{(2)}_{crit}$ decrease. We have also depicted the curves defined by the equations $\partial_xA(x,y;c_1, \alpha)=0$ and $\partial_y A(x,y;c_1, \alpha)=0$ by blue and red lines, respectively. As we have argued in the analysis of the quadrupole distortion, the points  in which they simultaneously cross the ergosurface $A(x,y;c_1, \alpha)=0$ correspond to the critical points.

\subsection{Ergoregion size and angular momentum}

In this section, we investigate the correlations between the size of the compact ergoregion in the vicinity of the horizon and two characteristics of the distorted black holes - the value the distortion parameter $c_2$, (or $c_1$), and the spin parameter $J/M^2$.  In all the analysis we keep the rotation parameter $\alpha$ and the mass parameter $m$  fixed to certain values and vary only the distortion parameter.  We consider the area of the cross-section  of the compact ergoregion with the $(x,y)$ - plane and denote it by $2A_E$. Consequently,  $A_E$ is the area of the compact part of the grey regions illustrated in the figures in section 3.1, for example in fig. \ref{c2gen}. We normalize it by the horizon area $A_H$

\begin{equation}
A_H = \int_H \sqrt{g_H}dyd\phi = \frac{16\pi m^2(1+\alpha^2)}{(1-\alpha^2)^2}\exp{\left(-2\sum^\infty_{n=1}c_{2n}\right)},
\end{equation}
and in this way we obtain the quantity
\begin{eqnarray}
\Delta A = \frac{A_E}{A_H},
\end{eqnarray}
\noindent
which we adopt as a measure of the size of the ergoregion in the vicinity of the horizon. In the static limit we have $\Delta A \rightarrow 0$.

\begin{figure}[htp]
\setlength{\tabcolsep}{ 0 pt }{\scriptsize\tt
		\begin{tabular}{ cc }
	\hspace{1.5cm} \includegraphics[width=6 cm]{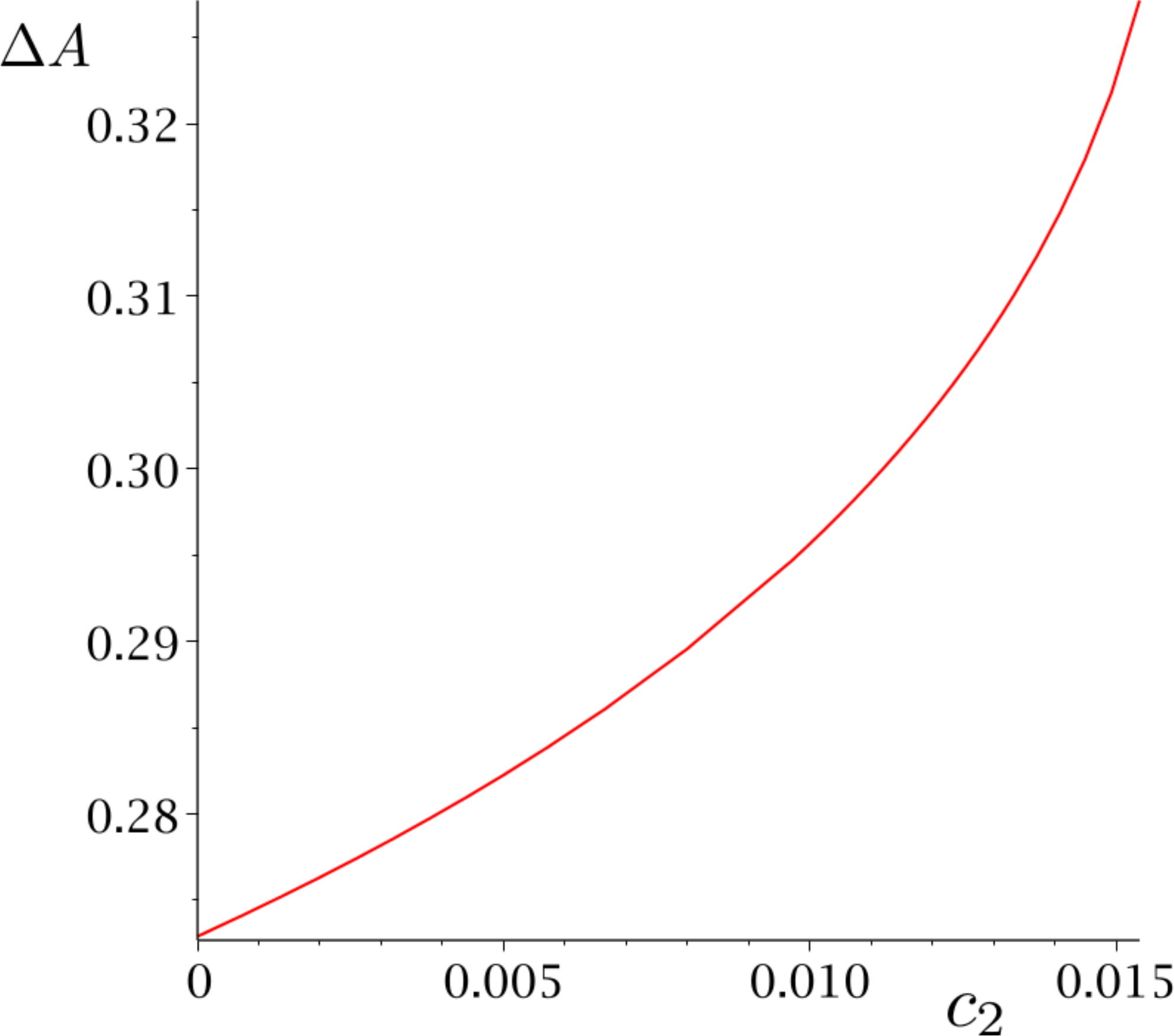}  &
      \hspace{1.5cm}      \includegraphics[width=6 cm]{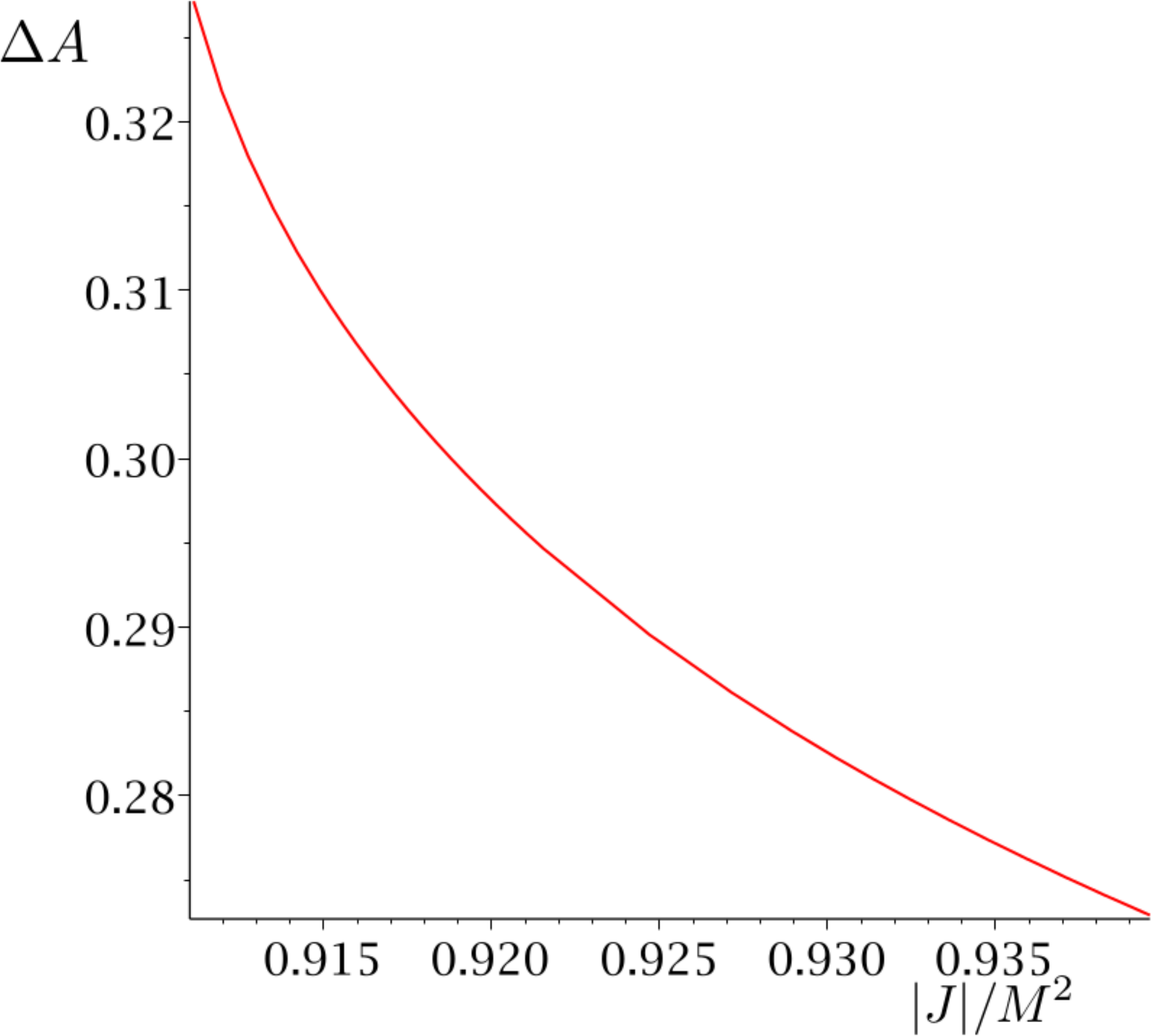} \\
                  \end{tabular}}
           \caption{\footnotesize{Dependence of the size of the compact ergoregion on the distortion parameter $c_2$ (left panel), and the spin parameter $J/M^2$ (right panel) for positive quadrupole distortion. The rotation parameter $\alpha$ is fixed to $\alpha = 0.7$.}}
		\label{areac2}
\end{figure}

\begin{figure}[htp]
\setlength{\tabcolsep}{ 0 pt }{\scriptsize\tt
		\begin{tabular}{ cc }
	\hspace{1.5cm} \includegraphics[width=6 cm]{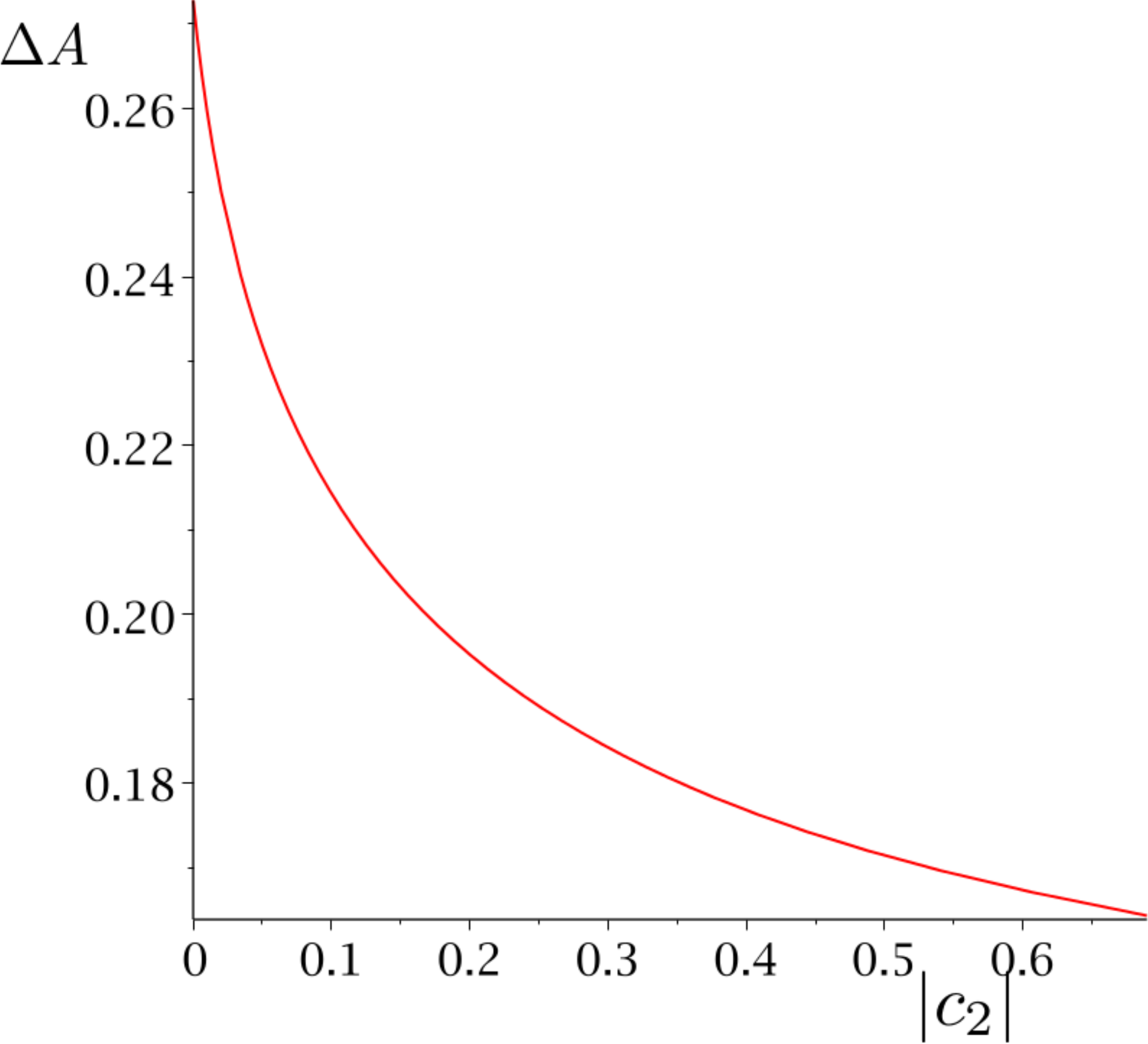}  &
      \hspace{1.5cm}      \includegraphics[width=6 cm]{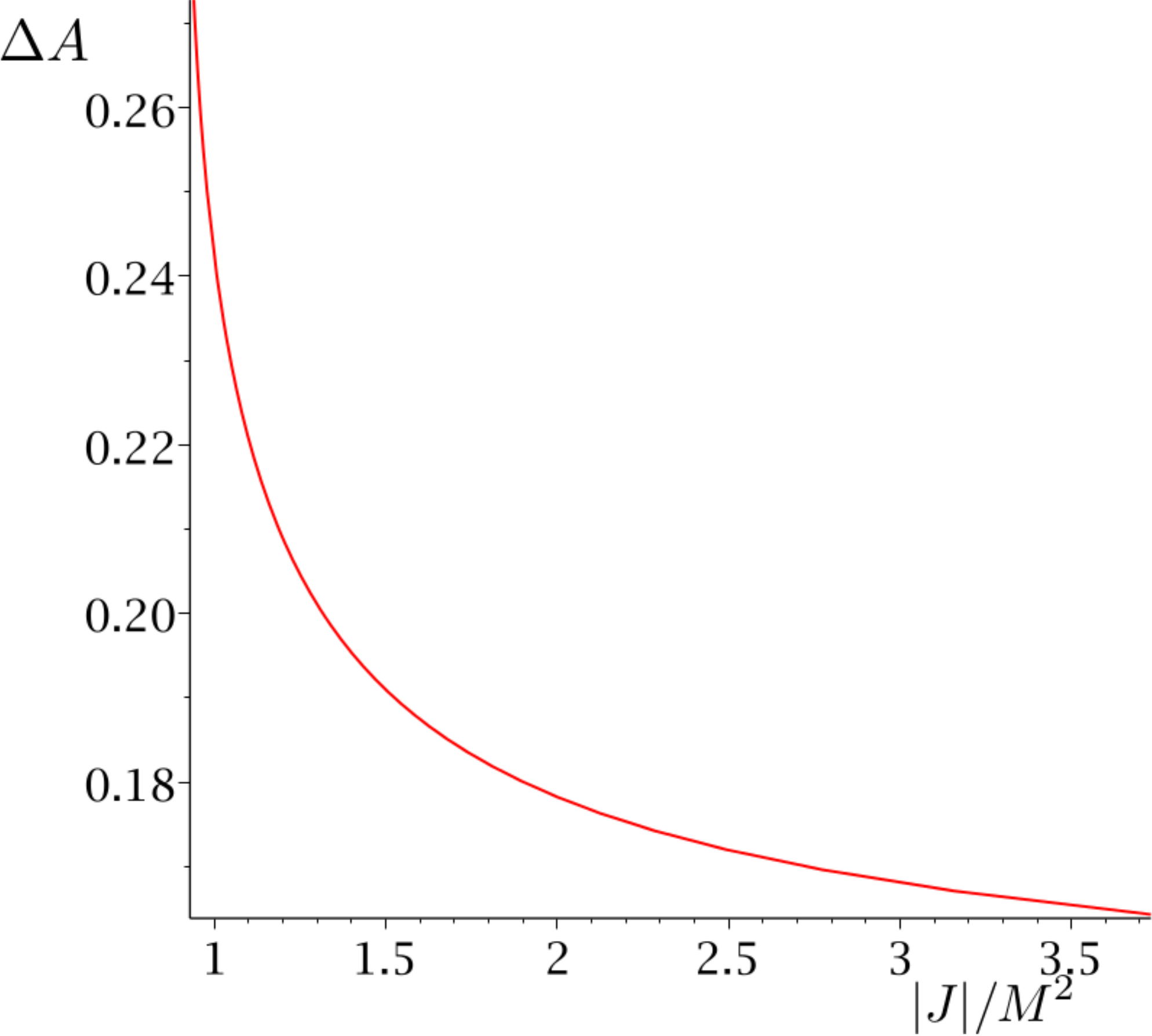} \\
                  \end{tabular}}
           \caption{\footnotesize{Dependence of the size of the compact ergoregion on the distortion parameter $c_2$ (left panel), and the spin parameter $J/M^2$ (right panel) for negative quadrupole distortion. The rotation parameter $\alpha$ is fixed to $\alpha = 0.7$.}}
		\label{areac2_neg}
\end{figure}

\begin{figure}[htp]
\setlength{\tabcolsep}{ 0 pt }{\scriptsize\tt
			\hspace{1.5cm} \includegraphics[width=6 cm]{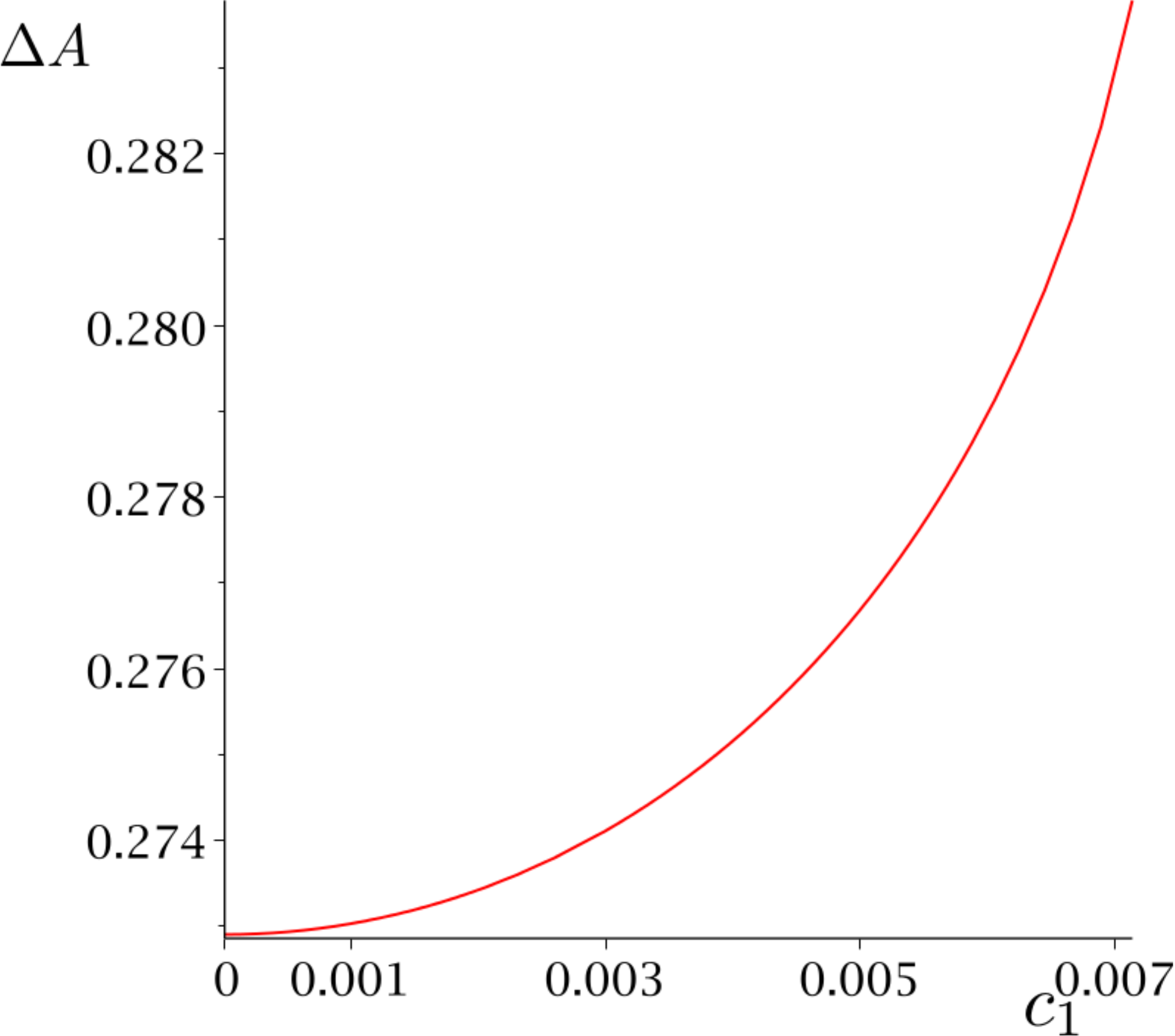}  }
           \caption{\footnotesize{Dependence of the size of the compact ergoregion on the distortion parameter $c_1$ for positive octupole distortion. The rotation parameter $\alpha$ is fixed to $\alpha = 0.7$.}}
		\label{areac1}
\end{figure}

The quantity $\Delta A$ is relevant only in the region in the parameter space where a compact ergoregion exists, i.e. for values of the distortion parameter below the critical value corresponding to each rotation parameter $\alpha$ and type of distortion. In Fig. \ref{areac2} - \ref{areac1} we illustrate the dependence of the size of the compact ergoregion $\Delta A$ on the distortion parameter in the case of positive and negative quadrupole, or octupole distortion. The rotation parameter $\alpha$ is fixed to $\alpha = 0.7$, while the parameter $m$ is set to $m=1$. The ergoregion does not depend on the scale parameter $m$ and the same qualitative behaviour is observed also for the other values of $\alpha\in (0,1)$.

For positive quadrupole or octupole distortion the ergoregion area increases as the distortion parameter $c_2$, or $c_1$ respectively, increases. For negative quadrupole distortion, we observe the opposite effect, the ergoregion area declines with the increase of the absolute value of the multipole moment. The size of the ergoregion for the isolated Kerr black hole for the same value of the rotation parameter corresponds to the limit $c_1=c_2=0$, and is equal to $\Delta A \approx 0.273$. Thus, for positive quadrupole or octupole distortion the compact ergoregion is always larger than the ergoregion of the isolated Kerr black hole. On the contrary, for negative quadrupole distortion the compact ergoregion area is always larger than the area in the non-distorted case. We do not consider separately the ergoregion area for negative octupole distortion. It coincides with the result for positive octupole distortion with the same absolute value of the distortion parameter $c_1$, due to the discrete symmetry $(y,c_1)\longrightarrow(-y,-c_1)$ between the two solutions.

We illustrate further the correlation between the ergoregion size and the value of the spin parameter $J/M^2$, where  $M$ and $J$ are the Komar mass and angular momentum on the horizon, for fixed value of the rotation parameter $\alpha$. For both positive and negative quadrupole distortion the ergoregion area decreases with the increase of the spin parameter. The isolated Kerr black hole corresponds to the point ($\Delta A \approx 0.273; J/M^2\approx 0.94$). For positive distortion the ergoregion area of the isolated Kerr black hole represents a lower limit, i.e. the ergoregion of a distorted black hole is always larger for any value of the spin parameter. On the contrary, for negative distortion the isolated case is an upper limit. In the case of octupole distortion, the ergoregion area for fixed $\alpha$ does not depend on the spin parameter $J/M^2$, i.e. infinitely many different configurations are possible for the same values of the Komar mass and angular momentum. This effect occurs in all the cases when only odd distortion parameters are present. Then, the Komar mass and angular momentum are independent of the multipole moments and coincide with those for the isolated Kerr black hole. On the other hand, the ergoregion configuration is sensitive to the values of the distortion parameters $c_n$.

\subsection{General distortion}

\begin{figure}[htp]
\setlength{\tabcolsep}{ 0 pt }{\scriptsize\tt
		\begin{tabular}{ cc }
	       \includegraphics[width=5 cm]{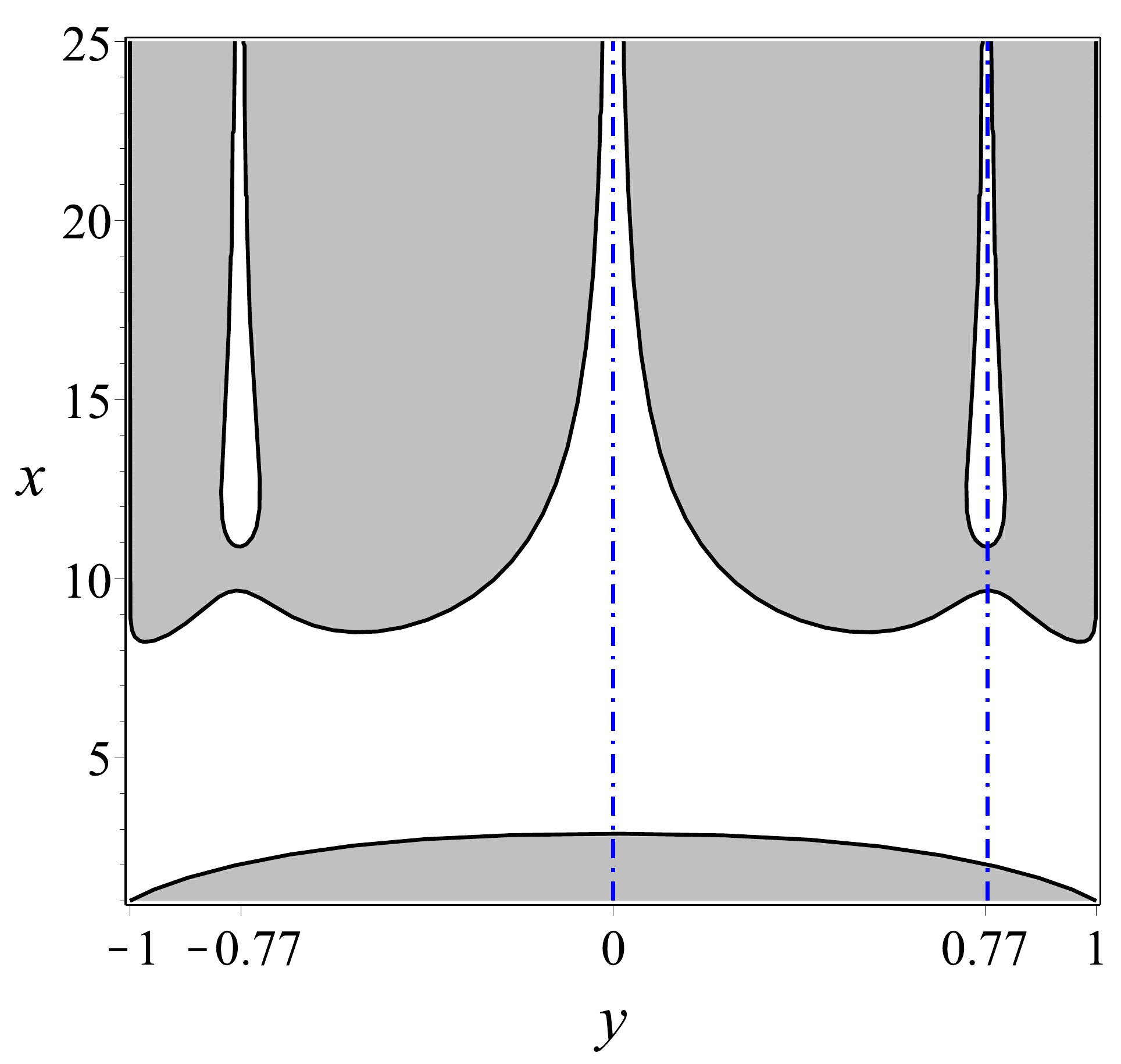} &
            \includegraphics[width=5 cm]{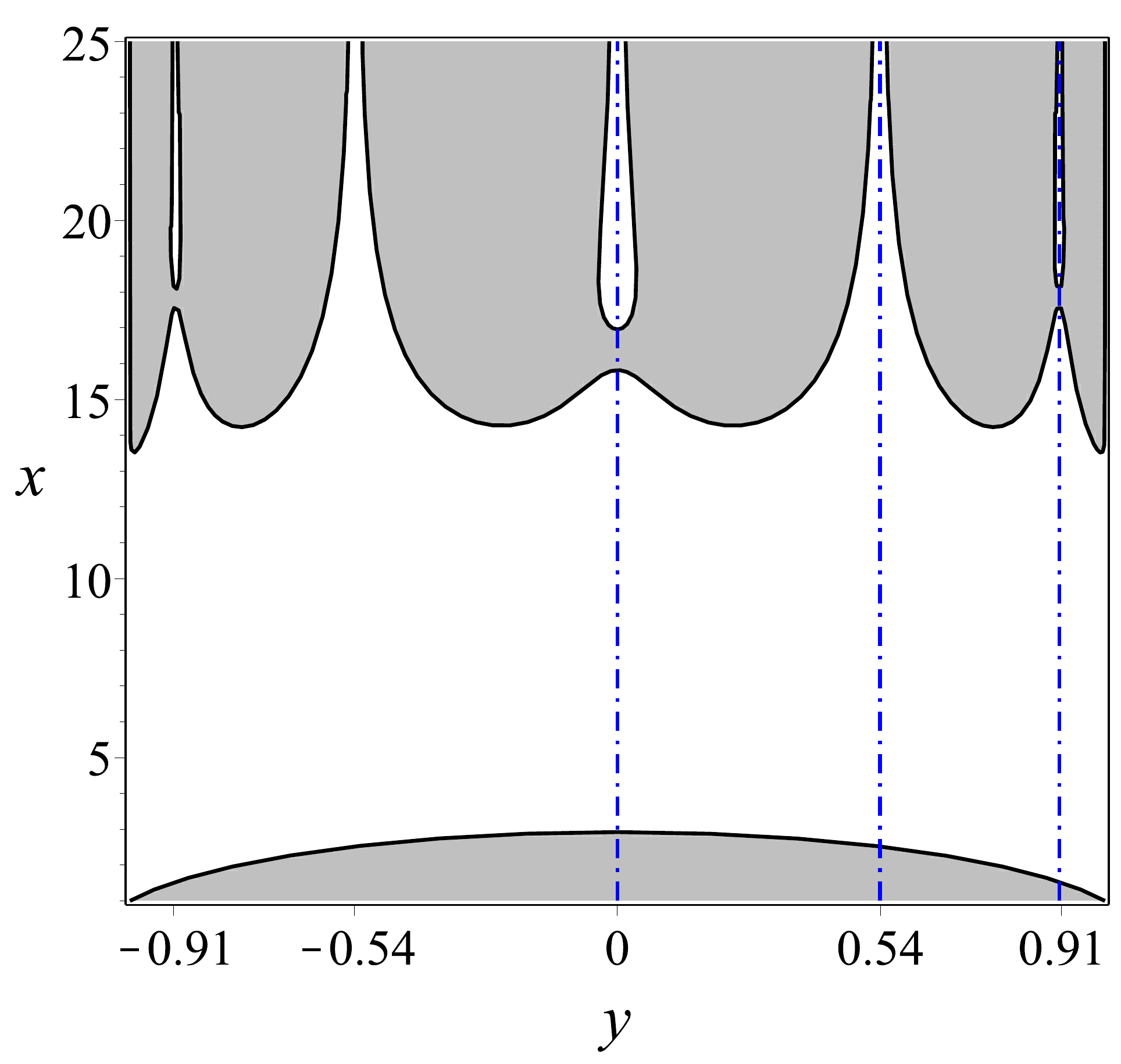} \\
            $c_4=0.4\times10^{-3}$ & $c_6=0.4\times10^{-6}$ \\
            $c_n=0$, $n\neq4$ & $c_n=0$, $n\neq6$ \\
            \includegraphics[width=5 cm]{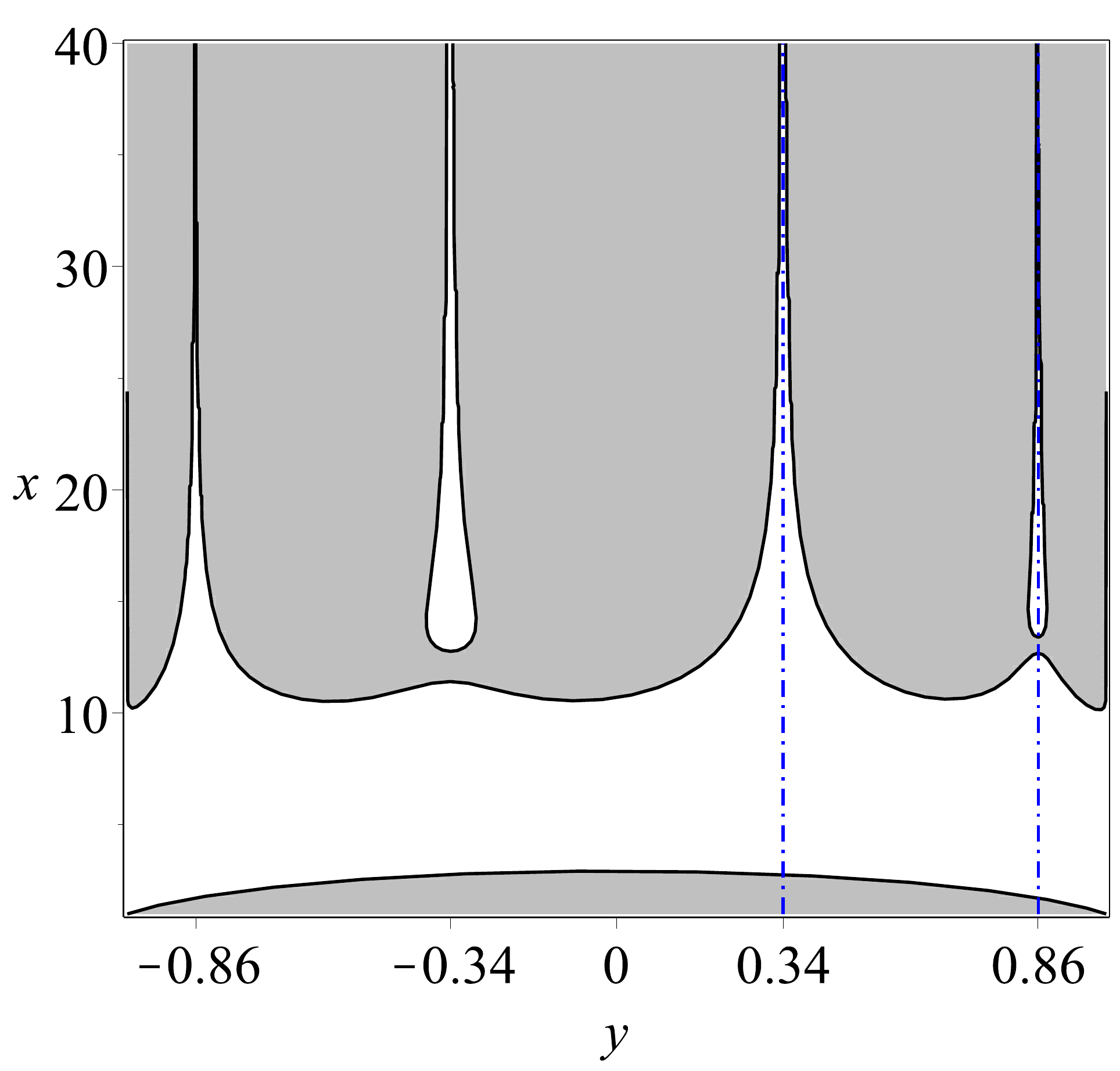} &
            \includegraphics[width=5 cm]{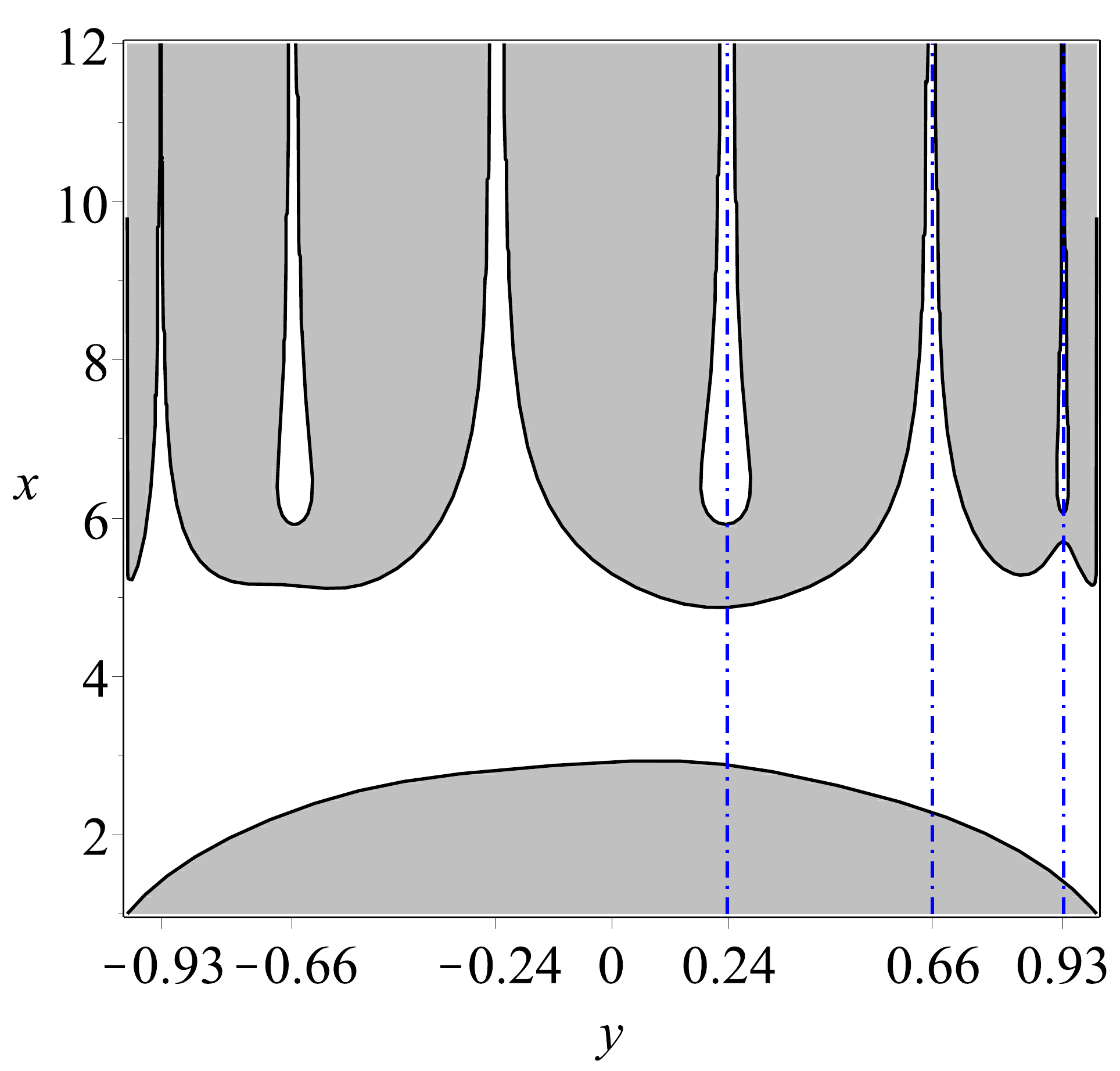}\\
             $c_5 = -c_1 = 0.2\times10^{-4}$ & $c_7=-c_1 = 0.2\times10^{-4} $ \\
             $c_n=0$, $n\neq 1,5$ & $c_n=0$, $n\neq 1,7$
                                   \end{tabular}}
           \caption{\footnotesize{Ergoregion (grey area) for even and odd distortion in cases when a compact ergoregion is present. The rotation parameter is fixed to $\alpha=0.7$.  The location of the cross-sections $y=y_i$ in the vicinity of which static regions are formed for large $x$ is specified on the $y$-axis.}}
		\label{ergo_gen}
\end{figure}

To gain intuition about the ergoregion of the distorted Kerr black hole we extend the analysis to some more general cases of distortion. In each case we choose again a single independent parameter $c_n\neq0$, which, however, corresponds to distortion of higher multipolarity. The solutions fall in three classes -  those characterized by even multipole moments, those characterized by odd multipole moments and the ones characterized by mixed, odd and even, multipole moments. Here, we do not consider the mixed case. In the case of even distortion we consider the special case defined by the following values of the distortion parameters $\{ c_{2n} \neq 0, \, c_{2k} = 0, k\neq n; \, c_{2k+1} =0, \forall k \in \mathcal{N}\}$, and in the case of odd distortion we consider the special case defined by the multipole moments $\{c_{1} = - c_{2n+1}\neq 0, \, c_{2k+1} =0, k\neq 0, n;\, c_{2k} = 0, \forall k\in\mathcal{N} \}$. The ergoregions for odd and even distortion behave in a similar way, respectively, as in the quadrupole and octupole cases we investigated. If we consider the behaviour of the ergoregion for fixed $\alpha$ as a function of the distortion parameter $c_n$, for high absolute values of $c_n$  we observe a single connected ergoregion extending to infinity. Again, there exists a critical value $c_n= c_{crit}$, depending on the value of $\alpha$, at which a transition in the type of the ergoregion occurs. For $c_n<c_{crit}$ the ergoregion consists of a compact part encompassing the horizon and a disconnected non-compact part. We have illustrated some configurations of the ergoregion of this type for the lowest multipole moments in Fig. \ref{ergo_gen}.

We can deduce some general properties of the ergoregion by analyzing the function $A(x,y)$, which determines its location. For general distortion parameters $c_n$ it can be presented in the form
\begin{eqnarray}
A(x,y) = (x^2-1)\left(1-\alpha^2e^{4\chi_1}\right)^2
 - 4\alpha^2(1-y^2)e^{4\chi_1}\cosh^2(2\chi_2),
\end{eqnarray}
where the functions $\chi_1$ and $\chi_2$ are given by
\begin{eqnarray}
 \chi_1&=& \sum^\infty_{n=0}c_{2n}\left[x\sum^{n-1}_{l=0}R^{2l}P_{2l}\left(\frac{xy}{R}\right) - y\sum^{n-1}_{l=0}R^{2l+1}P_{2l+1}\left(\frac{xy}{R}\right)\right]   \nonumber \\[2mm]
 &+& \sum^\infty_{n=0}c_{2n+1}\left[x\sum^{n-1}_{l=0}R^{2l+1}P_{2l+1}\left(\frac{xy}{R}\right) - y\sum^{n}_{l=0}R^{2l}P_{2l}\left(\frac{xy}{R}\right) \right], \nonumber \\[2mm]
 \chi_2 &=& \sum^\infty_{n=0}c_{2n}\left[ y\sum^{n-1}_{l=0}R^{2l}P_{2l}\left(\frac{xy}{R}\right) - x\sum^{n-1}_{l=0}R^{2l+1}P_{2l+1}\left(\frac{xy}{R}\right)\right]  \nonumber \\[2mm]
 &+& \sum^\infty_{n=0}c_{2n+1}\left[ y\sum^{n-1}_{l=0}R^{2l+1}P_{2l+1}\left(\frac{xy}{R}\right) - x\sum^{n}_{l=0}R^{2l}P_{2l}\left(\frac{xy}{R}\right)\right]\nonumber,
\end{eqnarray}
\noindent
and $R^2 = x^2+y^2-1$. For the special cases of distortion which we consider these functions reduce to the expressions
\begin{eqnarray}\label{chi_even}
 \chi_1&=& c_{2n}\left[x\sum^{n-1}_{l=0}R^{2l}P_{2l}\left(\frac{xy}{R}\right) - y\sum^{n-1}_{l=0}R^{2l+1}P_{2l+1}\left(\frac{xy}{R}\right)\right], \nonumber \\[2mm]
  \chi_2 &=& c_{2n}\left[y\sum^{n-1}_{l=0}R^{2l}P_{2l}\left(\frac{xy}{R}\right) - x\sum^{n-1}_{l=0}R^{2l+1}P_{2l+1}\left(\frac{xy}{R}\right)\right],
\end{eqnarray}
\noindent
when $\{ c_{2n} \neq 0, \, c_{2k} = 0, k\neq n; \, c_{2k+1} =0, \forall k \in \mathcal{N}\}$, and
\begin{eqnarray}\label{chi_odd}
 \chi_1&=& c_{2n+1}\left[x\sum^{n-1}_{l=0}R^{2l+1}P_{2l+1}\left(\frac{xy}{R}\right) - y\sum^{n}_{l=1}R^{2l}P_{2l}\left(\frac{xy}{R}\right) \right], \nonumber \\[2mm]
 \chi_2 &=&  c_{2n+1}\left[y\sum^{n-1}_{l=1}R^{2l+1}P_{2l+1}\left(\frac{xy}{R}\right) - x\sum^{n}_{l=0}R^{2l}P_{2l}\left(\frac{xy}{R}\right) \right],
\end{eqnarray}
\noindent
when $\{c_{1} = - c_{2n+1}\neq 0, \, c_{2k+1} =0, k\neq 0, n;\, c_{2k} = 0, \forall k\in\mathcal{N} \}$. The functions $\chi_1$ and $\chi_2$ are related by the shift $x\longleftrightarrow y$, i.e. if we perform the transformation $x\longleftrightarrow y$, it is satisfied that $\chi_1 \longleftrightarrow \chi_2$.

Examining the expressions (\ref{chi_even}) - (\ref{chi_odd}) we see that the ergoregion possesses similar properties as in the quadrupole and octupole cases, respectively. It does not contain the axis, and always extends to infinity, both for even and odd distortion. Indeed, for even distortion $c_{2n}\neq 0$ the functions $\chi_1$ and $\chi_2$ possess the asymptotic behaviour $\chi_1\sim x^{2n-1}$,  $\chi_2\sim x^{2n}$, up to the leading order in $x$. For odd distortion $c_{2n+1}\neq 0$ we obtain  $\chi_1\sim x^{2n}$ and $\chi_2\sim x^{2n+1}$. Consequently, the function $A(x,y)$ is negative for large values of $x$. The discrete symmetries in the cases of the quadrupole and octupole distortion are also retained. The ergoregion for even multipole moments is invariant with respect to the shift $y\longleftrightarrow -y$, and for odd multipole moments it is invariant with respect to the transformation $(y, c_{2n+1})\longleftrightarrow (-y, -c_{2n+1})$.

As in the case of quadrupole and octupole distortion, a number of static regions which extend to infinity are formed in the vicinity of some special cross-sections located at $y= y_i$. They correspond to the values of $y$, for which the leading order term in the asymptotic expansion of the function $\chi_2$ vanishes. The function $\chi_2$ possesses the following general form for even and odd distortion, respectively
\begin{eqnarray}
&&\chi_2 = c_{2n}\,y \,Q(x^2,y^2),\quad~~~c_{2n}\neq0  \nonumber\\[2mm]
&&\chi_2 = c_{2n+1}\,x \,\widetilde{Q}(x^2,y^2),  \quad~~~c_{2n+1} = -c_1\neq0\nonumber
\end{eqnarray}
where $Q(x^2,y^2)$ and $\widetilde {Q}(x^2,y^2)$ are polynomials in the variables $x^2$ and $y^2$. For example, in the simplest cases it is given by
\begin{eqnarray}
&&c_2\neq0, \quad~~~ \chi_2 = c_2\,y(1-x^2), \nonumber\\
&&c_4\neq0, \quad~~~ \chi_2 = \frac{1}{2}c_4\,y(1-x^2)\left[(5y^2-3)x^2 - y^2 + 3\right], \nonumber\\
&&c_6\neq0, \quad~~~ \chi_2 = \frac{1}{8}c_6\,y(1-x^2)\bigg[(252y^4 - 280y^2 + 60)x^4  \nonumber \\[1mm]
&&\quad~~~~~~~~~~~~-(168y^4-245y^2+75)x^2+12y^4-25y^2+15 \bigg]\nonumber \\[3mm]
&&c_3 \neq0, \quad~~~ \chi_2 = \frac{1}{2}c_3\,x(1-x^2)(3y^2-1), \nonumber\\
&&c_5 \neq0, \quad~~~ \chi_2 = \frac{1}{8}c_5\,x(1-x^2)\left[(35y^4-30y^2 + 3)x^2 - 15y^4+30y^2 -7\right], \nonumber\\
&&c_7 \neq0, \quad~~~ \chi_2 = \frac{1}{16}c_7\,x(1-x^2)\bigg[(231y^6 - 315y^4 + 105y^2 - 5)x^4  \nonumber \\[1mm]
&&\quad~~~~~~~~~~~~-(210y^6-420y^4+210y^2-16)x^2+35y^6-105y^4+105y^2-19 \bigg]\nonumber
\end{eqnarray}

For even multipole moments there is always a static region in the vicinity of the cross-section $y=0$. The location of the other cross-sections $y=y_i$ is determined by the real zeros of the polynomial coefficients in front of the highest degree of $x$, which is involved in the function $\chi_2$. For example, in the case $c_4\neq0$ static regions are formed for large $x$ around the cross-sections $y = \pm \sqrt{\frac{3}{5}}$. The location of the cross-sections $y=y_i$ for the simplest cases of distortion is presented in table \ref{static}, and illustrated in Fig. \ref{ergo_gen}.
\begin{table}[h]
\begin{center}
\begin{tabular}{ | c| c || c|c| }
\hline
Even distortion &  $y_i$& Odd distortion &  $y_i$\\
\hline
 $c_2\neq0$ &  0 &$c_3\neq0$ & $\pm$0.5773  \\
\hline
$c_4\neq0$ & 0, $\pm$ 0.7746 & $c_5\neq0$ & $\pm$0.3399, $\pm$0.8611 \\
\hline
$c_6\neq0$ & 0, $\pm$ 0.9062, $\pm$ 0.5385& $c_7\neq0$ & $\pm$0.2386 $\pm$0.6612 $\pm$0.9325\\
\hline
\end{tabular}
\end{center}
\caption{\footnotesize{Location of the cross-sections $y=y_i$ in the vicinity of which static regions are formed for large $x$.}}\label{static}
\end{table}

\section{Singularities}
In this section, we study the space-time singularities and whether there are singularities outside the horizon. Such analysis was partly done for quadrupole distortion in \cite{Breton:1997}, \cite{Shoom}. We extend it further, and consider also cases of more general distortion. The spacetime singularities are determined by the equation $B=0$, where $B(x,y)$ is the metric function (\ref{B}), since the Kretschmann invariant diverges for its solutions. It is equivalent to the following system of equations
\ba
&&x+1+(x-1)ab=0,\n{B1}\\
&& (1+y)a+(1-y)b=0. \n{B2}
\ea
We can see that, the set of solutions of (\ref{B1})-(\ref{B2}) is contained within the set of solutions of the equation $A(x,y)=0$. Consequently, all the curvature singularities which are located in the domain of outer communication lie on the boundary of the ergoregion.

\subsection{Quadrupole distortion}

For quadrupole distortion all the solutions of eq. (\ref{B2}) belong to the cross-section $y=0$. Therefore, the system of equations decouples and we can obtain the $x$-coordinate by solving (\ref{B1}).  If the distortion parameter $c_2$ is negative, it is satisfied only for $x<-1$. Therefore, for negative quadrupole distortion the spacetime singularities are always located in the equatorial plane behind the inner horizon. For positive $c_2$, there always exists a solution in the domain of outer communication $x>1$. As we already mentioned, it is located on the cross-section of the ergosurface with the equatorial plane. We will further argue that, the singularities do not lie on the boundary of the compact part of the ergoregion in the vicinity of the horizon, when such exists. Considering the analytical properties of eqs. (\ref{A_c2}) and (\ref{B1}) - (\ref{B2}), and the numerical investigation of the structure of the ergoregion in section 3.1, we can conjecture that, they are located on the boundary of the non-compact part. Indeed, the cross-section of the boundary of the ergoregion with the equatorial plane $y=0$ is determined by the equation

\begin{eqnarray}
A(x,y=0) &=& B_1\tilde{B}_1 =0,\label{A_singc2}
\end{eqnarray}
where
\begin{eqnarray}
B_1 &=& x+1+(x-1)ab = x+1 -\alpha^2(x-1)e^{4c_2x},\label{B1_c2} \\
\tilde{B}_1 &=& x-1+(x+1)ab = x-1 - \alpha^2(x+1)e^{4c_2x}.\label{B11_c2}
\end{eqnarray}

Let $x_1>1$ be a solution to eq. (\ref{B1_c2}) for some values of $c_2$ and $\alpha$, i.e. it corresponds to a ring singularity located at ($x = x_1, y=0$). For any $c_2>0$ and $\alpha$ there exists a single solution to (\ref{B1_c2}) such that $x_1>1$. Let $x_2>1$ be a solution to eq. (\ref{B11_c2}) for the same values of $c_2$ and $\alpha$, so it determines a non-singular point in the cross-section of the boundary of the ergoregion with the equatorial plane $y=0$. For any couple of such solutions, provided that $x_2$ exists, it is satisfied that $x_1>x_2$. Consequently, the singularity corresponds to the most distant point, with respect to the horizon, in the cross-section of the boundary of the ergoregion with the equatorial plane. Taking into account the behaviour of the ergoregion illustrated on Fig. \ref{c2gen}, we can conclude that this point belongs to the boundary of the static region which is formed for large $x$ in the vicinity of the cross-section $y=0$. The compact ergosurface, when it exists, delimits the part of the ergoregion closest to the horizon, therefore the singularity cannot lie on it.  We have illustrated, the location of the singularity for $\alpha=0.7$, $c_2 =0.008$ in Fig. \ref{sing}. The implicit functions defined by the equations $B_1=0$ and $\tilde{B}_1=0$ are presented with red and dark red lines, respectively. The cross-section of the former with the line $y=0$ corresponds to the singular point.

We should note that our argument relies on the fact that, the boundary of the non-compact static region around the cross-section $y=0$, intersects the equatorial plane for some value of $x$. This statement follows from our numerical investigation of the behaviour of the ergoregion performed in section 3.1, and is not proven rigorously by analytical techniques. Hence, the same level of precision applies for our results about the location of  the curvature singularities.

\begin{figure}[htp]
\setlength{\tabcolsep}{ 0 pt }{\scriptsize\tt
		\begin{tabular}{ cc }
	\hspace{1cm}\includegraphics[width=6 cm]{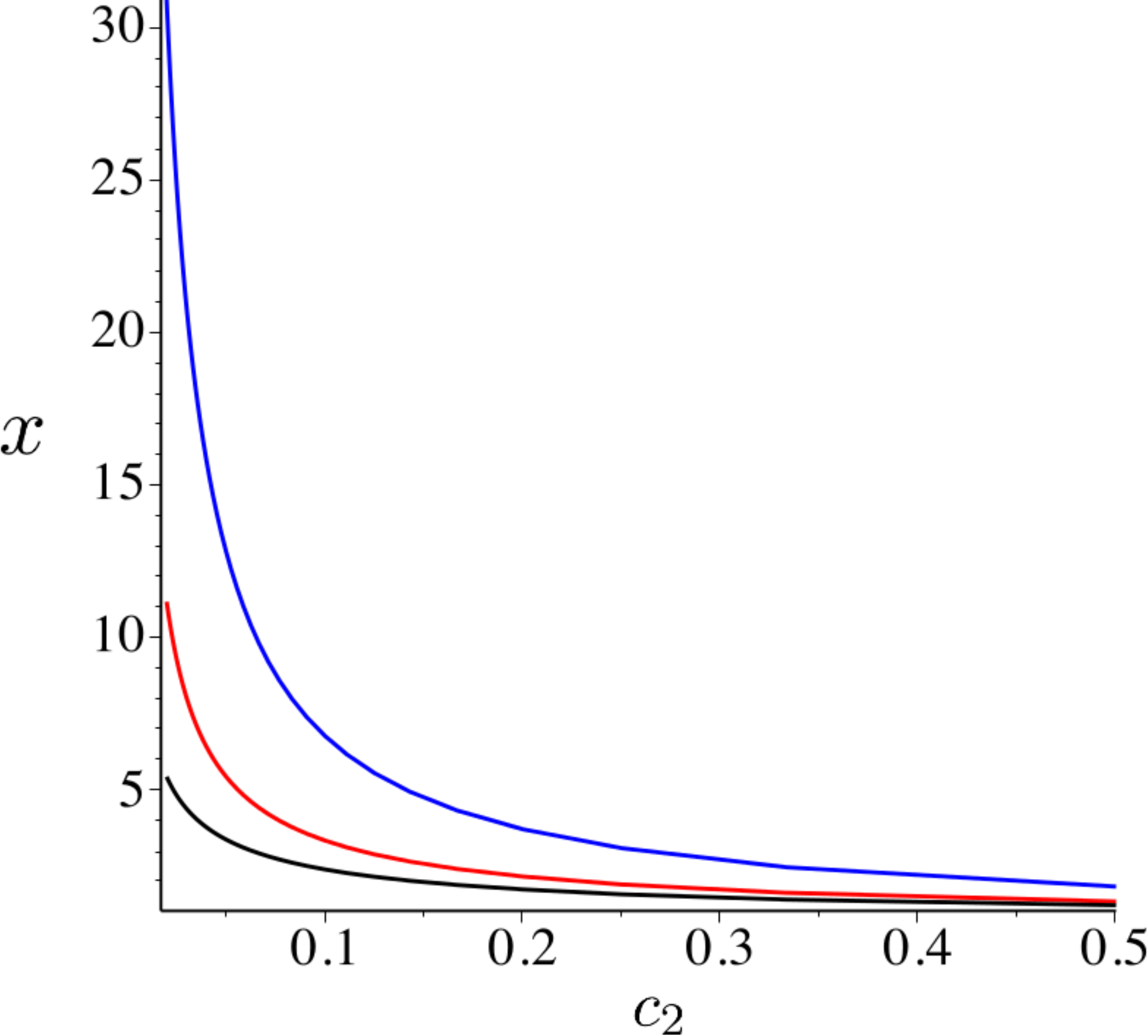} &\hspace{1cm}
            \includegraphics[width=6 cm]{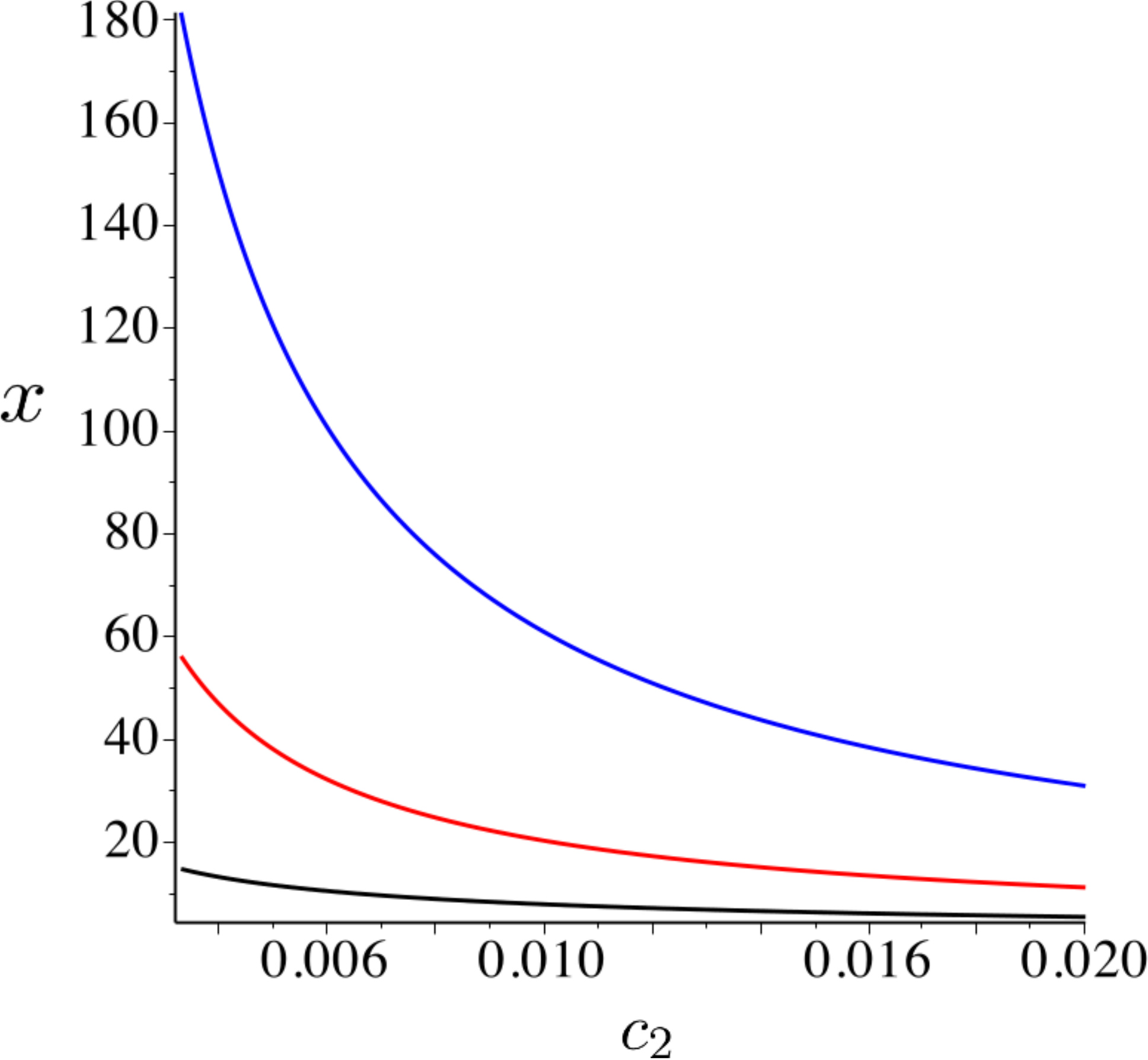} \\
            a & b
           \end{tabular}}
           \caption{\footnotesize{Location of the spacetime singularity on the equator for positive quadruple distortion for a) strong distortion $1/2<c_2<1/50$, and b) weak distortion $1/50<c_2<1/300$. Each curve corresponds to a different value of the rotation parameter $\alpha$. The blue  line is for $\alpha=0.3$, the red line is for $\alpha=0.7$, and the black line is for $\alpha=0.97$.  }}
		\label{Sing1}
\end{figure}
\begin{figure}[h]
\setlength{\tabcolsep}{ 0 pt }{\scriptsize\tt
		\begin{tabular}{ ccc }
	       \includegraphics[width=5 cm]{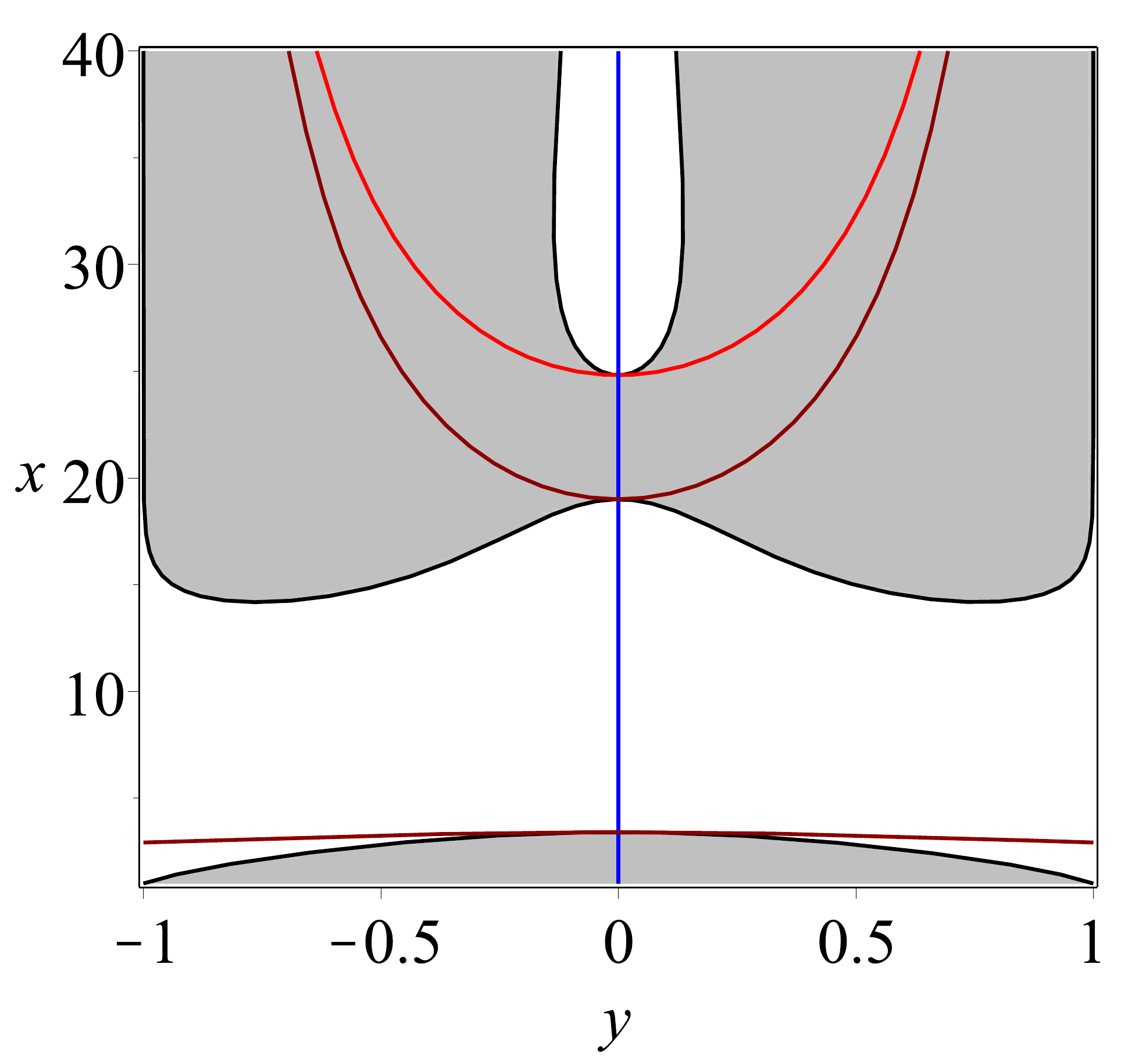} &
            \includegraphics[width=5 cm]{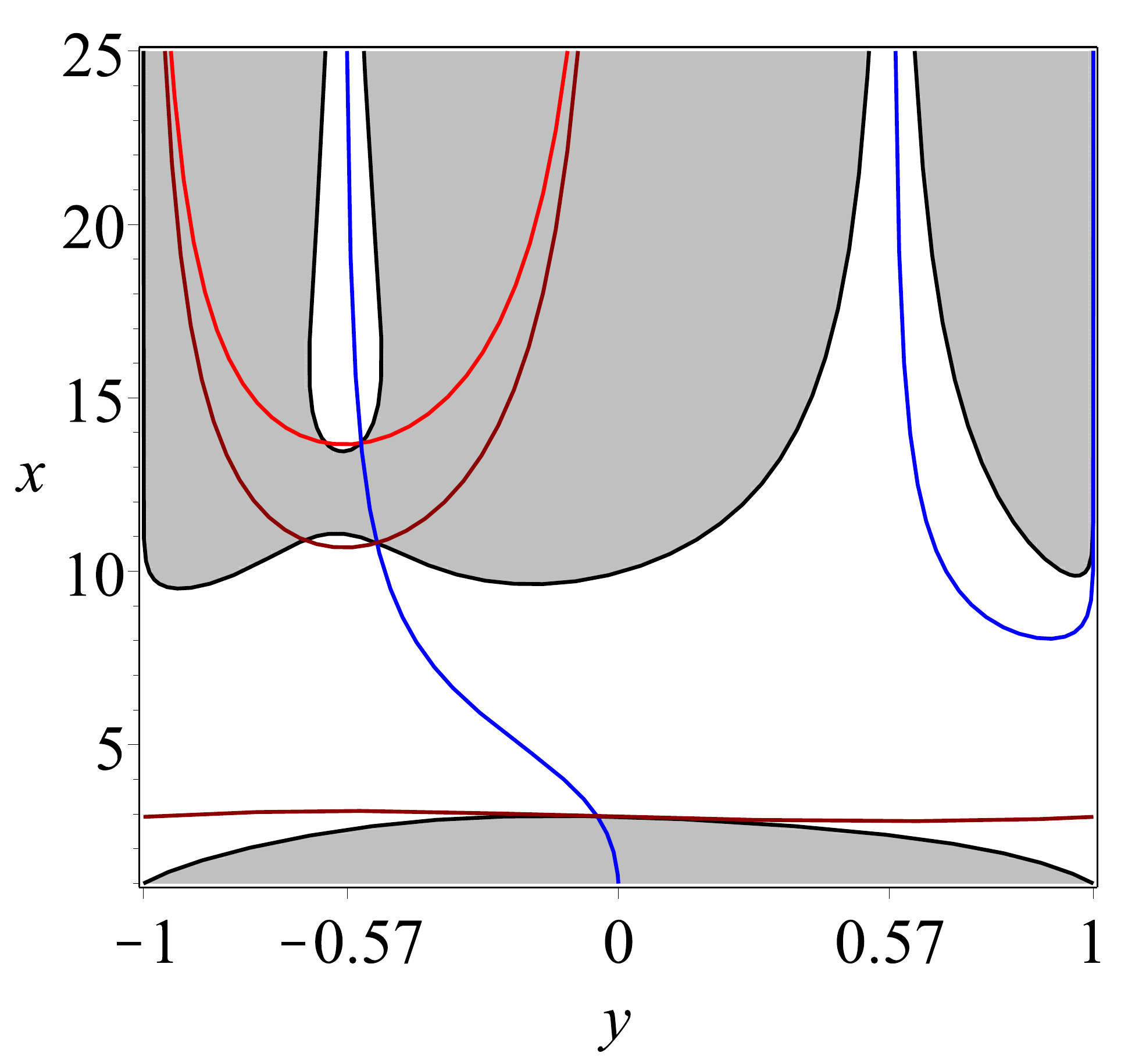} &
            \includegraphics[width=5 cm]{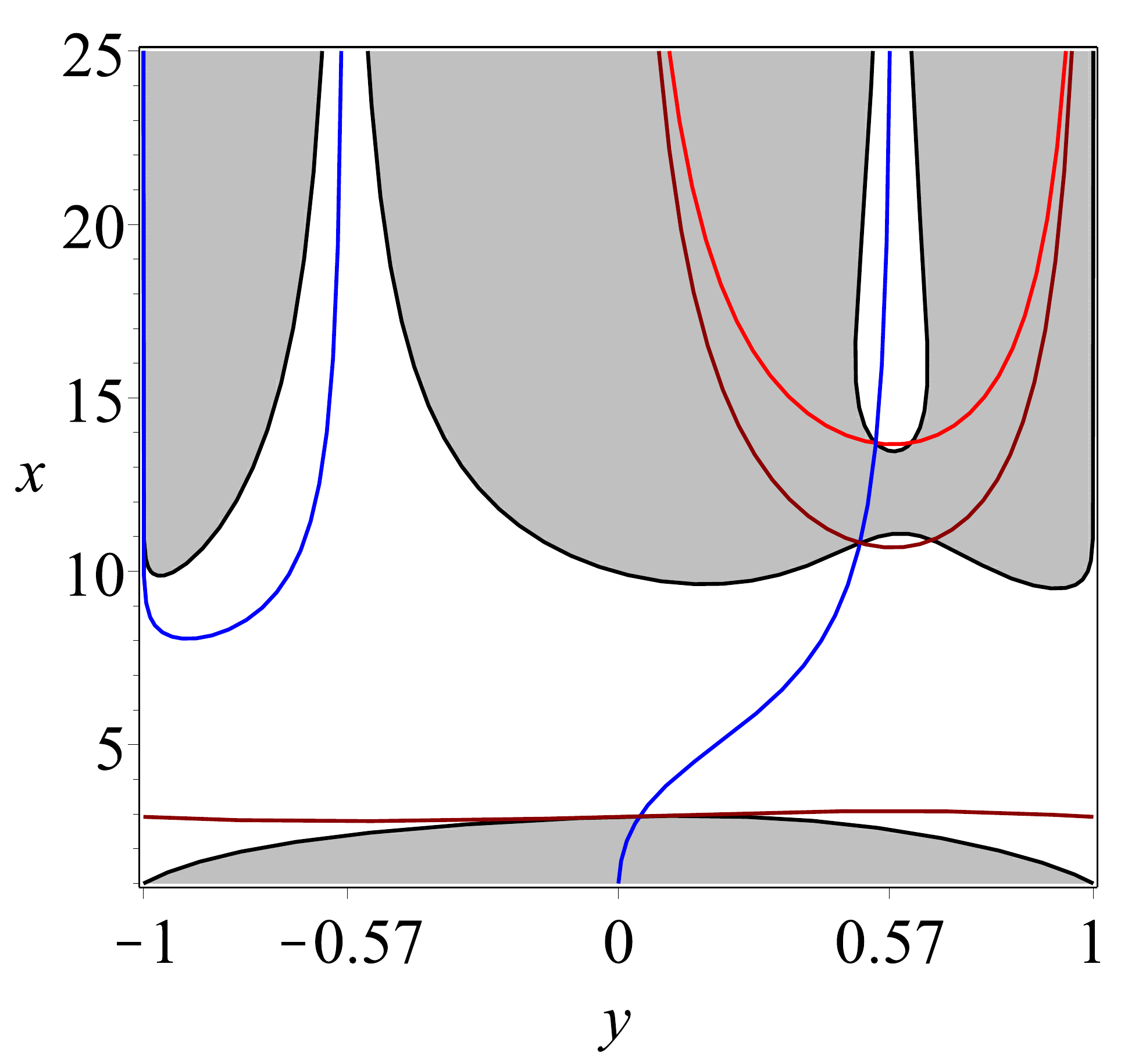} \\
            $c_2=0.008$ & $c_1 = -c_3=0.002$ & $c_1 = -c_3= -0.002$ \\
            $a)$& $b)$ &$c)$ \\
                                               \end{tabular}}
           \caption{\footnotesize{Location of the curvature singularities for a) positive quadrupole, b) positive octupole, and c) negative octupole distortion. The implicit functions defined by $B_1=0$, $B_2=0$, and $\tilde{B}_1=0$  are presented with red,  blue and dark red lines respectively. The singular points correspond to the cross-sections of the former two.}}
		\label{sing}
\end{figure}

As the value of the distortion parameter $c_2$ decreases the singularity moves further away from the horizon. This effect is consistent with the fact that the non-compact static region around the cross-section $y=0$ becomes more distant when $c_2$ decreases. Fig. \ref{Sing1} illustrates the location of the singularity for positive quadrupole distortion as a function of the multipole moment for different values of $\alpha$. The singularity is closer to the horizon for larger values of the rotation parameter $\alpha$, as compared to smaller values.

As we have mentioned before,  in this work, we consider the distorted Kerr black hole as a local solution, which is valid only in a certain neighbourhood of the horizon. A global solution can be constructed if (\ref{metric3}) is extended to an asymptotically flat solution by some sewing technique. The location of the outer ring singularity is a good measure for defining the range of the validity of the solution. The distorted black hole solution is supposed to describe the interior region with respect to some distribution of external sources. The interior solution will be characterized by a regular compact ergoregion in the vicinity of the horizon, in the region of the parameter space when such exists.

\begin{figure}[htp]
\setlength{\tabcolsep}{ 0 pt }{\scriptsize\tt
		\begin{tabular}{  cc }
	\hspace{1cm}\includegraphics[width=6 cm]{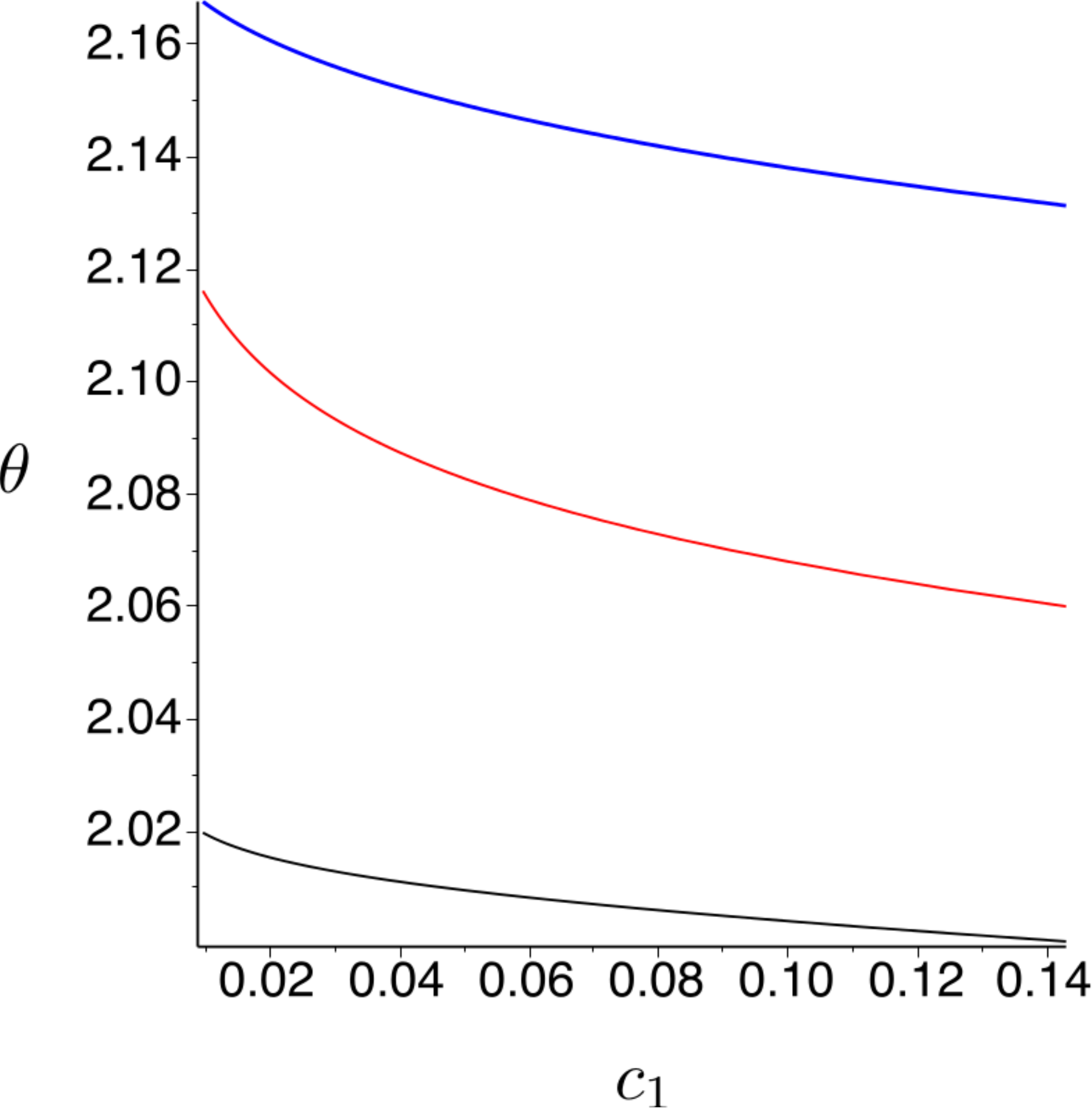} &\hspace{1cm}
            \includegraphics[width=6 cm]{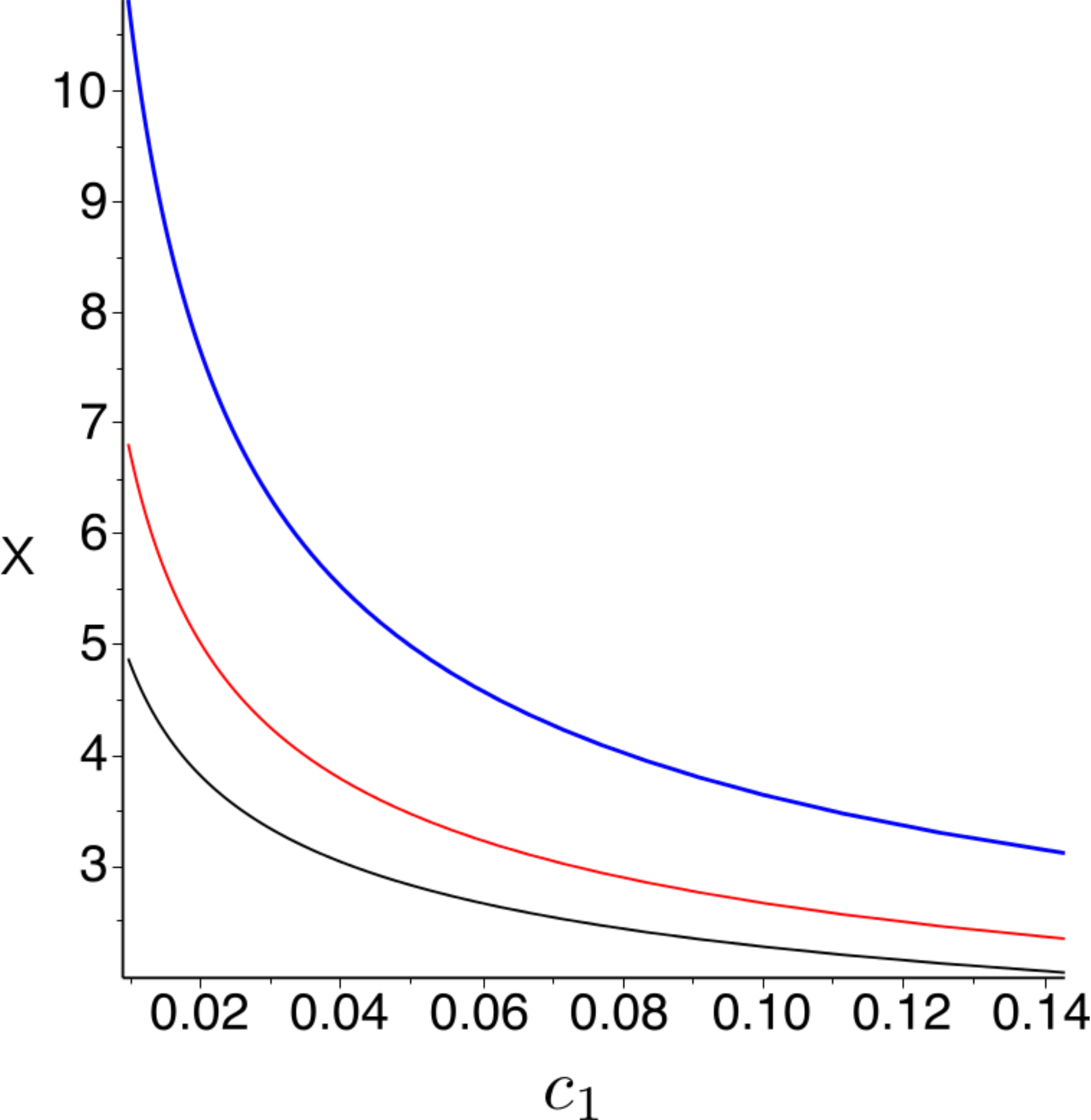} \\
            a & b
           \end{tabular}}
           \caption{\footnotesize{Location of the spacetime singularity for positive octupole distortions a) the $\theta$ position of the singularity and b) the $x$ position of the singularity. The blue line is for $\alpha=0.3$, the red line for $\alpha=0.7$, and the black line is for $\alpha=0.97$.  }}
		\label{Sing2}
\end{figure}

In the case of the isolated Kerr black hole the spin parameter should satisfy the condition
\be
a_{*}^2=\frac{J^2}{M^4}\leq1,
\ee
or otherwise, a naked singularity is present. If we consider the distorted Kerr solution, for which the sum $\sum_{n=1}^{\infty} c_{2n}$ is positive, $a_{*}^2$ is still less than one. However, if the distortion parameters satisfy  $\sum_{n=1}^{\infty} c_{2n}< 0$, $a_{*}^2$ can be greater than one. It is interesting that, even though there are no spacetime singularities outside the horizon for the case of negative quadrupole distortions, $a_{*}^2$ can be greater than one, in contrast to the case of the undistorted Kerr black hole, where $a_{*}^2>1$ means a naked singularity.

\subsection{Octupole distortion}

For octupole distortion, the singularities are not located on a particular cross-section of constant $y$. Therefore, eqs. (\ref{B1}) and (\ref{B2}) do not decouple but need to be solved simultaneously in order to find the $x$ and $y$ coordinates of the singularity. The singularity lies on the boundary of the ergoregion, and we argue again that, it is not located on the compact ergosurface when such exists. Similar to the quadrupole case, the reasoning is based on the analytical properties of eqs. (\ref{A_c1}) and (\ref{B1}) - (\ref{B2}), and the numerical investigation of the structure of the ergoregion in section 3.2. The equation which determines the boundary of the ergoregion can be represented in the form,

\begin{eqnarray}
A(x,y) &=& B_1\tilde{B}_1 - B_2\tilde{B}_2 =0,\label{A_singc1}
\end{eqnarray}
where,
\begin{eqnarray}
B_1 &=& x+1+(x-1)ab = x+1 -\alpha^2(x-1)e^{4\chi_1},\label{B1_c1} \\[1mm]
\tilde{B}_1 &=& x-1+(x+1)ab = x-1 - \alpha^2(x+1)e^{4\chi_1},\label{B11_c1} \\[2mm]
B_2 &=&(1+y)a + (1-y)b = 2\alpha e^{2\chi_1}\left(\sinh{2\chi_2} - y\cosh{2\chi_2}\right),\label{B2_c1} \\[1mm]
\tilde{B}_2 &=& (1-y)a + (1+y)b = 2\alpha e^{2\chi_1}\left(\sinh{2\chi_2} + y\cosh{2\chi_2}\right)\, , \label{B22_c1}
\end{eqnarray} and the functions $\chi_1$ and $\chi_2$ are defined by,
\begin{eqnarray}
\chi_1 &=& \frac{1}{2}c_1\,y(y^2-1)(3x^2-1), \nonumber\\[1mm]
\chi_2 &=& \frac{1}{2}c_1\,x(x^2-1)(3y^2-1).
\end{eqnarray}

The coordinates of the singular points  in the $(x,y)$ - plane for fixed values of the parameters $c_1$ and $\alpha$ are solution to the system $\{B_1=0, B_2 =0 \}$, which also belongs to the ergosurface. We consider for definiteness a positive distortion parameter $c_1>0$. Then, eq. (\ref{B1_c1}) is satisfied only in the range $y<0$, while eq. (\ref{B2_c1}) is satisfied in the ranges $ \frac{1}{\sqrt{3}}<y<1$ and $ -\frac{1}{\sqrt{3}}<y\leq0$. Let us consider the solution of eq. (\ref{B2_c1}) for $ -\frac{1}{\sqrt{3}}<y\leq0$. By the implicit function theorem  it defines a unique continuous function $x = f(y)$ in this interval. Moreover, the function $f(y)$ is monotonically decreasing and it is satisfied that  $x=1$ when $y=0$, and $x\rightarrow +\infty$ when $y\rightarrow-\frac{1}{\sqrt{3}}$. In a similar way, eq. (\ref{B1_c1}) defines a unique continuous function $x = g(y)$ in the interval $ -\frac{1}{\sqrt{3}}<y\leq0$, which is monotonically increasing and satisfies $x\rightarrow +\infty$ when $y\rightarrow 0$. Consequently, the system $\{B_1 = 0, B_2=0\}$ always has a single solution in the interval $ -\frac{1}{\sqrt{3}}<y<0$.

The crossing points of the curve $f(y)$ determined by the equation $B_2=0$ with the boundary of the ergoregion are divided into two types - such that equation $B_1=0$ is satisfied, and consequently corresponding to singularities, and regular points in which equation $\tilde{B}_1 = 0$ is satisfied. Let us denote by $(x_1, y_1)$ a solution to the system $\{B_1 = 0, B_2=0\}$, and by $(x_2,y_2)$ a solution to the system $\{ \tilde{B}_1=0, B_2=0 \}$ for the same values of the parameters $\alpha$ and $c_1$. Examining these equations we can prove that for any two such solutions, provided the latter exists, it is satisfied that $x_1>x_2$. Consequently, the singularity is located at the crossing point of the curve $f(y)$ with the boundary of the ergoregion which possesses the highest value of the coordinate $x$. Taking into account the properties of the function $f(y)$ and the behaviour of the ergoregion  illustrated in Fig. \ref{c1gen},  we can conclude that this point belongs to the boundary of the non-compact static region around $y=-\frac{1}{\sqrt{3}}$, which intersects the cross-section $y=-\frac{1}{\sqrt{3}}$ and extends to infinity. Consequently, the singularity does not lie on the compact ergosurface.

In Fig. \ref{sing}, we illustrate the location of the singularity for $\alpha= 0.7$, $c_1=0.002$. The implicit functions defined by the equations $B_1=0$,  $B_2=0$ , and $\tilde{B}_1=0$ are presented with red, blue, and dark red lines respectively. The singularity lies on the cross-section of the former two. Due to the discrete symmetry $(c_1, y)\longrightarrow(-c_1, -y)$ of the system of equations (\ref{B1})-(\ref{B2}) in the octupole case, the location of the singularity for $c_1<0$  is completely determined by the analysis for positive distortion parameter with the same absolute value. It is a mirror image of the location of the singularity for $c_1>0$ with respect to the equatorial plane $y=0$ (see Fig. \ref{sing}).

\begin{figure}[h]
\setlength{\tabcolsep}{ 0 pt }{\scriptsize\tt
		\begin{tabular}{ cc }
	       \includegraphics[width=5 cm]{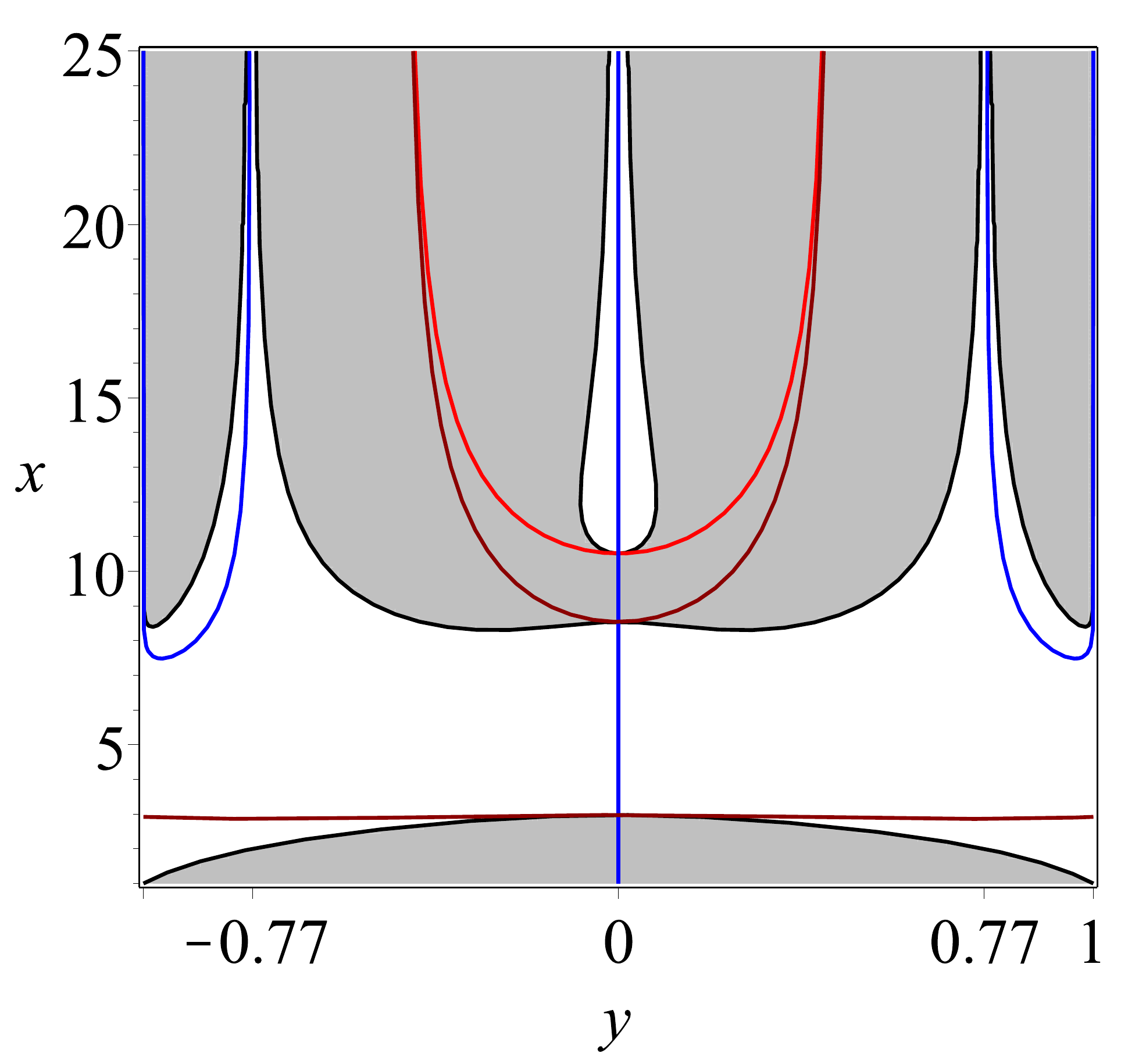} &
            \includegraphics[width=5 cm]{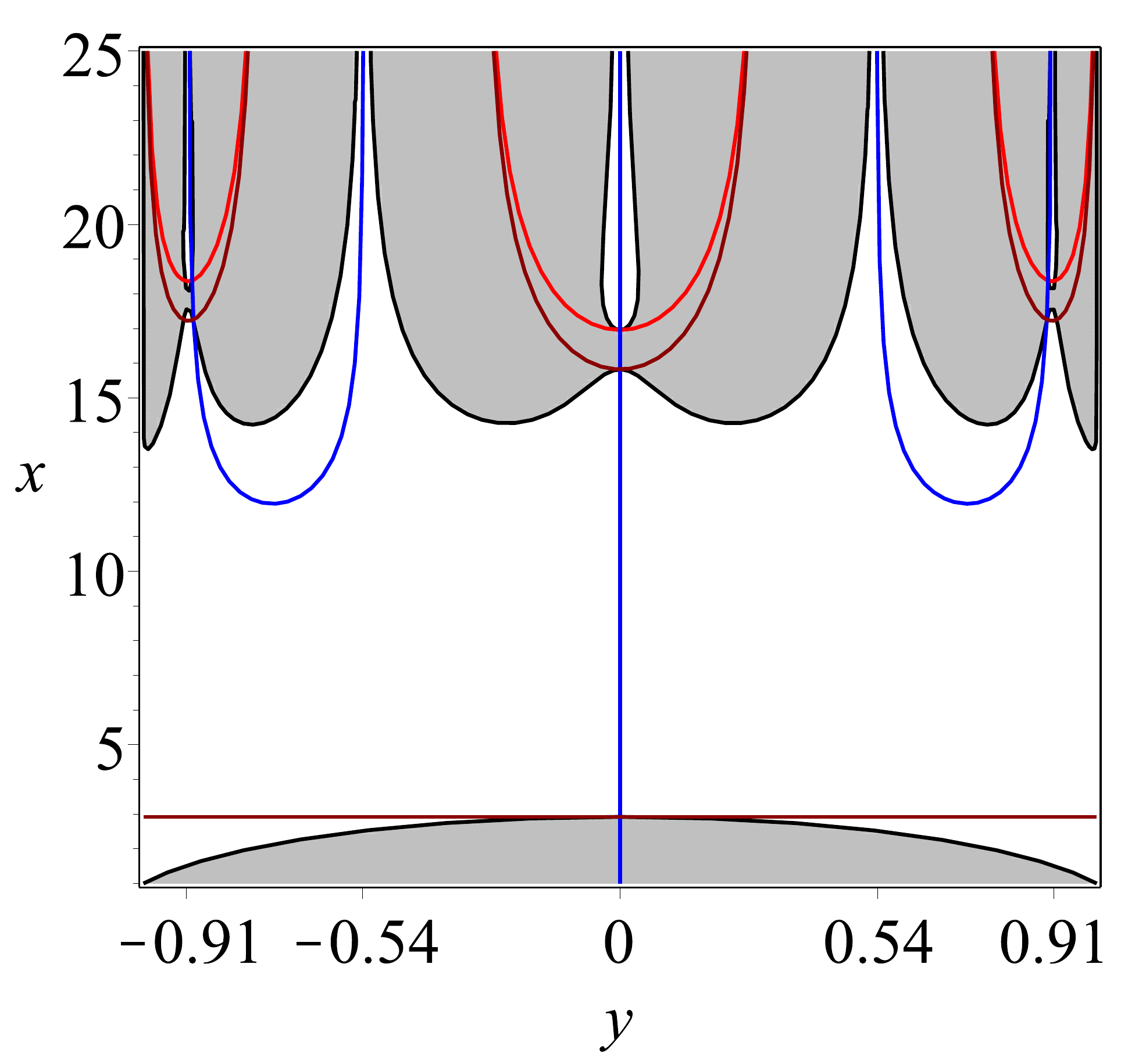} \\
            $c_4=-0.4\times10^{-3}$ & $c_6=0.4\times10^{-6}$ \\
            $c_n=0$, $n\neq4$ & $c_n=0, n\neq6$ \\
            \includegraphics[width=5 cm]{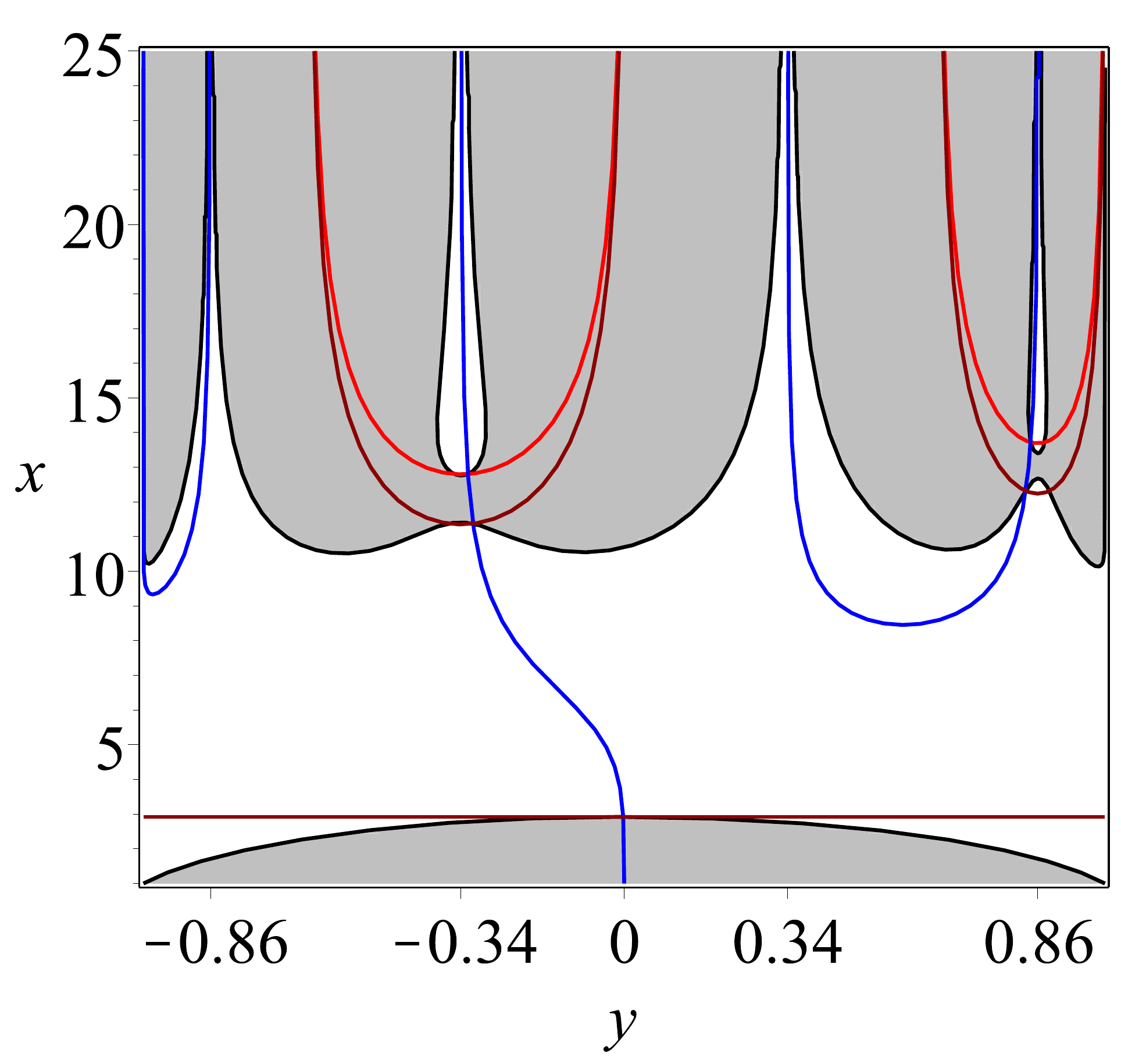} &
            \includegraphics[width=5 cm]{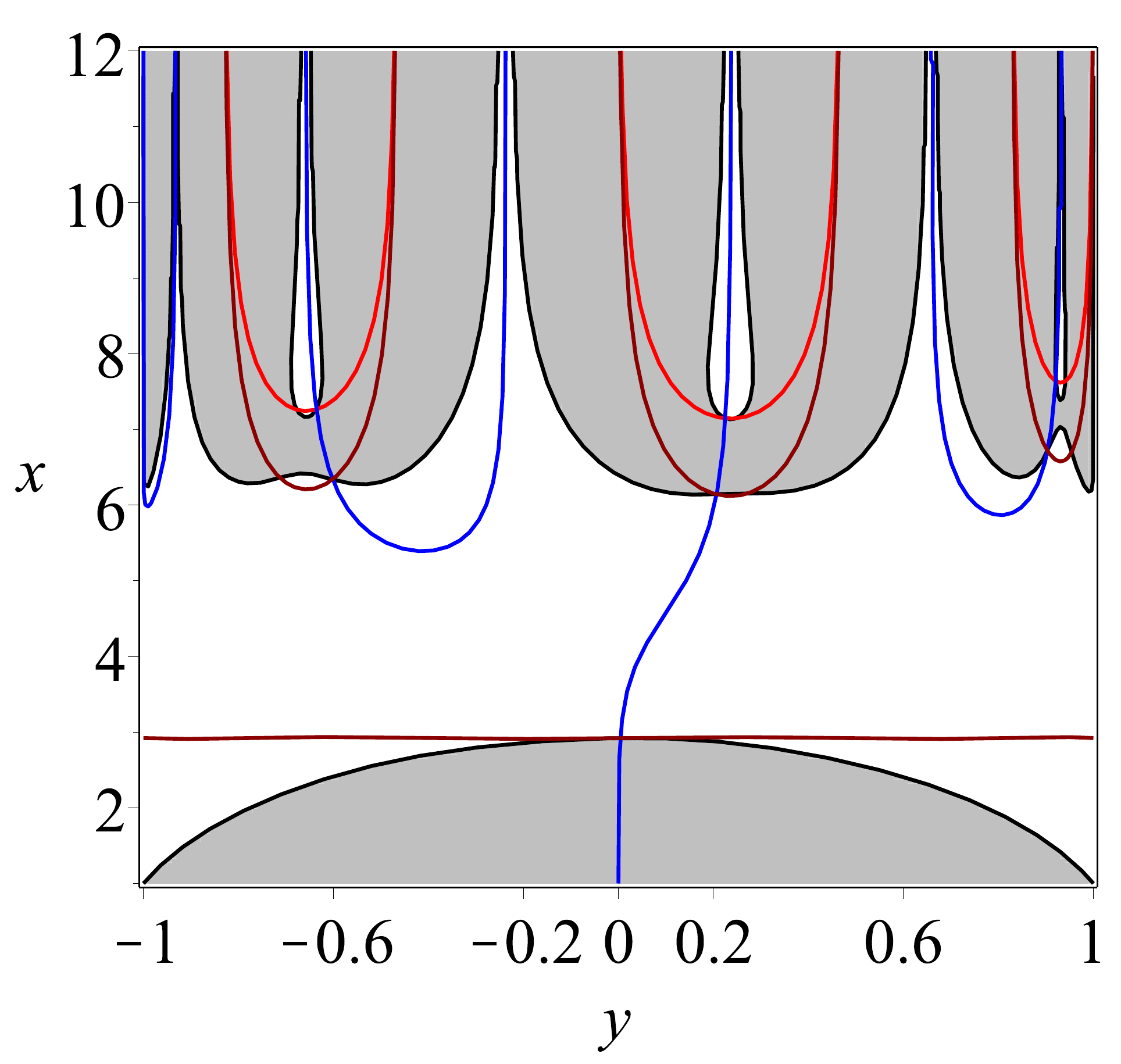}\\
             $c_5 = -c_1 = 0.2\times10^{-4}$ & $c_7=-c_1 = 0.6\times10^{-5} $ \\
             $c_n=0$, $n\neq 1,5$ & $c_n=0$, $n\neq 1,7$
                                   \end{tabular}}
           \caption{\footnotesize{Location of the curvature singularities for even and odd distortion. The rotation parameter is fixed to $\alpha=0.7$. The implicit functions defined by $B_1=0$, $B_2=0$, and $\tilde{B}_1=0$ are presented with red,  blue and dark red lines respectively. The singular points correspond to the cross-sections of the former two.}}
		\label{sing_gen}
\end{figure}

As in the quadrupole case, we  note that our argument relies on the numerical investigation of the possible ergoregion configurations performed in section 3.2. The described structure of the ergoregion and the position of singularities are not proven rigorously by analytical techniques.

In Fig. \ref{Sing2}, we investigate the location of the singularity in the $(x,y)$-plane in the case of positive octupole distortions as a function of the multipole moment for different values of $\alpha$. The location for negative distortion $c_1<0$ can be deduced from the presented results by keeping the $x$ coordinate  intact, and replacing the $y$ coordinate by a negative one with the same absolute value. Namely, if for positive values of the octupole distortion, the singularity is located on the upper plane $\frac{\pi}{2}< \theta< \pi$, for negative values it is located in the lower plane $0<\theta<\frac{\pi}{2}$. For positive values of the octupole moment, as the value of $\alpha$ decreases the singularity moves further away from the equatorial plane in the upper plane. Inversely, for negative values of the octupole moment, as the value of $\alpha$ decreases the singularity moves further away from the equatorial plane in the lower plane. We can see that, as the value of $\alpha$ increases the singularity is located closer to the outer horizon $x=1$. Once again, we can use the location of the singularity as a guideline for defining the position of the external sources and the region of validity of our solution.

If we consider more general even and odd distortion in the form $\{ c_{2n} \neq 0, \, c_{2k} = 0, k\neq n; \, c_{2k+1} =0, \forall k \in \mathcal{N}\}$, and  $\{c_{1} = - c_{2n+1}\neq 0, \, c_{2k+1} =0, k\neq 0, n;\, c_{2k} = 0, \forall k\in\mathcal{N} \}$, the representation of the metric function $A(x,y)$ in terms of the functions $B_1$ and $B_2$ provided in eqs. (\ref{A_singc1})-(\ref{B22_c1}) is still valid. However, the functions $\chi_1$ and $\chi_2$ are given by the expressions (\ref{chi_even})-(\ref{chi_odd}) for even and odd distortion, respectively.
We observe that there are certain similarities in the properties of the equations (\ref{B1_c1})-(\ref{B22_c1}) with the cases of quadrupole and octupole distortion. Therefore, it is interesting to see whether our results could be extended in these more general cases and whether the curvature singularities in the domain of outer communication possess a similar location. Namely, they could be also located on the boundaries of some of the static regions, which form for large $x$ in the vicinity of the special cross-sections $y=y_i$ (see section 3.4). We illustrate the location of the singularities for some of the most simple cases in these classes of distortion in Fig. \ref{sing_gen}.

\section{Principal invariants of the Riemann and Weyl tensors}
In this section, we consider the principal invariants of the Riemann and Weyl tensors - the Kretschmann $\mathcal{K}_K$, the Chern-Pontryagin $\mathcal{K}_C$, and the Euler $\mathcal{K}_E$ invariants,
\ba
&&\mathcal{K}_K=R_{\alpha\beta\gamma\delta}R^{\alpha\beta\gamma\delta},\\
&&\mathcal{K}_C={}^*R_{\alpha\beta\gamma\delta}R^{\alpha\beta\gamma\delta},\\
&&\mathcal{K}_E={}^*R^{*}_{\alpha\beta\gamma\delta}R^{\alpha\beta\gamma\delta}.
\ea
Here, $R_{\alpha\beta\gamma\delta}$ is the Riemann tensor, and the star denotes the left and right Hodge dual
\be
{}^*R_{\alpha\beta\mu\nu}=\frac{1}{2}\epsilon_{\alpha\beta\lambda\delta}R^{\lambda\delta}_{~~\mu\nu} , ~~~R^{*}_{\alpha\beta\mu\nu}=\frac{1}{2}\epsilon_{\mu\nu\lambda\delta}R_{\alpha\beta}^{~~\lambda\delta},
\ee
where $\epsilon_{\alpha\beta\lambda\delta}$ is the four-dimensional Levi-Civita pseudo-tensor. For the vacuum spacetime under consideration, we have $\mathcal{K}_E=-\mathcal{K}_K$. The scalar curvature invariants are considered in the context of quantum gravity and in effective theories of gravity applied to cosmology, playing an important role for the one-loop level renormalization of gravity \cite{Birrel}. The Kretschmann invariant is further used for calculating the energy density and stresses of a conformal scalar field in the Hartle-Hawking state outside a static black hole \cite{Don, Brown, Frolov}. We compute the Kretschmann and the Chern-Pontryagin invariants on the horizon for the isolated Kerr black hole in the cases of quadrupole and octupole distortion, and compare their behaviour.

\begin{figure}[htp]
\setlength{\tabcolsep}{ 0 pt }{\scriptsize\tt
		\begin{tabular}{ ccc }
	\includegraphics[width=4.5 cm]{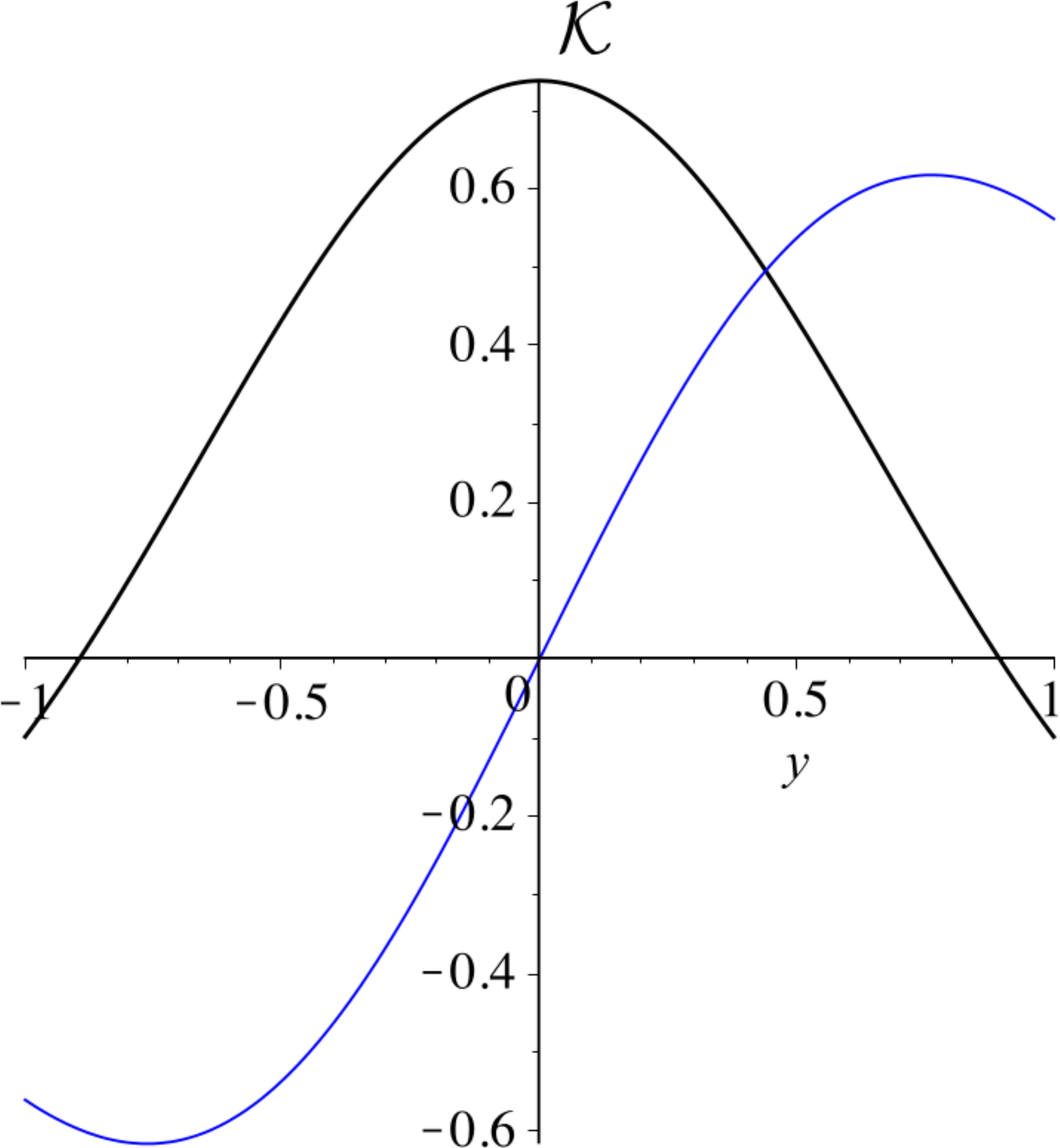}~~ &
            \includegraphics[width=4.5cm]{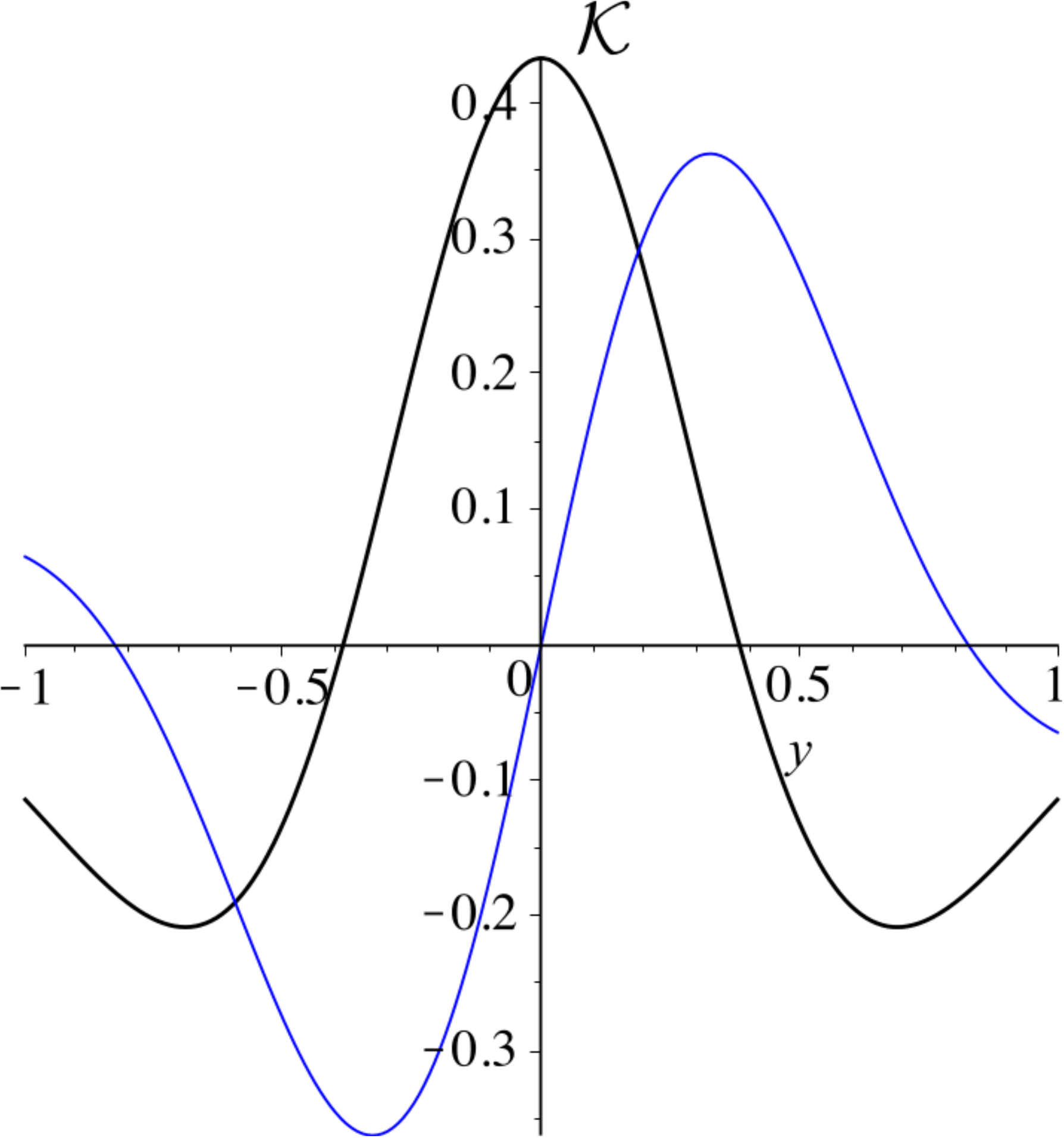}~~ &
            \includegraphics[width=4.5 cm]{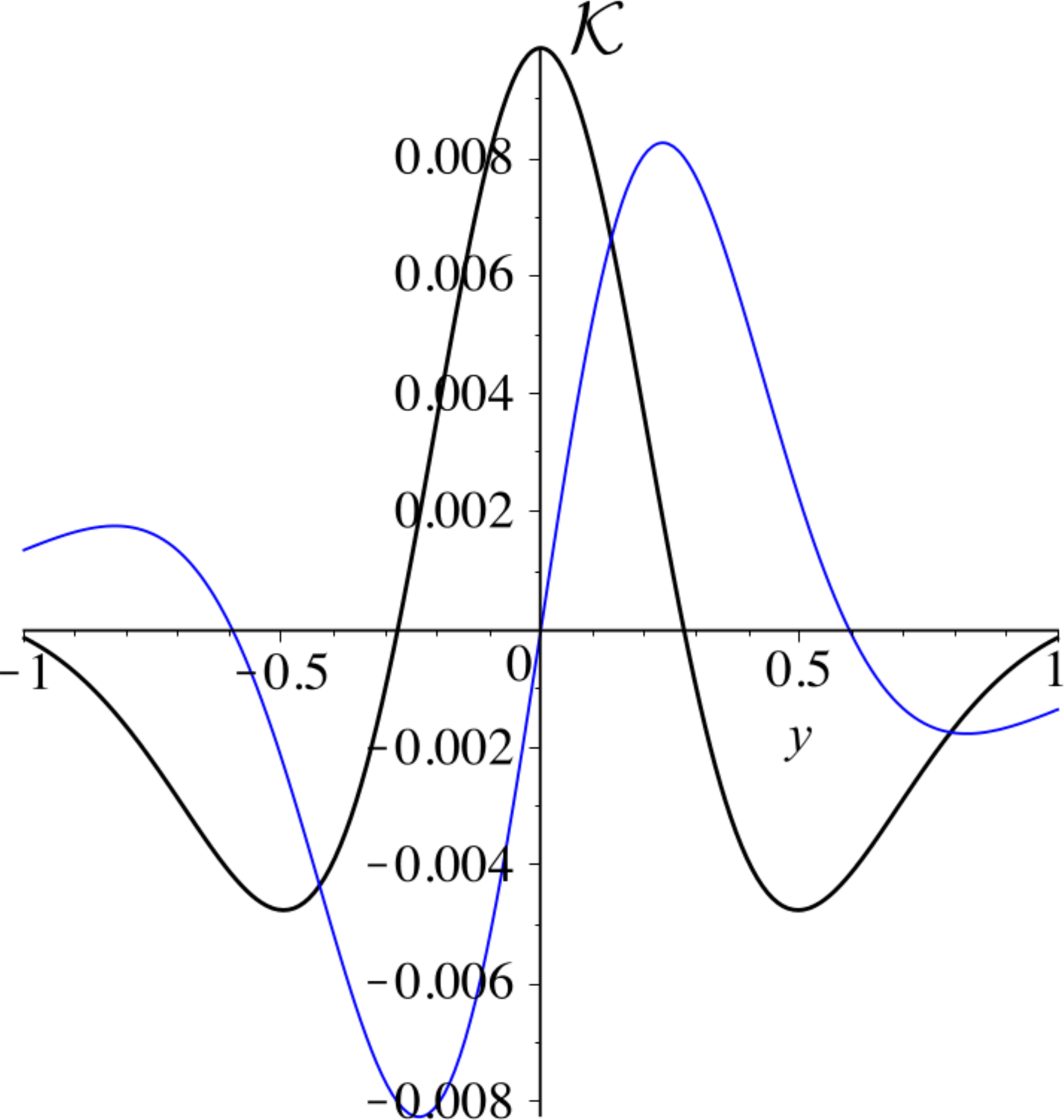} \\
            $\alpha=0.3$ & $\alpha=0.7$ &
            $\alpha=0.97$
           \end{tabular}}
           \caption{\footnotesize{ Scalar invariants vs. $y$ for the undistorted Kerr black hole. The black line is the Kretschmann invariant and the blue line is the Chern-Pontryagin invariant. }}
		\label{KerrCurvature}
\end{figure}

The behaviour of the scalar curvature invariants for the isolated Kerr black hole is illustrated in Fig. \ref{KerrCurvature}. The Kretschmann invariant on the horizon is everywhere positive for $\alpha<0.268$ and its maximum always lies in the equatorial plane $y=0$. For $\alpha>0.268$ the Kretschmann invariant is negative on the poles $y=1$ and $y=-1$ (or $\theta=0$ and $\theta=\pi$). In fact, for $\alpha>0.268$ the Kretschmann invariant gets negative at some angle $\theta=\theta_n$ and remains negative for all $\theta<\theta_n$ or $\theta>\pi-\theta_n$. As $\alpha\rightarrow 1$, the Kretschmann invariant on the axis vanishes. The Chern-Pontryagin invariant on the horizon calculated for $\alpha<0.577$ is positive for $\theta<\pi/2$, and negative for $\theta>\pi/2$.

\begin{figure}[htp]
\setlength{\tabcolsep}{ 0 pt }{\scriptsize\tt
		\begin{tabular}{ ccc }
	\includegraphics[width=4.2 cm]{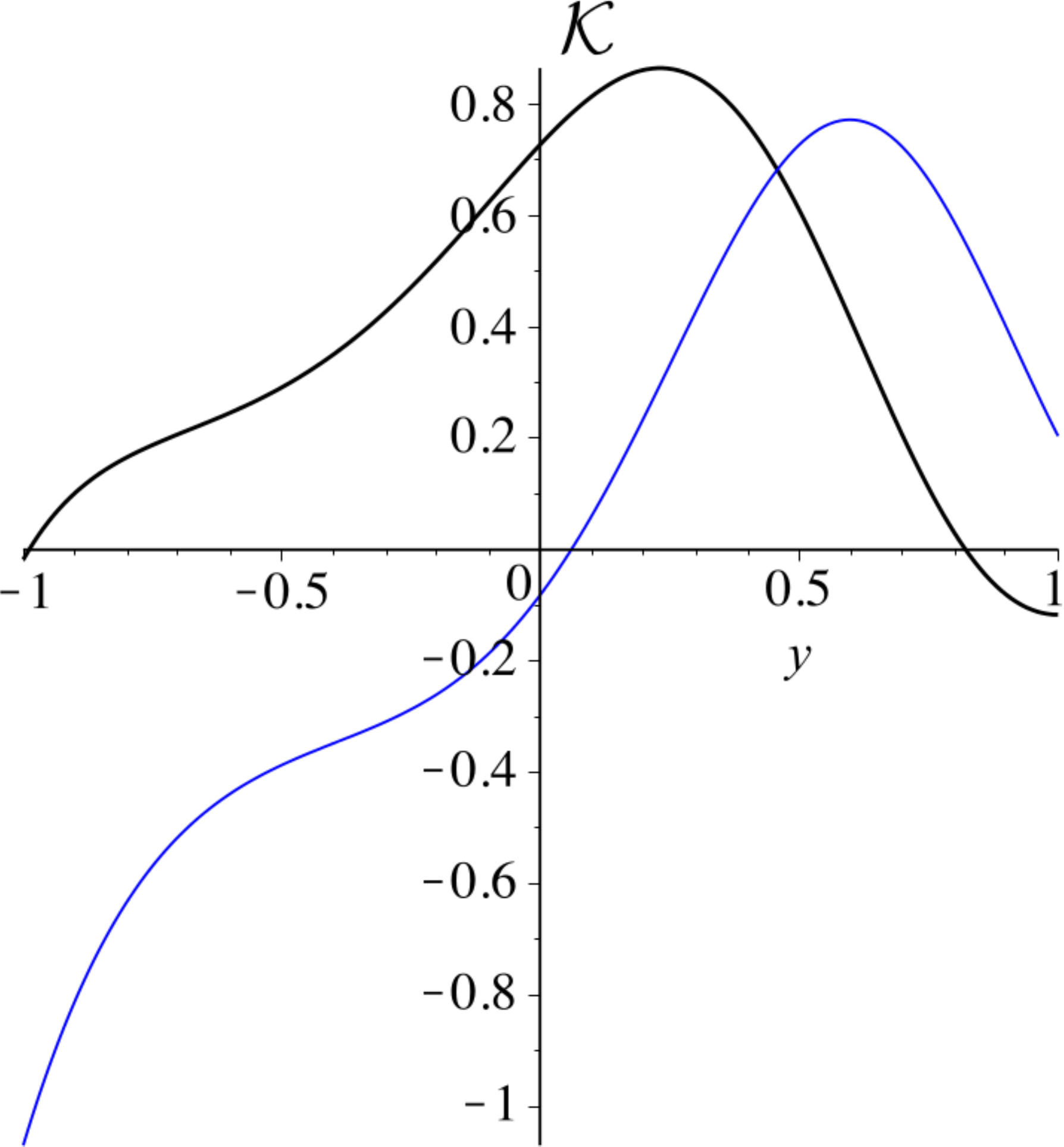}~~ &
            \includegraphics[width=4.2 cm]{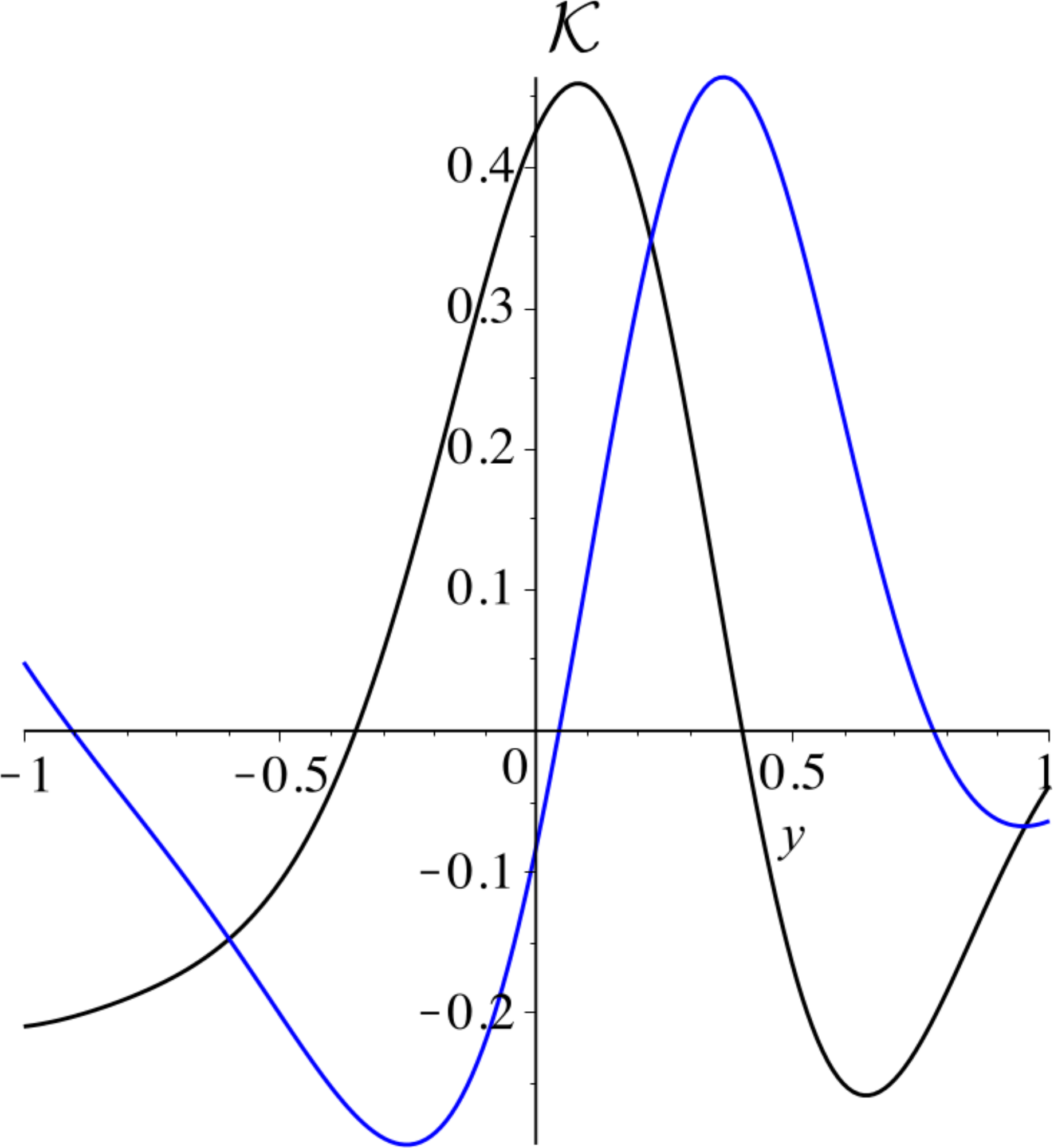}~~ &
            \includegraphics[width=4.2 cm]{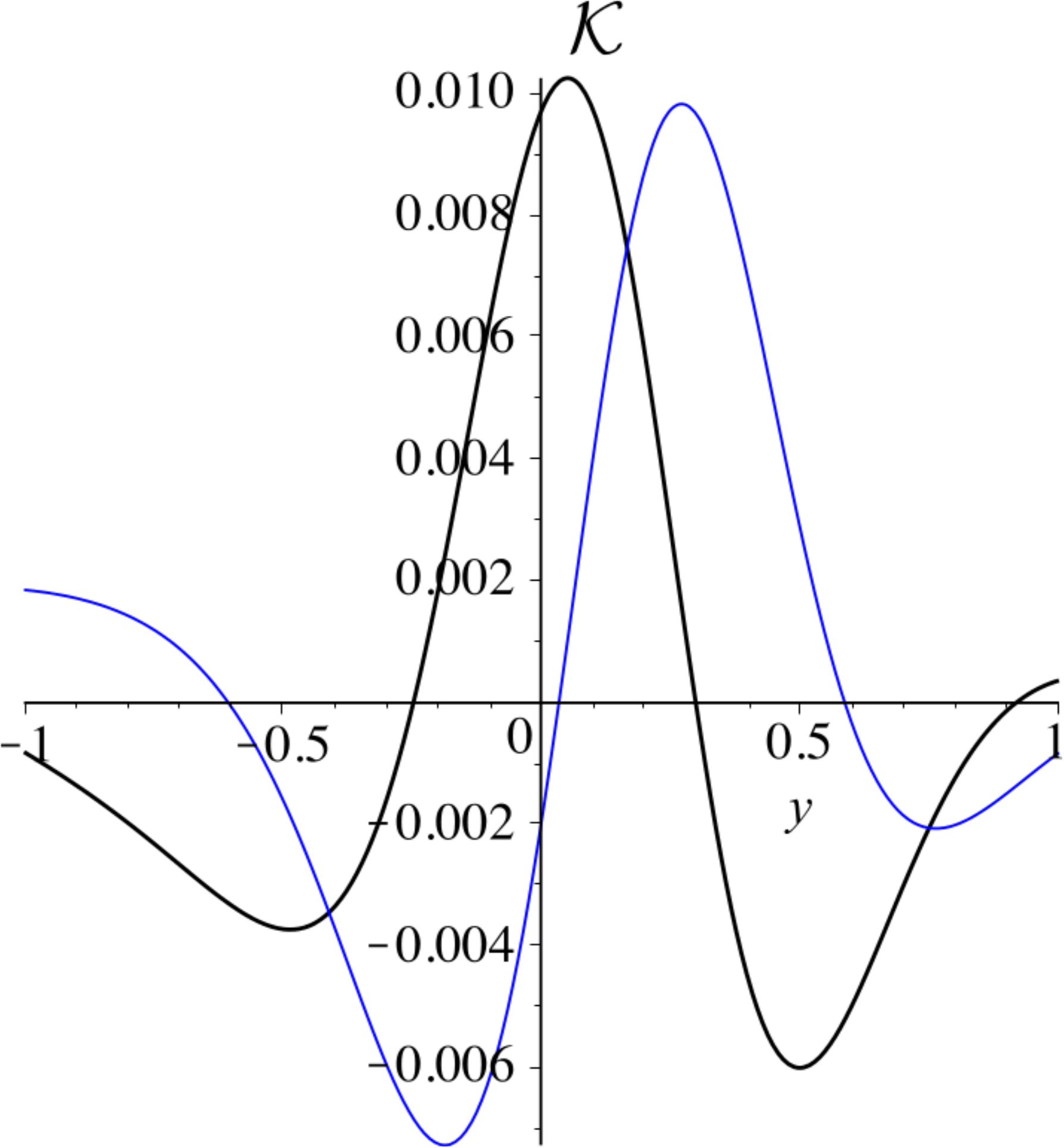} \\
            $\alpha=0.3,~c_1=-1/20$ & $\alpha=0.7,~c_1=-1/20$ &
            $\alpha=0.97,~c_1=-1/20$    \\ \\
            \includegraphics[width=4.2 cm]{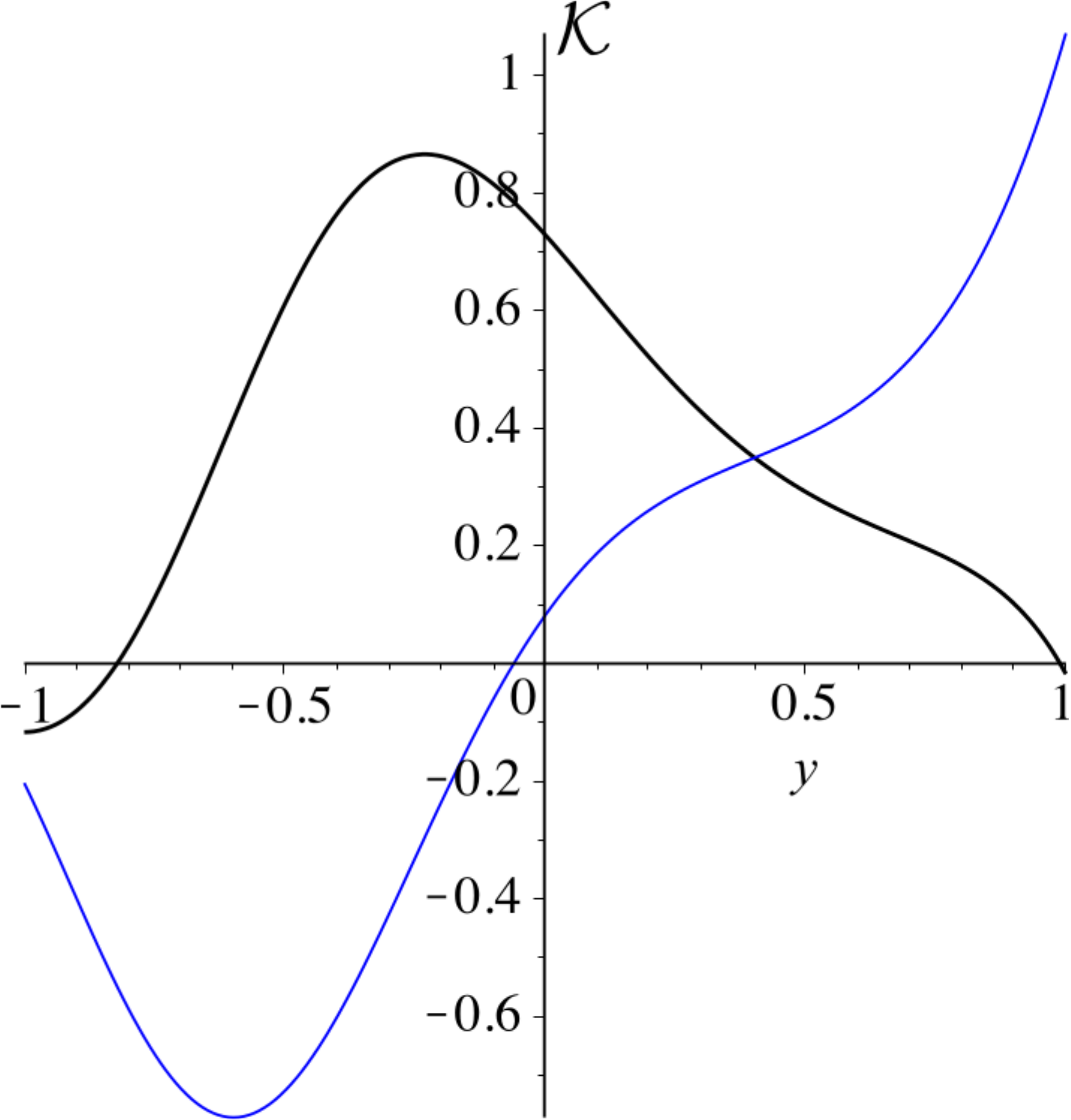} ~~&
            \includegraphics[width=4.2 cm]{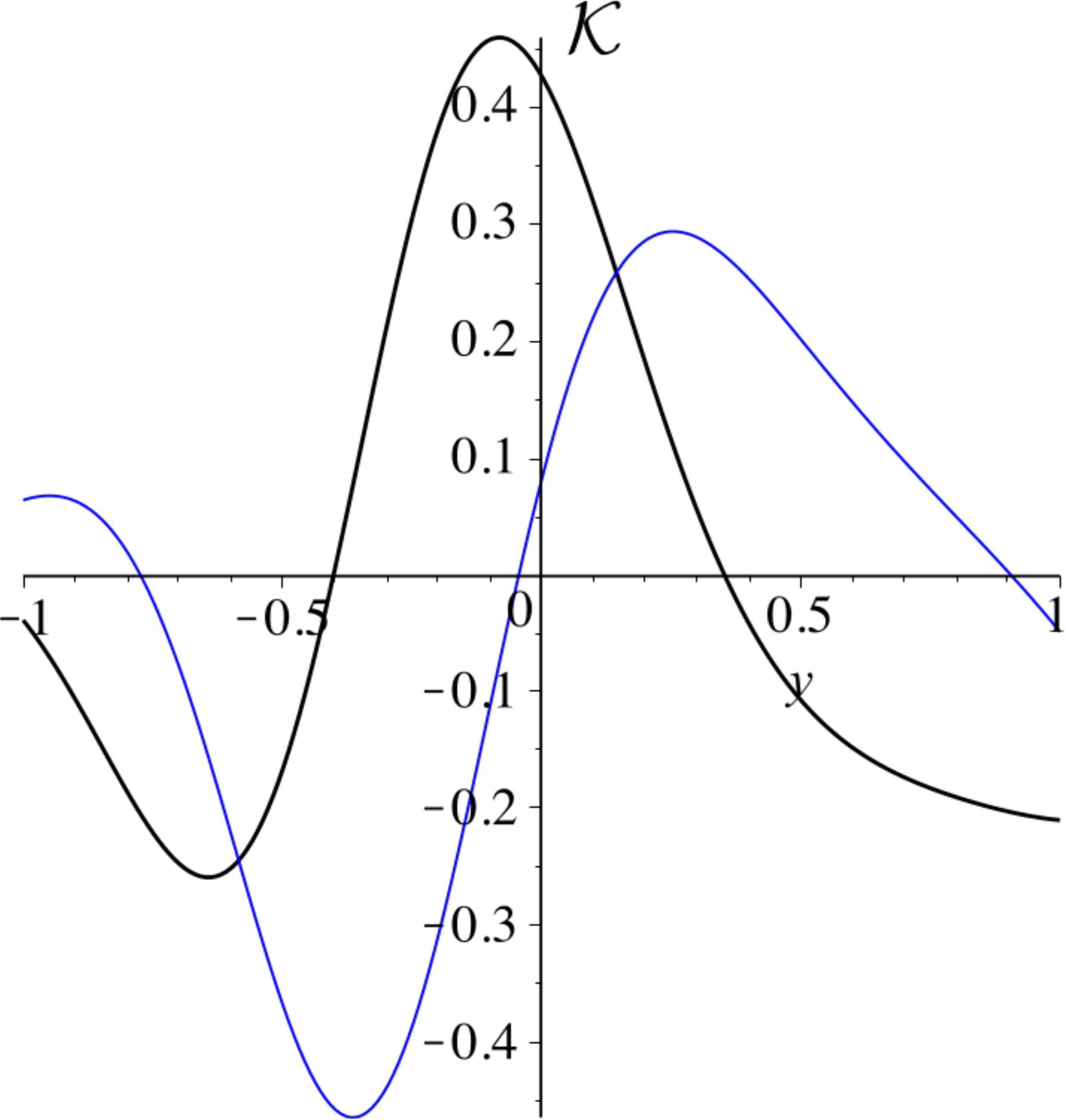}~~ &
            \includegraphics[width=4.2 cm]{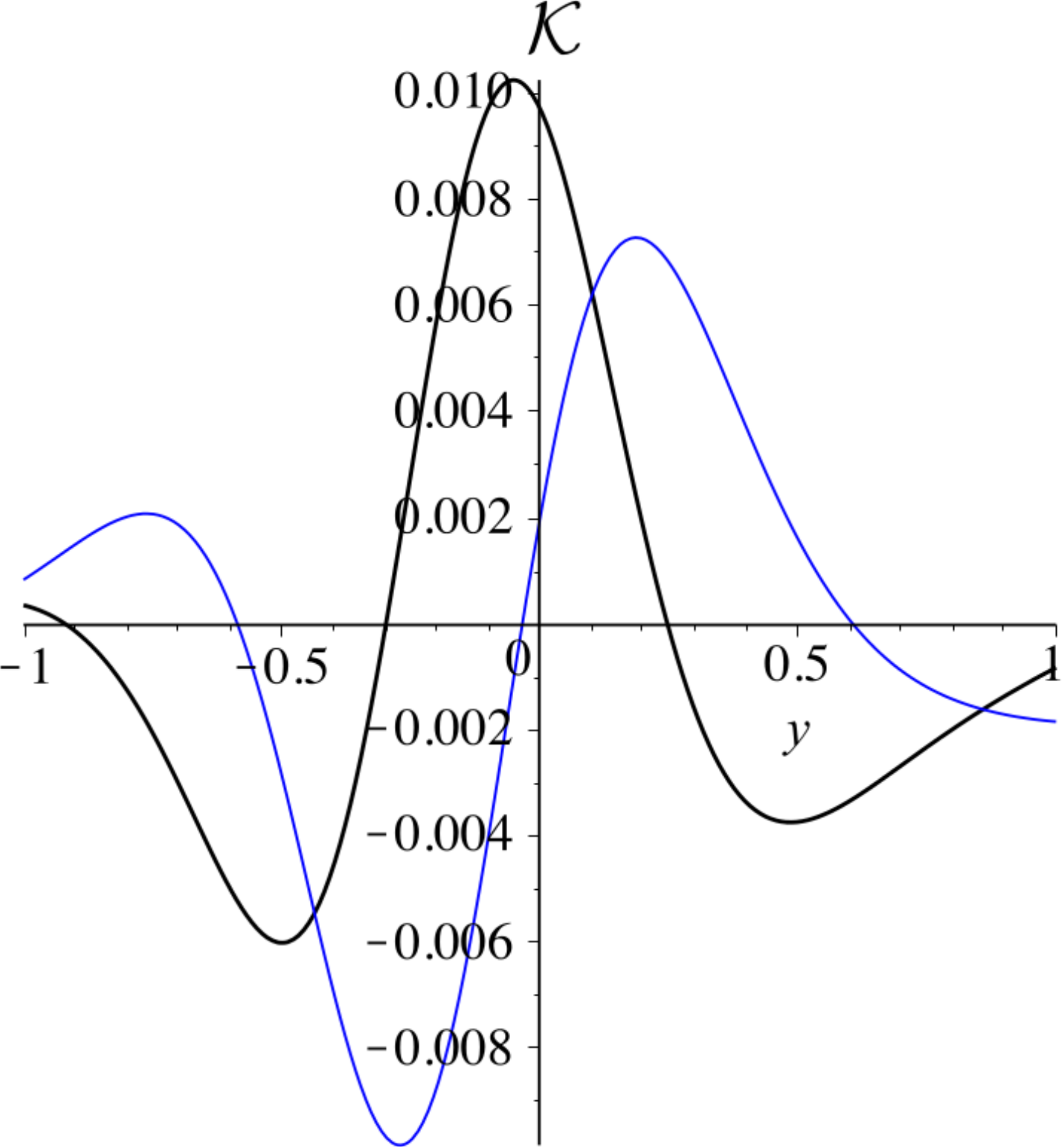} \\
                $\alpha=0.3,~c_1=1/20$ & $\alpha=0.7,~c_1=1/20$ &
            $\alpha=0.97,~c_1=1/20$    \\ \\
	\includegraphics[width=4.2 cm]{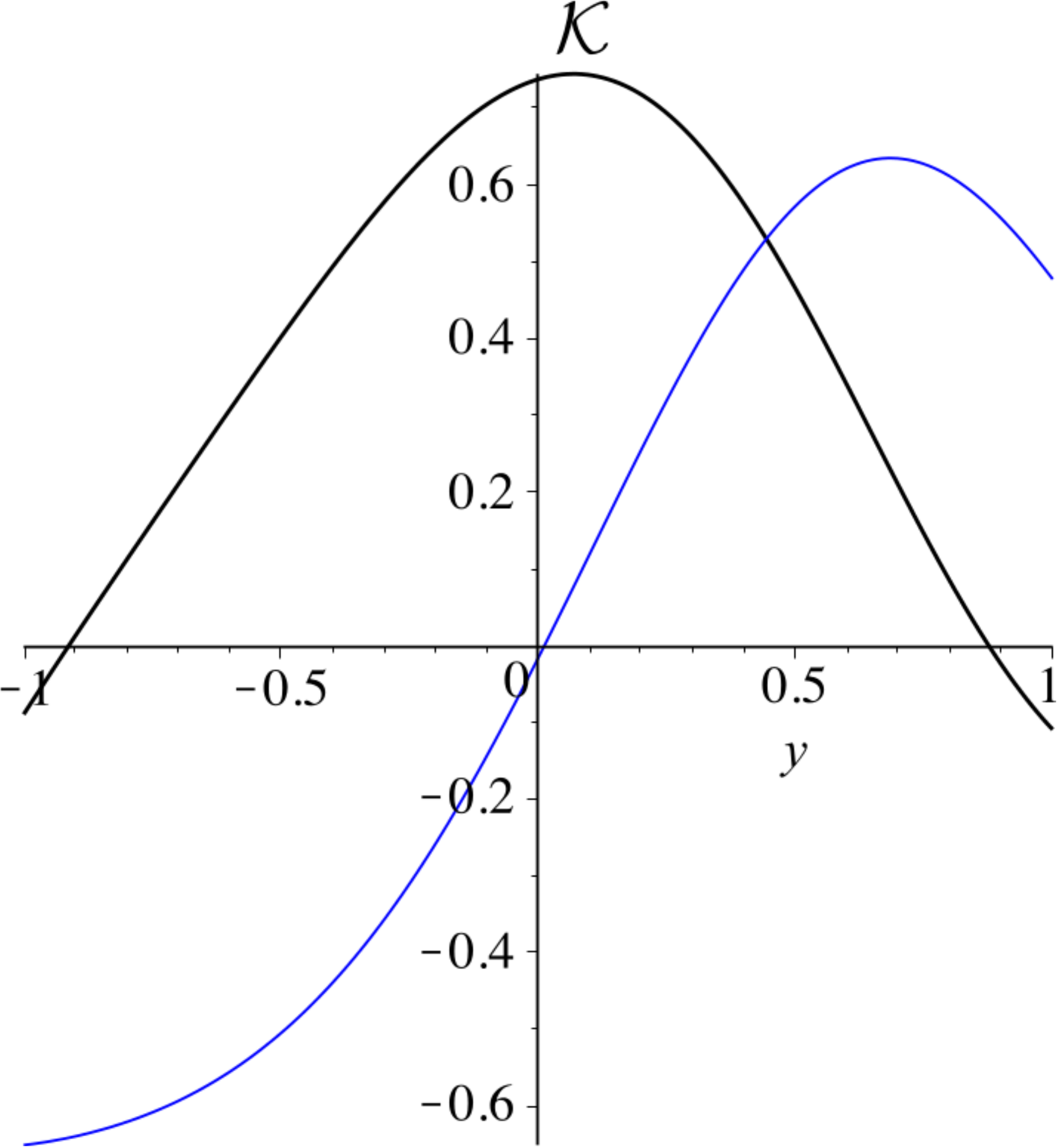}~~ &
            \includegraphics[width=4.2 cm]{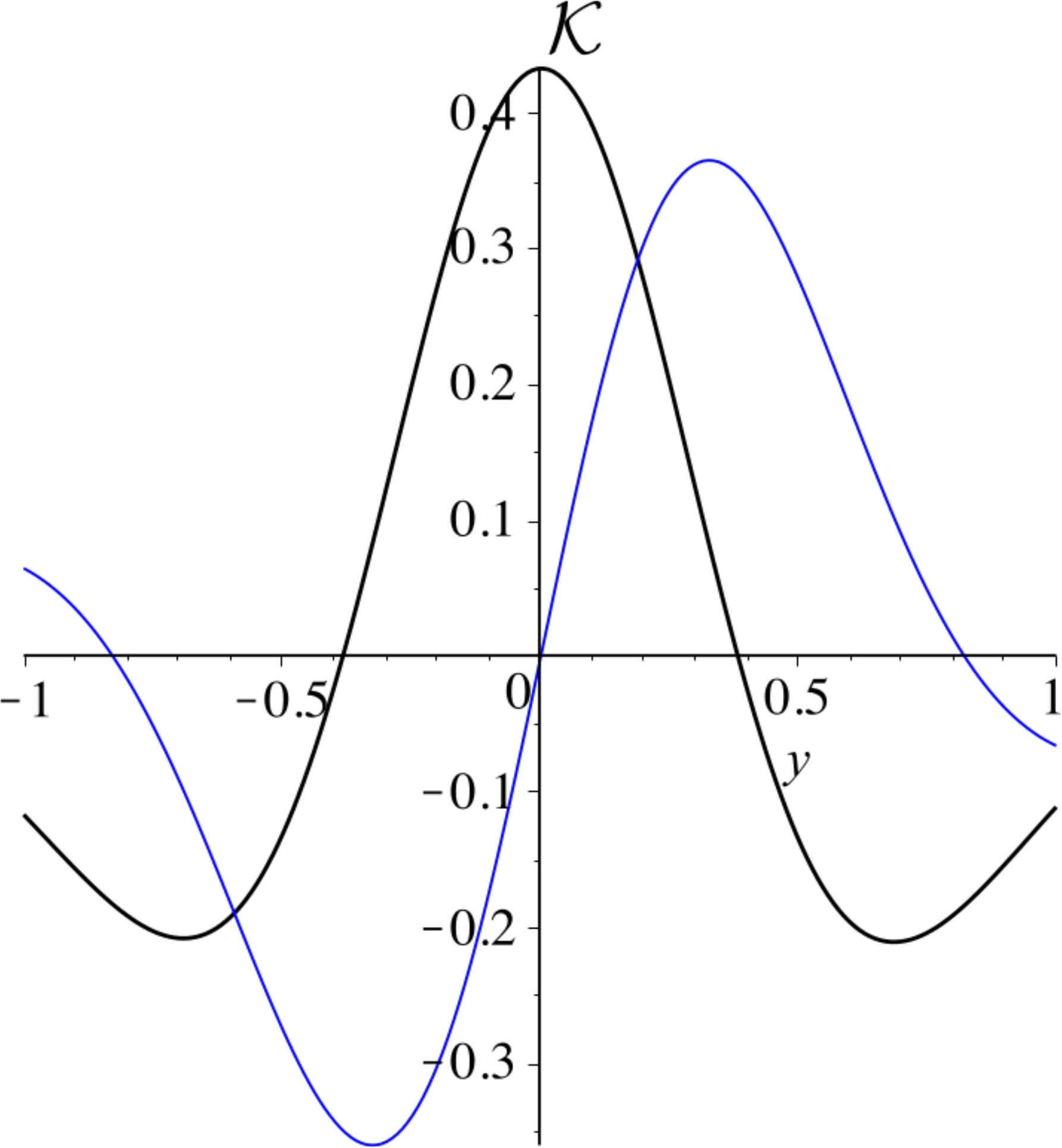}~~ &
            \includegraphics[width=4.2 cm]{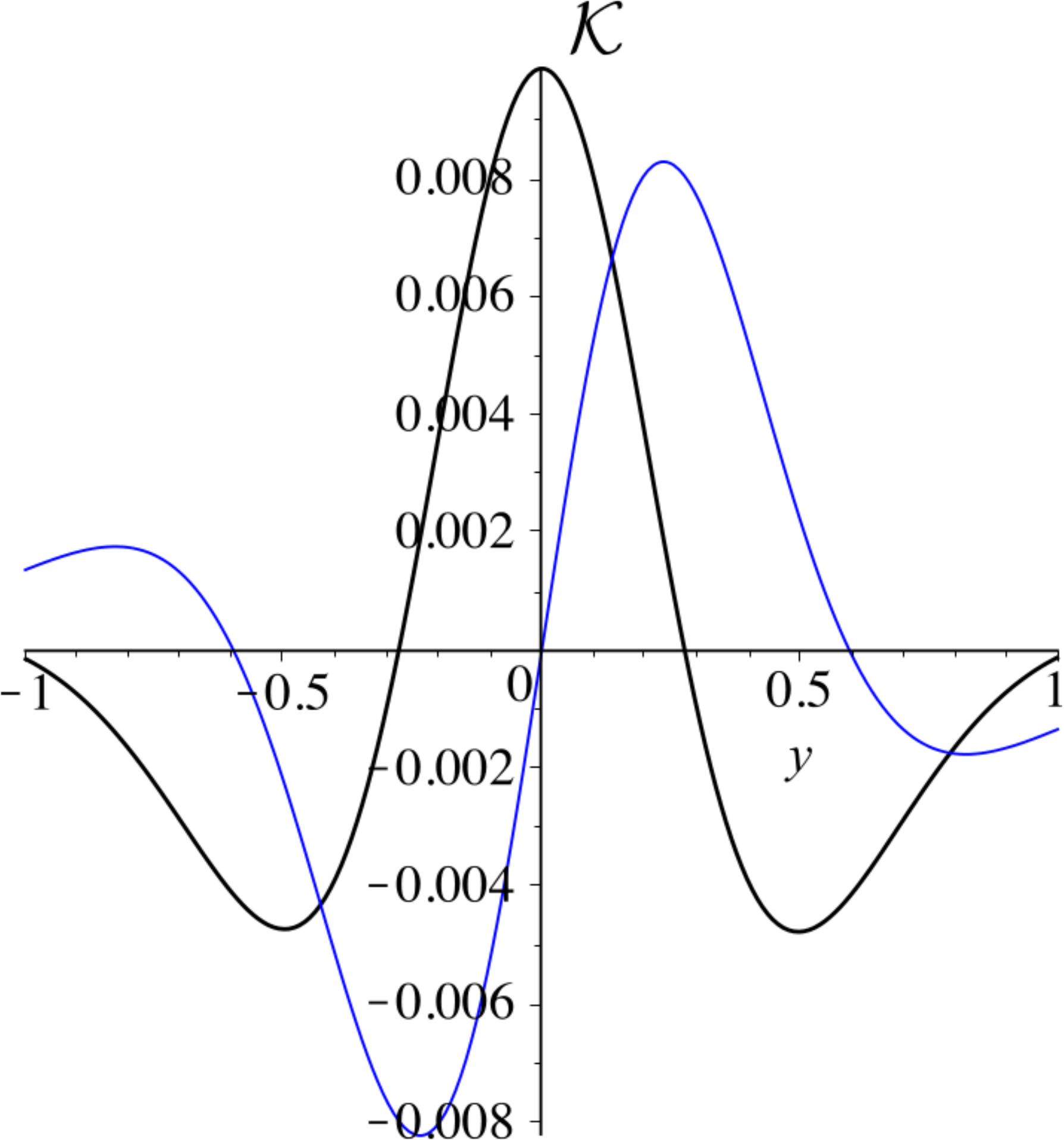} \\
            $\alpha=0.3,~c_1=-1/100$ & $\alpha=0.7,~c_1=-1/600$ &
            $\alpha=0.97,~c_1=-1/1000$    \\ \\
            \includegraphics[width=4.2 cm]{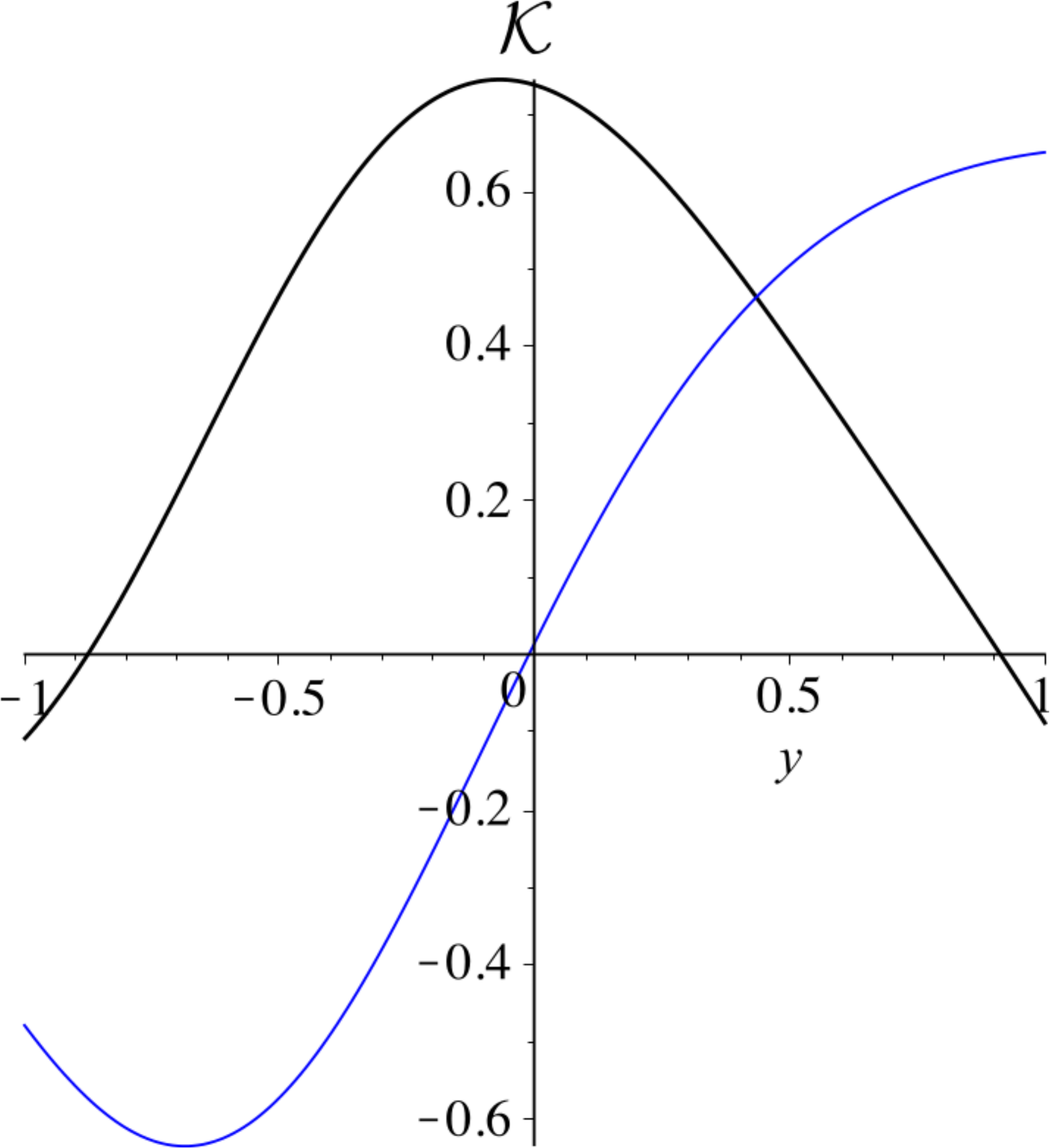} ~~&
            \includegraphics[width=4.2 cm]{07negOdd1600.pdf}~~ &
            \includegraphics[width=4.2 cm]{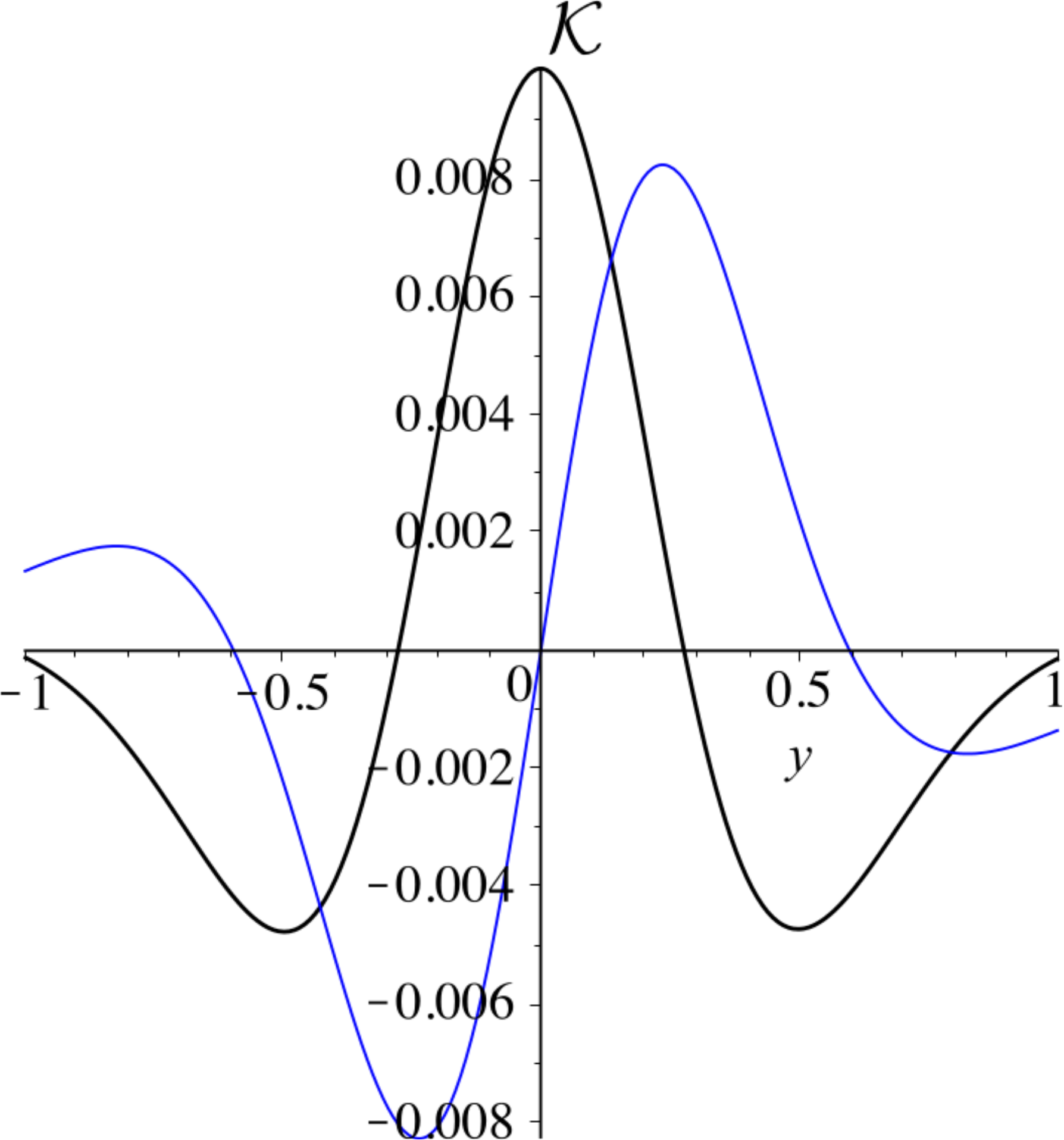} \\
                $\alpha=0.3,~c_1=1/100$ & $\alpha=0.7,~c_1=1/600$ &
            $\alpha=0.97,~c_1=1/1000$    \\
           \end{tabular}}
           \caption{\footnotesize{Scalar invariants vs. $y$ for octupole distortion. The black line is the Kretschmann invariant and the blue line is the Chern-Pontryagin invariant. }}
		\label{OddCurvature}
\end{figure}

In Fig. \ref{OddCurvature}, we illustrate the scalar curvature invariants in the case of octupole distortion.  The Kretschmann invariant on the horizon possesses similar behaviour to the case of the undistorted Kerr black hole. For negative  distortion parameter $c_1<0$ its maximum value lies in the upper plane $\theta<\pi/2$ close to the equator. The Chern-Pontryagin invariant for $c_1<0$  vanishes at some angle close to the equator in the upper plane instead on the equator as in the undistorted case. For small values of $\alpha$ and large absolute values of the negative multipole moment $c_1$, for example $\alpha=0.3$ and $c_1=-1/20$, we observe that the Chern-Pontryagin invariant decreases substantially on the axis $\theta=\pi$, compared to the case of undistorted Kerr black hole, where the maximum value of the Chern-Pontryagin invariant is close to the axis. For positive  distortion parameter $c_1>0$ the maximum value of the Kretschmann invariant lies in the lower plane  $\theta>\pi/2$ close to the equator. The Chern-Pontryagin invariant for $c_1>0$ vanishes at some angle close to the equator in the lower plane. For small value of $\alpha$ and large positive multipole moment such as $\alpha=0.3$ and $c_1=1/20$ the Chern-Pontryagin invariant increases substantially on the axis $\theta=0$ and its magnitude decreases substantially on the axis $\theta=\pi$. We have also considered the case of very weak distortion. Then, the behaviour of the scalar invariants is very similar to that of the undistorted Kerr black hole, as expected.

\begin{figure}[htp]
\setlength{\tabcolsep}{ 0 pt }{\scriptsize\tt
		\begin{tabular}{ ccc }
	\includegraphics[width=4.2 cm]{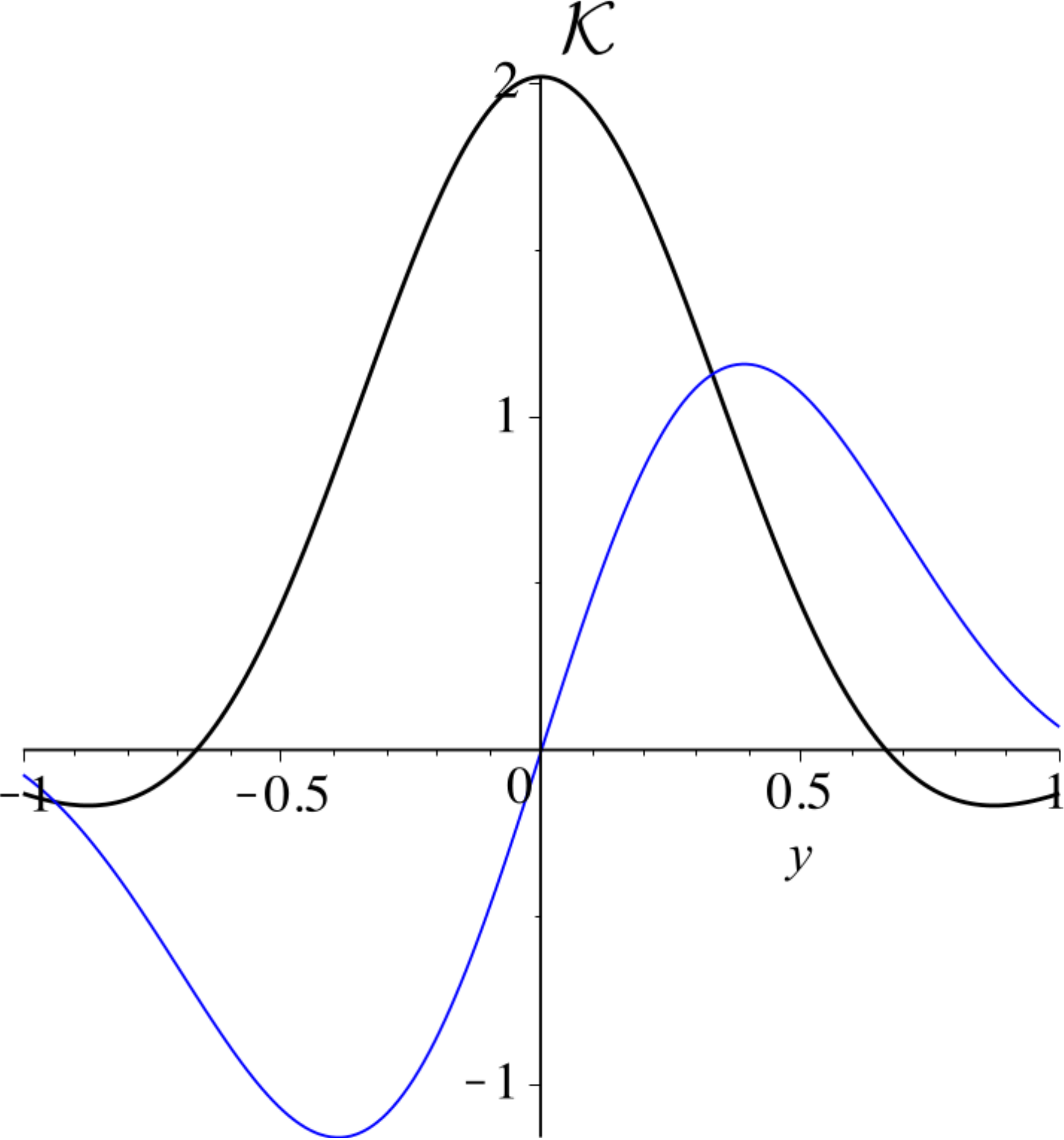}~~ &
            \includegraphics[width=4.2 cm]{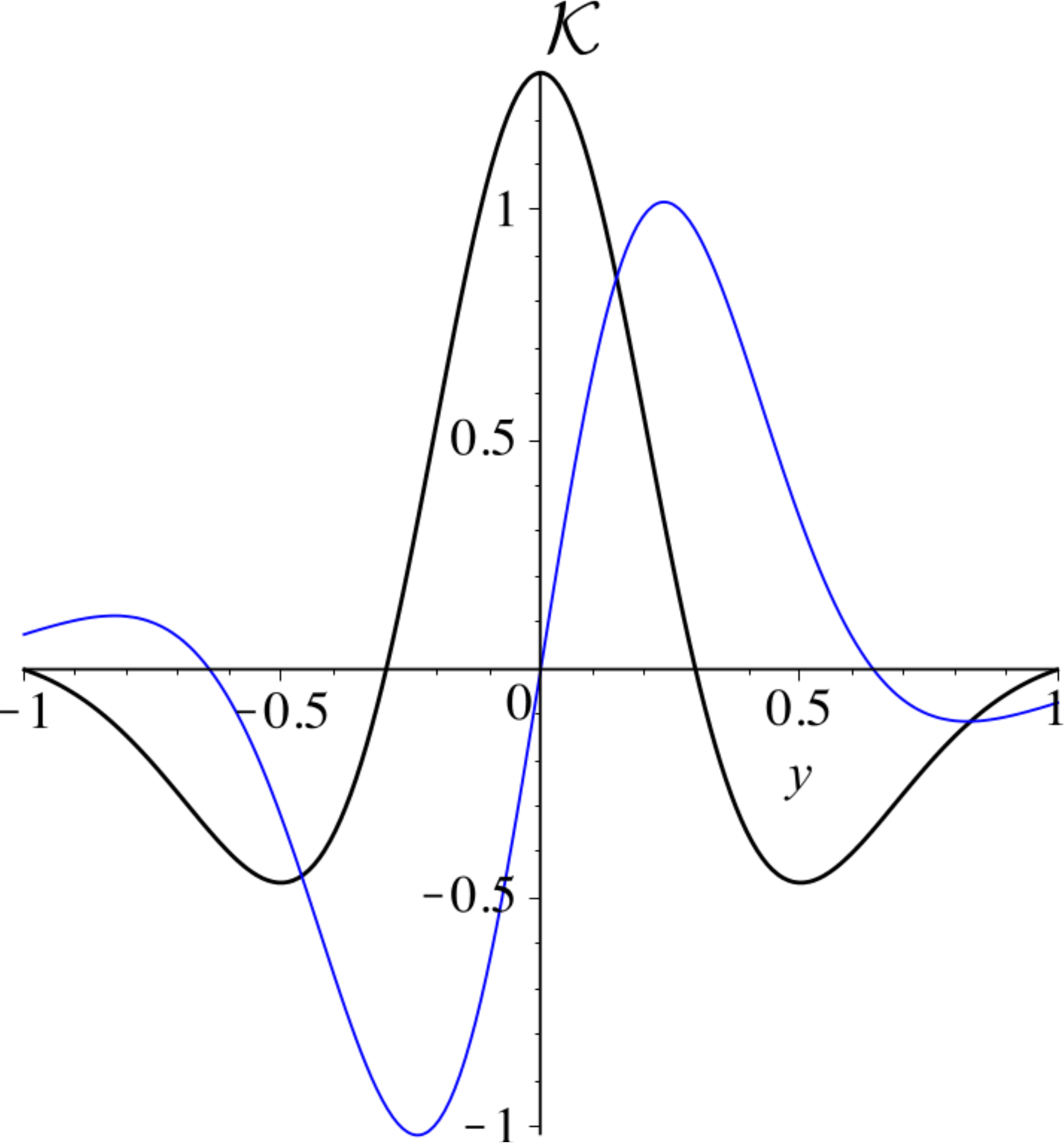}~~ &
            \includegraphics[width=4.2 cm]{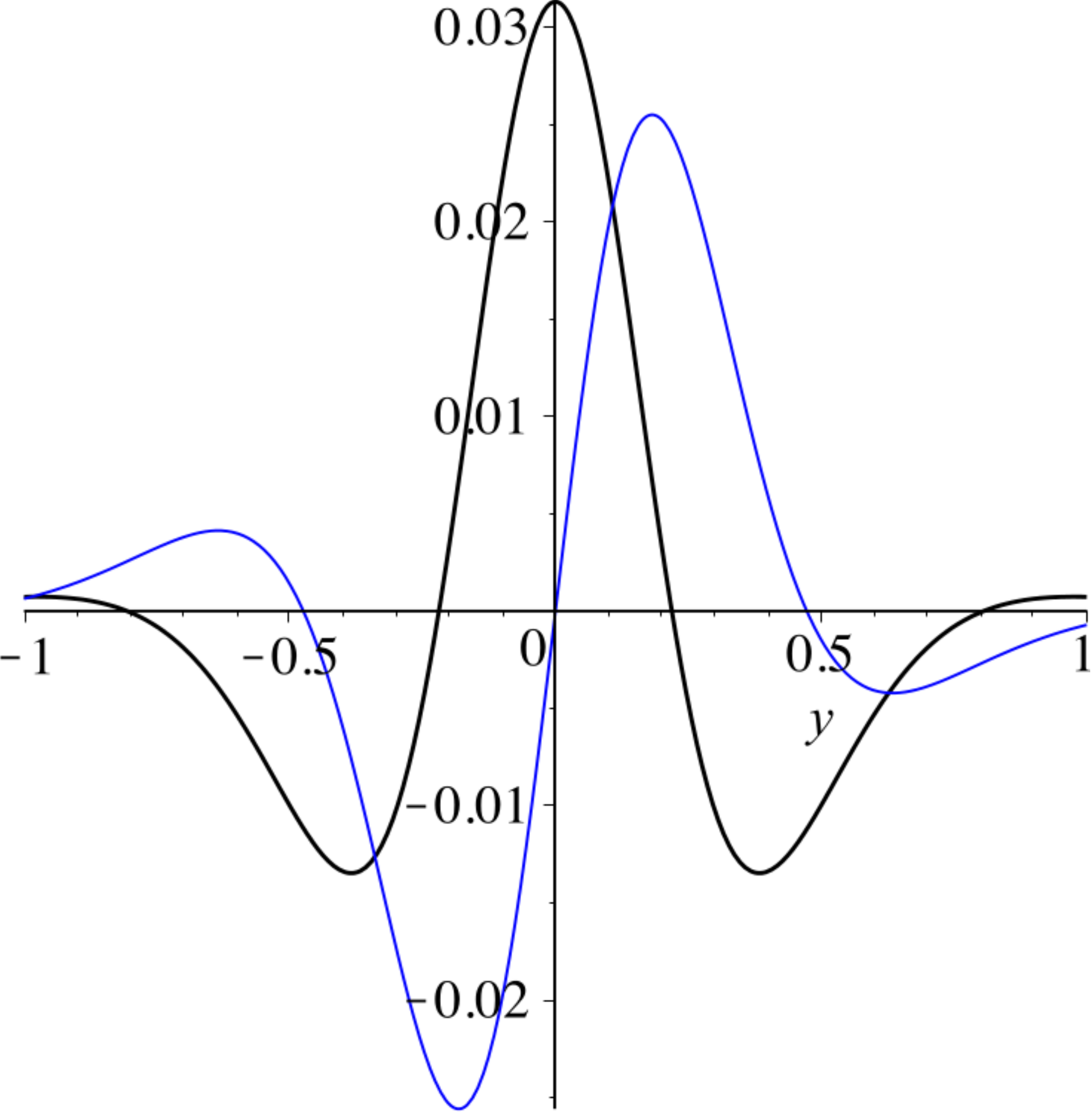} \\
            $\alpha=0.3,~c_2=1/12$ & $\alpha=0.7,~c_2=1/12$ &
            $\alpha=0.97,~c_2=1/12$    \\ \\
            \includegraphics[width=4.2 cm]{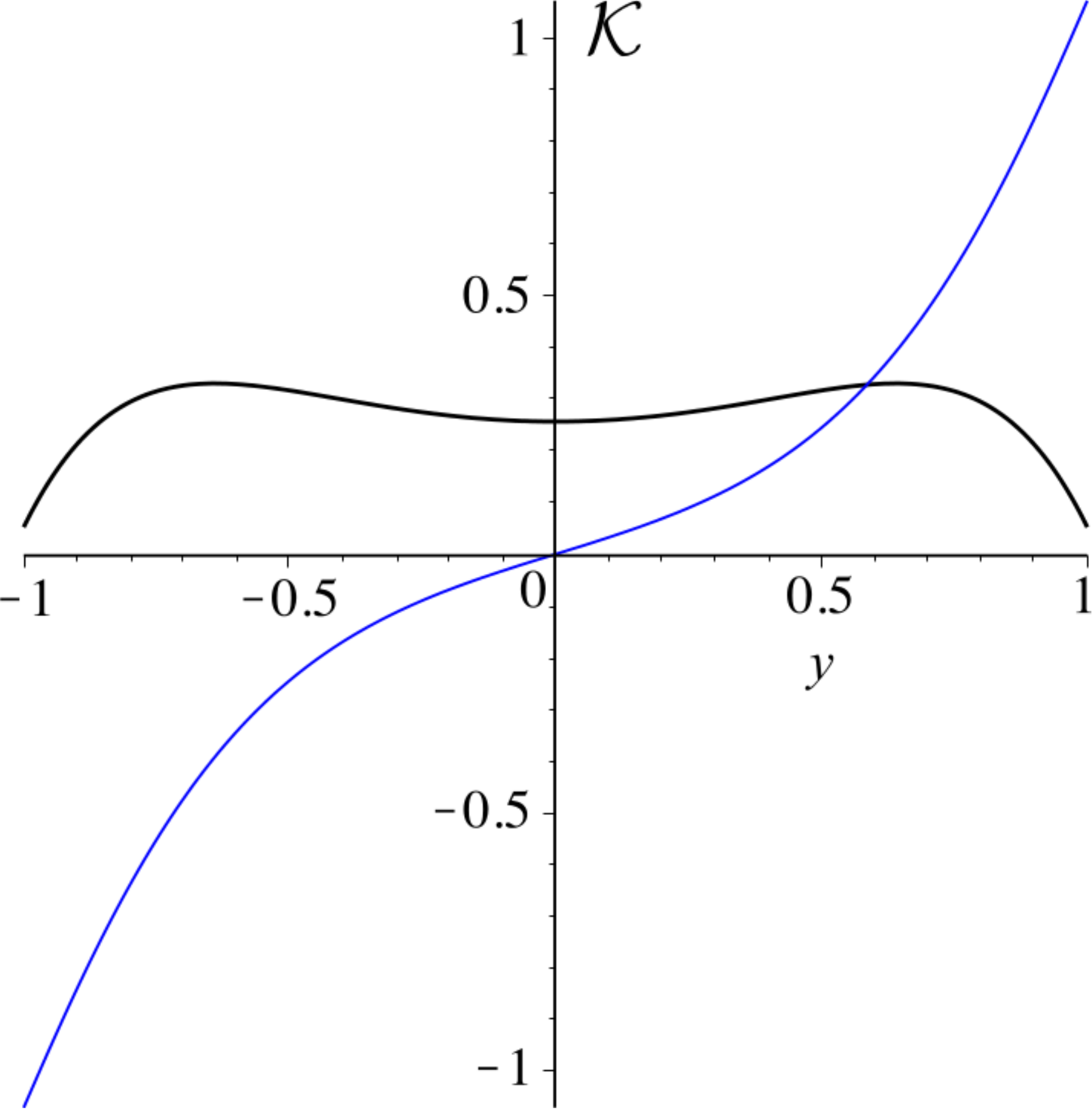} ~~&
            \includegraphics[width=4.2 cm]{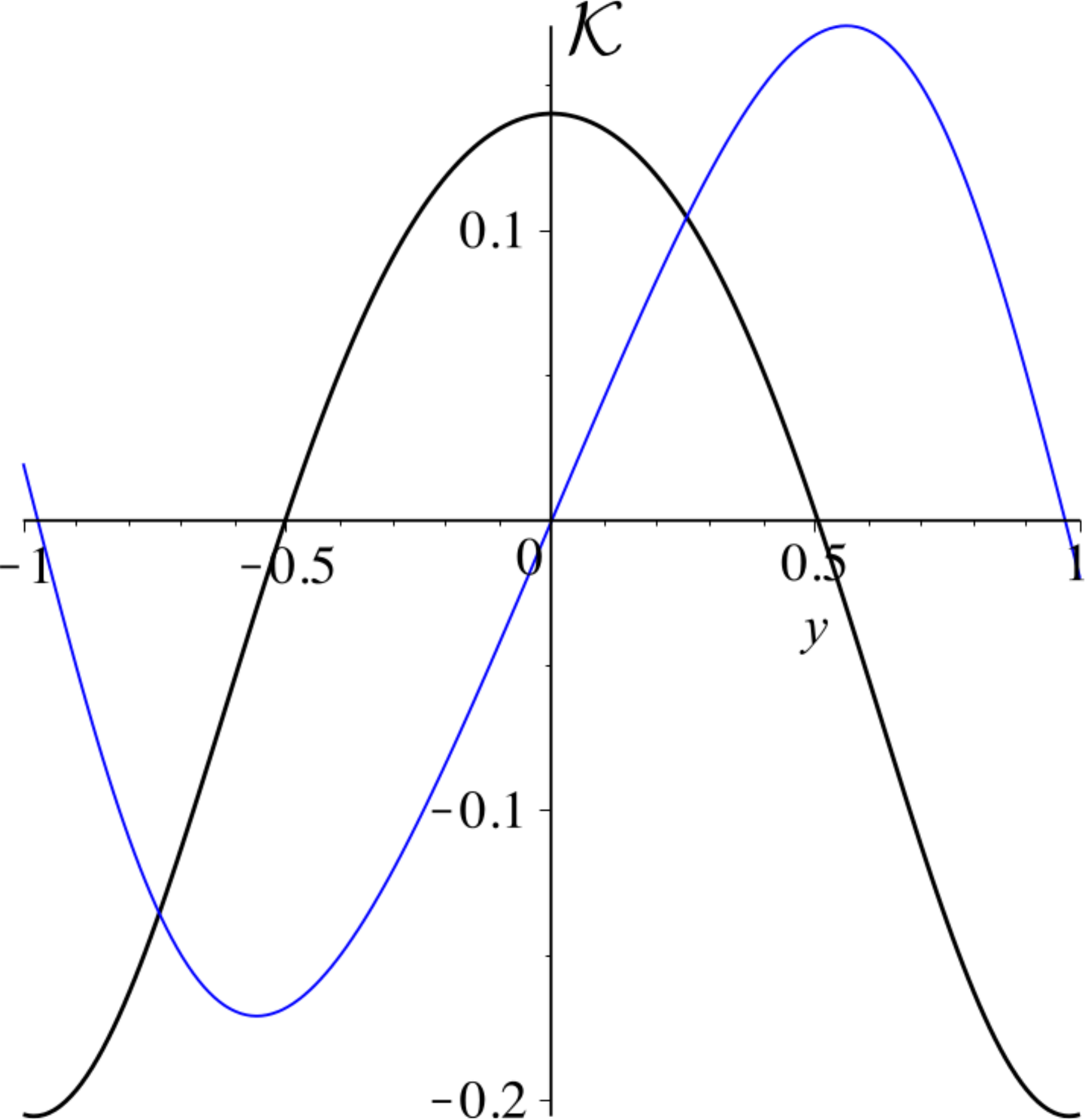}~~ &
            \includegraphics[width=4.2 cm]{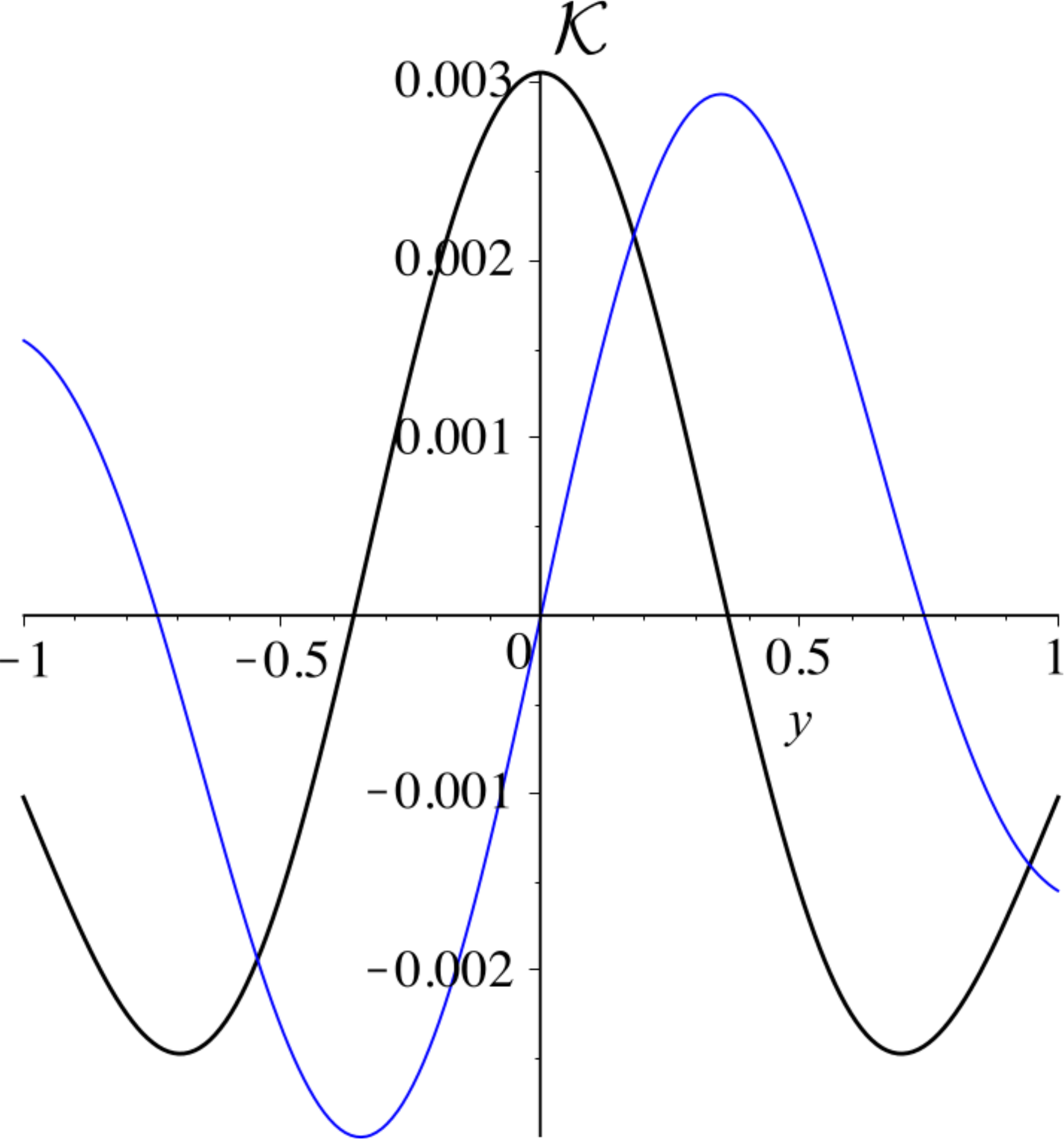} \\
                $\alpha=0.3,~c_2=-1/12$ & $\alpha=0.7,~c_2=-1/12$ &
            $\alpha=0.97,~c_2=-1/12$    \\ \\
            \includegraphics[width=4.2 cm]{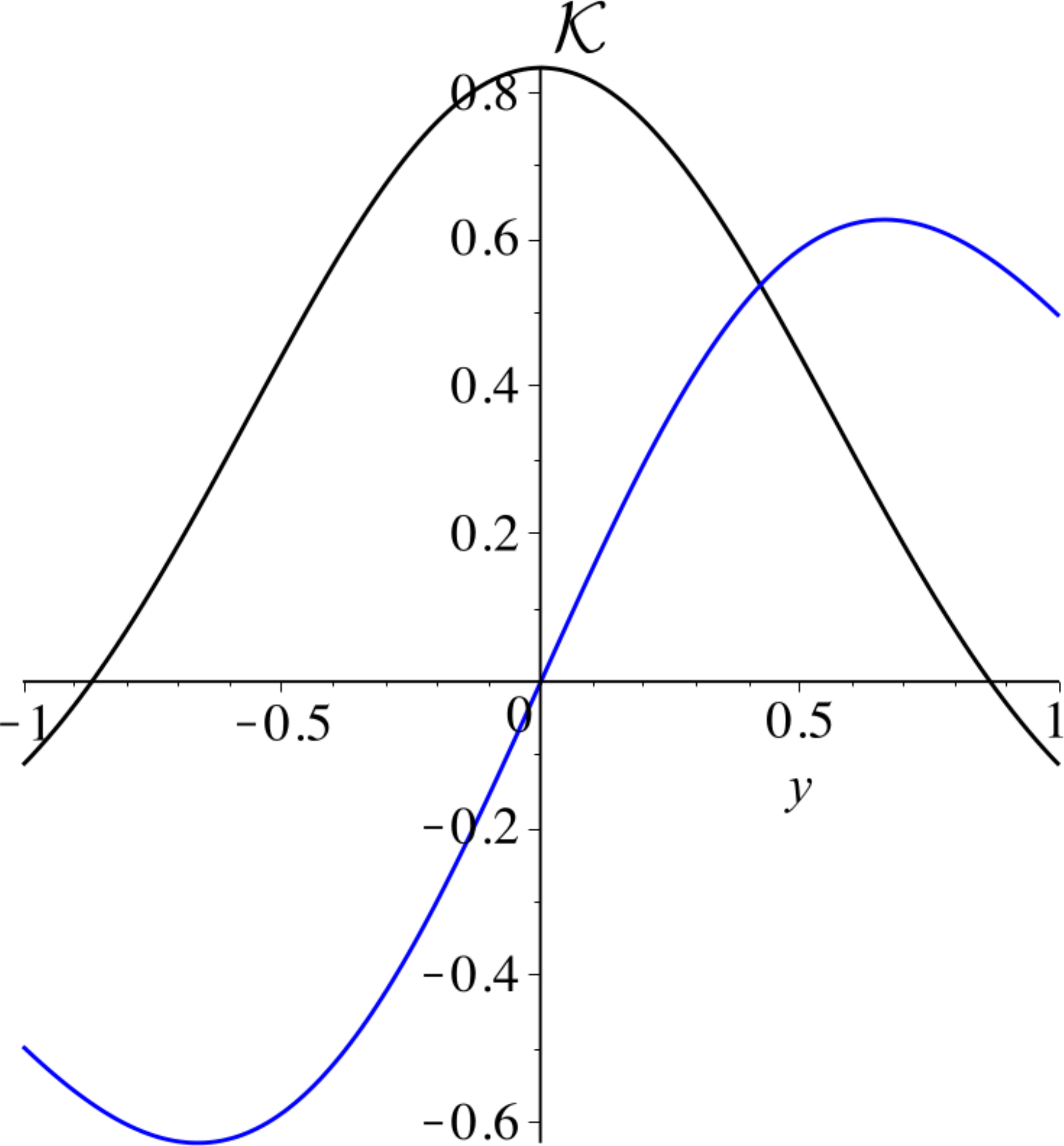} ~~&
            \includegraphics[width=4.2 cm]{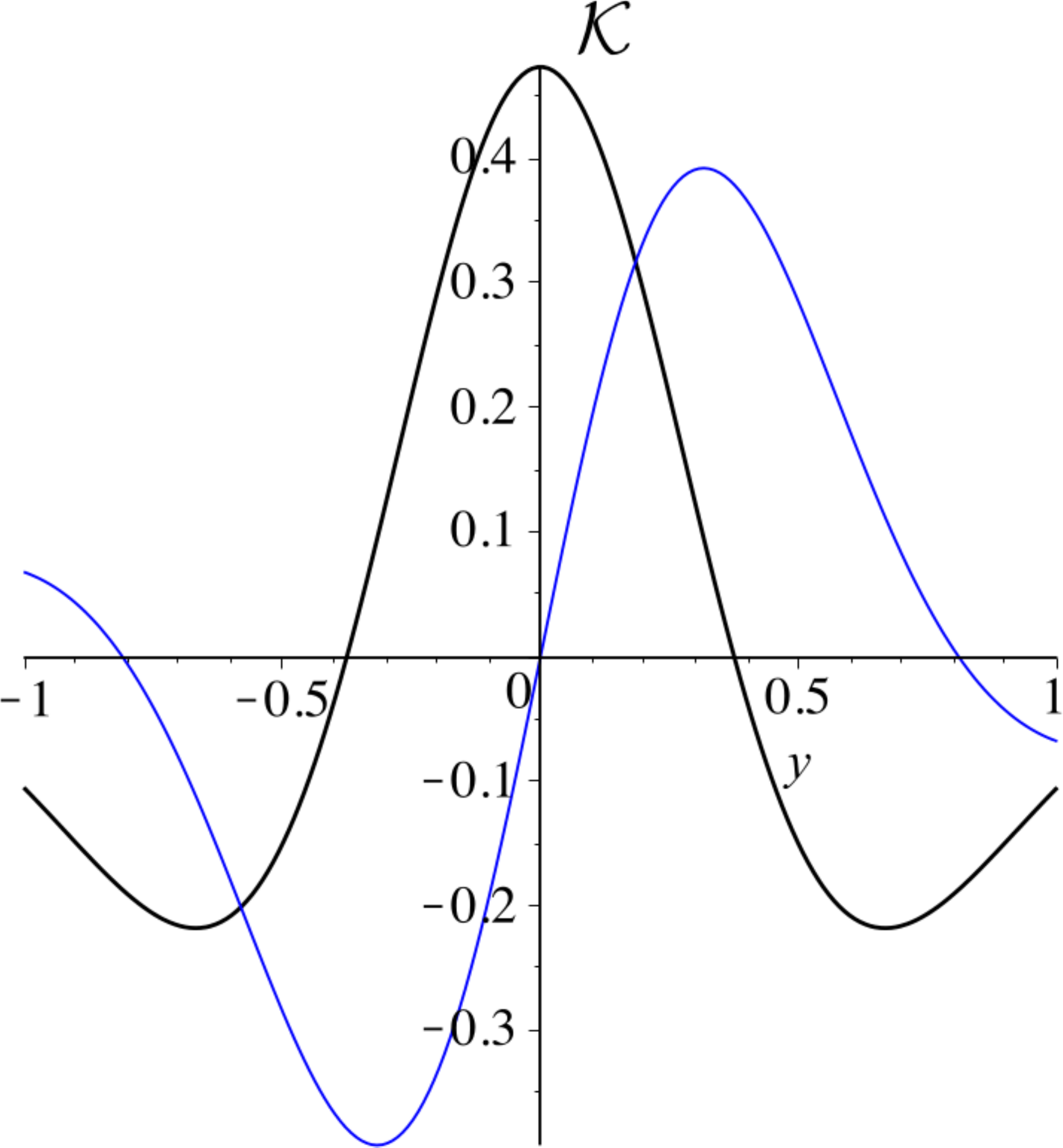}~~ &
            \includegraphics[width=4.2 cm]{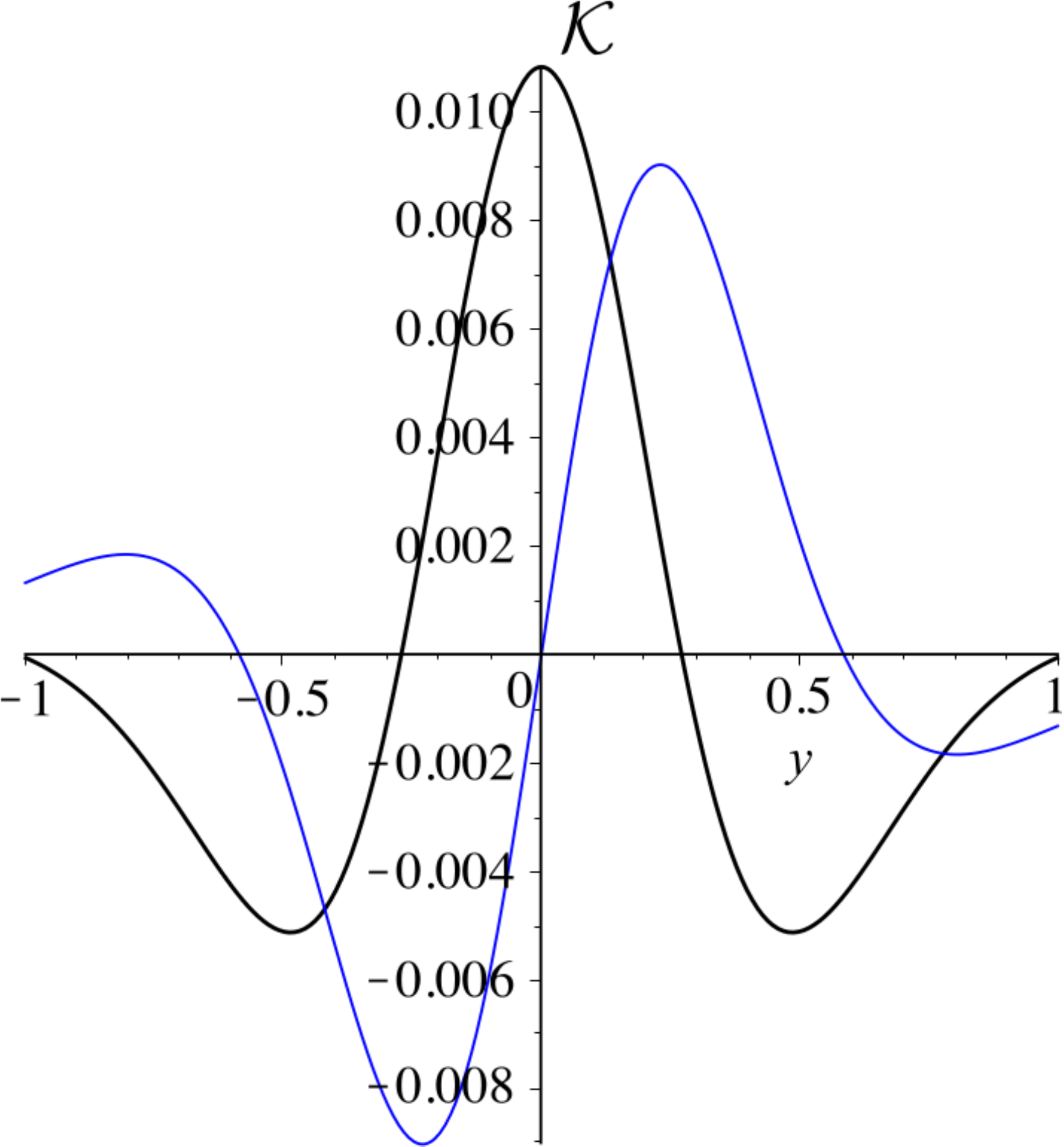} \\
                $\alpha=0.3,~c_2=1/100$ & $\alpha=0.7,~c_2=1/150$ &
            $\alpha=0.97,~c_2=1/150$   \\ \\
	\includegraphics[width=4.2 cm]{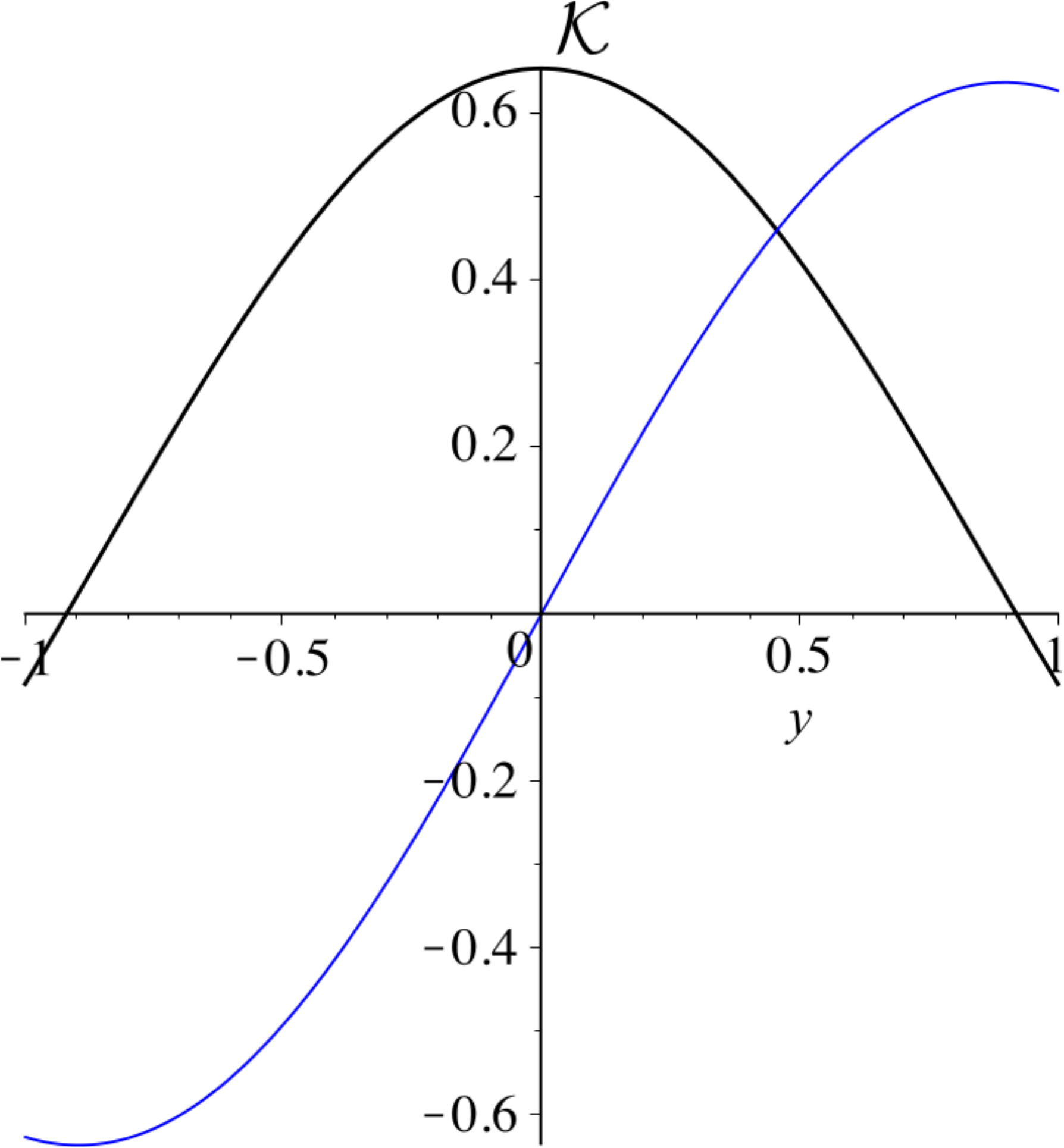}~~ &
            \includegraphics[width=4.2 cm]{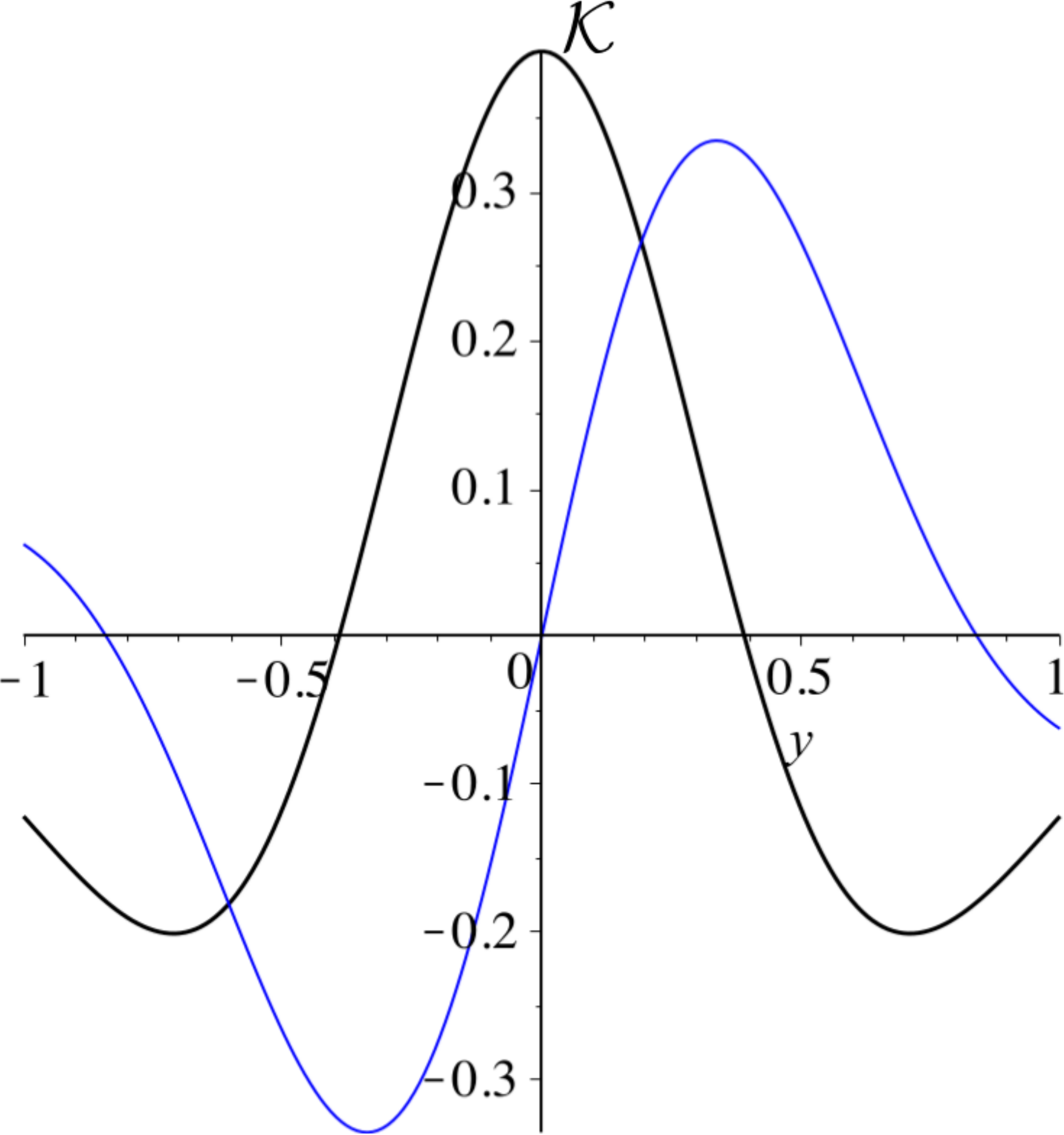}~~ &
            \includegraphics[width=4.2 cm]{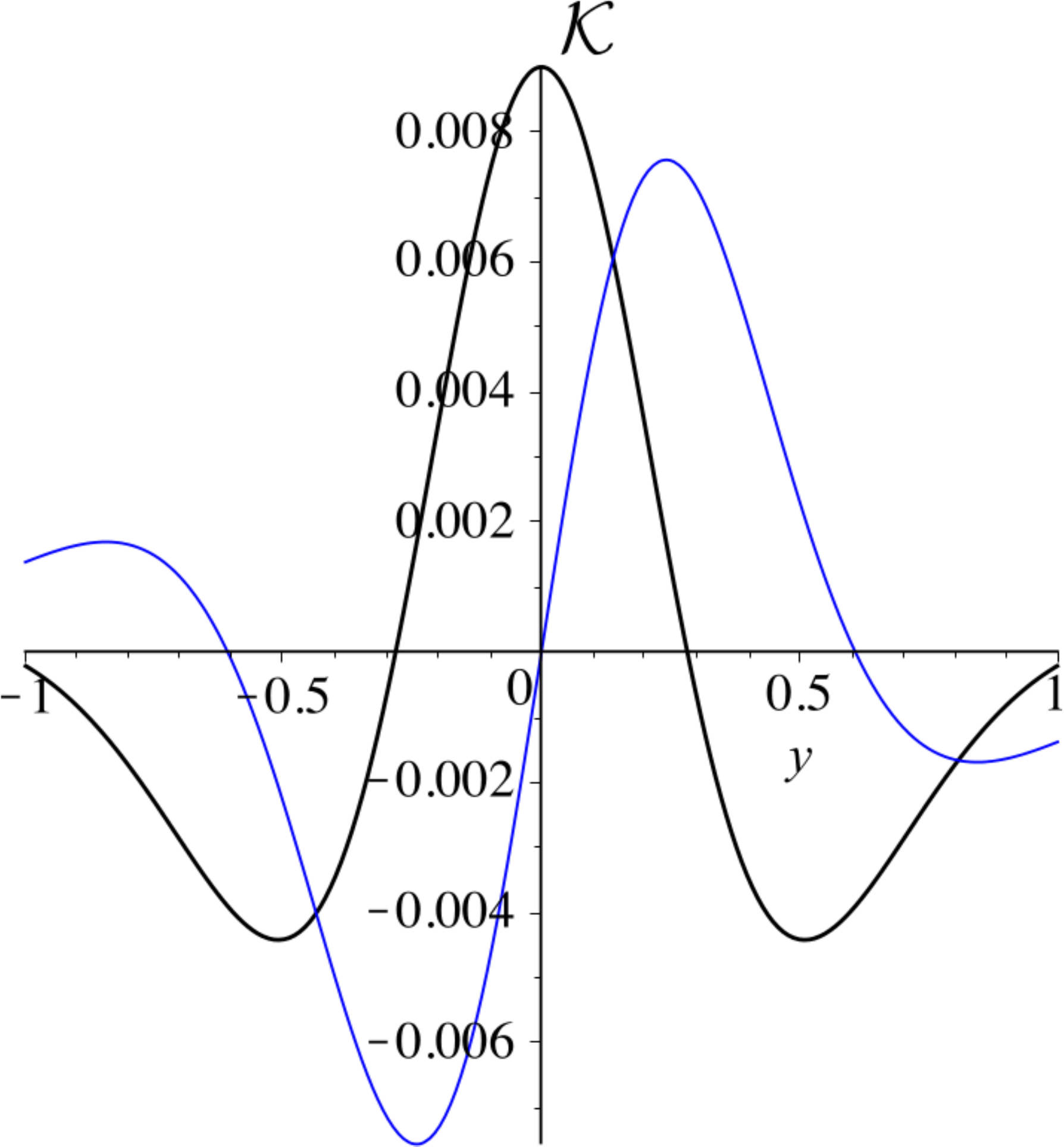} \\
            $\alpha=0.3,~c_2=-1/100$ & $\alpha=0.7,~c_2=-1/150$ &
            $\alpha=0.97,~c_2=-1/150$    \\
            \\
           \end{tabular}}
           \caption{\footnotesize{Scalar invariants vs. $y$ for quadrupole distortion. The black line is the Kretschmann invariant and the blue line is the Chern-Pontryagin invariant. }}
		\label{EvenCurvature}
\end{figure}
In Fig. \ref{EvenCurvature}, we illustrate the scalar curvature invariants for quadrupole distortion. The maximum value of the Kretschmann invariant lies on the equator in some of the cases, and it is greater than in the undistorted case for positive $c_2$, and smaller for negative $c_2$. However, there also exist values of the solution parameters when another location of its maximum is observed (see $\alpha=0.3$ and $c_2=-1/12$). The Kretschmann invariant drops to the value $\mathcal{K}_K=0.005$ at $y=0$ for $\alpha=0.3$ and $c_2=-1/3$. Therefore, we can see that the negative quadrupole distortion can effect the horizon in such a way that the Kretschmann invariant almost vanishes on the equator (compare to the case of $\alpha=0.3$ for the undistorted Kerr black hole see Fig. \ref{KerrCurvature}). For large values of $\alpha$ the behaviour of the Chern-Pontryagin invariant on the horizon is not affected much by the distortion. However, for small values of $\alpha$ and positive values of $c_2$ the Chern-Pontryagin invariant decreases considerably on the axis (see $\alpha=0.3$ and $c_2=1/12$). In contrast, for the isolated Kerr black hole the Chern-Pontryagin invariant reaches its maximum close to the axis.

\section{Summary}
Distorted black holes have been the subject of many studies \cite{Chandrasekhar}-\cite{Breton:1998}, \cite{ASF}-\cite{Ab}, which proved that they can show some remarkable and unique properties. Study of distorted back holes can be used as a theoretical guideline for categorizing the properties of black holes, namely for understanding which properties remain the same when the black hole is distorted and which ones change. In this paper, we consider a stationary and axisymmetric exact solution describing the Kerr black hole in the gravitational field of external sources.  We investigate the distorting effect of the sources in the particular cases of quadrupole and octupole external fields. In the papers \cite{Breton:1997} and \cite{Shoom} some properties of the distorted Kerr black hole have been studied.  In this work, we make the analysis more complete by investigating the ergoregion configurations and the behaviour of the curvature invariants on the horizon, which were not previously considered. Moreover, we have extended the analysis of the existence and location of curvature singularities.

In the ergoregion static observers are forced to corotate with the black hole. Given that effects such as the Penrose process or the super-radiance are associated to the ergoregion, it is one of the most important regions to be studied. The scalar curvature invariants are used in quantum gravity, the one-loop level renormalization of gravity \cite{Birrel}, and for calculating the energy density and stresses of a conformal scalar field in the Hartle-Hawking state \cite{Don, Brown, Frolov}. Therefore, it is important to investigate their behaviour on the horizon of a distorted black hole.

In the case of the isolated Kerr black hole the ergosurface is always a compact two dimensional surface which touches the horizon on the axis. However, this is not valid in the case of the distorted Kerr black hole. We can categorize the behaviour of the ergoregion in two classes.  In the first case, we observe a connected non-compact ergoregion. In the second case, the ergoregion consists of two parts -  a compact region which encompasses the black hole horizon and touches it on the axis, and a non-compact one disconnected from it. The first class of ergoregions is characteristic for high values of the multipole moments, while the second one is observed at low values when the properties of the solution are more similar to the isolated Kerr black hole. We have studied the transition between the connected and disconnected type of ergoregion. It occurs as a result of touching of the static regions which leads to the formation of the compact ergoregion in the vicinity of the horizon. The transition is realized at certain critical values of the distortion parameters which depend on the rotation parameter $\alpha$. In both the cases of quadrupole and octupole distortion the critical values decrease when $\alpha$ increases. In addition, we observe the formation of an interesting butterfly static region for octupole distortion. It is formed in the region of the parameter space when the connected non-compact ergoregion is realized.

It is interesting to discuss what is the physical meaning of the disconnected parts of the ergoregion. We know that certain relativistic matter configurations, such as compact stars and relativistic discs possess a toroidal ergoregion \cite{Butterworth}-\cite{Meinel}. We also observe some black hole and matter systems which are characterized by a Saturn-like ergoregion \cite{Herdeiro:2014a}. Therefore, we can make an analogy, and interpret the ergoregion which encompasses the horizon as associated with the black hole, and the other disconnected part as resulting from the external matter. By continuing the distorted black hole solution to an asymptotically flat global solution, the non-compact ergoregion disconnected from the horizon might be naturally truncated to a compact one. In this way it can be possible to achieve a Saturn-like ergoregion configuration for certain types of external matter. In this perspective, the continuation of the distorted solution to an asymptotically flat one can be performed on a hypersurface, which does not lie in the static region separating the two disconnected ergoregions. The matching hypersurface could also intersect the part of the ergoregion which is disconnected from the horizon. Considering external matter, which possesses an intrinsic ergoregion, however can lead to instability connected with it \cite{Kleihaus},\cite{Friedman}-\cite{Cardoso}. In the case that, with some sewing technique we achieve a compact ergoregion associated to the black hole, we observe the deformation of this ergoregion along with the deformation of the horizon due to the presence of external sources.

We have further investigated the influence of the external sources on the size of the compact ergoregion in the domain in the parameter space where such exists. For the purpose we examined the correlation of the compact ergoregion area with the value of the distortion parameter and the spin parameter $J/M^2$ for fixed $\alpha$. The ergoregion area is positively correlated with the value of the distortion parameter, and in the case of quadrupole distortion negatively correlated with the value of the spin parameter. For octupole distortion the ergoregion area does not depend on the spin parameter for fixed $\alpha$. Thus, infinitely many different configurations are possible for the same values of the Komar mass and angular momentum on the horizon.

The distorted Kerr black hole can possess curvature singularities in the domain of outer communication. When such exist, they are always located on the boundary of the ergoregion. We also presented arguments that in the cases of the quadrupole and octupole distortion the curvature singularities do not belong to the compact ergosurface in the vicinity of the horizon. Our analysis gives the following results for the location of the curvature singularities in the cases of quadrupole and octupole distortion. For negative quadrupole distortion there are no curvature singularities in the domain of outer communication. They are always located behind the inner Cauchy horizon. The solution is regular outside the horizon even though the spin parameter can satisfy $J^2/M^4>1$. This is in contrast to the isolated Kerr black hole, where the ratio $J^2/M^4>1$ leads to a naked singularity. It is interesting to see whether astrophysical black holes with $J^2/M^4>1$ can exist due to their interaction with external gravitational matter. For positive quadrupole distortion and octupole distortion there is always a ring singularity in the domain of outer communication. In the former case it is located in the equatorial plane. In the latter case we observe a reflection symmetry of the singularity location for positive and negative multipole moments with respect to the equatorial plane. The location of the singularity depends on the value of the multipole moments and the rotation parameter $\alpha$. As their values decrease the singularity moves further away from the horizon. For octupole distortion the distance between the singularity position and the equatorial plane also increases simultaneously. The location of the singularity can be used as a guideline for defining the region of validity of the local solution, and where the sewing technique for construction of a global solution may be carried out.

As a result of the analysis of the curvature invariants calculated on the horizon we see that the distortion can drastically change the position of their maxima and minima. In certain regions of the parameter space the Kretschmann invariant almost vanishes on the equator for the distorted Kerr black hole, while for the isolated Kerr black hole with the same rotation parameter it reaches its maximum value on the equator.

\section*{Acknowledgment}
S. A. and C. T. are grateful to the Natural Sciences and Engineering Research Council of Canada for financial support. J.K. and P.N. gratefully acknowledge support by the DFG Research Training Group 1620 ``Models of Gravity''. P.N. is partially supported by the Bulgarian NSF Grant $\textsl{DFNI T02/6}$.


\end{document}